\documentstyle[aps,twocolumn,epsf]{revtex}

\begin{document}
\draft
\title{Shot Noise in Mesoscopic Conductors}
\author{Ya. M. Blanter and M. B\"uttiker}
\address{D\'epartement de Physique Th\'eorique, Universit\'e 
de Gen\`eve, CH-1211, Gen\`eve 4, Switzerland \bigskip\\
\parbox{14cm}{\rm Theoretical and experimental work concerned with dynamic fluctuations
has developed into a very active and fascinating subfield of
mesoscopic physics. We present a review of this development focusing
on shot noise in small electric conductors. Shot noise is a
consequence of the quantization of charge. It can be used to obtain
information on a system which is not available through conductance
measurements. In particular, shot noise experiments can determine the
charge and statistics of the quasiparticles relevant for transport,
and reveal information on the potential profile and internal energy
scales of mesoscopic systems. Shot noise is generally more sensitive
to the effects of electron-electron interactions than the average
conductance. We present a discussion based on the conceptually
transparent scattering approach and on the classical Langevin and
Boltzmann-Langevin methods; in addition a discussion of results which
cannot be obtained by these methods is provided. We conclude the
review by pointing out a number of unsolved problems and an outlook on
the likely future development of the field.

}} 
\bigskip
\date{October 18, 1999}

\maketitle
\tableofcontents

\section{Introduction} \label{intro}

\subsection{Purpose of this Review} \label{int011}

During the past two decades mesoscopic physics has developed into a
fascinating subfield of condensed matter physics. In this article, we
review a special topic of this field: We are concerned with the
dynamical noise properties of mesoscopic conductors. After a
modest start, a little more than a decade ago, both theories and
experiments have matured. There is now already a substantial
theoretical literature and there are a number of interesting
experiments with which theoretical predictions can be compared. Some
experiments ask for additional theoretical work. The field has thus
reached a stage in which a review might be useful as a first
orientation for researchers who wish to enter the field. Also
researchers which are already active in the field might appreciate a
review to help them keep an overview over the rapid development which
has occurred. Any review, of course, reflects the authors' preferences
and prejudices and in any case cannot replace the study of the
original literature.  

Presently, there are no reviews covering the actual state of
development of the field. The only article which provides a
considerable list of references, and gives a description of
many essential features of shot noise in mesoscopic systems, has been
written in 1996 by de Jong and Beenakker \cite{dJBrev}. It is useful
as a first introduction to the subject, but since then the field has
developed considerably, and a broader review is clearly desirable. An
additional brief review has been written by Martin
\cite{MartinLesArcs}. The subject has been touched in books with
broader scopes by Kogan \cite{Koganbook} (Chapter 5) and Imry
\cite{Imrybook} (Chapter 8). These reviews, and in particular, the
work of de Jong and Beenakker \cite{dJBrev}, provided a considerable
help in starting this project. 

\subsection{Scope of the Review} \label{int012}

Our intention is to present a review on {\em shot noise in mesoscopic
conductors}. An effort is made to collect a {\em complete} list of 
references, and if not comprehensively re-derive, then at least to
mention results relevant to the field. We do not cite conference
proceedings and brief commentaries, unless we feel that they contain
new results or bring some understanding which cannot be found
elsewhere. Certainly, it is very possible that some papers, for
various reasons, have not come to our attention. We apologize to the
authors whose papers we might have overlooked.  

Trying to classify the already large literature, we chose to divide
the Review into sections according to the {\em methods} by which the
results are derived and not according to the systems we describe. Many
results can be obtained in the framework of the scattering approach
and/or by classical methods. We deliberately avoid an explanation of
the Green's function method, the master equation approach, and the
bosonization technique. An attempt to explain how any one of these
approaches work would probably double the size of this Review, which
is already long enough. Consequently, we make an effort to re-derive
the results existing in the literature by either the scattering or one
of the classical (but not the master equation) approaches, and to
present a unifying description. Certainly, for some systems these
simple methods just do not work. In particular, this concerns Section
\ref{strcorr}, which describes shot noise in strongly correlated
systems. Results obtained with more sophisticated methods are
discussed only briefly and without an attempt to re-derive them. We
incorporate a number of original results in the text; they are usually
extensions of results available in the literature, and are not
specially marked as original.   

A good review article should resemble a textbook useful to learn the
subject but should also be a handbook for an expert. From this
perspective this Review is more a textbook than a handbook. We
encourage the reader who wants to enter the field to read the Review
from the beginning to the end. On the other hand, we also try to help
the experts, who only take this Review to look for the results
concerning some particular phenomenon. It is for this reason that we
have included the Table I with references to the different subsections
in which the subject is discussed. The Review is concluded by a brief
summary.  

\subsection{Subjects not addressed in this Review} \label{int013}

First of all, we emphasize that this is not a general review of
mesoscopic physics. In many cases, it is necessary to describe briefly
the systems before addressing their noise properties. Only a few
references are provided to the general physical background. These
references are not systematic; we cite review articles where this is
possible, and give references to the original papers only in cases
such reviews are not readily available. This Review is not intended to
be a tool to study mesoscopic physics, though possibly many important
ideas are present here. 

A multitude of sources typically contribute to the noise of electrical
conductors. In particular, some of the early experimental efforts in
this field suffered from $1/f$-noise which is observed at low
frequencies. There is still to our knowledge no established theory of
$1/f$-noise. The present article focuses on fundamental, unavoidable
sources of noise, which are the thermal (Nyquist-Johnson-noise) and
the shot noise due to the granularity of charge. The reader may find
an account of other sources of noise in the book by Kogan
\cite{Koganbook}.    

We mention also that, though we try to be very accurate in references,
this Review cannot be regarded as a historical chronicle of shot
noise research. We re-derive a number of results by other methods than 
were used in the original papers, and even though we make an effort to
mention all achievements, we do not emphasize historical priorities.  

\subsection{Fundamental sources of noise} \label{int014}

{\bf Thermal noise}. At non-zero temperature, thermal fluctuations are
an unavoidable source of noise. Thermal agitation causes the
occupation number of the states of a system to fluctuate. 

Typically, in electric conductors, we can characterize the occupation
of a state by an occupation number $n$ which is either zero or
one. The thermodynamic average of the occupation number $\langle n
\rangle$ is determined by the Fermi distribution function $f$ and we
have simply $\langle n \rangle  = f$. In an equilibrium state the
probability that the state is empty is on the average given by $1-f$,
and the probability that the state is occupied is on the average given
by $f$. The fluctuations away from this average occupation number are
$(n-\langle n \rangle)^{2} = n^{2} - 2 n \langle n \rangle + \langle n
\rangle^{2}$. Taking into account that for a Fermi system $n^{2} =
n$, we find immediately that the fluctuations of the occupation number
at equilibrium away from its thermal average are given by  
\begin{equation} \label{occf1}
\left\langle ( n - \langle n \rangle)^{2} \right\rangle = f(1-f). 
\end{equation}  
The mean squared fluctuations vanish in the zero temperature limit. At
high temperatures and high enough energies the Fermi distribution
function is much smaller than one and thus the factor $1 - f$ in
Eq. (\ref{occf1}) can be replaced by $1$. The fluctuations are
determined by the (Maxwell) -- Boltzmann distribution.  

The fluctuations in the occupation number give rise to equilibrium
current fluctuations in the external circuit which are via the
fluctuation-dissipation theorem related to the conductance of the
system. Thus investigation of equilibrium current fluctuations
provides the same information as investigation of the
conductance. This is not so with the shot noise of electrical
conductors which provides information which cannot be obtained from a
conductance measurement. We next briefly discuss the source of shot
noise.  

{\bf Shot noise}. Shot noise in an electrical conductor is a
consequence of the quantization of the charge. Unlike for thermal
noise, to observe shot noise, we have to investigate the
non-equilibrium ({\em transport}) state of the system.  

To explain the origin of shot noise we consider first a fictitious
experiment in which only one particle is incident upon a barrier. At
the barrier the particle is either transmitted with probability $T$ or
reflected with probability $R = 1 - T$. We now introduce the
occupation numbers for this experiment. The incident state is
characterized by an occupation number $n_{in}$, the transmitted state
is characterized by an occupation number $n_{T}$ and the reflected
state  by an occupation number $n_{R}$. If we repeat this experiment
many times, we can investigate the average of these occupation numbers
and their fluctuations away from the average behavior. In our
experiments the incident beam is occupied with probability $1$ and
thus $\langle n_{in} \rangle = 1$. However the transmitted state is
occupied only with probability $T$ and is empty with probability
$R$. Thus $\langle n_{T} \rangle = T$ and  $\langle n_{R} \rangle =
R$. The fluctuations away from the average occupation number, can be
obtained very easily in the following way: The mean squared
fluctuations in the incident beam vanish, $(n_{in} - \langle n_{in}
\rangle)^{2} = 0$. To find the mean squared fluctuations in the
transmitted and reflected state, we consider the average of the
product of the occupation numbers of the transmitted and reflected
beam $\langle n_T n_R \rangle$. Since in each event the particle is
either transmitted or reflected, the product $n_T n_R$ vanishes for
each experiment, and hence the average vanishes also, $\langle n_T
n_R\rangle = 0$. Using this we find easily that the mean squares of
the transmitted and reflected beam and their correlation is given by  
\begin{equation} \label{occf2}
\langle (\Delta n_{T})^{2} \rangle = \langle (\Delta n_{R})^{2}
\rangle =  - \langle \Delta n_{T} \Delta n_{R} \rangle = TR,
\end{equation} 
where we have used the abbreviations $\Delta n_{T} = n_{T} - \langle
n_{T} \rangle$ and $\Delta n_{R} = n_{R} - \langle n_{R}
\rangle$. Such fluctuations are called {\em partition noise} since the
scatterer divides the incident carrier stream into two streams. The
partition noise vanishes both in the limit when the transmission
probability is $1$ and in the limit when the transmission probability
vanishes, $T = 0$. In these limiting cases no partitioning takes
place. The partition noise is maximal if the transmission probability
is $T = 1/2$. 

Let us next consider a slightly more sophisticated, but still
fictitious, experiment. We assume that the incident beam is
now occupied only with probability $f$. Eventually $f$ will just be
taken to be the Fermi distribution function. The initial state is
empty with probability $1-f$. Apparently, in this experiment the
average incident occupation number is $\langle n_{in} \rangle =f$, and
since the particle is transmitted only with probability $fT$ and
reflected with probability $fR$, we have $\langle n_{T} \rangle = fT$
and  $\langle n_{R} \rangle = fR$. Since we are still only considering
a single particle, we have as before that in each event the product
$n_{T}n_{R}$ vanishes. Thus we can repeat the above calculation to
find  
\begin{equation} \label{occf3}
\langle (\Delta n_{T})^{2} \rangle = Tf (1-Tf),
\end{equation} 
\begin{equation} \label{occf4}
\langle (\Delta n_{R})^{2} \rangle = Rf (1-Rf), 
\end{equation}
\begin{equation} \label{occf5}
\langle \Delta n_{T} \Delta n_{R} \rangle = -TR f^{2}.
\end{equation} 
If we are in the zero temperature limit, $f =1$, we recover the
results discussed above. Note that now even in the limit $T =1$ the
fluctuations in the transmitted state do not vanish, but fluctuate
like the incident state. For the transmitted stream, the factor
$(1-Tf)$ can be replaced by $1$, if either the transmission
probability is small or if the occupation probability of the incident
carrier stream is small.  

We can relate the above results to the fluctuations of the current in
a conductor. To do this we have to put aside the fact that in a
conductor we deal not as above just with events of a single carrier
but with a state which may involve many (indistinguishable)
carriers. We imagine a perfect conductor which guides the incident
carriers to the barrier, and imagine that we have two additional
conductors which guide the transmitted and reflected carriers away
from the conductor, such that we can discuss, as before, incident,
transmitted and reflected currents separately. Furthermore, we want to
assume that we have to consider only carriers moving in one direction
with a velocity $v(E)$ which is uniquely determined by the energy $E$
of the carrier. Consider next the average incident current. In a
narrow energy interval $dE$, the incident current is $dI_{in} (E) =
ev(E) d\rho(E)$, where $d\rho(E)$ is the density of carriers per unit
length in this energy range. The density in an energy interval $dE$ is
determined by the density of states (per unit length) $\nu (E) =
d\rho/dE$ times the occupation factor $n_{in}(E)$ of the state at
energy $E$. We thus have $d\rho(E) = n_{in}(E) \nu (E) dE$. The
density of states in our perfect conductors is  $\nu (E) =
1/(2\pi\hbar v(E))$. Thus the incident current in a narrow energy
interval is simply  
\begin{equation} \label{curin}
dI_{in} (E) = \frac{e}{2\pi\hbar} n_{in}(E) dE.
\end{equation} 
This result shows that there is a direct link between currents and the
occupation numbers. The total incident current is $I_{in} =
(e/2\pi\hbar) \int n_{in}(E) dE$ and on the average is given by
$\langle I_{in} \rangle = (e/2\pi\hbar) \int f(E) dE$. Similar
considerations give for the average transmitted current $\langle
I_{T}\rangle = (e/2\pi\hbar) \int f(E) T dE$ and for the reflected
current $\langle I_{R} \rangle = (e/2\pi\hbar) \int f(E) R
dE$. Current fluctuations are dynamic phenomena. The importance of the
above consideration is that it can now easily be applied to
investigate time-dependent current fluctuations. For occupation
numbers which vary slowly in time, Eq. (\ref{curin}) still holds. The
current fluctuations in a narrow energy interval are at long times
determined by $dI_{in} (E,t) = (e/2\pi\hbar) n_{in}(E,t) dE$ where
$n_{in}(E,t)$ is the time dependent occupation number of states with
energy $E$. A detailed derivation of the connection between currents
and occupation numbers is the subject of an entire Section of this
Review. We are interested in the low frequency current noise and thus
we can Fourier transform this equation. In the low frequency limit we
obtain $I(\omega) = (e/2\pi\hbar) \int dE  n(E, E+ \hbar \omega)$. As
a consequence the fluctuations in current and the fluctuations in
occupation number are directly related. In the zero frequency limit
the current noise power is $S_{II} = e^{2} \int dE  S_{nn}(E)$. In
each small energy interval particles arrive at a rate $dE/(2\pi\hbar)$
and contribute, with a mean square fluctuation, as given by one of the
equations (\ref{occf3}) -- (\ref{occf5}), to the noise power. We have
$S_{nn} (E) = (1/\pi\hbar) \langle \Delta n \Delta n \rangle $. Thus
the fluctuation spectra of the incident, transmitted, and reflected
currents are    
\begin{equation}  
S_{I_{in}I_{in}} = 2 \frac{e^2}{2\pi\hbar} \int dE f (1-f);
\end{equation} 
\begin{equation} \label{occf6}
S_{I_{T}I_{T}} = 2 \frac{e^2}{2\pi\hbar} \int dE \ Tf (1-Tf);
\end{equation} 
\begin{equation} \label{occf7}
S_{I_{R}I_{R}} = 2 \frac{e^2}{2\pi\hbar} \int dE \ Rf (1-Rf).
\end{equation}
The transmitted and reflected current are correlated,  
\begin{equation} \label{occf8} 
S_{I_{T}I_{R}} = - 2 \frac{e^2}{2\pi\hbar} \int dE \ Tf \ Rf.  
\end{equation} 
In the limit that either $T$ is very small or $f$ is small, the factor
$(1-Tf)$ in Eq. (\ref{occf6}) can be replaced by one. In this limit,
since the average current through the barrier is $\langle I \rangle =
(e/2\pi\hbar) \int dE Tf$, the spectrum, Eq. (\ref{occf6}), is 
Schottky's result \cite{Schottky} for shot noise,  
\begin{equation} \label{Schottky3}
S_{I_{T}I_{T}} = 2e \langle  I \rangle.
\end{equation} 
Schottky's result corresponds to the uncorrelated arrival of
particles with a distribution function of time intervals between
arrival times which is Poissonian,  $P(\Delta t) = \tau^{-1}
\exp(-\Delta t/\tau)$, with $\tau$ being the mean time interval
between carriers. Alternatively, Eq. (\ref{Schottky3}) is also
referred to in the literature as the {\em Poisson value} of shot
noise.      

The result Eq. (\ref{occf6}) is markedly different from
Eq. (\ref{Schottky3}) since it contains, in comparison to Schottky's
expression, the extra factor $(1-Tf)$. This factor has the consequence
that the shot noise (\ref{occf6}) is always smaller than the Poisson
value. For truly ballistic systems ($T = 1$) the shot noise even
vanishes in the zero temperature limit.  As the temperature increases,
in such a conductor ($T = 1$) there is shot noise due to the
fluctuation in the incident beam arising from the thermal
fluctuations. Eventually at high temperatures the factor $1 - f$ can
be replaced by $1$, and the ballistic conductor exhibits Poisson
noise, in accordance with Schottky's formula,
Eq. (\ref{Schottky3}). The full Poisson noise given by Schottky's
formula is also reached for a scatterer with very small transparency
$T \ll 1$. We emphasize that the above statements refer to the
transmitted current. In the limit $T \ll 1$ the reflected current
remains nearly noiseless up to high temperatures when  $(1-Rf)$ can be
replaced by $1$. We also remark that even though electron motion in
vacuum tubes (the Schottky problem \cite{Schottky}) is often referred
to as ballistic, it is in fact a problem in which carriers have been
emitted by a source into vacuum either through thermal activation over
or by tunneling through a barrier with very small transparency.  

Our discussion makes it clear that out of equilibrium, and at finite
temperatures, the noise described by Eq. (\ref{occf6}) contains the
effect of both the fluctuations in the incident carrier beam as well
as the partition noise. In a transport state, noise in mesoscopic
conductors has two distinct sources which manifest themselves in the
fluctuations of the occupation numbers of states: (i) thermal
fluctuations;  (ii) partition noise due to the discrete nature of
carriers\footnote{Note that this terminology, common in mesoscopic
physics, is different from that used in the older literature
\protect\cite{Ziel55}, where shot noise (due to random injection of
particles into a system) and partition noise (due to random division
of the particle stream between different electrodes, or by potential
barriers) are two distinct independent sources of fluctuations.}.   

Both the thermal and shot noise at low frequencies and low voltages
reflect in many situations independent quasi-particle
transport. Electrons are, however, interacting entities and both the
fluctuations at finite frequencies and the fluctuation properties far
from equilibrium require in general a discussion of the role of the
long-range Coulomb interaction. A quasi-particle picture is no longer
sufficient and collective properties of the electron system come into
play.    
 
The above considerations are, of course, rather simplistic and should
not be considered as a quantitative theory. Since statistical effects
play a role, one would like to see a derivation which relates the
noise to the symmetry of the wave functions. Since we deal with many
indistinguishable particles, exchange effects can come into play. The
following Sections will treat these questions in detail.   

\subsection{Composition of the Review} \label{int015}

The review starts with a discussion of the scattering approach
(Section \ref{scat}). This is a fully quantum-mechanical theory which
applies to phase-coherent transport. It is useful to take this
approach as a starting point because of its conceptual clarity. The
discussion proceeds with a number of specific examples, like quantum
point contacts, resonant double barriers, metallic diffusive wires,
chaotic cavities, and quantum Hall conductors. We are interested not
only in current fluctuations at one contact of a mesoscopic sample but
also in the correlations of currents at different
contacts. Predictions and experiments on such correlations are
particularly interesting since current correlations are sensitive to
the statistical properties of the system. Comparison of such
experiments with optical analogs is particularly instructive.    

Section \ref{freq} describes the frequency dependence of noise via the
scattering approach. The main complication is that generally one has
to include electron-electron interactions to obtain fluctuation 
spectra which are current conserving. For this reason, not many
results are currently available on frequency-dependent noise, though
the possibility to probe in this way the inner energy scales and
collective response times of the system looks very promising.  

We proceed in Section \ref{hybr} with the description of
superconducting and hybrid structures, to which a generalized
scattering approach can be applied. New noise features appear from the
fact that the Cooper pairs in superconductors have the charge $2e$.   

Then in Sections \ref{Langevin} and \ref{BoltzmannL} we review recent
discussions which apply more traditional classical approaches to the
fluctuations of currents in mesoscopic conductors. Specifically, for a
number of systems, as far as one is concerned only with ensemble
averaged quantities, the Langevin and Boltzmann-Langevin approaches
provide a useful discussion, especially since it is known how to
include inelastic scattering and effects of interactions.  

Section \ref{strcorr} is devoted to shot noise in strongly correlated
systems. This Section differs in many respects from the rest of the 
Review, mainly because strongly correlated systems are mostly too
complicated to be described by the scattering or Langevin
approaches. We resort to a brief description and commentary of the
results, rather than to a comprehensive demonstration how they are
derived.  

\begin{table} \label{tableone}
\begin{tabular}{ll}
Subject & Subdivision \\
\hline
{\bf Ballistic conductors} & \\ 
$\bullet$ Electron-phonon interactions & \ref{bol63} \\
$\bullet$ Electron-electron interactions in & \\
\ \ \ non-degenerate ballistic conductors & \ref{bol65}\\ 
$\bullet$ Hanbury Brown -- Twiss effects & \ref{scat0268},
\ref{scat0269}, \\
& \ref{hybr41}, \ref{con083} \\
$\bullet$ Aharonov-Bohm effect & \ref{scat0260}\\
\hline
{\bf Tunnel barriers} & \\
$\bullet$ Normal barriers & \ref{scat0261} \\
$\bullet$ Barriers in diffusive conductors & \ref{scat0264}\\
$\bullet$ Coulomb blockade regime & \ref{scorr71} \\
$\bullet$ Frequency dependence of noise & \ref{freq032}, \ref{freq033}
\\
$\bullet$ Barriers of oscillating random height & \ref{freq033} \\ 
$\bullet$ NS interfaces & \ref{hybr41} \\ 
$\bullet$ Josephson junctions & \ref{hybr42} \\
$\bullet$ Barriers in Luttinger liquids & \ref{scorr73}\\
$\bullet$ Counting statistics for normal & \\
\ \ \ and NS barriers & \ref{con082} \\
\hline
{\bf Quantum point contacts} & \\
$\bullet$ Normal quantum point contacts & \ref{scat0262} \\
$\bullet$ SNS contacts & \ref{hybr43} \\
\hline
{\bf Double-barrier structures} & \\
$\bullet$ Resonant tunneling; linear regime & \ref{scat0263} \\
$\bullet$ Double-barrier suppression & \ref{scat0263}, \ref{lang52}\\ 
$\bullet$ Counting statistics & \ref{con082} \\
$\bullet$ Double wells and crossover to the & \\
\ \ \ diffusive regime & \ref{scat0263}, \ref{lang52} \\
$\bullet$ Quantum wells in the non-linear regime: & \\
\ \ \ super-Poissonian shot noise enhancement & \ref{lang53} \\
$\bullet$ Interaction effects in quantum wells &
\ref{scat027}, \ref{lang52},\\
& \ref{lang53} \\
$\bullet$ Coulomb blockade in quantum dots (normal, & \\
\ \ \ superconducting, or ferromagnetic electrodes) & \ref{scorr71}\\ 
$\bullet$ Frequency dependence of noise & \ref{lang52},
\ref{lang53},\\
& \ref{scorr71} \\
$\bullet$ Resonant tunneling through localized states; & \\
\ \ \ Anderson and Kondo models & \ref{scat0263}, \ref{scorr72}\\
$\bullet$ NINIS junctions & \ref{hybr41} \\
\hline
{\bf Disordered conductors} & \\
$\bullet$ Noise suppression in metallic diffusive wires &
\ref{scat0264}, \ref{bol62}\\
$\bullet$ Counting statistics & \ref{con082}\\
$\bullet$ Multi-terminal generalization and & \\
\ \ \ Hanbury Brown -- Twiss effects & \ref{scat0269}, \ref{bol62}\\
$\bullet$ Interaction effects & \ref{scat027}, \ref{bol63}\\
$\bullet$ Frequency dependence of noise & \ref{freq032}, \ref{bol64}
\\  
$\bullet$ Disordered contacts and interfaces & \ref{scat0264} \\
$\bullet$ Disordered NS and SNS contacts & \ref{hybr41},
\ref{hybr43}\\ 
$\bullet$ Crossover to the ballistic regime & \ref{scat0264} \\  
$\bullet$ Localized regime & \ref{scat0264} \\
$\bullet$ Non-degenerate diffusive conductors & \ref{bol65}\\
$\bullet$ Composite fermions with disorder & \ref{scorr74}\\
$\bullet$ Noise induced by thermal transport & \ref{con084}\\
\hline
{\bf Chaotic cavities} & \\
$\bullet$ Noise suppression in two-terminal & \\
\ \ \ chaotic cavities & \ref{scat0265}, \ref{bol66},\\
& \ref{bol67} \\  
$\bullet$ Cavities with diffusive boundary scattering & \ref{bol66} \\
$\bullet$ Multi-terminal generalization and & \\
\ \ \ Hanbury Brown -- Twiss effects & \ref{scat0265}, \ref{bol67}\\ 
$\bullet$ Counting statistics & \ref{con082} \\
\hline
{\bf Quantum Hall effect} & \\
$\bullet$ IQHE edge channels & \ref{scat0266}, \ref{scat0260}\\
$\bullet$ Hanbury Brown -- Twiss effects with & \\
\ \ \ IQHE edge channels & \ref{scat0267}, \ref{scat0269} \\
$\bullet$ FQHE edge channels & \ref{scorr73}\\
$\bullet$ Composite fermions & \ref{scorr74}\\
\hline
{\bf Systems with purely capacitive coupling} & \\
$\bullet$ Frequency dependence of noise & \ref{freq034},
\ref{freq035}\\  
\end{tabular}
\vspace{0.5cm}
\caption{The results reviewed in this article arranged by subject.} 
\end{table}

We conclude the Review (Section \ref{conclus}) with an outlook. We
give our opinion concerning possible future directions along which the
research on shot noise will develop. In the concluding Section, we
provide a very concise summary of the state of the field and we
list some (possibly) important unsolved problems.  

Some topics are treated in a number of Appendices mostly to provide a
better organization of the manuscript. The Appendices report important
results on topics which are relatively well rounded, are relevant for
the connections between different sub-fields, and might very well
become the subject of much further research.

\section{Scattering theory of thermal and shot noise} \label{scat}

\subsection{Introduction} \label{scat021}

In this Section we present a theory of thermal and shot noise
for fully phase-coherent mesoscopic conductors. The discussion is
based on the scattering approach to electrical conductance. This
approach, as we will show, is conceptually simple and transparent. A 
phase-coherent description is needed if we consider an individual
sample, like an Aharonov-Bohm ring, or a quantum point contact. Often,
however, we are interested in characterizing not a single sample but
rather an ensemble of samples in which individual members differ only
in the microscopic arrangement of impurities or small variations in
the shape. The ensemble averaged conductance is typically, up to a
small correction, determined by a classical expression like a Drude
conductance formula. Similarly, noise spectra, after ensemble
averaging, are, up to small corrections, determined by purely
classical expressions. In these case, there is no need to keep
information about phases of wavefunctions, and shot noise expressions
may be obtained by classical methods. Nevertheless, the generality of
the scattering approach and its conceptual clarity, make it the
desired starting point of a discussion of noise in electrical
conductors.   

Below we emphasize a discussion based on second quantization. 
This permits a concise treatment of the many particle problem. 
Rather than introducing the Pauli principle by hand, in this approach
it is a consequence of the underlying symmetry of the wave
functions. It lends itself to a discussion of the effects related to
the quantum mechanical indistinguishability of identical particles. In
fact it is an interesting question to what extend we can directly
probe the fact that exchange of particles leaves the wave function
invariant up to a sign. Thus an important part of our discussion will
focus on {\it exchange effects} in current-current correlation
spectra.   

We start this Section with a review of fluctuations in idealized {\it
one-} and {\it two-}particle scattering problems. This simple
discussion highlights the connection between symmetry of the 
wave functions (the Pauli principle) and the fluctuation
properties. It introduces in a simple manner some of the basic
concepts and it will be interesting to compare the results of the one-
and two-particle scattering problems with the many-particle problem
which we face in mesoscopic conductors. 

\subsection{The Pauli principle} \label{scat022}

The investigation of the noise properties of a system is interesting
because it is fundamentally connected with the statistical properties
of the entities which generate the noise. We are concerned with
systems which contain a large number of {\it indistinguishable}
particles. The fact that in quantum mechanics we cannot label
different particles implies that the wave function must be invariant,
up to a phase, if two particles are {\it exchanged}. The invariance of
the wave function under exchange of two particles implies that we
deal with wavefunctions which are either symmetric or antisymmetric
under particle exchange. (In strictly two dimensional systems more
exotic possibilities are permitted). These symmetry statements are
known as the Pauli principle. Systems with symmetric (antisymmetric)
wavefunctions are described by Bose-Einstein (Fermi-Dirac) statistics,
respectively. 

\begin{figure}
\narrowtext
{\epsfxsize=6cm\centerline{\epsfbox{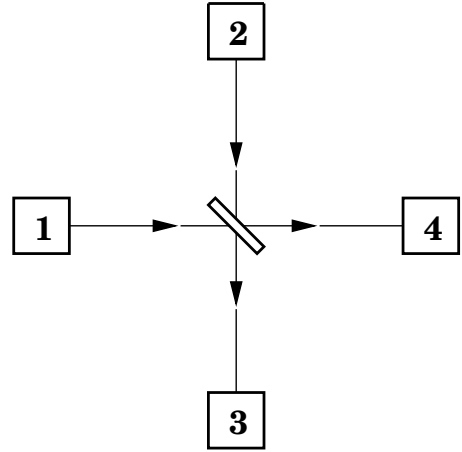}}}
\vspace{0.5cm}
\caption{An arrangement of scattering experiment with two sources ($1$
and $2$) and two detectors ($3$ and $4$).}
\protect\label{figloudon}
\end{figure}

Prior to the discussion of the noise properties
in electrical conductors, which is our central subject, in this
subsection we illustrate in a simple manner the fundamental connection
between the symmetry of the wave function and the statistical
properties of scattering experiments. We deal with open systems
similarly to a scattering experiment in which particles are incident
on a target at which they are scattered. The simplest case in which
the symmetry of the wave function matters is the case of two identical
particles. Here we present a discussion of idealized two-particle
scattering experiments. We consider the arrangement shown in
Fig. \ref{figloudon}, which contains two sources $1$ and $2$ which can
emit particles and two detectors $3$ and $4$ which respond ideally
with a signal each time a particle enters a detector. An
arrangement similar to that shown in Fig. \ref{figloudon} is used in
optical experiments. In this field experiments which invoke one or two
incoming particle streams (photons) and two detectors are known
as Hanbury Brown -- Twiss experiments \cite{HBT}, after the pioneering
experiments carried out by these two researchers to investigate the
statistical properties of light.   

In Fig. \ref{figloudon} the scattering is, much as in an optical table
top experiment, provided by a half-silvered mirror (beam-splitter),
which permits transmission from the input-channel $1$ through the 
mirror with probability amplitude $s_{41} = t$ to the detector at arm
$4$ and generates reflected particles with amplitude $s_{31} = r$ into
detector $3$. We assume that particles from source $2$ are scattered
likewise and have probability amplitudes $s_{32} = t$ and $s_{42} =
r$. The elements  $s_{ij}$, when written as a matrix, form the
scattering matrix $s$. The elements of the scattering matrix satisfy 
$|r|^{2}+|t|^{2} = 1$ and $tr^{\ast} + rt^{\ast} = 0$, stating that
the scattering matrix is unitary. A simple example of a scattering
matrix often employed to describe scattering at a mirror in optics is
$ r = - i/\sqrt{2}$ and $t = 1/\sqrt{2}$.  
 
We are interested in describing various input states emanating 
from the two sources. To be interesting, these input states contain
one or two particles. We could describe these states in terms of
Slater determinants, but it is more elegant to employ the second
quantization approach. The incident states are described by
annihilation operators $\hat a_i$ or creation operators
$\hat{a}^{\dagger}_i$ in arm $i$, $i = 1, 2$. The outgoing states are,
in turn, described by annihilation  operators $\hat b_i$ and creation
operators $\hat{b}^{\dagger}_i$, $i = 3, 4$. The operators of the
input states and output states are not independent but are related by
a unitary transformation which is just the scattering matrix,   
\begin{equation} \label{scatmatr0}
\left( \matrix{ \hat b_{3} \cr 
\hat{b}_{4}} \right) = s \left( \matrix{
\hat{a}_{1} \cr \hat{a}_{2} } \right). 
\end{equation}
Similarly, the creation operators $\hat a^{\dagger}_i$ and $\hat
b^{\dagger}_i$ are related by the adjoint of the scattering matrix
$s^{\dagger}$. Note that the mirror generates quantum mechanical
superpositions of the input states. The coefficients of these
superpositions are determined by the elements of the scattering
matrix.  

For bosons, the ${\hat{a}_i}$ obey the commutation relations
\begin{equation} \label{comm1}
[\hat{a}_{i},\hat{a}^{\dagger}_{j}] = \delta_{ij} .
\end{equation}
Since the scattering matrix is unitary, the $\hat{b}_{i}$ obey the
same commutation relations. In contrast, for fermions, the
$\hat{a}_{i}$ and $\hat{b}_{i}$ obey anti-commutation relations
\begin{equation} \label{comm2}
\{\hat{a}_{i},\hat{a}^{\dagger}_{j} \} = \delta_{ij} .
\end{equation}
The different commutation relations for fermions and bosons assure
that multi-particle states reflect the underlying symmetry of the
wave function.  

The occupation numbers of the incident and transmitted states are
found as $\hat{n}_{i} = \hat{a}^{\dagger}_{i}  \hat{a}_{i}$ and 
$\hat{n}_{i} = \hat{b}^{\dagger}_{i} \hat{b}_{i}$,
\begin{equation} \label{cons}
\left( \matrix{ \langle \hat{n}_{3} \rangle \cr 
 \langle\hat{n}_{4}\rangle }\right) = 
\left( \matrix{ R & T \cr T & R} \right)\left( \matrix{
\langle \hat{n}_{1} \rangle \cr \langle \hat{n}_{2} \rangle } \right),
\end{equation} 
where we introduced transmission and reflection probabilities, $T =
|t|^{2}$ and $R = |r|^{2}$.  

{\bf Single, independent particle scattering}. Before treating
two-particle states it is useful to consider briefly a series of 
scattering experiments in each of which only one particle is incident
on the mirror. Let us suppose that a particle is incident in arm
$1$. Since we know that in each scattering experiment there is one
incident particle, the average occupation number in the incident arm
is thus $ \langle n_{1} \rangle =1.$ The fluctuations away from the
average occupation number $\Delta n_{1} = n_{1} -  \langle n_{1}
\rangle$ vanish identically (not only on the average). In particular,
we have $\langle (\Delta n_{1})^{2} \rangle = 0.$ Particles are
transmitted into arm $4$ with probability $T$, and thus the mean
occupation number in the transmitted beam is $\langle n_{4} \rangle =
T$. Similarly, particles are reflected into arm $3$ with probability
$R$, and the average occupation number in our series of experiments is
thus $ \langle n_{3} \rangle = R$. Consider now the correlation
$\langle n_{3}n_{4}\rangle$ between the occupation numbers in arm $3$
and $4$. Since each particle is either reflected or transmitted, it
means that in this product one of the occupation numbers is always $1$
and one is zero in each experiment. Consequently the correlation of
the occupation numbers vanishes, $\langle n_{3}n_{4} \rangle =
0$. Using this result, we obtain for the fluctuations of the
occupation numbers $\Delta n_{3} = n_{3} - \langle n_{3} \rangle$ in
the reflected beam and $\Delta n_{4} = n_{4} - \langle n_{4} \rangle$
in the transmitted beam, 
\begin{equation} \label{part}
\langle (\Delta n_{3})^{2} \rangle = \langle (\Delta n_{4})^{2}
\rangle = - \langle \Delta n_{3} \Delta n_{4} \rangle = TR. 
\end{equation} 
The fluctuations in the occupation numbers in the transmitted and 
reflected beams and their correlation are a consequence of the fact 
that a carrier can finally only be either reflected or transmitted. 
These fluctuations are known as {\it partition noise}. The partition
noise vanishes for a completely transparent scatterer $T  =1$ and for
a completely reflecting scatterer $R =1$ and is maximal for $T = R =
1/2$. We emphasize that the partition noise is the same, whether we
use a fermion or boson in the series of experiments. To detect the
sensitivity to the symmetry of the wave function, we need to consider
at least two particles.  

{\bf Two-particle scattering}. We now consider two particles incident
on the mirror of Fig. 1, focusing on the case that one particle is
incident in each arm.  We follow here closely a discussion by Loudon
\cite{Loudon} and refer the reader to this work for additional
information. The empty (vacuum) state of the system (input arms and
output arms) is denoted by $|0\rangle$. Consider now an input state
which consists of two particles with a definite momentum, one incident
in arm $1$ and one incident in arm $2$. For simplicity, we assume that
the momentum of both particles is the same. With the help of the
creation operators given above we can generate the input state
$|\Psi \rangle= \hat{a}^{\dagger}_{1}\hat{a}^{\dagger}_{2}|0 \rangle$.
The probability that both particles appear in output arm $3$ is
$P(2,0) = \langle \Psi|\hat n_{3} \hat n_{3}|\Psi \rangle$, the
probability that in each output arm there is one particle is $P(1,1) =
\langle \Psi|\hat n_{3} \hat n_{4}|\Psi \rangle$, and the probability
that both particles are scattered into arm $4$ is  $P(0,2) = \langle
\Psi|\hat n_{4} \hat n_{4}|\Psi \rangle$. Considering specifically the
probability $P(1,1)$, we have to find 
\begin{equation} \label{twob}
P(1,1) = \langle \Psi|\hat n_{3} \hat n_{4}|\Psi \rangle =
\langle 0|\hat{a}_{2}\hat{a}_{1} \hat{b}^{\dagger}_{3}\hat{b}_{3} \hat
b^{\dagger}_{4} \hat b_{4}
\hat{a}^{\dagger}_{1}\hat{a}^{\dagger}_{2}|0 \rangle. 
\end{equation}
First we notice that in the sequence of $\hat b$-operators 
$\hat{b}_{3}$ and $\hat{b}^{\dagger}_{4}$ anti-commute (commute), and
we can thus write $\hat{b}^{\dagger}_{3}\hat{b}_{3} \hat
b^{\dagger}_{4} \hat b_{4}$ also in the sequence $\mp
\hat{b}^{\dagger}_{3} \hat b^{\dagger}_{4} \hat{b}_{3} \hat
b_{4}$. Then, by inserting a complete set of states with fixed number
of particles $|n\rangle\langle n|$ into this product, $\mp
\hat{b}^{\dagger}_{3} \hat b^{\dagger}_{4}|n\rangle\langle n|
\hat{b}_{3} \hat b_{4}$, it is only the state with $n =0$ which
contributes since to the right of $\langle n|$ we have two creation
and two annihilation operators. Thus the probability $P(1,1)$ is given
by the absolute square of a probability amplitude  
\begin{equation} \label{p11amp}
P(1,1) = | \langle 0|\hat{b}_{3}\hat
b_{4}\hat{a}^{\dagger}_{1}\hat{a}^{\dagger}_{2}|0 \rangle |^{2}.
\end{equation}
To complete the evaluation of this probability, we express
$\hat{a}^{\dagger}_{1}$ and $\hat{a}^{\dagger}_{2}$ in terms of the
output operators $\hat{b}^{\dagger}_{3}$ and $\hat{b}^{\dagger}_{4}$
using the adjoint of Eq. (\ref{scatmatr0}). This gives 
\begin{equation} \label{aa}
\hat{a}^{\dagger}_{1}\hat{a}^{\dagger}_{2} =
rt \hat{b}^{\dagger}_{3}\hat{b}^{\dagger}_{3} + 
r^{2} \hat{b}^{\dagger}_{3}\hat{b}^{\dagger}_{4}  
+ t^{2} \hat{b}^{\dagger}_{4}\hat{b}^{\dagger}_{3}
+ rt\hat{b}^{\dagger}_{4}\hat{b}^{\dagger}_{4}.
\end{equation}
Now, using the commutation relations, we pull the annihilation
operators to the right until one of them acts on the vacuum and the
corresponding term vanishes. A little algebra gives 
\begin{equation} \label{p11}
P(1,1) = ( T \pm R )^{2}.
\end{equation}
Eq. (\ref{p11}) is a concise statement of the Pauli principle. 
For bosons $P(1,1)$ depends on the transmission and reflection
probability of the scatterer, and vanishes for an ideal mirror $T = R
= 1/2$. The two particles are preferentially scattered into the same
output branch. For fermions $P(1,1)$ is independent of the
transmission and reflection probability and given by  $P(1,1) =
1$. Thus fermions are scattered with probability one into the
different output branches. It is instructive to compare these result
with the one for classical particles, $P(1,1) = T^2 + R^2$. We see
thus that the probability to find two bosons (fermions) in two
different detectors is suppressed (enhanced) in comparison with the
same probability for classical particles. 

A similar consideration also gives for the probabilities 
\begin{equation} \label{p20}
P(2,0) = P(0,2) = 2 R T
\end{equation}
for bosons, whereas for fermions the two probabilities vanish $P(2,0)
= P(0,2) =0$. For classical partition of carriers the probability to
find the two particles in the same detector is $RT$, which is just 
one half of the probability for bosons.

The average occupation numbers are $ \langle n_{3} \rangle =
\langle n_{4} \rangle =1$, since we have now two particles in branch
$3$ with probability $P(2,0)$ and one particle with probability
$P(1,1)$. Consequently, the correlations of the fluctuations in the
occupation numbers  $\Delta \hat n_{i} = \hat n_{i} -  \langle n_{i}
\rangle$ are given by  
\begin{equation} \label{foccb}
\langle \Delta \hat n_{3} \Delta \hat n_{4} \rangle  = -4RT ,
\end{equation}
for bosons and by $\langle \Delta \hat n_{3} \Delta \hat n_{4}\rangle
= 0$ for fermions. For bosons the correlation is negative due to the
enhanced probability that both photons end up in the same output
branch. For fermions there are no fluctuations in the occupation
number and the correlation function thus vanishes.  

{\bf Two-particle scattering: Wave packet overlap}. The discussion
given above implicitly assumes that both "particles" or "waves" arrive
simultaneously at the mirror and "see" each other. Clearly, if the two
particles arrive at the mirror with a time-delay which is large enough
such that there is no overlap, the outcome of the experiments
described above is entirely different. If we have only a sequence of
individual photons or electrons arriving at the mirror we have for the
expectation values of the occupation numbers $\langle n_{3} \rangle =
\langle n_{4} \rangle = R + T = 1$, and the correlation of the
occupation number $\langle n_{3}n_{4} \rangle  = 0$
vanishes. Consequently, the correlation of the fluctuations of the
occupation number  is $ \langle \Delta n_{1} \Delta n_{2} \rangle =
-1$. Without any special sources at hand it is impossible to time the
carriers such that they arrive simultaneously at the mirror, and we
should consider all possibilities.  

To do this, we must consider the states at the input in more
detail. Let us assume that a state in input arm $i$ can be written
with the help of plane waves $\Psi_{i} (k, x_{i}) = \exp(-ikx_{i})$
with $x_{i}$ the coordinate along arm $i$ normalized such that it
grows as we move away from the arm toward the source. Similarly, let
$y_{i}$ be the coordinates along the output arms such that $y_{i}$
vanishes at the mirror and grows as we move away from the splitter. A
plane wave $\Psi_{1}(k, x_{1}) = \exp(-ikx_{1})$  incident from arm
$1$ thus leads to a reflected wave in output arm $3$ given by
$\Psi_{1} (k, y_{3}) = r \exp(-iky_{3})$ and to a transmitted wave
$\Psi_{1} (k, y_{4}) = t \exp(-iky_{4})$ in output arm $4$. We call
such a state a "scattering state". It can be regarded as the limit of
a wave packet with a spatial width that tends towards infinity and an
energy width that tends to zero. To built up a particle that is
localized in space at a given time we now invoke superpositions of
such scattering states. Thus, let the incident particle in arm $i$ be
described  by $\Psi_{i}(x_{i}, t ) = \int dk \alpha_{i} (k)
\exp(-ikx_{i}) \exp(-iE(k) t/\hbar),$ where $\alpha_{i} (k)$ is a
function such that 
\begin{equation} \label{norm}
\int_{0}^{\infty} dk 
|\alpha_{i}(k)|^{2} = 1,  
\end{equation}
and $E(k)$ is the 
energy of the carriers as a function of the wave vector $k$. In second
quantization the incident states are written with the help of the
operators 
\begin{eqnarray} \label{Aop}
\hat A^{\dagger}_{i} (x_{i}, t ) & = & \int_{0}^{\infty} dk_{i} 
\alpha_{i}(k_{i}) \Psi_{i} (k, x_{i}) \hat a^{\dagger}_{i}(k_{i})
\nonumber \\ 
& \times & \exp{(-iE(k_{i})t/\hbar)},
\end{eqnarray}
and the initial state of our two-particle scattering experiment is
thus $\hat A^{\dagger}_{1} (x_{1}, t ) \hat A^{\dagger}_{2} (x_{2},
t ) |0 \rangle$. We are again interested in determining
the probabilities that two particles appear in an output branch or
that one particle appears in each output branch, $P(2,0) =
\int_{0}^{\infty} dk_{3} dk_{4} \langle \hat n_{3}(k_{3})\hat
n_{3}(k_{4}) \rangle$, $P(1,1) = \int_{0}^{\infty} dk_{3} dk_{4}$
$\langle \hat n_{3}(k_{3})\hat n_{4}(k_{4}) \rangle$, and $P(0,2) =
\int_{0}^{\infty} dk_{3} dk_{4} \langle \hat n_{4}(k_{3})\hat
n_{4}(k_{4}) \rangle$. Let us again consider $P(1,1)$. Its evaluation
proceeds in much the same way as in the case of pure scattering
states. We re-write $P(1,1)$ in terms of the absolute square of an
amplitude,  
\begin{equation} \label{Ap11} P(1,1) = \int_{0}^{\infty} dk_{3} dk_{4} 
|\langle 0|\hat b_{3}(k_{3}) \hat b_{4}(k_{4}) |\Psi \rangle|^{2}. 
\end{equation}
We then write the $\hat a$ operators in the $\hat A$ in terms of 
the output operators $\hat b$. Instead of Eq. (\ref{p11}), we obtain 
\begin{equation} \label{pro11}
P(1,1) = T^{2} + R^{2} \pm 2 T R |J|^{2}, 
\end{equation}
where 
\begin{equation} \label{over}
J = \int_{0}^{\infty} dk \alpha^{\ast}_{1}(k) \alpha_{2}(k)
\exp(ik(x_{1}-x_{2}))
\end{equation}
is the {\it overlap integral} of the two particles. For the case of
complete overlap $|J| =1$ we obtain  Eq. (\ref{p11}). For the case
that we have no overlap we obtain the classical result $P(1,1) = T^{2}
+ R^{2}$ which is independent of whether a boson or fermion is
incident on the scatterer. In the general case, the overlap depends on
the form of the wave packet. If two Gaussian wave packets of spatial
width $\delta$ and central velocity $v$ are timed to arrive at time
$\tau_1$ and $\tau_2$ at the scatterer, the overlap integral is 
\begin{equation} \label{overg}
|J|^{2} = \exp[-v^{2}(\tau_1 - \tau_2)^{2}/2\delta^{2}] .
\end{equation}
A significant overlap occurs only during the time $\delta /v$. For
wave packets separated in time by more than this time interval the
Pauli principle is not effective.  

Complete overlap occurs in two simple cases. We can assume that 
the two wave packets are identical and are timed to arrive exactly 
at the same instant at the scatterer. Another case, in which we have
complete overlap, is in the basis of scattering states. In this case
$|\alpha_{i} (k)|^{2} = \delta (k -k_{i})$ for a scattering state with
wave vector $k_{i}$ and consequently for the two particles with $k_i$
and $k_j$ we have $J = \delta_{ij}$. The first option of timed wave
packets seems artificial for the thermal sources which we want to
describe.  Thus in the following we will work with scattering states.

\begin{table} \label{tabloudon}
\begin{tabular}{llll}
Probability & Classical & Bosons & Fermions\\
\hline
$P(2,0)$ & $RT$ & $RT (1 + \vert J \vert^2)$ & $RT (1 - \vert J
\vert^2)$ \\
$P(1,1)$ & $R^2 + T^2$ & $R^2 + T^2 - 2RT\vert J \vert^2$ & $R^2 + T^2
+ 2RT\vert J \vert^2$ \\
$P(0,2)$ & $RT$ & $RT (1 + \vert J \vert^2)$ & $RT (1 - \vert J
\vert^2)$ \\
\end{tabular}
\caption{Output probabilities for one particle incident in each
input arm (From Ref. \protect\cite{Loudon}).} 
\end{table}

The probabilities $P(2,0)$, $P(1,1)$ and $P(0,2)$ for these 
scattering experiments are shown in Table II. The
considerations which lead to these results now should be extended to
take the polarization of photons or spin of electrons into account.  

{\bf Two-particle scattering: Spin}. Consider the case of fermions and
let us investigate a sequence of experiments in each of which we have
an equal probability of having electrons with spin up or down in an
incident state. Assuming that the scattering matrix is independent of
the spin state of the electrons, the results discussed above describe
the two cases when both spins point in the same direction (to be
denoted as $P(1\uparrow, 1\uparrow)$ and $P(1\downarrow,
1\downarrow)$). Thus what remains is to consider the case in which one
incident particle has spin up and one incident particle has spin
down. If the detection is also spin sensitive, the probability which
we determine is $P(1\uparrow,1\downarrow)$. But in such an experiment
we can tell which of the two particles went which way at the scatterer
and there is thus no interference. The outcome is classical:
$P(1\uparrow,1\downarrow) = R^{2}$ for an initial state  with a spin
up in arm $1$ and a spin down in arm $2$. For the same state we have
$P(1\downarrow,1\uparrow) = T^{2}$.  

Now let us assume that there is no way of detecting the spin state of
the outgoing particles. For a given initial state we have $P(1,1) =
\sum_{\sigma,\sigma^{\prime}} P(1\sigma,1\sigma^{\prime})$ where
$\sigma$ and $\sigma^{\prime}$ are spin variables. If we consider all
possible incident states with equal probability, we find $P(1,1) =
T^{2} + R^{2} + T R |J|^{2}$ i. e. a result with an interference
contribution which is only half as large as given in
Eq. (\ref{pro11}). For further discussion, see Appendix \ref{con083}.  

The scattering experiments considered above assume that we can 
produce one or two particle states either in a single mode or 
by exciting many modes. Below we will show that thermal sources,
the electron reservoirs which are of the main interest here,  cannot
be described in this way.


\subsection{The scattering approach} \label{scat023}

The idea of the scattering approach (also referred to as Landauer
approach) is to relate transport properties of the system (in
particular, current fluctuations) to its scattering properties, which
are assumed to be known from a quantum-mechanical calculation. In its
traditional form the method applies to {\em non-interacting} systems
in the {\em stationary} regime\footnote{To avoid a possible
misunderstanding, we stress that the long range Coulomb interaction
needs to be taken into account when one tries to apply the scattering
approach for the description of systems in time-dependent external
fields, or finite-frequency fluctuation spectra in stationary fields.
On the other hand, for the description of {\em zero-frequency}
fluctuation spectra in {\em stationary fields}, a consistent theory
can be given without including Coulomb effects, even though the
fluctuations themselves are, of course, time-dependent and random.}. 
The system may be either at {\em equilibrium} or in a {\em
non-equilibrium} state; this information is introduced through the
distribution functions of the contacts of the sample. To be clear, we
consider first a two-probe geometry and particles obeying Fermi
statistics (having in mind electrons in mesoscopic
systems). Eventually, the generalization to many probes and Bose
statistics is given; extensions to interacting problems are discussed
at the end of this Section. In the derivation we essentially follow
Ref. \cite{Buttiker92}. 
\begin{figure} 
{\epsfxsize=8.5cm\centerline{\epsfbox{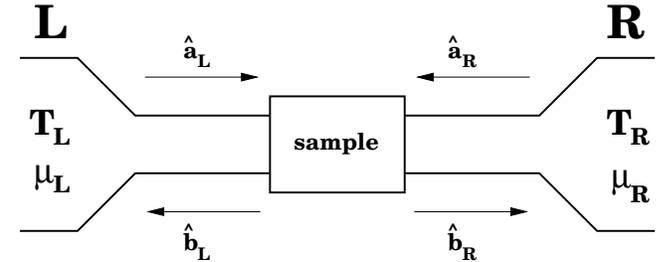}}}
\vspace{0.3cm}
\caption{Example of two-terminal scattering problem for the case of
one transverse channel.}
\label{setupscat}
\end{figure}

{\bf Two-terminal case; current operator.} We consider a mesoscopic
sample connected to two reservoirs (terminals, probes), to be referred
to as ``left'' (L) and ``right'' (R). It is assumed that the
reservoirs are so large that they can be characterized by a
temperature $T_{L,R}$ and a chemical potential $\mu_{L,R}$; the
distribution functions of electrons in the reservoirs, defined via
these parameters, are then Fermi distribution functions
\begin{displaymath}
f_{\alpha} (E) = [\exp[(E - \mu_{\alpha})/k_{B}T_{\alpha}]+1]^{-1}, \
\ \ \alpha = L,R
\end{displaymath}
(see Fig.~\ref{setupscat}). We must note at this stage, that, although
there are no inelastic processes in the sample, a strict equilibrium
state in the reservoirs can be established only via inelastic
processes. However, we consider the reservoirs (the leads) to be 
wide compared to the typical cross-section of the mesoscopic
conductor. Consequently, as far as the reservoirs are concerned, the
mesoscopic conductor represents only a small perturbation, and
describing their local properties in terms of an equilibrium state is
thus justified. We emphasize here, that even though the dynamics of
the scattering problem is described in terms of a Hamiltonian, the
problem which we consider is irreversible. Irreversibility is
introduced in the discussion, since the processes of a carrier leaving
the mesoscopic conductor and entering the mesoscopic conductor are
unrelated, uncorrelated events. The reservoirs act as sources of
carriers determined by the Fermi distribution but also act as perfect
sinks of carriers irrespective of the energy of the carrier that is
leaving the conductor.   

Far from the sample, we can, without loss of generality, assume that
transverse (across the leads) and longitudinal (along the leads)
motion of electrons are separable. In the longitudinal (from left to
right) direction the system is open, and is characterized by the
continuous wave vector $k_l$. It is advantageous to separate incoming
(to the sample) and outgoing states, and to introduce the longitudinal
energy $E_l = \hbar^2k_l^2/2m$ as a quantum number. Transverse motion
is quantized and described by the discrete index $n$ (corresponding to
transverse energies $E_{L,R;n}$, which can be different for the left
and right leads). These states are in the following referred to as
{\em transverse (quantum) channels}. We write thus $E = E_n +
E_l$. Since $E_l$ needs to be positive, for a given total energy $E$
only a finite number of channels exists. The number of incoming
channels is denoted $N_{L,R}(E)$ in the left and right lead,
respectively.  
  
We now introduce {\em creation} and {\em annihilation operators} of
electrons in the scattering states. In principle, we could have used
the operators which refer to particles in the states described by the
quantum numbers $n, k_l$. However, the scattering matrix which we
introduce below, relates current amplitudes and not wave function
amplitudes. Thus we introduce operators $\hat a^{\dagger}_{Ln} (E)$
and $\hat a_{Ln} (E)$ which create and annihilate electrons with total
energy $E$ in the transverse channel $n$ in the left lead, which are
incident upon the sample. In the same way, the creation $\hat
b^{\dagger}_{Ln} (E)$ and annihilation $\hat b_{Ln} (E)$ operators
describe electrons in the outgoing states. They obey anticommutation
relations    
\begin{displaymath} 
\hat a^{\dagger}_{Ln} (E) \hat a_{Ln'} (E') + \hat a_{Ln'} (E') \hat
a^{\dagger}_{Ln} (E) = \delta_{nn'} \delta(E - E'), 
\end{displaymath}
\begin{displaymath} 
\hat a_{Ln} (E) \hat a_{Ln'} (E') + \hat a_{Ln'} (E') \hat a_{Ln} (E)
= 0, 
\end{displaymath} 
\begin{displaymath} 
\hat a^{\dagger}_{Ln} (E) \hat a^{\dagger}_{Ln'} (E') + \hat
a^{\dagger}_{Ln'} (E') \hat a^{\dagger}_{Ln} (E) = 0.
\end{displaymath} 
Similarly, we introduce creation and annihilation operators $\hat
a^{\dagger}_{Rn} (E)$ and $\hat a_{Rn}(E)$ in incoming states and
$\hat b^{\dagger}_{Rn} (E)$ and $\hat b_{Rn}(E)$ in outgoing states in
the right lead (Fig.~\ref{setupscat}).   

The operators $\hat a$ and $\hat b$ are related via
the {\em scattering matrix} $s$,
\begin{equation} \label{scatmatr}
\left( \matrix{ \hat b_{L1} \cr \dots \cr \hat b_{LN_L} \cr \hat
b_{R1} \cr \dots \cr \hat b_{RN_R} } \right) = s \left( \matrix{
\hat a_{L1} \cr \dots \cr \hat a_{LN_L} \cr \hat a_{R1} \cr \dots \cr
\hat a_{RN_R} } \right). 
\end{equation}
The creation operators $\hat a^{\dagger}$ and $\hat b^{\dagger}$ obey
the same relation with the hermitian conjugated matrix $s^{\dagger}$. 

The matrix $s$ has dimensions $(N_L + N_R) \times (N_L +
N_R)$. Its size, as well as the matrix elements, depends on the total
energy $E$. It has the block structure
\begin{eqnarray} \label{scatmatr1}
s = \left( \matrix{ r & t' \cr t & r' } \right).
\end{eqnarray}
Here the square diagonal blocks $r$ (size $N_L \times N_L$) and $r'$
(size $N_R \times N_R$) describe electron reflection back to the left
and right reservoirs, respectively. The off-diagonal, rectangular
blocks $t$ (size $N_R \times N_L$) and $t'$ (size $N_L \times N_R$)
are responsible for the electron transmission through the sample. The
flux conservation in the scattering process implies that the matrix
$s$ is quite generally unitary. In the presence of time-reversal
symmetry the scattering matrix is also symmetric.      

The current operator in the left lead (far from the sample) is
expressed in a standard way,  
\begin{eqnarray*} 
\hat I_L (z,t) & = & \frac{\hbar e}{2im} \int d\bbox{r}_{\perp}
\left[ \hat \Psi^{\dagger}_L (\bbox{r}, t) \frac{\partial}{\partial z}
\hat \Psi_L (\bbox{r}, t) \right. \nonumber \\ 
& - & \left. \left( \frac{\partial}{\partial z}
\hat \Psi^{\dagger}_L (\bbox{r}, t) \right) \hat \Psi_L
(\bbox{r}, t) \right],    
\end{eqnarray*}
where the field operators $\hat \Psi$ and $\hat \Psi^{\dagger}$ are
defined as 
\begin{eqnarray*}
\hat \Psi_L (\bbox{r},t) & = & \int dE e^{-iEt/\hbar}
\sum_{n=1}^{N_L(E)} \frac{\chi_{Ln}
(\bbox{r}_{\perp})}{(2\pi\hbar v_{Ln}(E))^{1/2}} \nonumber \\
& \times & \left[ \hat a_{Ln} e^{ik_{Ln}z} + \hat b_{Ln}
e^{-ik_{Ln} z} \right] 
\end{eqnarray*}
and
\begin{eqnarray*}
\hat \Psi^{\dagger}_L (\bbox{r},t) & = & \int dE e^{iEt/\hbar}
\sum_{n=1}^{N_L(E)} \frac{\chi^*_{Ln}
(\bbox{r}_{\perp})}{(2\pi\hbar v_{Ln}(E))^{1/2}} \nonumber \\
& \times & \left[ \hat a^{\dagger}_{Ln} e^{-ik_{Ln}z} + \hat
b^{\dagger}_{Ln} e^{ik_{Ln} z} \right]. 
\end{eqnarray*}
Here $\bbox{r}_{\perp}$ is the transverse coordinate(s) and $z$
is the coordinate along the leads (measured from left to right);
$\chi_n^{L}$ are the transverse wave functions, and we have introduced
the wave vector, $k_{Ln} = \hbar^{-1}[2m(E - E_{Ln})]^{1/2}$ (the
summation only includes channels with real $k_{Ln}$), and 
the velocity of carriers $v_n(E) =
\hbar k_{Ln}/m$ in the $n$-th transverse channel.   

After some algebra, the expression for the current can be cast into
the form 
\begin{eqnarray} \label{cur0}
\hat I_L (z,t) & = & \frac{e}{4\pi\hbar} \sum_n \int dE dE'
e^{i(E-E')t/\hbar} \frac{1}{\sqrt{v_{Ln}(E) v_{Ln}(E')}} \nonumber \\
& \times & \left\{ \left[ v_{Ln} (E) + v_{Ln} (E') \right]
\right. \nonumber \\
& \times & \left. \left[ \exp
\left[ i \left(k_{Ln} (E') - k_{Ln} (E) \right) z \right] \hat
a^{\dagger}_{Ln} (E) \hat a_{Ln} (E') \right. \right. \nonumber \\
& - & \left. \left. \exp \left[ i \left(k_{Ln} (E) - k_{Ln} (E')
\right) z \right] \hat b^{\dagger}_{Ln} (E) \hat b_{Ln} (E') \right]
\right. \nonumber \\
& + & \left. \left[ v_{Ln} (E) - v_{Ln} (E') \right] \right. 
\\
& \times & \left. \left[ \exp
\left[ -i \left(k_{Ln} (E) + k_{Ln} (E') \right) z \right] \hat
a^{\dagger}_{Ln} (E) \hat b_{Ln} (E') \right. \right. \nonumber \\
& - & \left. \left. \exp \left[ i \left(k_{Ln} (E) + k_{Ln} (E')
\right) z \right] \hat b^{\dagger}_{Ln} (E) \hat a_{Ln} (E') \right]
\right\}. \nonumber
\end{eqnarray}
This expression is cumbersome, and, in addition, depends explicitly on
the coordinate $z$. However, it can be considerably simplified. The
key point  is that for all observable quantities (average current,
noise, or higher moments of the current distribution) the energies $E$
and $E'$ in Eq. (\ref{cur0}) either coincide, or are close to each
other. On the other hand, the velocities $v_n(E)$ vary with energy
quite slowly, typically on the scale of the Fermi energy. Therefore,
one can neglect their energy dependence, and reduce the expression
(\ref{cur0}) to a much simpler form\footnote{A discussion of the
limitations of Eq. (\ref{cur1}) is given in
Ref. \protect\cite{Pretre96}.},  
\begin{eqnarray} \label{cur1}
\hat I_L (t) & = & \frac{e}{2\pi\hbar} \sum_n \int dE dE'
e^{i(E-E')t/\hbar} \nonumber \\
& \times & \left[ \hat a^{\dagger}_{Ln} (E) \hat a_{Ln} (E') -
\hat b^{\dagger}_{Ln} (E) \hat b_{Ln} (E') \right].
\end{eqnarray}
Note that $\hat n^{+}_{Ln} (E) = \hat a^{\dagger}_{Ln} (E) \hat a_{Ln}
(E)$ is the operator of the occupation number of the incident carriers
in lead $L$ in channel $n$. Similarly, $\hat n^{-}_{Ln} (E) = \hat
b^{\dagger}_{Ln} (E) \hat b_{Ln} (E)$ is the operator of the
occupation number of the out-going carriers in lead $L$ in channel
$n$. Setting $E' = E + \hbar \omega$ and carrying out the integral over 
$\omega$ gives 
\begin{equation} \label{ab}
\hat I_L (t) = \frac{e}{2\pi\hbar} \sum_n \int dE 
\left[\hat n^{+}_{Ln} (E,t) - \hat n^{-}_{Ln} (E,t)\right].
\end{equation}
Here  $\hat n^{\pm}_{Ln} (E,t)$ are the time-dependent occupation
numbers for the left and right moving carriers at energy $E$. Thus
Eq. (\ref{ab}) states that the current at time $t$ is simply
determined by the difference in occupation number between the left and
right movers in each channel. We made use of this intuitively
appealing result already in the introduction. Using
Eq. (\ref{scatmatr}) we can express the current in terms of the $\hat
a$ and $\hat a^{\dagger}$ operators alone,  
\begin{eqnarray} \label{cur2}
\hat I_L (t) & = & \frac{e}{2\pi\hbar} \sum_{\alpha\beta} \sum_{mn}
\int dE dE' e^{i(E-E')t/\hbar} \nonumber \\
& \times & \hat a^{\dagger}_{\alpha m} (E)
A_{\alpha\beta}^{mn} (L; E,E') \hat a_{\beta n} (E').  
\end{eqnarray}
Here the indices $\alpha$ and $\beta$ label the reservoirs and may
assume values $L$ or $R$. The matrix $A$ is defined as
\begin{equation} \label{curmatr}
A_{\alpha\beta}^{mn} (L; E, E') = \delta_{mn} \delta_{\alpha L}
\delta_{\beta L} - \sum_{k} s^{\dagger}_{L\alpha; mk}(E) s_{L\beta;
kn} (E'),
\end{equation}
and $s_{L \alpha; mk} (E)$ is the element of the scattering matrix
relating $\hat b_{Lm} (E)$ to $\hat a_{\alpha k} (E)$. Note that
Eq. (\ref{cur2}) is independent of the coordinate $z$ along the lead.  

{\bf Average current.} Before we proceed in the next subsection with
the calculation of current-current correlations, it is instructive to
derive the average current from Eq. (\ref{cur2}). For a system at
thermal equilibrium the quantum statistical average of the product of
an electron creation operator and annihilation operator of a Fermi gas
is   
\begin{equation} \label{av1}
\left\langle \hat a^{\dagger}_{\alpha m} (E) \hat a_{\beta n} (E')
\right\rangle = \delta_{\alpha\beta} \delta_{mn} \delta (E-E')
f_{\alpha} (E). 
\end{equation} 
Using Eq. (\ref{cur2}) and  Eq. (\ref{av1}) 
and taking into account the unitarity of the scattering matrix 
$s$, we obtain
\begin{equation}  \label{av11}
\left\langle I_L \right\rangle = \frac{e}{2\pi\hbar} \int dE\ {\rm Tr}
\left[ t^{\dagger} (E) t(E) \right] \left[ f_L (E) - f_R(E) \right]. 
\end{equation}
Here the matrix $t$ is the off-diagonal block of the scattering matrix
(\ref{scatmatr1}), $t_{mn} = s_{RL; mn}$. 
In the zero-temperature limit and for a small applied 
voltage Eq. (\ref{av11}) gives a conductance 
\begin{equation} \label{gtr}
G = \frac{e^2}{2\pi\hbar} {\rm Tr}\left[ t^{\dagger} (E_{F}) t(E_{F})
\right]. 
\end{equation} 
Eq. (\ref{gtr}) establishes the relation between the scattering 
matrix evaluated at the 
Fermi energy and the conductance. It is a basis invariant expression. 
The matrix $t^{\dagger} t$
can be diagonalized; it has a real set of eigenvalues ({\em
transmission probabilities}) $T_n (E)$ (not to be confused with
temperature), each of them assumes a value between zero and one. In
the basis of {\em eigen channels} we have instead of Eq. (\ref{av11}) 
\begin{equation}  \label{avcur1}
\left\langle I_L \right\rangle = \frac{e}{2\pi\hbar} \sum_n \int dE\
T_n(E) \left[ f_L (E) - f_R(E) \right].  
\end{equation} 
and thus for the conductance 
\begin{equation} \label{condbasic}
G = \frac{e^2}{2\pi\hbar} \sum_n T_n,
\end{equation} 
Eq. (\ref{condbasic}) is known as a multi-channel generalization of
the Landauer formula. Still another version of this result expresses the 
conductance in terms of the transmission probabilities $ T_{RL,mn} =
|s_{RL,mn}|^{2}$ for carriers incident in channel $n$ in the left lead
$L$ and transmitted into channel $m$ in the right lead $R$. In this
basis the Hamiltonians of the left and right lead (the reservoirs) are
diagonal and the conductance is given by  
\begin{equation} \label{gnat}
G = \frac{e^2}{2\pi\hbar} \sum_{mn} T_{mn}.
\end{equation} 
We refer to this basis as the {\em natural basis}. 
We remark already here that, independently of the 
choice of basis,
the conductance can
be expressed in terms of transmission probabilities
only. 
This is not case for the shot noise to be discussed subsequently.
Thus the scattering matrix rather then transmission probabilities 
represents the fundamental object governing the kinetics of carriers.

{\bf Multi-terminal case}. We consider now a sample connected by ideal
leads to a number of reservoirs labeled by an index $\alpha$, with the
Fermi distribution functions $f_{\alpha} (E)$. At a given energy $E$
the lead $\alpha$ supports $N_{\alpha} (E)$ transverse channels. We
introduce, as before, creation and annihilation operators of electrons
in an incoming $\hat a^{\dagger}_{\alpha n}$, $\hat a_{\alpha n}$ and
outgoing $\hat b^{\dagger}_{\alpha n}$, $\hat b_{\alpha n}$ state of
lead $\alpha$ in the transverse channel $n$. These operators are again
related via the scattering matrix. We write down this relation,
similar to Eq. (\ref{scatmatr}), in components, 
\begin{equation} \label{scatmatr2}
\hat b_{\alpha m} (E) = \sum_{\beta n} s_{\alpha \beta; mn} (E) \hat
a_{\beta n} (E). 
\end{equation}
The matrix $s$ is again unitary, and, in the presence of time-reversal
symmetry, symmetric. 

Proceeding similarly to the derivation presented above, we obtain 
the multi-terminal generalization of Eq. (\ref{cur2}) for the current
through the lead $\alpha$,
\begin{eqnarray} \label{cur3}
\hat I_{\alpha} (t) & = & \frac{e}{2\pi\hbar} \sum_{\beta\gamma}
\sum_{mn} \int dE dE' e^{i(E-E')t/\hbar} \nonumber \\
& \times & \hat a^{\dagger}_{\beta m}
(E) A_{\beta\gamma}^{mn} (\alpha; E,E') \hat a_{\gamma n} (E'),
\end{eqnarray}
with the notation
\begin{equation} \label{curmatr3}
A_{\beta\gamma}^{mn} (\alpha; E, E') = \delta_{mn}
\delta_{\alpha\beta} \delta_{\alpha\gamma} - \sum_{k}
s^{\dagger}_{\alpha\beta; mk}(E) s_{\alpha\gamma; kn} (E').
\end{equation}
The signs of currents are chosen to be positive for incoming
electrons. 

Imagine that a voltage $V_{\beta}$ is applied to the reservoir
$\beta$, that is, the electro-chemical potential is $\mu_{\beta} = \mu
+ eV_{\beta}$, where $\mu$ can be taken to be the equilibrium chemical
potential. From
Eq. (\ref{cur3}) we find the average current,   
\begin{eqnarray} \label{condmult}
\left\langle I_{\alpha} \right\rangle & = & \frac{e^2}{2\pi\hbar}
\sum_{\beta} V_{\beta} \int dE \left( -\frac{\partial f}{\partial E}
\right) \nonumber \\
& \times & \left[ N_{\alpha} \delta_{\alpha\beta} - {\rm Tr} \left(
s^{\dagger}_{\alpha\beta} s_{\alpha\beta} \right) \right], 
\end{eqnarray}
where the trace is taken over channel indices in lead $\alpha$. 
As usual, we define the conductance matrix $G_{\alpha\beta}$ via
$G_{\alpha\beta} = d\langle I_{\alpha} \rangle /dV_{\beta}|_{V_{\beta}
=0}$. In the linear regime this gives 
\begin{equation} \label{kirk}
\left\langle I_{\alpha} \right\rangle = \sum_{\beta} G_{\alpha\beta}
V_{\beta} 
\end{equation}
with  
\begin{equation} \label{condmult1}
G_{\alpha\beta} = \frac{e^2}{2\pi\hbar}
\int dE \left( -\frac{\partial f}{\partial E} \right) \left[
N_{\alpha} \delta_{\alpha\beta} - {\rm Tr} \left(
s^{\dagger}_{\alpha\beta} s_{\alpha\beta} \right) \right]. 
\end{equation}
The scattering matrix is evaluated at the Fermi energy. 
Eq. (\ref{condmult1}) has been successfully applied to a wide range of
problems from ballistic transport to the quantum Hall effect. 

{\bf Current conservation, gauge invariance, and reciprocity}. Any
reasonable theory of electron transport must be current-conserving and
gauge invariant. {\em Current conservation} means that the sum of
currents entering the sample from all terminals is equal to zero {\em
at each instant of time}. For the multi-terminal geometry discussed
here this  means $\sum_{\alpha} I_{\alpha} = 0$. The current is taken
to be positive if it flows from the reservoir towards the mesoscopic
structure. For the average current in the two-terminal geometry, we
have $I_L + I_R = 0$. We emphasize that current conservation must hold
not only on the average but at each instant of time. In particular,
current conservation must also hold for the fluctuation spectra which
we discuss subsequently. In general, for time dependent currents, we
have to consider not only contacts which permit carrier exchange with
the conductor, but also other nearby metallic structures, for instance
gates, against which the conductor can be polarized. The requirement
that the results are {\em gauge invariant} means in this context, that
no current arises if voltages at all reservoirs are simultaneously
shifted by the same value (and no temperature gradient is
applied). For the average currents (see Eqs. (\ref{avcur1}),
(\ref{condmult1})) both properties are a direct consequence of the
unitarity of the scattering matrix.   

For the conductance matrix  $G_{\alpha\beta}$
current conservation and gauge invariance require that 
the elements of this matrix in each row and in each column 
add up to zero, 
\begin{equation}\label{curcon}
\sum_{\alpha} G_{\alpha\beta} = \sum_{\beta} G_{\alpha\beta} = 0 . 
\end{equation}
Note that for the two terminal case this implies
$G \equiv G_{LL} =  G_{RR} = - G_{LR} = - G_{RL}$. 
In the two terminal case, it is thus sufficient to evaluate one
conductance to determine the conductance matrix. In multi-probe
samples the number of elements of the one has to determine to find the
conductance matrix is given by the constraints (\ref{curcon}) and by
the fact that the conductance matrix is a susceptibility and obeys the
Onsager-Casimir symmetries 
\begin{displaymath}
G_{\alpha\beta} (B) = G_{\beta\alpha} (-B) .
\end{displaymath}
In the scattering approach the Onsager-Casimir symmetries 
are again a direct consequence of the reciprocity 
symmetry of the scattering matrix under field reversal. 

In the stationary case, the current conservation and the 
gauge invariance of the results are a direct consequence of the 
unitarity of the scattering matrix. In general, for non-linear and
non-stationary problems, current conservation and gauge invariance are
not automatically fulfilled. Indeed, in ac-transport a direct
calculation of average particle currents does not yield a current
conserving theory. Only the introduction of displacement currents,
determined by the long range Coulomb interaction, leads to a theory
which satisfies these basic requirements. We will discuss these issues
for noise problems in Section \ref{freq}.

\subsection{General expressions for noise} \label{scat024}

We are concerned with fluctuations of the current away 
from their average value. We thus introduce the operators
$\Delta \hat I_{\alpha} (t) \equiv \hat I_{\alpha} (t)
- \langle I_{\alpha} \rangle$.  
We define the correlation function $S_{\alpha\beta}(t-t')$
of the current in contact
$\alpha$ and the current in contact $\beta$ 
as\footnote{Note that several definitions,
differing by numerical factors, can be found in the literature. The
one we use corresponds to the general definition of time-dependent
fluctuations found in Ref. \cite{LL}. We define the Fourier transform
with the coefficient $2$ in front of it, then our normalization yields
the equilibrium (Nyquist-Johnson) noise $S = 4k_{B}TG$ and is in
accordance with Ref. \cite{dJBrev}, see below. The standard
definition of Fourier transform would yield the Nyquist-Johnson noise
$S = 2k_{B}TG$. Ref. \cite{Buttiker92} defines the spectral function
which is multiplied by the width of the frequency interval where noise
is measured.}     
\begin{equation} \label{noisedef}
S_{\alpha\beta}(t - t') \equiv \frac{1}{2} \left\langle \Delta \hat
I_{\alpha} (t) \Delta \hat I_{\beta} (t') + \Delta \hat I_{\beta} (t')
\Delta \hat I_{\alpha} (t) \right\rangle.
\end{equation}
Note that in the absence of time-dependent external fields the
correlation function must be function of only $t - t'$. Its Fourier
transform,  
\begin{displaymath}
2\pi \delta(\omega + \omega') S_{\alpha\beta} (\omega) \equiv
\left\langle \Delta \hat I_{\alpha} (\omega) \Delta \hat I_{\beta}
(\omega') + \Delta \hat I_{\beta} (\omega') \Delta \hat I_{\alpha}
(\omega) \right\rangle,
\end{displaymath}
is sometimes referred to as {\em noise power}. 

To find the noise power we need the quantum statistical expectation
value of products of four operators $\hat a$. For a Fermi gas (or a
Bose gas) at equilibrium this expectation value is 
\begin{eqnarray} \label{averfour}
& & \left\langle \hat a^{\dagger}_{\alpha k} (E_1) \hat a_{\beta l}
(E_2) \hat a^{\dagger}_{\gamma m} (E_3) \hat a_{\delta n} (E_4)
\right\rangle \nonumber \\
& - & \left\langle \hat a^{\dagger}_{\alpha k} (E_1) \hat a_{\beta l}
(E_2) \right\rangle \left\langle \hat a^{\dagger}_{\gamma 
m} (E_3) \hat a_{\delta n} (E_4) \right\rangle \nonumber \\
& &  = \delta_{\alpha\delta} \delta_{\beta\gamma} \delta_{kn}
\delta_{ml} \delta(E_1 - E_4) \delta(E_2 - E_3) \nonumber \\
& \times & f_{\alpha} (E_1) \left[ 1 \mp f_{\beta} (E_2) \right]. 
\end{eqnarray}
(The upper sign corresponds to Fermi statistics, and 
the lower sign corresponds to Bose statistics. This convention 
will be maintained whenever we compare systems with differing
statistics. It is also understood that for Fermi statistics
$f_{\alpha}(E)$ is a Fermi distribution and for Bose statistics
$f_{\alpha}(E)$ is a Bose distribution function). Making use of
Eq. (\ref{cur3}) and of the expectation value (\ref{averfour}), we
obtain the expression for the noise power \cite{Buttiker92}, 
\begin{eqnarray} \label{noisefr}
S_{\alpha\beta} (\omega)  & = & \frac{e^2}{2\pi \hbar}
\sum_{\gamma\delta} \sum_{mn} \int dE A_{\gamma\delta}^{mn} (\alpha;
E, E + \hbar\omega) \nonumber \\
& \times & A_{\delta\gamma}^{nm} (\beta; E + \hbar\omega, E)
\nonumber \\
& \times & \left\{ f_{\gamma} (E) \left[ 1 \mp f_{\delta} (E +
\hbar\omega) \right] \right. \nonumber \\
& + & \left. \left[ 1 \mp f_{\gamma} (E) \right] f_{\delta}
(E + \hbar\omega) \right\}.   
\end{eqnarray}
Note that with respect 
to frequency, it has the symmetry properties $S_{\alpha\beta} (\omega)
= S_{\beta\alpha} (-\omega)$. For arbitrary frequencies and 
an arbitrary $s$-matrix Eq. (\ref{noisefr}) is neither current
conserving nor gauge invariant and additional considerations are
needed to obtain a physically meaningful result. 

In the reminder of this Section, we will only be interested in the
{\em zero-frequency} noise. For the noise power at $\omega = 0$ 
we obtain \cite{Buttiker92}
\begin{eqnarray} \label{noise0}
S_{\alpha\beta} \equiv S_{\alpha\beta} (0) & = & \frac{e^2}{2\pi
\hbar} \sum_{\gamma\delta} \sum_{mn} \int dE A_{\gamma\delta}^{mn}
(\alpha; E, E) \nonumber \\
& \times & A_{\delta\gamma}^{nm} (\beta; E, E) \\
& \times & \left\{ f_{\gamma} (E) \left[ 1 \mp f_{\delta} (E) \right]
+ \left[ 1 \mp f_{\gamma} (E) \right] f_{\delta} (E)
\right\}. \nonumber   
\end{eqnarray}
Eqs. (\ref{noise0}) are current conserving and gauge invariant. 
Eq. (\ref{noise0}) can now be used to predict the low frequency noise
properties of arbitrary multi-channel and multi-probe, phase-coherent
conductors. We first elucidate the general properties of this result,
and later on analyze it for various physical situations. 

{\bf Equilibrium noise}. If the system is in thermal equilibrium at
temperature $T$, the distribution functions in all reservoirs coincide
and are equal to $f(E)$. Using the property $f(1 \mp f) =
-k_{B}T\partial f/\partial E$ and employing the unitarity of the
scattering matrix, which enables us to write 
\begin{displaymath}
\sum_{\gamma\delta} {\rm Tr} \left( s^{\dagger}_{\alpha\gamma}
s_{\alpha\delta} s^{\dagger}_{\beta\delta} s_{\beta\gamma} \right) = 
\delta_{\alpha\beta} N_\alpha 
\end{displaymath} 
(where as before the trace is taken over transverse channel indices, 
and $N_{\alpha}$ is the number of channels in the lead $\alpha$),
we find 
\begin{eqnarray} \label{noiseeq1}
S_{\alpha\beta} & = & \frac{e^2 k_{B}T}{\pi\hbar} \int dE \left( -
\frac{\partial f}{\partial E}\right) \nonumber \\
& \times & \left[ 2N_{\alpha}
\delta_{\alpha\beta} - {\rm Tr} \left( s^{\dagger}_{\alpha\beta}
s_{\alpha\beta} + s^{\dagger}_{\beta\alpha}  s_{\beta\alpha} \right)
\right].  
\end{eqnarray}
This is the equilibrium, or {\em Nyquist-Johnson} noise. In the
approach discussed here it is a consequence of the the thermal
fluctuations of occupation numbers in the reservoirs. Comparing
Eqs. (\ref{condmult}) and (\ref{noiseeq1}), we see that  
\begin{equation} \label{FDT1}
S_{\alpha\beta} = 2k_{B}T \left( G_{\alpha\beta} + G_{\beta\alpha}
\right). 
\end{equation}
This is the manifestation of the fluctuation-dissipation theorem:
equilibrium fluctuations are proportional to the corresponding
generalized susceptibility, in this case to the conductance. For the
time-reversal case (no magnetic field) the conductance matrix is
symmetric, and Eq. (\ref{FDT1}) takes the form 
\begin{displaymath}
S_{\alpha\beta} = 4k_{B}T G_{\alpha\beta}, 
\end{displaymath}
which is familiar for the two-terminal case, $S = 4k_{B}T G$, with
$G$ being the conductance. From Eq. (\ref{FDT1}) we see that 
the fluctuation spectrum of the mean squared current at a contact 
$\alpha$ is positive (since $G_{\alpha \alpha} >0$)
but that the current-current correlations of the fluctuations 
at different probes are negative  (since $G_{\alpha\beta} <0$).
The sign of the equilibrium current-current fluctuations is
independent of statistics: Intensity-intensity fluctuations for a
system of bosons in which the electron reservoirs are replaced by
black body radiators are also negative. We thus see that equilibrium
noise does not provide any information of the system beyond that
already known from conductance measurements.  

Nevertheless, the equilibrium noise is important, if only to calibrate
experiments and as a simple test for theoretical discussions. 
Experimentally, a careful study of thermal noise in a multi-terminal
structure (a quantum Hall bar with a constriction) was recently
performed by Henny {\em et al} \cite{henny}. Within the experimental
accuracy, the results agree with the theoretical predictions. 

{\bf Shot noise. Zero temperature}. We now consider noise in a 
system of fermions\footnote{For bosons at zero temperature one needs
to take into account Bose condensation effects.} 
in a transport state. In the zero temperature limit the Fermi
distribution  in each reservoir is a step function $f_{\alpha} (E) =
\theta(\mu_{\alpha} - E)$.  Using this we can rewrite
Eq. (\ref{noise0}) as  
\begin{eqnarray} \label{noiset0}
S_{\alpha\beta} & = & \frac{e^2}{2\pi\hbar} \sum_{\gamma \ne \delta}
\int dE \ {\rm Tr} \left[ s^{\dagger}_{\alpha\gamma} s_{\alpha\delta}
s^{\dagger}_{\beta\delta} s_{\beta\gamma} \right] \nonumber \\
& & \times \left\{ f_{\gamma} (E) \left[ 1 - f_{\delta} (E) \right] +
f_{\delta} (E) \left[ 1 - f_{\gamma} (E) \right] \right\}. 
\end{eqnarray} 
We are now prepared to make two general statements. First,
correlations of the current at the {\em same} lead,
$S_{\alpha\alpha}$, are {\em positive}. This is easy to see, since
their signs are determined by positively defined
quantities\footnote{These quantities are called ``noise conductances''
in Ref. \protect\cite{Buttiker92}.} ${\rm Tr}
[s^{\dagger}_{\alpha\gamma} s_{\alpha\delta}
s^{\dagger}_{\alpha\delta} 
s_{\alpha\gamma}]$. The second statement is that the correlations at
{\em different} leads, $S_{\alpha\beta}$ with $\alpha \ne \beta$, are
{\em negative}. This becomes clear if we use the property
$\sum_{\delta} s_{\alpha\delta} s^{\dagger}_{\beta\delta} = 0$ and
rewrite Eq. (\ref{noiset0}) as     
\begin{eqnarray*}
S_{\alpha\beta} & = & -\frac{e^2}{\pi\hbar} \int dE \nonumber \\ 
& \times & {\rm Tr} \left[
\left( \sum_{\gamma} s_{\beta\gamma} s^{\dagger}_{\alpha\gamma}
f_{\gamma}(E) \right) \left( \sum_{\delta} s_{\alpha\delta}
s^{\dagger}_{\beta\delta} f_{\delta}(E) \right) \right].
\end{eqnarray*}
The integrand is now positively defined. Of course, current
conservation implies that if all cross-correlations $S_{\alpha\beta}$
are negative for all $\beta$ different from $\alpha$, the spectral
function $S_{\alpha\alpha}$ must be positive.  

Actually, these statements are even more general. One can prove that
cross-correlations in the system of fermions are generally negative at
any temperature, see Ref. \cite{Buttiker92} for details. On the other
hand, this is not correct for a system of bosons, where under
certain conditions cross-correlations can be positive.  

{\bf Two-terminal conductors}. Let us now consider the
zero-temperature shot noise of a two-terminal conductor. Again we
denote the leads as left (L) and right (R). Due to current
conservation, we have $S \equiv S_{LL} = S_{RR} = - S_{LR} = -
S_{RL}$. Utilizing the representation  of the scattering matrix
(\ref{scatmatr1}), and taking into account that the unitarity of the
matrix $s$ implies $r^{\dagger}r + t^{\dagger}t = 1$, we obtain after
some algebra 
\begin{eqnarray} \label{shottr}
S& = & \frac{e^2}{\pi\hbar} {\rm Tr} \ (r^{\dagger} r t^{\dagger} t) \
e \vert V \vert,   
\end{eqnarray}
where the scattering matrix elements are evaluated at the Fermi
level. This is the basis invariant relation between the scattering
matrix and the shot noise at zero temperature. Like the expression of
the conductance, Eq. (\ref{gtr}), we can express this result in the
basis  of eigen channels with the help of the transmission
probabilities $T_{n}$ and reflection probabilities $R_{n} =1-T_{n}$,
\begin{equation} \label{shot2term}
S_{LL} = \frac{e^3 \vert V \vert}{\pi\hbar} \sum_n T_n \left( 1 -
T_n \right). 
\end{equation}
We see that the non-equilibrium (shot) noise is not simply determined
by the conductance of the sample. Instead, it is determined by a sum
of products of transmission and reflection probabilities of the eigen
channels. Only in the limit of low-transparency $T_n \ll 1$ in {\em
all} eigen channels is the shot noise given by the {\em  Poisson
value}, discussed by Schottky,  
\begin{equation} \label{Poisson1}
S_P = \frac{e^3 \vert V \vert}{\pi\hbar} \sum_n T_n = 2e \langle I
\rangle.
\end{equation}
It is clear that zero-temperature shot noise is always suppressed in
comparison with the Poisson value\footnote{This statement is only
valid for non-interacting systems. Interactions may cause
instabilities in the system, driving the noise to super-Poissonian
values. Noise in systems with multi-stable current-voltage
characteristics (caused, for example, by a non-trivial structure of
the energy bands, like in the Esaki diode) may also be
super-Poissonian. These features are discussed in Section
\ref{Langevin}.}. In particular, neither closed ($T_n = 0$) nor open
($T_n = 1$) channels contribute to shot noise; 
the maximal contribution comes from channels with $T_n = 1/2$. The
suppression below the Poissonian limit given by Eq. (\ref{Poisson1})
was one of the aspects of noise in mesoscopic systems which triggered
many of the subsequent theoretical and experimental works. A
convenient measure of {\em sub-Poissonian shot noise} is the {\em Fano
factor} $F$ which is the ratio of the actual shot noise and the
Poisson noise that would be measured if the system produced noise due
to single independent electrons,  
\begin{equation} \label{Fano00} 
F = \frac{S_{LL}}{S_P}. 
\end{equation}
For energy independent transmission and/or in the linear regime the
Fano factor is  
\begin{equation} \label{Fano}
F = \frac{\sum_n T_n(1-T_n)}{\sum_n T_n}.
\end{equation}
The Fano factor assumes values between zero (all channels are
transparent) and one (Poissonian noise). In particular, for one
channel it becomes $(1-T)$.   

Unlike the conductance, which can be expressed in terms of
(transmission) probabilities independent of the choice of basis, the
shot  noise, even for the two terminal conductors considered here, can
not be expressed in terms of probabilities. The trace of
Eq. (\ref{shottr}) is a sum over $k,l,m,n$ of terms
$r^{*}_{kn}r_{km}t^{*}_{lm}t_{ln}$, which by themselves
are not real valued if $m \ne n$ (in contrast to the Eq. (\ref{gnat})
for the conductance). This is a signature that carriers from different
quantum channels interfere and must remain indistinguishable. It is 
very interesting to examine whether it is possible to find
experimental arrangements which directly probe such {\em exchange
interference} effects, and we return to this question later on. In the
remaining part of this subsection we will use the eigen channel basis
which offers the most compact representation of the results.  

The general result for the noise power of the current fluctuations 
in a two-terminal conductor is 
\begin{eqnarray} \label{noistwo1}
S & = & \frac{e^2}{\pi\hbar} \sum_n \int dE \left\{ T_n(E) 
\left[ f_L (1 \mp f_L) + f_R (1 \mp f_R) \right]
\right. \nonumber \\
& \pm & \left. T_n(E) \left[ 1 - T_n(E) \right] \left( f_L - f_R
\right)^2 \right\}.   
\end{eqnarray}
Here the first two terms are the equilibrium noise contributions, 
and the third term, which changes sign if we change statistics from
fermions to bosons, is the non-equilibrium or shot noise 
contribution to the power spectrum. Note that this term is second 
order in the distribution function. At high energies, in the range 
where both the Fermi and Bose distribution function are well 
approximated by a Maxwell-Boltzmann distribution, it is negligible 
compared to the equilibrium noise described by the first two terms. 
According to Eq. (\ref{noistwo1}) the shot noise term 
enhances the noise power compared to the equilibrium noise for 
fermions but diminishes the noise power for bosons. 
 
In the practically important case, when the scale of the energy
dependence of transmission coefficients $T_n (E)$ is much larger than
both the temperature and applied voltage, these quantities in
Eq. (\ref{noistwo1}) may be replaced by their values taken at the
Fermi energy. We obtain then (only fermions are considered henceforth)
\begin{eqnarray} \label{noistwo2}
S & = & \frac{e^2}{\pi\hbar} \left[ 2k_{B}T \sum_n T_n^2
\right. \nonumber \\ 
& + & \left. eV \coth \left( \frac{eV}{2k_{B}T} \right) \sum_n T_n
\left( 1 - T_n \right) \right], 
\end{eqnarray}
where $V$ is again the voltage applied between the left and right
reservoirs. The full noise is a complicated function of temperature
and applied voltage rather than a simple superposition of equilibrium
and shot noise\footnote{The full noise can be divided into
equilibrium-like and transport parts, see Ref. \cite{Buttiker92}. This
division is, of course, arbitrary.}.  For low voltages $eV \ll k_BT$ we
obtain $S= 4k_{B}TG$, in accordance with the general result
(\ref{FDT1}).  

Note that, since $\coth x > 1/x$ for any $x > 0$, the actual noise
(\ref{noistwo2}) for any voltage is {\em higher} than the equilibrium
noise. This is not generally correct if the transmission coefficients
are strongly energy dependent. As pointed out by Lesovik and Loosen
\cite{Loosen93}, in certain situations (for instance, when the
transmission coefficients sharply peak as functions of energy) the
total non-equilibrium noise may be actually {\em lower} than the
equilibrium noise at the same temperature. 

We conclude this subsection with some historical remarks. Already
Kulik and Omel'yanchuk \cite{Kulik84} noticed that the shot noise in
ballistic contacts (modeled as an orifice in an insulating layer
between two metallic reservoirs) vanishes if there is no elastic
impurity scattering. Subsequently, Khlus \cite{Khlus} considered such
a point contact with elastic scattering and derived
Eq. (\ref{noistwo2}) by means of a Keldysh Green's function
technique. The papers by Kulik and 
Omel'yanchuk and by Khlus remained unknown, they were either not or
only poorly cited even in the Russian literature. Later Lesovik
\cite{Lesovik89} derived Eq. (\ref{noistwo1}) in the framework of the
scattering approach (for the case of fermions). Independently, Yurke
and Kochanski \cite{Yurke} investigated the momentum noise of a
tunneling microscopic tip, also based on the scattering approach,
treating only the one-channel case; Ref. \cite{Lesovik89} treated the
multi-channel case, but assumed at the outset that the scattering
matrix is diagonal and that the diagonal channels are independent. A
generalization for many-channel conductors described by an arbitrary
scattering matrix (without assumption of independence) and for the
many-terminal case was given in Refs. \cite{Buttiker90,Buttiker91}. 
The same results were later discussed by Landauer and Martin 
\cite{LM91} and Martin and Landauer \cite{ML92} appealing to wave
packets. The treatment of wave-packet overlap (see subsection
\ref{scat022}) is avoided by assuming that wave packets are 
identical and timed to arrive at the same instant. Ref. 
\cite{Buttiker92}, which we followed in this subsection, is a long
version of the papers \cite{Buttiker90,Buttiker91}.  

\subsection{Voltage Fluctuations} \label{scat025}

\begin{figure}
{\epsfxsize=6.cm\centerline{\epsfbox{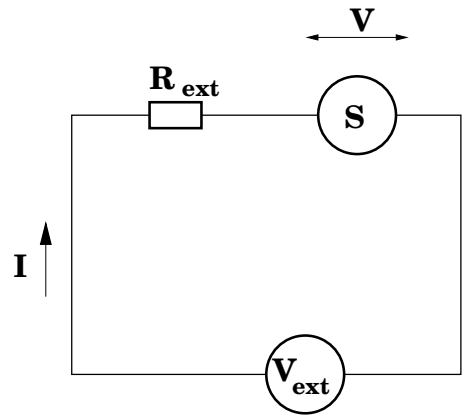}}}
\vspace{0.3cm}
\caption{Noise measurements in an external circuit. The sample is
denoted $S$.}
\protect\label{circuitf}
\end{figure}

{\bf Role of external circuit}. 
Thus far all the results which we have presented are based on the 
assumption that the sample is part of an external circuit 
with zero impedance. In this case the voltage (voltages) applied to
the sample can be viewed to be a fixed non-fluctuating quantity
and the noise properties are determined by the current correlations
which we have discussed. The idealized notion of a zero-impedance
external circuit does often not apply. Fig. \ref{circuitf}
shows a simple example: The sample S is part of an electrical
circuit with resistance $R_{ext}$ and a voltage source which generates 
a voltage $V_{ext}$. (In general the external circuit is described by
a frequency dependent impedance $Z_{ext}(\omega))$. As a consequence,
in such a circuit, we deal with both current fluctuations and voltage
fluctuations.  

The current fluctuations through the sample are now governed by the
fluctuations $ \Delta V(t)$ of the voltage across the sample, which
generate a fluctuating current $\Delta I_{V}(\omega) = G(\omega)
\Delta V(\omega)$, where $G(\omega)$ is frequency-dependent
conductance (admittance) of the sample. In addition, there is the
contribution of the spontaneous current fluctuations $\delta
I(\omega)$ themselves. The total fluctuating current through the
sample is thus given by   
\begin{equation}\label{l1}
\Delta I(\omega) = G(\omega) \Delta V(\omega) +\delta I(\omega).
\end{equation} 
Eq. (\ref{l1}) has the form of a Langevin equation with a fluctuating
source term given by the spontaneous current fluctuations determined by
the noise power spectrum $S$ of a two terminal conductor. To complete
these equations we must now relate the current through the sample to
the external voltage. The total current $I$ is related to the external
voltage $V_{ext}$ and the voltage across the sample by the Kirchoff
law $V + R_{ext}I = V_{ext}$. Here $V_{ext}$ is a constant, and the
voltage and current fluctuations are thus related by $\Delta V +
R_{ext} \Delta I =0$, or  
\begin{equation}\label{l2}
\Delta V(\omega) = - R_{ext} \Delta I(\omega).
\end{equation}
For $R_{ext} = 0$ (zero external impedance) we have the case of a {\em
voltage controlled} external circuit, while for $R_{ext} \to
\infty$ (infinite external impedance) we have the case of a {\em
current controlled} external circuit. The Langevin approach assumes
that the mesoscopic sample and the external circuit can be treated as
separate entities, each of which might be governed by quantum effects,
but that there are no phase coherent effects which would require the
treatment of the sample and the circuit as one quantum mechanical
entity. In such a case the distinction between sample and external
circuit would presumably be meaningless. Eliminating the voltage
fluctuations in Eqs. (\ref{l1}) and (\ref{l2}) gives for the current
fluctuations $\Delta I(\omega) (1+G(\omega)R_{ext}) =  \delta
I(\omega)$ and with the resistance of the sample $R = 1/G(0)$ and the
noise power spectrum $S$ we obtain in the zero-frequency
limit\footnote{The quantity $S_{II}$ is defined by
Eq.(\protect\ref{noisedef}), as before. The quantity $S_{VV}$ is
defined by the same expression where current fluctuations $\Delta I$
are replaced by the voltage fluctuations $\Delta V$.}
\begin{equation}\label{SIext}
S_{II} = \frac{S}{(1+ R_{ext}/R)^{2}} .
\end{equation}
Eq. (\ref{SIext}) shows that the external impedance becomes 
important if it is comparable or larger than the resistance of the
sample. Eliminating the current we obtain $(1/R_{ext} + G(\omega))
\Delta V(\omega) = - \delta I(\omega)$, and thus a voltage fluctuation
spectrum given by  
\begin{equation}\label{SVext}
S_{VV} = \frac{S}{(\frac{1}{R_{ext}} + \frac{1}{R})^{2}}. 
\end{equation}
At equilibrium, where the current noise power is given by $S=4k_{B}T
G$, Eq. (\ref{SVext}) gives 
$S_{VV} =$ $4k_{B}T/R(1/R_{ext}+1/R)^{2}$ which reduces in the limit 
$R_{ext} \to \infty$
to the familiar Johnson-Nyquist result 
$S_{VV} = 4k_{B}T R$ for the voltage fluctuations in an infinite
external impedance circuit. The procedure described above can also be
applied to shot noise as long as we are only concerned with effects
linear in the voltage $V$. Far from equilibrium, this approach applies
if we replace the conductance (resistance) by the differential
conductance (resistance) and if a linear fluctuation theory is
sufficient.  

{\bf External circuit: multi-probe conductors}. For a multi-probe
geometry the consideration of the external circuit is similarly based
on the Langevin equation \cite{Buttiker92},
\begin{equation}\label{lm}
\Delta I_{\alpha} = \sum_{\beta} G_{\alpha\beta} \Delta V_{\beta} + 
\delta I_{\alpha},
\end{equation}
where $G_{\alpha\beta}$ is an element of the conductance matrix
and $\delta I_{\alpha}$ is a fluctuating current with 
the noise power spectrum $S_{\alpha\beta}$. 
The external circuit loops connecting to a multi-probe conductor
can have different impedances: the external impedance is thus
also represented by a matrix which connects voltages and 
currents at the contacts of the multi-probe conductor. 
Ideally, the current source and sink contacts are connected
to a zero impedance external circuit, whereas the voltage probes 
are connected to an external loop with infinite impedance. 

At equilibrium, in the limiting case that all probes are connected 
to infinite external impedance loops, 
the voltage fluctuations can be expressed in terms 
of multi-probe resistances. Consider first a four-probe conductor. 
A four-probe resistance is obtained 
by injecting current in contact $\alpha$ taking it out at contact
$\beta$ and using two additional contacts $\gamma$ and $\delta$ 
to measure the voltage difference $V_\gamma -V_\delta$. 
The four-probe resistance is defined as 
$R_{\alpha\beta,\gamma\delta} = (V_\gamma -V_\delta)/I$. 
Using the conductance matrix of a four-probe conductor, a little 
algebra shows that, 
\begin{equation}\label{Rfour}
R_{\alpha\beta,\gamma\delta} = D^{-1}\left( G_{\gamma \alpha}
G_{\delta \beta} - G_{\gamma \beta} G_{\delta \alpha} \right), 
\end{equation}
where $D$ is any sub-determinant of rank three of the conductance
matrix. (Due to current conservation and gauge invariance all possible
subdeterminants of rank three of the conductance matrix are identical
and are even functions of the applied magnetic field). 
Eqs. (\ref{Rfour}) 
can be applied to a conductor with any number of contacts larger than
four, since the conductance matrix of any dimension can be reduced to
a conductance matrix of dimension four, if the additional contacts not
involved in the measurement are taken to be connected to infinite
external impedance loops. Similarly, there exists an effective
conductance matrix of dimension three which permits to define a
three-probe measurement. In such a measurement one of the voltages is 
measured at the current source contact or current sink contact and
thus two of the indices in $R_{\alpha\beta,\gamma\delta}$ are
identical, $\alpha = \gamma$ or $\beta = \delta$. Finally, if the
conductance matrix is reduced to a $2\times 2$ matrix, we obtain a
resistance for which two pairs of indices are identical,
$R_{\alpha\beta,\alpha\beta}$ or $R_{\alpha\beta,\beta\alpha} = -
R_{\alpha\beta,\alpha\beta}$. With these resistances we can now
generalize the familiar Johnson-Nyquist relation $S_{VV} = 4k_{B}TR$
for two-probe conductors, to the case of a multi-probe conductor.  
For the correlation of a voltage difference $V_{\alpha}-V_{\beta}$
measured between contacts $\alpha$ and $\beta$ with a voltage
fluctuation $V_{\gamma}-V_{\delta}$ measured between contacts $\gamma$
and $\delta$ Eq. (\ref{lm}) leads to  
\begin{equation}\label{fldisext}
\langle (V_{\alpha}-V_{\beta}) (V_{\gamma}-V_{\delta})\rangle = 
2k_{B}T (R_{\alpha\beta,\gamma\delta} + R_{\gamma\delta,\alpha\beta}).
\end{equation}
The mean squared voltage fluctuations $\alpha = \gamma$ and 
$\beta = \delta$ are determined by the two-terminal resistances 
$R_{\alpha\beta,\alpha\beta}$
of the multi-probe conductor. The correlations of voltage fluctuations
(in the case when all four indices differ) are related to symmetrized
four-probe resistances.  

If shot noise is generated, for instance, by 
a current incident at contact $\alpha$ and taken out at contact 
$\beta$ (in a zero external impedance loop)
and with all other contacts connected to an infinite impedance
circuit, the voltage fluctuations are \cite{PB} 
\begin{equation}\label{peder}
\langle (\delta V_{\gamma} - \delta V_{\delta})(\delta V_{\epsilon} - 
\delta V_{\zeta})\rangle = 
\sum_{\alpha\eta} R_{\alpha\beta,\epsilon\zeta}
R_{\eta\beta,\gamma\delta} S_{\alpha\eta}, 
\end{equation}
where $S_{\alpha\eta}$ is the noise power spectrum of the current 
correlations at contacts $\alpha$ and $\eta$, and $\beta$ is an
arbitrary index. 

These examples demonstrate that the fluctuations in a conductor are in
general a complicated expression of the noise power spectrum
determined for the zero-impedance case, the resistances (or far from
equilibrium the differential resistances) and the external impedance
(matrix). These considerations are of importance since in experiments
it is the voltage fluctuations which are actually measured and which
eventually are converted to current fluctuations.

\subsection{Applications} \label{scat026}

In this subsection, we give some simple applications of the general
formulae derived above, and illustrate them with experimental results.
We consider only zero frequency limit. As we explained in the
Introduction, we do not intend to give here a review of all results
concerning a specific system. Instead, we focus on the application of
the scattering approach. For results derived for these systems with
other methods, the reader is addressed to Table 1. 

\begin{figure}
{\epsfxsize=8cm\centerline{\epsfbox{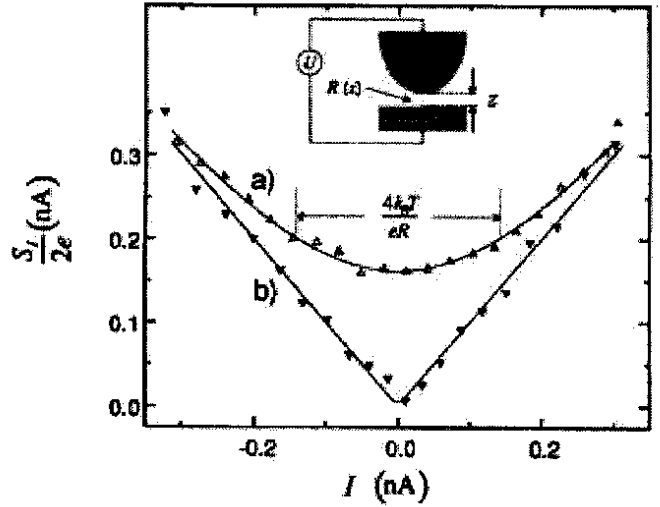}}}
\vspace{0.3cm}
\caption{Crossover from thermal to shot noise measured by Birk, de
Jong, and Sch\"onenberger \protect\cite{Birk}. Solid curves correspond
to Eq. (\ref{tunbar1}); triangles show experimental data for the two
samples, with lower (a) and higher (b) resistance.}
\label{fig220}
\end{figure}

\subsubsection{Tunnel barriers} \label{scat0261}

For a tunnel barrier, which can be realized, for example, as a layer
of insulator separating two normal metal electrodes, all the
transmission coefficients $T_n$ are small, $T_n \ll 1$ for any
$n$. Separating terms linear in $T_n$ in Eq. (\ref{noistwo2}) and
taking into account the definition of the Poisson noise,
Eq.(\ref{Poisson1}), we obtain
\begin{equation} \label{tunbar1}
S = \frac{e^3 V}{\pi \hbar} \coth \left( \frac{eV}{2k_{B}T}
\right) \sum_n T_n = \coth \left( \frac{e\vert V \vert }{2k_{B}T}
\right) S_P.
\end{equation}
At a given temperature, Eq. (\ref{tunbar1}) describes the crossover
from thermal noise at voltages  $e \vert V \vert \ll k_{B}T$ to shot
noise at voltages  $e \vert V \vert \gg k_{B}T$. The transition is
independent from any details of the tunnel barrier and occurs at $e
\vert V \vert = k_{B}T$.   

Eq. (\ref{tunbar1}) is also obtained in the zero-frequency, zero
charging energy limit of microscopic theories of low transparency
normal tunnel junctions or Josephson junctions
\cite{LO68,Dahm,Rogovin71,Rogovin74,Schoen,LeeLev96}. These theories
employ the tunneling Hamiltonian approach and typically only keep the
terms of lowest non-vanishing order in the tunneling
amplitude. Poissonian shot noise was measured experimentally in
semiconductor diodes, see {\em e.g.} Ref. \cite{LiuVdZiel}; these
devices, however, could hardly be called mesoscopic, and it is not
always easy to separate various sources of noise. More recently, Birk,
de Jong, and Sch\"onenberger \cite{Birk} presented measurements of
noise in a tunnel barrier formed between an STM tip and a metallic
surface. Specifically addressing the crossover between the thermal and
shot noise, they found an excellent agreement with
Eq. (\ref{tunbar1}). Their experimental results are shown in
Fig. \ref{fig220}. 

\subsubsection{Quantum point contacts} \label{scat0262}

A point contact is usually defined as a constriction between two
metallic reservoirs. Experimentally, it is typically realized by
depleting of a two-dimensional electron gas formed with the help of a
number of gates. Changing the gate voltage $V_g$ leads to the
variation of the width of the channel, and consequently of the
electron concentration. All the sizes of the constriction are assumed
to be shorter than the mean free path due to any type of scattering,
and thus transport through the point contact is ballistic. In a {\em
quantum} point contact the width of the constriction is comparable to
the Fermi wavelength.
\begin{figure}
{\epsfxsize=6.cm\centerline{\epsfbox{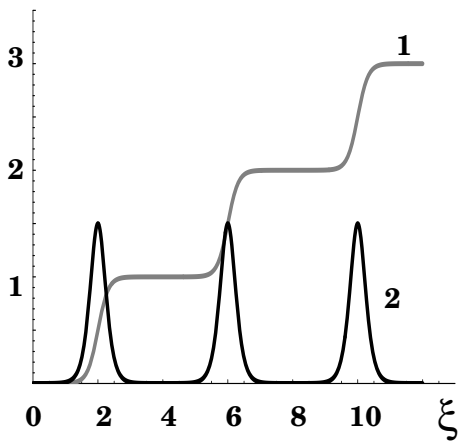}}}
\vspace{0.3cm}
\caption{Conductance in units of $e^2/2\pi\hbar$ (curve 1) and
zero-frequency shot noise power in units of $e^3 \vert V \vert
/6\pi\hbar$ (curve 2) for a quantum point contact with $\omega_y =
4\omega_x$ as a function of the gate voltage. Here $\xi = (E_F -
V_0)/\hbar\omega_x$ is a dimensionless energy.} 
\protect\label{qpt01}
\end{figure}

\begin{figure}
\epsfxsize=8.5cm{\centerline{\epsfbox{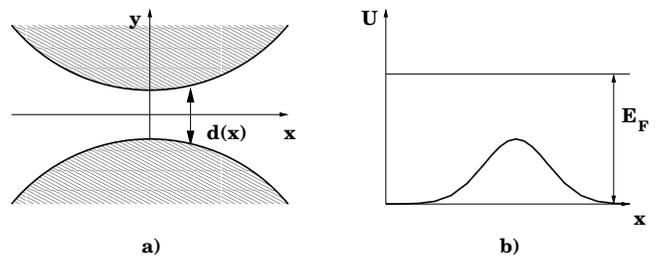}}}
\vspace{0.3cm}
\caption{Geometry of the quantum point contact in the hard-wall model
(a) and the effective potential for one-dimensional motion (b).}
\label{qpt02}
\end{figure}

Quantum point contacts have drawn wide attention after experimental
investigations \cite{qptexp1,qptexp2} showed steps in the dependence
of the conductance on the gate voltage. This stepwise dependence is
illustrated in Fig.~\protect\ref{qpt01}, curve 1. An explanation
was provided by Glazman {\em et al} \cite{GLKS}, who modeled the
quantum point contact as a ballistic channel between two infinitely
high potential walls (Fig.~\ref{qpt02}a). If the distance between the
walls $d(x)$ (width of the contact) is changing slowly in comparison
with the wavelength, transverse and longitudinal motion can be
approximately separated. The problem is then effectively reduced to
one-dimensional motion in the adiabatic potential $U(x) = \pi^2 n^2
\hbar^2 /2m d^2(x)$, which depends on the width profile and the 
the transverse channel number $n$. Changing the gate voltage leads to
the modification of the potential profile. Theoretically it is easier
to fix the geometry of the sample, {\em i.e} the form of the
potential, and vary the Fermi energy in the channel $E_F$
(Fig.~\ref{qpt02}b). The external potential is smooth, and therefore
may be treated semi-classically. This means that the channels with $n
< k_Fd_{min}/\pi$ (here $\hbar k_F \equiv (2mE_F)^{1/2}$, and
$d_{min}$ is the minimal width of the contact) are open and
transparent, $T_n = 1$, while the others are closed, $T_n = 0$. The
conductance (\ref{condbasic}) is proportional to the number of open
channels and therefore exhibits plateaus as a function of the gate
voltage. At the plateaus, shot noise is equal to zero, since all the
channels are either open or closed. The semi-classical description
fails when the Fermi energy lies close to the top of the potential in
one of the transverse channels. Then the transmission coefficient for
this channel increases from zero to one due to quantum tunneling
through the barrier and quantum reflection at the barrier. The
transition from one plateau to the next is associated with a spike in
the shot noise as we will now discuss.  

A more realistic description of the quantum point contact takes
into account that the potential in the transverse direction $y$ is
smooth \cite{Buttikerqpt}. The constriction can then be thought of as
a bottleneck with an electrostatic potential 
of the form of a saddle. Quite generally the potential
can be expanded in the directions away from the center of the
constriction, 
\begin{displaymath}
V(x,y) = V_0 - \frac{1}{2} m\omega_x^2 x^2 + \frac{1}{2} m\omega_y^2
y^2,
\end{displaymath}
where the constant $V_0$ denotes the potential at the saddle point. 
Experimentally it is a function of the gate voltage.
The transmission probabilities are given by Ref. \cite{Buttikerqpt},
\begin{eqnarray} \label{qpttrans}
T_n(E) & = & \left[1 + \exp \left( -\pi \epsilon_n \right)
\right]^{-1}, \nonumber \\ 
\epsilon_n & \equiv & 2\left[ E - \hbar\omega_y \left( n + \frac{1}{2}
\right) - V_0 \right] /\hbar\omega_x.
\end{eqnarray}
The transmission probability $T_n(E)$ exhibits a crossover from zero
to one as the energy $E$ passes the value $V_0 + \hbar\omega_y
(n+1/2)$. The resulting zero temperature shot noise as a function of
$V_{0}$, using Eq. (\ref{shot2term}), is illustrated in
Fig.~\ref{qpt01} for the case $\omega_y = 4\omega_x$ (curve 2). The
conductance of this quantum point contact is shown in Fig.~\ref{qpt01}
as curve 1. As expected, the shot noise dependence is a set of
identical spikes between the plateaus. The height of each spike is
$e^3 \vert V \vert/4\pi\hbar$ up to exponential accuracy. At the
plateaus shot noise is exponentially suppressed. This behavior of shot
noise in a quantum point contact was predicted by Lesovik
\cite{Lesovik89}. The shot noise of a saddle point model of a quantum
point contact was presented in Ref. \cite{dJBrev}. Scherbakov {\em et
al} \cite{Bogachek} thoroughly analyze and compare shot noise for
various models of quantum point contacts. Using a classical (master
equation) approach, shot noise suppression was also confirmed by Chen
and Ying \cite{ChenYing}.  

If a magnetic field is applied in the transverse direction, the
energies $\epsilon_n$ are pushed up. Shot noise is thus an oscillating
function of the magnetic field for a fixed gate voltage. Strong
magnetic fields may even drive the quantum point contact to the regime
$E_F < \epsilon_0$, suppressing the shot noise completely
\cite{Bogachek}.  

Using Eq. (\ref{noistwo1}), it is easy also to study shot noise in the
non-linear regime as the function of the applied bias voltage $V$. We
obtain 
\begin{equation} \label{noiseqpt2621}
S \approx \frac{e^2 \omega_x}{(2\pi)^2} N_V, \ \ \ N_V =
\frac{e \vert V \vert}{\hbar \omega_y},
\end{equation}  
where $N_V$ is the number of channels which open in the energy
interval between zero and $e\vert V \vert$. Eq. (\ref{noiseqpt2621})
applies when this number is large, $N_V \gg 1$. For even higher
voltages $e \vert V \vert > V_0$, the noise becomes voltage
independent, as found by Larkin and Reznikov \cite{Reznikov96}. They
also discuss self-consistent interactions and found that the
non-linear shot noise is {\em suppressed} as compared to the
non-interacting value.  

As the number of open channels becomes large, so that the width of the
constriction is much wider than the Fermi wavelength ({\em classical}
point contact), the shot noise stays the same, while the conductance
grows proportional to the number of channels. Thus, shot noise
becomes small in comparison with the Poisson value (at the top
of the $n$-th spike this suppression equals $(4n)^{-1}$), and in this
sense shot vanishes for a classical point contact, as found by Kulik 
and Omel'yanchuk \cite{Kulik84}. Note, however, that the shot noise
really disappears only when inelastic scattering becomes significant
(see subsection \ref{scat027}). 
\begin{figure}
\epsfxsize=7.cm{\centerline{\epsfbox{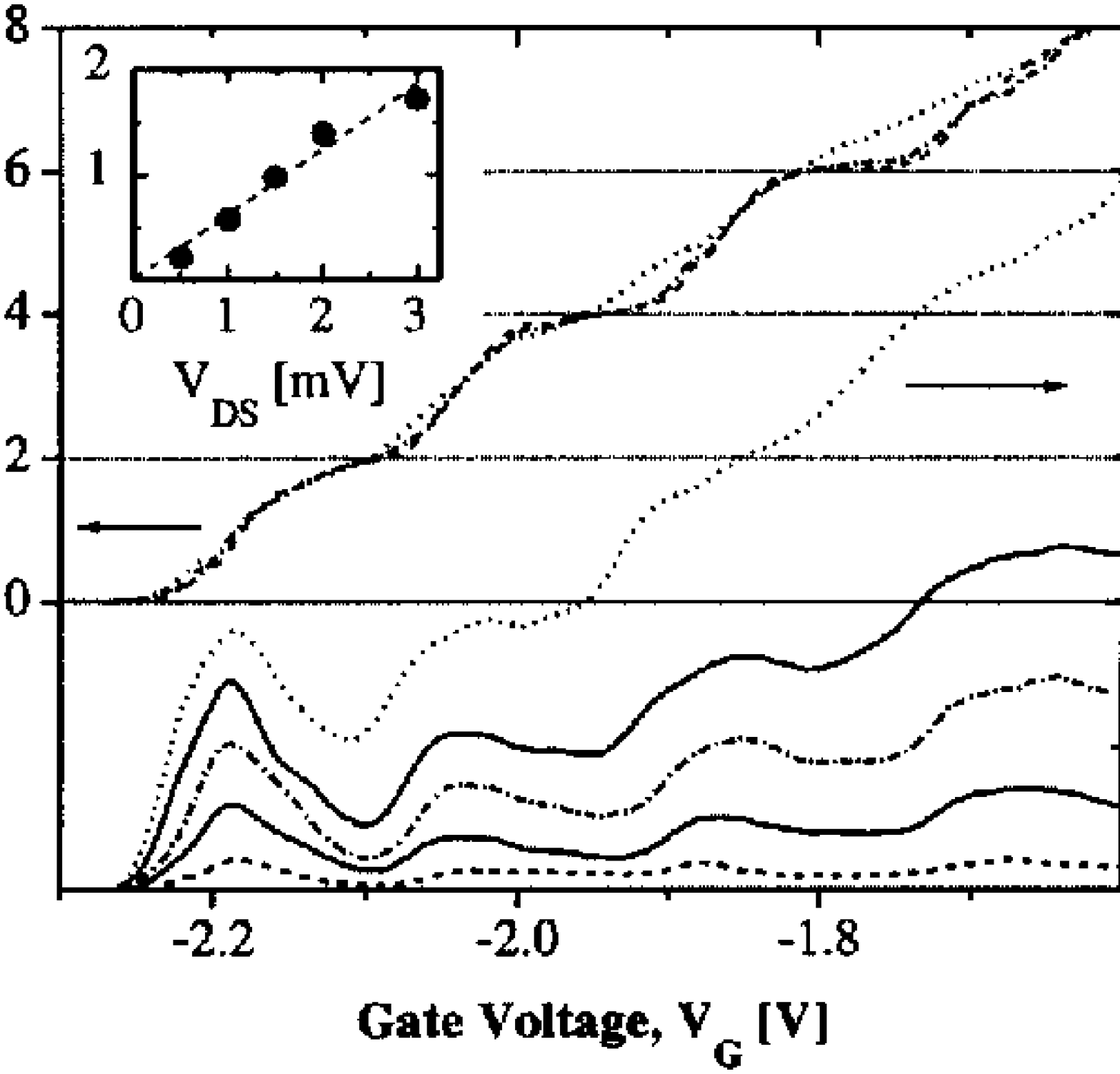}}}
\vspace{0.3cm}
\caption{Conductance (upper plot) and shot noise (lower plot) as
functions of the gate voltage, as measured by Reznikov {\em et al}
\protect\cite{Reznikov95}. Different curves correspond to five
different bias voltages.}
\label{qpt03}
\end{figure}

Experimentally, sub-Poissonian shot noise suppression was 
observed by Akimenko, Verkin, and Yanson \cite{Akimenko} in a
slightly different system, a metallic quantum point contact, which is
essentially an orifice in a thin insulating layer between two metallic
reservoirs. In this system, however, it is difficult to separate
different sources of noise. In ballistic quantum point contacts
sub-Poisson suppression was observed in an early experiment by Li {\em
et al} \cite{LiQPT}, and later by Dekker {\em et al}
\cite{Dekker91,Liefrink91}. Reznikov {\em et al} \cite{Reznikov95}
found clearly formed peaks in the shot noise as a function of the gate
voltage. A considerable improvement in the experimental technique
was obtained by measuring noise in the MHz range at frequencies
far above the range where $1/f$-noise contributes. The results of
Reznikov {\em et al} \cite{Reznikov95} are shown in
Fig.~\ref{qpt03}. Compared to the theory the experimental noise peaks
exhibit a slight asymmetry around the transition 
point. Kumar {\em et al} \cite{Kumar96} developed a different low
frequency technique based on voltage correlation measurements to
filter out unwanted noise. They found that ``{\em the agreement}
$\langle$of experimental results$\rangle$ {\em with theoretical
expectations, within the calculable statistical deviations, is nearly
perfect}''. Recently, van den Brom and van Ruitenbeek \cite{Brom}
demonstrated that shot noise measurements can be used to extract
information on the transmission probabilities of the eigen channels of
nanoscopic metallic point contacts. Subsequently, B\"urki and Stafford
\cite{Stafford99} were able to reproduce their results quantitatively
based on a simple theoretical model which takes into account only two
features of the contacts, the confinement of electrons and the
coherent backscattering from imperfections.  

\subsubsection{Resonant tunnel barriers} \label{scat0263}

\begin{figure}
\epsfxsize=8.cm{\centerline{\epsfbox{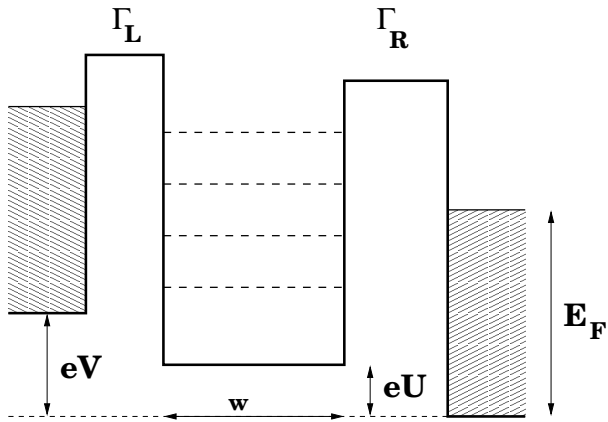}}}
\vspace{0.3cm}
\caption{Resonant double barriers. The case of low voltage is
illustrated; resonant levels inside the well are indicated by dashed
lines.}
\label{rtw1}
\end{figure}

The transport through two consecutive tunnel barriers allows already
to discuss many aspects of shot noise suppression. Let us first
consider the case of purely one-dimensional electron motion through
two potential barriers with transmission probabilities $T_L$ and
$T_R$, separated by a distance $w$, as shown in
Fig.~\ref{rtw1}. Eventually, we will assume that the transmission of
each barrier is low, $T_L \ll 1$ and $T_R \ll 1$. An exact expression
for the transmission coefficient of the whole structure is
\begin{eqnarray} \label{rtwtc1}
& & T(E) \\
& = & \frac{T_LT_R}{1 + (1-T_L)(1-T_R) -
2\sqrt{(1-T_L)(1-T_R)}\cos \phi(E)}, \nonumber 
\end{eqnarray}
with $\phi(E)$ being the phase accumulated during motion between the
barriers; in our particular case $\phi(E) =
2w(2mE)^{1/2}/\hbar$. Eq. (\ref{rtwtc1}) has a set of maxima at the
resonant energies $E_n^r$ such that the phase $\phi(E_n^r)$ equals
$2\pi n$. Expanding the function $\phi(E)$ around $E_n^r$, and
neglecting the energy dependence of the transmission coefficients, we
obtain the Breit-Wigner formula \cite{StoneLee85,ButtikerIBM}
\begin{equation} \label{breitw}
T(E) = T_n^{\max} \frac{\Gamma_{n}^{2}/4}{(E - E_n^r)^2 +
\Gamma_n^2/4}, \ \ \ T_n^{\max} =
\frac{4\Gamma_{Ln}\Gamma_{Rn}}{\Gamma_{n}^{2}}.  
\end{equation}
$T_n^{\max}$ is the maximal transmission probability at resonance. We
have introduced the partial decay widths $\Gamma_{Ln,Rn} = \hbar
\nu_{n} T_{L,R}$. The attempt frequency  $\nu_{n}$ of the $n$-th
resonant level is given by $\nu_{n}^{-1} = (\hbar/2)(d\phi/dE_n^r) =
w/v_n$, $v_n = (2E_n^r/m)^{1/2}$. $\Gamma_{n} \equiv \Gamma_{Ln} +
\Gamma_{Rn}$ is the total decay width of the resonant level. Eq. 
(\ref{breitw}) is, strictly speaking, only valid when the energy $E$
is close to one of the resonant energies\footnote{We assume that the
resonances are well separated, $\Gamma \ll \hbar^2/2mw^2$.}
$E_n^r$. In many situations, however, the Lorentz tails of $T(E)$ far
from the resonances are not important, and one can write 
\begin{displaymath}
T(E) = \sum_n T_n^{\max}\frac{\Gamma_{n}^2/4}{(E - E_n^r)^2 +
\Gamma_{n}^2/4}.
\end{displaymath}

Despite the fact that the transparencies of both barriers are
low, we see that the total transmission coefficient shows sharp peaks
around resonant energies. This effect is a consequence of constructive
interference  and is known as {\em resonant tunneling}.
The transmission coefficient at the top of a peak equals
$T_n^{\max}$; for a symmetric resonance $\Gamma_{L n} = \Gamma_{R n}$
the transmission is ideal, $T_n^{\max} = 1$. This dependence may be
probed\footnote{For discussion of experimental realizations, see
below.} by applying a gate voltage. The gate voltage moves the
positions of the resonant levels, and the conductance exhibits peaks
around each resonance. 

In the {\em linear regime} the shot noise is determined by the
transmission coefficient evaluated at the Fermi level, and is
thus an oscillating function of the gate voltage, vanishing almost 
completely between the peaks.  The Fano factor (\ref{Fano})
at the top of each peak is equal to $F = (\Gamma_{Ln} -
\Gamma_{Rn})^2/\Gamma_{n}^2$. It  vanishes for a symmetric barrier.
For a resonance with $T_n^{\max} > 1/2$ the Fano factor reaches a
maximum each time when the transmission probability passes through $T
= 1/2$; for a resonance with $T_n^{\max} <1/2$ the shot noise is
maximal at resonance. 

{\bf One-dimensional problem, non-linear regime}. For arbitrary
voltage, direct evaluation of the expressions (\ref{avcur1}) and
(\ref{noistwo1}) gives an average current,
\begin{equation} \label{conddbtnlin}
I = \frac{e}{\hbar} \sum_{n =1}^{N_{V}}
\frac{\Gamma_{Ln}\Gamma_{Rn}}{\Gamma_{n}},
\end{equation}
and a zero-temperature shot noise, \begin{equation}
\label{noisedbtnlin}
S = \frac{2e^2}{\hbar} \sum_{n =1}^{N_{V}}
\frac{\Gamma_{Ln}\Gamma_{Rn}(\Gamma_{Ln}^2 +
\Gamma_{Rn}^2)}{\Gamma_{n}^3}. 
\end{equation}
Here $N_V$ is the number of resonant levels in the energy strip
$e \vert V \vert$ between the chemical potentials of the left and
right reservoirs. Eqs. (\ref{conddbtnlin}) and (\ref{noisedbtnlin})
are only valid when this number is well defined -- the energy
difference between any resonant level and the chemical potential of
any reservoir must be much greater than $\Gamma$. Under this condition
both the current and the shot noise are independent of the applied
voltage. The  dependence of both the current and the shot noise on the
bias voltage $V$ is thus a set of plateaus, the height of each plateau
being proportional to the number of resonant levels through which
transmission is possible. Outside this regime, when one of the
resonant levels is close to the chemical potential of left and/or
right reservoir, a smooth transition with a width of order $\Gamma$
from one plateau to the next occurs.  

Consider now for a moment, a structure with a single resonance. If the
applied voltage is large enough, such that the resonance is between
the Fermi level of the source contact and that of the the sink
contact, the Fano factor is 
\begin{equation} \label{Fanodbt}
F = \frac{\Gamma_L^2 + \Gamma_R^2}{\Gamma^2}. 
\end{equation}
It varies between $1/2$ (symmetric barrier) and $1$ (very
asymmetric barrier). Expression (\ref{Fanodbt}) was 
obtained by Chen and Ting \cite{ChenTing91} using a nonequilibrium
Green's functions technique, and independently in Ref.
\cite{Buttiker91} using the scattering approach. It was confirmed in
Monte Carlo simulations performed by Reklaitis and Reggiani
\cite{Reklaitis97,Reklaitis971}. If the width of the resonance is
comparable to the applied voltage, Eq. (\ref{Fanodbt}) has to be
supplemented by correction terms due to the Lorentz tails of the
Breit-Wigner formula, as found in Ref. \cite{Buttiker91} and later by
Averin \cite{Averin93}. 

It is also worthwhile to point out that our
quantum-mechanical derivation assumes that the electron preserves full
quantum coherence during the tunneling process ({\em coherent
tunneling model}). Another limiting case occurs 
when the electron completely loses
phase coherence once it is inside the well ({\em sequential
tunneling model}). This latter situation can be described both
classically (usually, by means of a master equation) and
quantum-mechanically ({\em e.g.}, by connecting to the well one or
several fictitious voltage probes which serve as ``dephasing'' leads).
These issues are addressed in Section \ref{Langevin}, where we show
that the result for the Fano factor Eq. (\ref{Fanodbt}) remains
independent of whether we deal with a coherent process or a fully
incoherent process.  The Fano factor Eq. (\ref{Fanodbt}) is thus
insensitive to dephasing. 

{\bf Quantum wells}. The double-barrier problem is also relevant
for {\em quantum wells}, which are two- or three-dimensional
structures\footnote{For simplicity, we use a three-dimensional
notation. Specialization to two dimensions is trivial.},
consisting of two planar (linear in two dimensions) potential
barriers. Of interest is transport in the direction perpendicular to the
barriers (across the quantum well, axis $z$). These
systems have drawn attention already in the seventies, 
when resonant tunneling was investigated both theoretically
\cite{TsuEsaki} and experimentally \cite{Esaki1}.

If the area of the barriers (in the plane $xy$) ${\cal A}$ is very
large, the summation over the transverse channels in Eqs.
(\ref{avcur1}), (\ref{noistwo1}) can be replaced by
integration, and we obtain for the average current
\begin{eqnarray*}
I & = & \frac{e\nu_2{\cal A}}{2\pi\hbar} \int_{0}^{\infty} dE_{\perp}
dE_z T(E_z) \\
& \times & \left\{ f_L\left(E_z + E_{\perp} \right) - f_R\left(E_z +
E_{\perp} \right)\right\} 
\end{eqnarray*}
and the shot noise
\begin{eqnarray*}
S & = & \frac{e^2\nu_2{\cal A}}{\pi\hbar} \int_{0}^{\infty} dE_{\perp}
dE_z T(E_z) \left[ 1 - T(E_z)\right] \\
& \times & \left\{ f_L\left(E_z + E_{\perp} \right) - f_R\left(E_z +
E_{\perp} \right)\right\}, 
\end{eqnarray*}
with $\nu_2 = m/2\pi\hbar^2$ the density of states of the
two-dimensional electron gas (per spin). The key point is that the
transmission coefficient depends  only on the energy of the
longitudinal\footnote{Along the current, {\em not} along the well.}
motion $E_z$, and thus is given by the solution of the one-dimensional
double-barrier problem, discussed above. Denoting $\mu_L = E_F+eV$,
$\mu_R = E_F$, and integrating over $dE_{\perp}$, we write (the
temperature is set to zero)  
\begin{eqnarray} \label{qwc00}
I & = & \frac{e\nu_2{\cal A}}{2\pi\hbar} \left\{ eV \int_{0}^{E_F}
dE_z T(E_z) \right. \nonumber \\
& + & \left. \int_{E_F}^{E_F+eV} dE_z \left(E_F + eV - E_z
\right) T(E_z) \right\}
\end{eqnarray}
and
\begin{eqnarray} \label{qwn00}
S & = & \frac{e^2\nu_2{\cal A}}{\pi\hbar} \left\{ eV \int_{0}^{E_F}
dE_z T(E_z) \left[ 1 - T(E_z) \right] \right. \\
& + & \left. \int_{E_F}^{E_F+eV} dE_z
\left(E_F + eV - E_z \right) T(E_z) \left[ 1 - T(E_z)
\right] \right\}. \nonumber
\end{eqnarray}

Expressions (\ref{qwc00}) and (\ref{qwn00}) are valid
in the linear and non-linear regimes, provided interactions are not
important. We will consider the noise in the non-linear regime in
Section \ref{Langevin}, where it will be shown that effects of
charging of the well may play an important role. Here, specializing on
the regime {\em linear in the bias voltage $V$}, we obtain 
a current $\langle I \rangle = G V$ determined by the conductance
\begin{equation} \label{condqwlin}
G = \frac{e^2\nu_2{\cal A} N_F}{\hbar}
\frac{\Gamma_L\Gamma_R}{\Gamma}, 
\end{equation}
and a shot noise power
\begin{equation} \label{noiseqwlin}
S = 2e 
\frac{\left(\Gamma_L^2 +
\Gamma_R^{2}\right)}{\Gamma^2} \langle I \rangle.
\end{equation}
Here $N_F$ is the number of resonant states in the one-dimensional
problem, which lie below $E_F$. Eqs. (\ref{condqwlin}) and
(\ref{noiseqwlin}) are valid only when the distance between all 
resonant levels and the Fermi level is much greater that
$\Gamma$. This dependence may be probed again, like in a
one-dimensional structure, with the help of a gate. Both the current
and the shot noise exhibit plateaus as a function of gate voltage;
these plateaus are smoothly joined over a width of order
$\Gamma$. This dependence resembles that of the non-linear
one-dimensional regime, but it clearly is a consequence 
of different physics. Nevertheless, {\em in the plateau
regime} the Fano factor is the same, Eq. (\ref{Fanodbt}). This
fact was noted by Davis {\em et al} \cite{Davies92},
who presented both quantum and classical derivations of this
result. Classical theories of shot noise suppression in quantum
wells are discussed in Section \ref{Langevin}.

{\bf Averaging}. Another point of view was taken by Melsen and
Beenakker \cite{Melsen94} and independently by Melnikov
\cite{Melnikov}, who investigated not a single resonant tunneling
structure but an  ensemble of systems. Imagine an ensemble of
quasi-one-dimensional double-barrier systems, in which some parameter
is random. For definiteness, we assume that the systems are subject to
a random gate voltage. Then in some of them the Fermi level is close
to one of the resonant energies, and in others it lies between two
resonant levels. Therefore, on average, shot noise must be finite even
in the linear regime. To quantify this argument, we turn to the exact
expression (\ref{rtwtc1}) for the transmission
coefficient\footnote{The Breit-Wigner formula (\ref{breitw}) cannot 
be used for this purpose, since it is not exact far from the
resonance.} and assume that the phase $\phi$ is a random variable,
uniformly distributed on the interval $(0, 2\pi)$. We only consider
the linear regime.

First, we calculate the conductance,
\begin{equation} \label{condbeen}
G = \frac{e^2}{2\pi\hbar} \sum_n T_n = \frac{e^2
N}{(2\pi)^2\hbar} \int_0^{2\pi} T(\phi) d\phi = \frac{e^2
N}{2\pi\hbar} \frac{T_L T_R}{T_L + T_R},
\end{equation}
where $N$ is the number of transverse channels, and we have
taken into account $T_L \ll 1$, $T_R \ll 1$. The same
calculation for the shot noise yields \cite{Melnikov,dJBrev}
\begin{eqnarray} \label{noisbeen}
S & = & \frac{e^3 \vert V \vert N}{2\pi^2\hbar} \int_0^{2\pi} T(\phi)
\left[ 1 - T(\phi) \right] d\phi \nonumber \\
& = & \frac{e^3 \vert V \vert N}{\pi\hbar} \frac{T_L T_R \left( T_L^2
+ T_R^2 \right)}{\left( T_L + T_R \right)^3}. 
\end{eqnarray}
The Fano factor is given again by Eq. (\ref{Fanodbt}).

We can learn two lessons from this simple model. First, the Fano
factor (\ref{Fanodbt}) $F = (\Gamma_L^2 + \Gamma_R^2)/\Gamma^2$
appears each time when there is some kind of averaging
in the system which involves one-dimensional motion
across two barriers. In the two examples we considered, the
one-dimensional non-linear problem, and the quantum well in the linear
regime, this averaging is provided by the summation over all the
levels between the chemical potentials of the two reservoirs (one
dimension), or the summation over the transverse channels at the given
total energy (quantum well). Thus, both problems prove to be
self-averaging. At the same time, this averaging is absent in the
one-dimensional linear problem, and the Fano factor has nothing to do
with Eq. (\ref{Fanodbt}) even between the resonances.

The second lesson is provided by the distribution function of the 
transmission coefficients of the eigen channels 
in the one-dimensional problem \cite{Melsen94,Melnikov}. 
Without giving details, we mention only that it 
has a {\em bimodal} form. The transmission
coefficients assume values between $T_{\min} = T_LT_R/4 \ll 1$ and
$T_{\max}$; those close to $T_{\min}$ and $T_{\max}$ have higher
probability than those lying in between. The Fano factor is very
sensitive to the appearance of transmission coefficients close to one,
since it is these values which cause the sub-Poisson
suppression. Thus, for a symmetric barrier $T_{\max} = 1$, and the
probability to find the transmission coefficient close to $1$ is
high. This yields the lowest possible Fano factor $1/2$
which is possible in this situation. We will return in more detail 
to the distribution of transmission probabilities for
metallic diffusive conductors and chaotic cavities.

{\bf Related work}. Here we mention briefly additional 
theoretical results on noise in double-barrier and similar
structures. 

Runge \cite{Runge} investigates  noise in double-barrier quantum
wells, allowing for elastic scattering inside the well. He employs
a non-equilibrium Green's function technique and a coherent potential
approximation, and arrives at rather cumbersome expressions for the
average current and noise power. In the limit of zero temperature,
however, his results yield the same Fano factor (\ref{Fanodbt}),
despite the fact that both current and noise are sensitive to impurity
scattering.

Lund B\o \ and Galperin \cite{Bo,Bo1} consider a resonant quantum
well in a strong magnetic field perpendicular to the interfaces (along
the axis $z$). They find that the shot noise power (in the linear and
nonlinear regimes, but without charging effects taken into account)
shows peaks each time when the new Landau level in the well crosses
the chemical potential in the right reservoir.

Xiong \cite{Xiong} analyzes noise in superlattices of finite
size (several consecutive barriers) using the transfer matrix
method. His numerical results clearly show shot noise suppression with
respect to the Poisson value, but, unfortunately, the Fano factor
is not plotted.

{\bf Resonant tunneling through localized states}. The following
problem was discussed by Nazarov and Struben \cite{Struben}. Consider
now non-linear transport through {\em one}, one-dimensional, symmetric
barrier, situated in the region $-w/2 < z < w/2$. We assume that there
are resonant states which are randomly distributed {\em inside} the
barrier and strongly localized. Applying the model suggested for this
situation by Larkin and Matveev \cite{Larkin87}, we assume that these
resonant states are provided by impurities inside the barrier; the
localization radius of each state is denoted by $\xi$, $\xi \ll
w$. Transition rates are exponentially sensitive to the position of
these impurities inside the barrier. In the regime of low impurity
concentration, only those situated close to the center of the barrier
contribute to the transport properties. Thus, the problem is
effectively mapped onto a double-barrier problem, where the impurity
region near the center of the barrier serves as a potential well, and
is separated by two ``barriers'' from the left and right
reservoirs. The tunneling rates through these ``barriers'' to a
resonant state depend on the position $z$ of the impurity which
provides this resonant state. We have \cite{Larkin87}
\begin{displaymath}
\Gamma_{L,R} (z) = \Gamma_0 \exp\left[ \pm z/\xi \right], \ \ \ \vert
z \vert < w/2.
\end{displaymath}
We have assumed that this amplitude is energy independent, and
that impurities are uniformly distributed in energy and space. Thus,
our expressions (\ref{conddbtnlin}) and (\ref{noisedbtnlin}) hold, and
must be averaged over impurity configurations. We write
\begin{displaymath}
I = \frac{e}{\hbar} n_0 \Gamma_0 \int_{-\infty}^{\infty}
\frac{dz}{2\cosh (z/\xi)}
\end{displaymath}
and
\begin{displaymath}
S = \frac{2e^2}{\hbar} n_0 \Gamma_0 \int_{-\infty}^{\infty}
dz \frac{\cosh(2z/\xi)}{4\cosh^3 (z/\xi)},
\end{displaymath}
where $n_0$ is the spatial concentration of impurities. By extending
the integration to infinity, we have taken into account $\xi \ll w$.
Performing the average and calculating the Fano factor, we find
\cite{Struben}
\begin{displaymath}
F = 3/4,
\end{displaymath}
which is markedly different form the usual double-barrier suppression
$F = 1/2$. 

The model can be generalized\footnote{This example concerns
interacting systems, and is included in this Section only as an
exception.} to include Coulomb correlations \cite{Glazman88}. Imagine
that each resonant center has two degenerate electron states available
for tunneling, corresponding to two different spin states. However, if
one of the states is filled, the other one is shifted up by the
Coulomb energy $U$. We assume that the Coulomb energy is very large,
so that once one electron has tunneled, the tunneling of the second
one is suppressed. Then the effective tunneling rate through the
``left'' barrier is $2\Gamma_L$ (we assume that voltage is applied
from left to the right), and instead of Eqs. (\ref{conddbtnlin}) and
(\ref{noisedbtnlin}) we write for the current \cite{Glazman88} and
shot noise power per spin \cite{Struben}  
\begin{displaymath}
I = \frac{eN_V}{\hbar} \frac{\Gamma_L\Gamma_R}{2\Gamma_L + \Gamma_R},
\ \ \ S = \frac{2e^2N_V}{\hbar} \frac{\Gamma_L\Gamma_R (4\Gamma_L^2 +
\Gamma_R^2)}{(2\Gamma_L + \Gamma_R)^3}. 
\end{displaymath}
Averaging over impurity configurations and calculating the Fano
factor, we find again $F = 3/4$ \cite{Struben}. 

{\bf Experiments}. The simplest experimental system one can imagine
which should exhibit the features of a two-barrier structure is just
a one-dimensional channel constrained by two potential barriers. If
the barriers are close to each other, the region between the 
two barriers can be considered as a zero-dimensional system and is
called a {\em quantum dot}. In addition, one usually places one more
electrode ({\em gate}), which couples only capacitively to the
dot. Roughly speaking, the voltage applied to the gate shifts all
electron levels in the dot with respect to the chemical potential of
the reservoirs, and may tune them to the resonance position. However,
typically quantum dots are so small that Coulomb interaction effects
(Coulomb blockade) become important, and the theoretical picture
described above is no longer valid. If the space between the barriers
is large and one-dimensional (one channel), interaction effects are
also important, and a Luttinger liquid state is formed. For a more
extensive discussion of noise in interacting systems, the reader is
addressed to Section \ref{strcorr}.  
\begin{figure}
\epsfxsize=8.cm{\centerline{\epsfbox{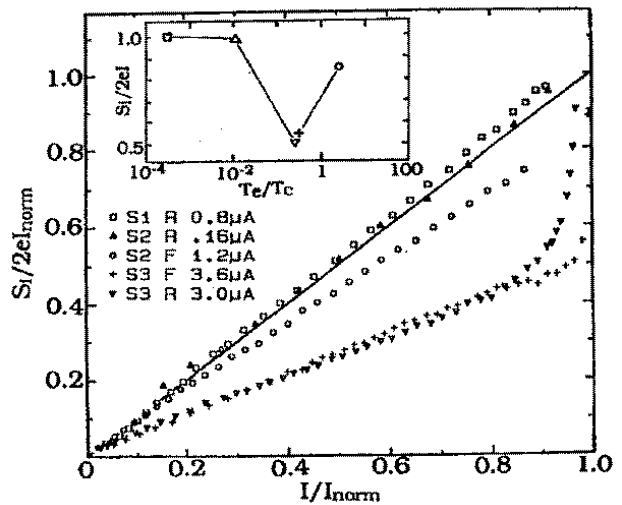}}}
\vspace{0.3cm}
\caption{The Fano factor observed experimentally by Li {\em et al}
\protect\cite{LiRTQW} as a function of current for three quantum
wells, which differ by their asymmetry. The solid line represents the
Poisson shot noise value.}   
\label{rtw2}
\end{figure}

Quantum wells, however, are macroscopic objects, and hence are less
sensitive to interactions. Thus, experiments carried out on quantum
wells may probe the non-interacting theory of noise suppression 
in a double-barrier system. Sub-Poissonian shot noise suppression 
in quantum wells was observed by Li {\em et al} \cite{LiRTQW} even
before a theory of this suppression was available. Li {\em et al}
noted that the suppression is maximal for symmetric barriers, and is
insignificant for very asymmetric structures (Fig.~\ref{rtw2}). This
suppression was later observed by van de Roer {\em et al} \cite{Roer},
Ciambrone {\em et al} \cite{Ciambrone}, Liu {\em et al} \cite{Liu95},
and Przadka {\em et al} \cite{Przadka}. Liu {\em et al} compared their
experimental data with the results of numerical simulations attempting
to take into account specific features of their sample, and found
that theory and experiment are in a reasonable agreement. Yau {\em et
al} \cite{Yau} observed shot noise suppression in double quantum wells
(triple barrier structures). We should note, however, that in all
experimental data available, the Fano factor depends considerably on
the applied voltage in the whole range of voltages. Apparently, this
happens because already relatively low voltages drive the system out
of the linear regime. To the best of our knowledge, this issue has not
been addressed systematically, although some results, especially
concerning the negative differential resistance range, exist. They are
summarized in Section \ref{Langevin}. 

\subsubsection{Metallic diffusive wires} \label{scat0264}

{\bf $1/3$-suppression}. We consider now transport in multi-channel
diffusive wires in the metallic regime. This means that, on one hand,
the length of the wire $L$ is much longer than the mean free path $l$
due to disorder. On the other hand, in a quasi-one-dimensional
geometry all electron states are localized in the presence of
arbitrarily weak disorder; the localization length equals $L_{\xi} =
N_{\perp} l$, where $N_{\perp}$ is the number of transverse
channels\footnote{We use below two-dimensional notations: for a strip
of width $w$ the number of transverse channels is equal to $N_{\perp}
= p_Fw/\pi\hbar$. All results expressed through $N_{\perp}$ remain
valid also for a three-dimensional (wire) geometry.}. Thus for a wire
to be metallic we must have $L \ll L_{\xi}$ (which of course implies
$N_{\perp} \gg 1$). As everywhere so far, we ignore inelastic
processes. 

Comparison between the Drude-Sommerfeld formula for
conductance\footnote{The factor $2/\pi$ which might look 
unusual to some readers only reflects a different definition of the
mean free path, and is not essential for any results which we describe
below.}, 
\begin{equation} \label{Drude}
G = \frac{2}{\pi} \frac{e^2 n\tau}{m} \frac{w}{L},
\end{equation}
($n$ is the electron concentration, and $\tau = l/v_F$ is the momentum
relaxation time), and the Landauer formula (\ref{condbasic}) yields
the expression for the {\em average transmission coefficient},
\begin{equation} \label{Taver}
\langle T \rangle = \frac{l}{L}.
\end{equation}
In the diffusive regime we have $\langle T \rangle \ll 1$.

A naive point of view would be to assume that {\em all} the
transmission coefficients of the wire are of the order of 
the average transmission eigenvalue $\langle T
\rangle$ and thus, that all transmission probabilities are small. From
our previous consideration it would then follow that the Fano factor
is very close to one: a metallic diffusive wire would exhibit full
Poissonian shot noise. On the other hand, it is well known that a
macroscopic metallic conductor exhibits no shot noise. Using this
information as a guide one might equally naively expect that a
mesoscopic metallic diffusive conductor also exhibits no shot noise. 

\begin{figure}
\epsfxsize=7.cm{\centerline{\epsfbox{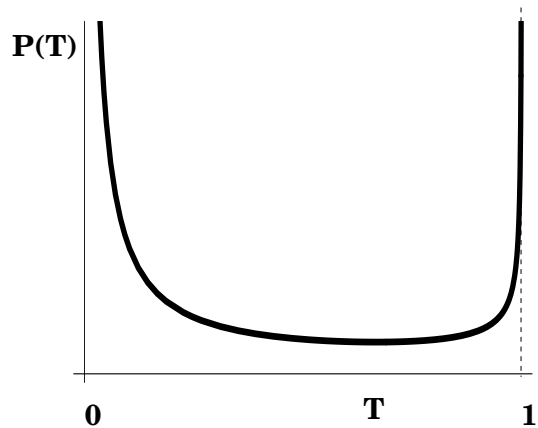}}}
\vspace{0.3cm}
\caption{Distribution function of transmission coefficients
(\ref{Tdistrib}) for $L/l = 10$.}
\label{diffus1}
\end{figure}

In fact, these naive assumptions are incorrect. In particular, the
fact that the transmission eigenvalues of a metallic conductor are not
all small has long been recognized\footnote{This 
statement has a long history, and we only cite two early papers on the
subject by Dorokhov \cite{Dorokhov} and Imry \cite{Imry86}. For a
modern discussion we refer to Ref. \cite{BeenRMP}.}: In the metallic
regime, for any energy open channels (with $T \sim 1$) coexist with
closed ones ($T \ll 1$). The distribution function of transmission
coefficients has in fact a bimodal form. This bimodal distribution
leads to sub-Poissonian shot noise. Quantitatively, this situation can
be described by random matrix theory of one-dimensional transport. It
implies \cite{SMMP} that the channel-dependent inverse localization
lengths $\zeta_n$, related to the transmission coefficients by $T_n =
\cosh^{-2} (L/\zeta_n)$, are uniformly distributed between $0$ and
$l^{-1}$. This statement can be transformed into the following
expression for the distribution function of transmission coefficients,
\begin{eqnarray} \label{Tdistrib}
P(T) = \frac{l}{2L} \frac{1}{T\sqrt{1-T}}, \ \ \ & & T_{\min} < T < 1,
\nonumber \\
T_{\min} & = & 4\exp(-2L/l),
\end{eqnarray}
and $P(T) = 0$ otherwise. As discussed, it has a bimodal form: almost
open and almost close channels are preferred. The dependence $P(T)$ is
illustrated in Fig.~\ref{diffus1}.

The distribution function $P(T)$ must be used now to average
expressions (\ref{condbasic}) and (\ref{shot2term}) over impurity
configurations. Direct calculation confirms Eq. (\ref{Taver}), and,
thus, the distribution function (\ref{Tdistrib}) yields the
Drude-Sommerfeld formula (\ref{Drude}) for the average
conductance. Furthermore, we obtain
\begin{displaymath}
\langle T (1-T) \rangle = \frac{l}{3L},
\end{displaymath}
which implies that the zero-temperature shot noise power is
\begin{equation} \label{shotdiffus}
S = \frac{e^3 \vert V \vert}{3\pi\hbar} \frac{N_{\perp}l}{L} =
\frac{1}{3} S_P.
\end{equation}
The shot noise suppression factor for metallic diffusive wires is
equal to $F = 1/3$. The remarkable feature is that this result is {\em
universal}: As long as the geometry of the wire is
quasi-one-dimensional and $l \ll L \ll L_{\xi}$ (metallic diffusive
regime), the Fano factor does not depend on the degree of
disorder\footnote{It seems that the question whether the Fano factor
depends on the {\em type} of disorder has never been addressed. In all
cases disorder is assumed to be Gaussian white noise, {\em i.e.}
$\langle U(\bbox{r}) U(\bbox{r'}) \rangle \propto \delta
(\bbox{r} - \bbox{r'})$.}, the number of transverse channels,
and any other individual features of the sample. This result was first
obtained by Beenakker and one of the authors \cite{Been92} using the
approach described above. Independently, Nagaev \cite{Nagaev92}
derived the same suppression factor $1/3$ by using a 
classical theory based on a Boltzmann equation with Langevin
sources. This theory and subsequent developments are described in
Section \ref{BoltzmannL}. 

Later on, the $1/3$ suppression of shot noise became a subject of a
number of microscopic derivations. Altshuler, Levitov, and Yakovets
\cite{Yakovets} recovered the Fano factor $1/3$ by direct microscopic
calculation using the Green's function technique. Nazarov
\cite{Nazarov94} proved that the distribution (\ref{Tdistrib}) holds
for an arbitrary (not necessarily quasi-one-dimensional) geometry;
thus, the $1/3$-suppression is ``super-universal''. He used a slightly
different technique, expressing scattering matrices through Green's
functions and then performing disorder averages. The same technique,
in more elaborated form, was used in Ref. \cite{Blanter97}, which also
obtains the $1/3$-suppression. 

It is clear that the quantum-mechanical theories of Refs.
\cite{Been92,Yakovets,Nazarov94,Blanter97} are equivalent for
the quasi-one-dimensional geometry, since they deal with disorder
averages basically in the same way. On the other hand, their
equivalence to the classical consideration of Ref. \cite{Nagaev92} is
less obvious.  

\begin{figure}
\epsfxsize=5.cm{\centerline{\epsfbox{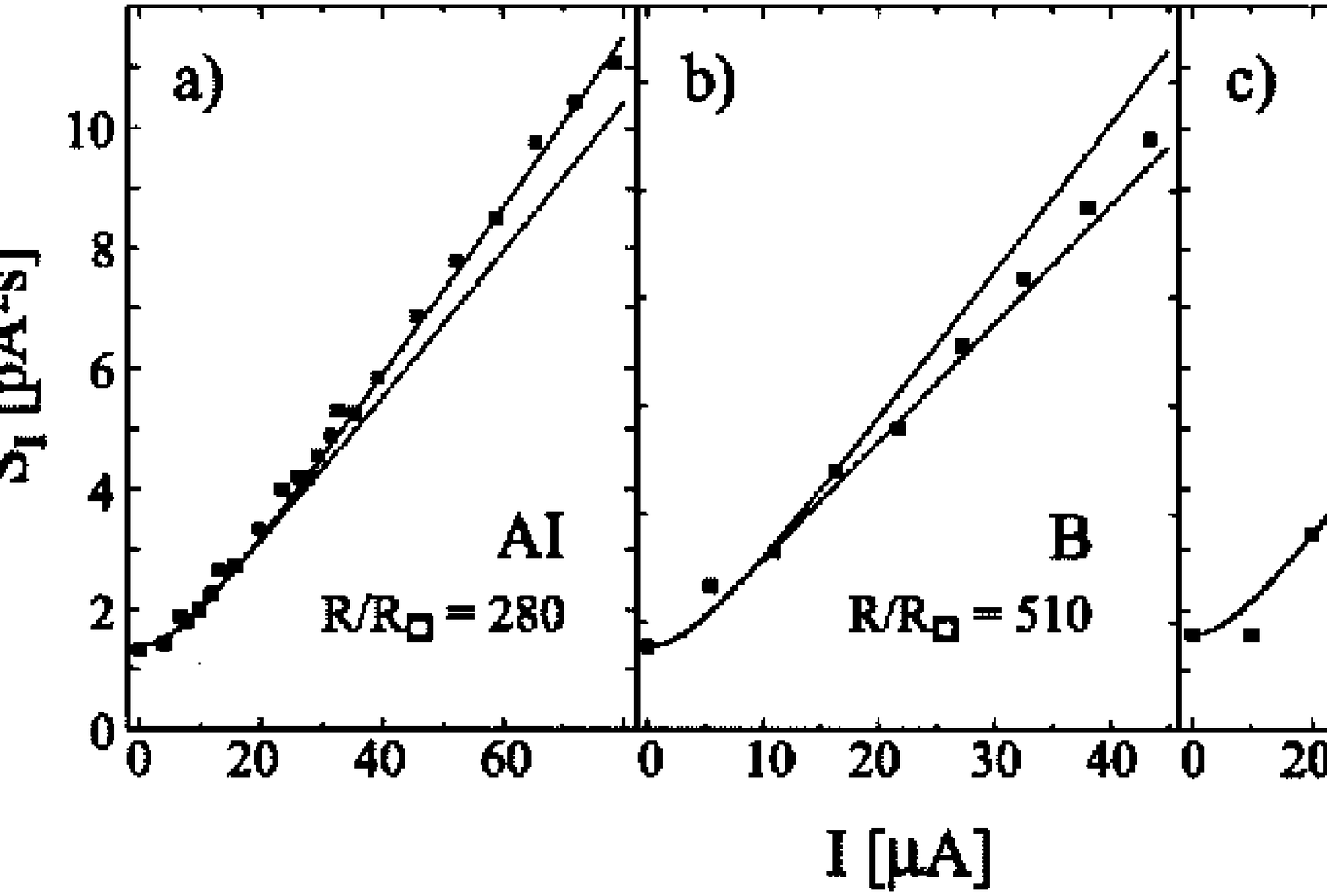}}}
\vspace{0.3cm}
\caption{Shot noise measurements by Henny {\em et al}
\protect\cite{Henny} on three different samples. The lower solid line
is $1/3$--suppression, the upper line is the hot-electron result $F =
\sqrt{3}/4$ (see Section \protect\ref{BoltzmannL}). The samples (b)
and (c) are short, and clearly display $1/3$--suppression. The sample
(a) is longer (has lower resistance), and the shot noise deviates from
the non-interacting suppression value due to inelastic processes.}
\label{diffus2}
\end{figure}

Experimentally, shot noise in metallic diffusive wires was
investigated by Liefrink {\em et al} \cite{Liefrink94}, who observed
that it is suppressed with respect to the Poisson value. The
suppression  factor in this experiment lies between $0.2$ and $0.4$
(depending on gate voltage). More precise experiments were performed
by Steinbach, Martinis, and Devoret \cite{Steinbach} who analyzed
silver wires of different length. In the shortest wires examined they
found a shot noise slightly larger than $1/3$ and explained this
larger value as due to electron-electron interaction\footnote{In long
wires, interaction effects play a role; this is addressed in Section
\ref{BoltzmannL}.}. A very accurate measurement of the $1/3$ noise
suppression was performed by Henny {\em et al} \cite{Henny}. Special
care was taken to avoid electron heating effects by attaching very
large reservoirs to the wire. Their results are displayed in
Fig.~\ref{diffus2}.   

{\bf Localized regime}. In {\em quasi-one-dimensional} wires with
length $L \gg L_{\xi}$ the transmission coefficients are
``crystallized'' around exponentially small values \cite{BeenRMP}. 
This leads to a conductance and a shot noise power which decay
exponentially with the length $L$. Shot noise is not suppressed with
respect to the Poissonian value: $F = 1$.  

The shot noise in the {\em one-dimensional} case was analyzed by
Melnikov \cite{Melnikov}, who obtained different suppression factors
for various models of disorder in a one-channel wires. In particular,
the model of Gaussian delta-correlated one-dimensional disorder leads
to the suppression factor $3/4$. 

{\bf Weak localization and mesoscopic fluctuations}. In the metallic
regime, quantum interference effects due to disorder, which eventually
drive the system into the localized regime, manifest themselves in the
form of weak localization corrections. For the shot noise, the weak
localization correction was studied by de Jong and Beenakker
\cite{Jong92}, and later by Mac\^edo \cite{Macedo1}, Mac\^edo and
Chalker \cite{Chalker}, and Mac\^edo \cite{Macedo2}. They found 
\begin{equation}
\label{wldiffus} S =
\frac{e^3\vert V \vert}{\pi\hbar} \left[ \frac{N_{\perp}l}{3L} -
\frac{1}{45} \right].
\end{equation}
Comparison with a similar expression for the conductance,
\begin{displaymath}
G = \frac{e^2}{2\pi\hbar} \left[ \frac{N_{\perp}l}{L} - \frac{1}{3}
\right],
\end{displaymath}
yields the Fano factor
\begin{equation} \label{Fanowl}
F = \frac{1}{3} + \frac{4}{45} \frac{L}{N_{\perp}l}.
\end{equation}
The second terms represent weak localization corrections ($L \ll
N_{\perp}l$). These expressions are valid for the case of preserved
time-reversal symmetry (orthogonal symmetry). In the case of broken
time-reversal symmetry (unitary symmetry; technically, this means that
a weak magnetic field is applied) weak localization corrections are
absent and the Fano factor stays at $1/3$. Thus, we see that weak
localization effects {\em suppress} noise, but {\em enhance} the Fano
factor, in agreement with the general expectation that it lies above
$1/3$ in the localized regime. The crossover from the metallic to the
localized regime for shot noise has not been investigated.  

De Jong and Beenakker \cite{Jong92}, Mac\^edo \cite{Macedo1}, and
Mac\^edo and Chalker \cite{Chalker} studied also mesoscopic
fluctuations of shot noise\footnote{This requires the knowledge of the
joint distribution function of two transmission eigenvalues.}, which
are an analog of the universal conductance fluctuations. 
For the root mean square of the shot noise power, they found
\begin{displaymath}
{\rm r.m.s.}\ S = \frac{e^3 \vert V \vert}{\pi\hbar}
\sqrt{\frac{46}{2835 \beta}}, 
\end{displaymath}
where the parameter $\beta$ equals $1$ and $2$ for the orthogonal and
unitary symmetry, respectively. These fluctuations are independent of 
the number of transverse channels, length of the wire, or degree of
disorder, and may be called \cite{Jong92} "universal noise
fluctuations".

The picture which emerges is, therefore, that like the
conductance, the ensemble averaged shot noise is a classical
quantity. Quantum effects in the shot noise manifest themselves only
if we include weak localization effects or if we ask about
fluctuations away from the average. With these results it is thus no
longer surprising that the $1/3$ noise suppression factor derived
quantum mechanically and from a classical Boltzmann equation for the
fluctuating distribution are in fact the same. The same picture holds
of course not only for metallic diffusive wires but whenever we
ensemble average. We have already discussed this for the resonant
double barrier and below will learn this very same lesson again for
chaotic cavities.  

{\bf Chiral symmetry}. Mudry, Brouwer, and Furusaki
\cite{Mudry1,Mudry2} studied the transport properties of disordered
wires with chiral symmetry ({\em i.e.} when the system consists of two
or several sublattices, and only transitions between different
sublattices are allowed). Examples of these models include
tight-binding hopping models with disorder or the random magnetic flux
problem. Chiral models exhibit properties usually different
from those of standard disordered wires\footnote{Extensive list of
references is provided by Ref. \cite{Mudry1}.}, for instance, the
conductance at the band center scales not exponentially with the
length of the wire $L$, but rather as a power law
\cite{Gade}. Ref. \cite{Mudry2}, however, finds that in the diffusive
regime ($l \ll L \ll N_{\perp} l$) the Fano factor equals precisely
$1/3$, like for ordinary symmetry. The only feature which appears due
to the chiral symmetry is the absence of weak localization
corrections in the zero order in $N_{\perp}$. Weak localization
corrections, both for conductance and shot noise, scale as
$L/(lN_{\perp})$ and discriminate between chiral unitary and chiral
orthogonal symmetries. 

{\bf Transition to the ballistic regime}. In the zero-temperature
limit, a perfect wire does not exhibit shot noise. For this reason,
one should anticipate that in the 
ballistic regime, $l > L$, shot noise is suppressed below
$1/3$. The crossover between metallic and ballistic regimes in
disordered wires was studied by de Jong and Beenakker \cite{Jong92}
(see their Eq. (A10)), who found for the noise suppression factor
\begin{equation} \label{fanotransball}
F = \frac{1}{3} \left( 1 - \frac{1}{(1 + L/l)^3} \right).
\end{equation}
It, indeed, interpolates between $F=1/3$ for $l \ll L$ and $F = 0$
for $L \gg l$. Later, they \cite{Jong95,Jong96} illustrated
Eq. (\ref{fanotransball}) by using the classical (Boltzmann-Langevin)
approach for single-channel wires\footnote{This must be considered as
a toy model, since the classical theory ignores localization effects.
Single-channel wires are in reality either ballistic or localized, but
never metallic.}. Liu, Eastman, and Yamamoto \cite{Eastman} performed
Monte Carlo simulations of shot noise for the same situation, and
found agreement with Eq. (\ref{fanotransball}).

Nazarov \cite{Nazarov94} and, independently, Beenakker and Melsen
\cite{MelsenPRB94} addressed the metallic -- ballistic crossover in a
disordered quantum point contact, {\em i.e.} a constriction between
two quasi-one-dimensional metallic diffusive conductors (of identical
width). The diffusive conductors have a mean free path $l$, 
a combined length $L$ and a total resistance  $R_N$.
The constriction has a resistance $R_T = (h/e^{2})N_0$
in the presence of $N_0$ open channels. 
For this system the Fano factor 
is a function of the ratio $\gamma = N_0L/(lN) = R_N/R_T$,
and is given by 
\begin{equation} \label{Fanomelsen}
F = \frac{1}{3} \left( 1 - \frac{1}{(1 + \gamma)^3} \right).
\end{equation}
Eq.(\ref{Fanomelsen}) describes a crossover from $F = 1/3$ (the
metallic regime) to $F = 1$ (classical point contact between metallic
diffusive banks) and actually follows \cite{MelsenPRB94} from
Eq. (\ref{fanotransball}).  

{\bf Disordered interfaces}. Schep and Bauer \cite{Schep} considered
transport through disordered interfaces, modeled as a configuration of
short-ranged scatterers randomly distributed in the plane
perpendicular to the direction of transport. In the limit $g \ll
N_{\perp}$, with $g$ and $N_{\perp}$ being the dimensionless
conductance and the number of transverse channels, respectively, they
found the following distribution function of transmission
coefficients,
\begin{eqnarray} \label{disinterf}
P(T) & = & \frac{g}{\pi N_{\perp}} \frac{1}{T^{3/2} \sqrt{1 - T}},
\nonumber \\ 
& & \left[ 1 + \left( \frac{\pi N_{\perp}}{2g} \right)^2 \right]^{-2}
< T < 1, 
\end{eqnarray}
and zero otherwise. Eq. (\ref{disinterf}) accidentally has the same
form as the distribution function of transmission coefficients for the
symmetric opaque double-barrier structure. The noise suppression factor
for this system equals $1/2$, {\em e.g.} the suppression is weaker
than for metallic diffusive wires.  

\subsubsection{Chaotic cavities} \label{scat0265}

\begin{figure}
\epsfxsize=8.5cm{\centerline{\epsfbox{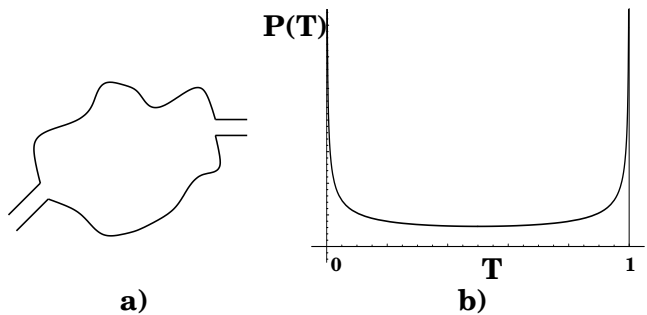}}}
\vspace{0.3cm}
\caption{(a) An example of a chaotic cavity. (b) Distribution function
of transmission eigenvalues (\ref{chaodistr}).}
\label{chaot}
\end{figure}

{\bf $1/4$-suppression}. Chaotic cavities are quantum systems which in
the classical limit would exhibit chaotic electron motion. We consider
{\em ballistic} chaotic systems without any disorder inside the
cavity; the chaotic nature of classical motion is a consequence of the
shape of the cavity or due to surface disorder. The results presented
below are {\em averages} over ensembles of cavities. The ensemble 
can consist of a collection of cavities with slightly
different shape or a variation in the surface disorder, or 
it can consist of cavities investigated at slightly different 
energies. Experimentally, chaotic cavities are usually realized as
{\em quantum dots}, formed in the 2D electron gas by back-gates. They
may be open or almost closed; we discuss first the case of open
chaotic quantum dots, shown in Fig.~\ref{chaot}a. We neglect charging
effects\footnote{In the case when the cavity is open, {\em i.e.}
connected by ideal leads to the electron reservoirs, charging effects
may still play a role. This effect, called {\em mesoscopic charge
quantization} \cite{AGmes}, was recently shown to affect very 
weakly the conductance of open chaotic cavities \cite{Brouwer99}. 
Results for shot noise are currently unavailable.}. One more standard
assumption, which we use here, is that there is no direct
transmission: Electrons incident from one lead cannot enter another
lead without being reflected from the surface of the cavity (like
Fig.~\ref{chaot}a). 

The description of transport properties of open chaotic cavities based
on the random matrix theory was proposed independently by Baranger and
Mello \cite{Baranger} and Jalabert, Pichard, and Beenakker
\cite{Jalabert}. They assumed that the scattering matrix of the
chaotic cavity is a member of Dyson's circular ensemble of random
matrices, uniformly distributed over the unitary group. For the cavity
where both left and right leads support the same number of transverse
channels $N_{\perp} \gg 1$, this conjecture implies the following
distribution function of transmission eigenvalues, 
\begin{equation} \label{chaodistr}
P(T) = \frac{1}{\pi \sqrt{T(1-T)}},
\end{equation}
shown in Fig.~\ref{chaot}b. 
As a consequence of the assumption underlying random matrix theory,
this distribution is universal: It does not depend on {\em any}
features of the system. Taking into account that $\langle T \rangle =
1/2$ and $\langle T (1-T) \rangle = 1/8$, we obtain for conductance,
\begin{displaymath}
G = \frac{e^2N_{\perp}}{4\pi\hbar},
\end{displaymath}
and for the zero-temperature shot noise,
\begin{displaymath}
S = \frac{e^3 \vert V \vert N_{\perp}}{8\pi\hbar} = \frac{1}{4} S_P. 
\end{displaymath}
The Fano factor equals $F = 1/4$, and is, of course, also universal.
For the one-channel case, the whole distribution function of shot
noise may be found analytically \cite{PvLB}, 
\begin{eqnarray}
& & P(S)\\
& = & \left\{ \matrix{ \frac{\sqrt{1 + \sqrt{1-4\eta}} + \sqrt{1 -
\sqrt{1 - 4\eta}}}{\sqrt{16\eta(1-4\eta)}}, & \ \ \ \mbox{\rm
orthogonal symmetry} \cr 
\frac{2}{\sqrt{1 - 4\eta}}, &\ \ \ \mbox{\rm unitary symmetry} }
\right. , \nonumber 
\end{eqnarray}
where we defined $\eta = \pi\hbar S/(e^3 \vert V \vert)$. 

In the general case when the numbers of transverse channels in the
left $N_L$ and right $N_R$ lead are not equal, but still $N_L \gg 1$
and $N_R \gg 1$, the distribution function of transmission eigenvalues
has been calculated by Nazarov \cite{Nazarov95}. Using this result, a
calculation of the shot noise gives for the Fano factor
\cite{Nazarov95,BeenRMP}   
\begin{equation} \label{Fanochaot}
F = \frac{N_LN_R}{(N_L + N_R)^{2}}.
\end{equation}
This suppression factor has $1/4$ as its {\em maximal} value for the
symmetric case $N_L = N_R$, and shot noise is suppressed down to zero
in the very asymmetric case $N_L \ll N_R$ or $N_L \gg N_R$. Indeed,
for $N_L \ll N_R$ the transport properties are determined by the less
transparent (left) contact; however, since the contact is still ideal,
the noise in this situation is totally suppressed. 

Eq. (\ref{Fanochaot}) results from an ensemble average, and, as we
discussed, must be a classical result. Indeed, it has been derived
\cite{Blanter99} by purely classical means, see Section
\ref{BoltzmannL}. 

{\bf Crossover to double-barrier behavior}. We assume now that the
cavity is separated from the leads by tunnel barriers. Brouwer
and Beenakker \cite{Brouwer96} were able to calculate the distribution
function of transmission coefficients in this system for the symmetric
case, when the number of channels supported by the left and the right
lead are equal, $N_L = N_R = N$, and the transmission coefficients
$\tilde T_i$ in each channel $i$ are same for the left and the right
barriers. Assuming in addition $N \tilde T_i \gg 1$ for all channels,
they found  
\begin{equation} \label{distribchaotun}
P(T) = \frac{1}{N} \sum_{i=1}^N \frac{\tilde T_i (2 - \tilde T_i)}{\pi
(\tilde T_i^2 -4\tilde T_i T + 4T) \sqrt{T(1-T)}}.
\end{equation}  
Calculating the averages
\begin{displaymath}
\langle T \rangle = \frac{1}{2N} \sum_i \tilde T_i,
\end{displaymath}
and
\begin{displaymath}
\langle T (1-T) \rangle = \frac{1}{8N} \sum_i \tilde T_i (2 - \tilde
T_i),
\end{displaymath}
we find the Fano factor,
\begin{equation} \label{Fanochaotun}
F = \frac{\langle T (1-T) \rangle}{\langle T \rangle} = \frac{1}{4}
\frac{ \sum_i \tilde T_i (2 - \tilde T_i)}{\sum_i \tilde T_i}.
\end{equation}
In particular, if all the transmission probabilities $\tilde T_i$ are
the same and equal to $\tilde T$, we obtain $F = (2 - \tilde
T)/4$. This expression reproduces the limiting cases $F = 1/4$ for
$\tilde T = 1$ (no barriers -- open quantum cavity) and $F = 1/2$ for
$\tilde T \to 0$ (double-barrier suppression in symmetric
system). Thus, Eq. (\ref{Fanochaotun}) describes crossover between the
behavior characteristic for an open cavity and the situation when the
barriers are so high that the dynamics inside the cavity does not play
any role. 

\subsubsection{Edge channels in the quantum Hall effect regime}
\label{scat0266} 

Now we turn to the description of effects which are inherently
multi-terminal. The calculation of the scattering matrix is in general
a difficult problem. However, in some special situations the
scattering  matrix can be deduced immediately even for multi-terminal
conductors.  

\begin{figure}
\epsfxsize=8.cm{\centerline{\epsfbox{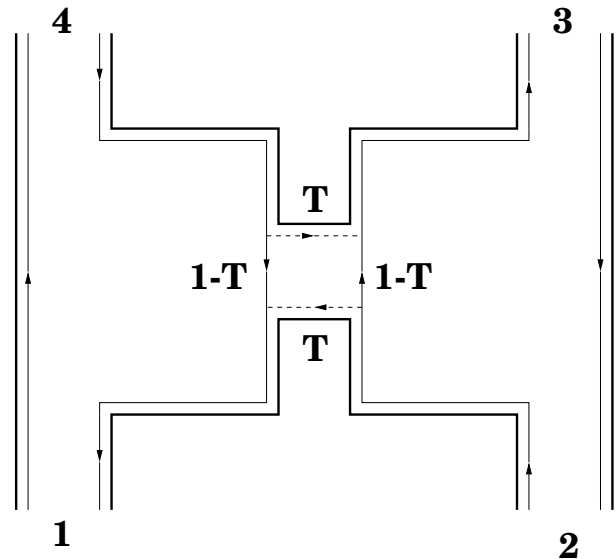}}}
\vspace{0.3cm}
\caption{Four-probe quantum Hall conductor. Solid lines indicate edge
channels; dashed lines show the additional scattering probability $T$ 
through the quantum point contact.}
\label{edge1}
\end{figure}

We consider a four-terminal conductor (Fig.~\ref{edge1}) made by
patterning a two-dimensional electron gas. The conductor is brought
into the quantum Hall regime by a strong transverse magnetic field.
In a region with integer filling of Landau levels the only extended
states at the Fermi energy which connect contacts \cite{bu88} are edge
states\footnote{We do not give a microscopic description of edge
states. Coulomb effects in the integer quantum Hall effect regime
lead to a spatial decomposition into compressible and incompressible
regions. Edge channels in the fractional quantum Hall effect regime
will be discussed in Section \ref{strcorr}.}, the quantum mechanical
equivalent of classical skipping orbits. Since the net current at a
contact is determined by the states near the Fermi surface, transport
in such a system can be described by considering the edge states. Note
that this fact makes no statement on the spatial
distribution of the current density. In particular a description based
on edge states does not mean that the current density vanishes away
from the edges. This point which has caused considerable confusion and
generated a number of publications is well understood, and we refer
the reader here only to one particularly perceptive discussion
\cite{hirai}. Edge states are uni-directional; if the
sample is wide enough, backscattering from one edge state to another
one is suppressed \cite{bu88}. In the plateau regime of the integer 
quantum Hall effect, the number of edge channels is equal to the number 
of filled Landau levels. For the discussion given here, we assume, for
simplicity, that we have only one edge state. In a quantum Hall
conductor wide enough so that there is no backscattering, there is no
shot noise \cite{Buttiker90}. Hence we introduce a constriction
(Fig.~\ref{edge1}) and allow scattering between different edge states
at the constriction \cite{Buttiker90}: the probability of scattering
from contact $4$ to the contact $3$ is $T$, while that from $4$ to $1$
is $1-T$. In the following, we will focus on the situation when the
chemical potentials of all the four reservoirs are arranged so that
$\mu_2 = \mu_3 = \mu$, $\mu_1 = \mu_4 = \mu + eV$.  

The scattering matrix of this system has the form
\begin{eqnarray} \label{edgescat}
s = \left( \matrix{ 0 & s_{12} & 0 & s_{14} \cr 0 & 0 & s_{23} & 0
\cr 0 & s_{32} & 0 & s_{34} \cr s_{41} & 0 & 0 & 0 }
\right).
\end{eqnarray}
Here the elements $s_{14} = r$, $s_{32} = r'$, $s_{12} = t$, $s_{34} =
t'$ form the $2\times 2$ scattering matrix at the constriction: $\vert
t\vert^2 = \vert t' \vert^2 = T$, $\vert r\vert^2 = \vert r' \vert^2 =
1-T$, $r^*t' + t^*r' = r^*t + r't'^* = 0$. The two remaining
elements, $s_{41} = \exp(i\theta_1)$ and $s_{23} = \exp(i\theta_2)$
describe propagation along the edges without scattering. It is
straightforward to check that the matrix (\ref{edgescat}) is unitary.

We consider first the shot noise at zero temperature
\cite{Buttiker90,Buttiker92}. Using the general Eq. (\ref{noiset0}),
we see immediately that the only non-zero components of the shot noise
power tensor are $S_{11} = S_{33} = -S_{13} = -S_{31}$, with
\begin{equation} \label{edge2}
S_{31} = -  \frac{e^3\vert V \vert}{\pi\hbar} T(1-T).
\end{equation}
Indeed, if there is no scattering between the edge states, there is no
shot noise in the system. The same is true for the case when this
scattering is too strong: all the current from $2$ flows to $1$, and
from $4$ to $3$.

For finite temperature all components become non-zero (except for
$S_{24}$) and can be found \cite{Buttiker92} from
Eq. (\ref{noise0}). We only give the result for $S_{13} = S_{31}$,
\begin{equation} \label{edge3}
S_{13} = -\frac{e^2}{\pi\hbar} \int dE \ T(1-T) (f_1 - f_3)^2,
\end{equation}
where $f_1$ and $f_3$ are Fermi distribution functions in the
reservoirs $1$ and $3$, respectively. This result is remarkable since
it vanishes for $T=1$ at any temperature: The correlation function
(\ref{edge3}) is always ``shot-noise-like''.

One more necessary remark is that all the shot noise components in
this example are actually expressed only through absolute values of
the scattering matrix elements: Phases are not important for noise in
this simple edge channel problem. 

Early experiments on noise in quantum Hall systems were oriented to
other sources of noise (see {\em e.g.} Refs. \cite{Kil,BKane}), and
are not discussed here. Shot noise in the quantum Hall regime was
studied by Washburn {\em et al} \cite{Washburn91} who measured voltage 
fluctuations in a six-terminal geometry with a constriction, which, in
principle, allows for direct comparison with the above theory. They
obtained results in two magnetic fields, corresponding to the filling
factors $\nu=1$ and $\nu=4$. Although their results were dominated 
by $1/f$-noise, Washburn {\em et al} were able to find that shot noise
is very much reduced below the Poisson value, and the order of
magnitude corresponds to theoretical results. 

\subsubsection{Hanbury Brown -- Twiss Effects with edge channels}
\label{scat0267} 

The conductor of Fig.~\ref{edge1} is an electrical analog of
the scattering of photons at a half-silvered mirror (see
Fig.~\ref{figloudon}). Like in the table top experiment of Hanbury
Brown and Twiss \cite{HBT,Loudon}, there is the possibility of two 
sources which send particles to an object (here the quantum point
contact) permitting scattering into transmitted and reflected
channels which can be detected separately. Bose statistical effects 
have been exploited by Hanbury Brown and Twiss \cite{HBT}
to measure the diameter of stars. The electrical geometry of
Fig.~\ref{edge1} was implemented by Henny {\em et al} \cite{henny},
and the power spectrum of the current correlation between contact $1$
(reflected current) and contact $3$ (transmitted current) was measured
in a situation where current is incident from contact $4$
only. Contact $2$ was closed such that effectively only a
three-terminal structure resulted. For the edge-channel situation
considered here, in the zero-temperature limit, this does not affect
the correlation between transmitted and reflected current. (At finite
temperature, the presence of the fourth contact would even be
advantageous, as it avoids, as described above, the "contamination"
due to thermal noise of the correlation function of reflected and
transmitted currents, see Eq. (\ref{edge3})). The experiment by Henny
{\em et al} finds good agreement with the predictions of
Ref. \cite{Buttiker90}, {\em i.e.} Eq. (\ref{edge2}). With
experimental accuracy the current correlation $S_{31}$ is negative and
equal in magnitude to the mean square current fluctuations $S_{11} =
S_{33}$ in the transmitted and reflected beam. The experiment finds
thus complete anti-correlation. This outcome is related to the Fermi
statistics only indirectly: If the incident carrier stream is
noiseless, current conservation alone leads to Eq. (\ref{edge2}). As
pointed out by Henny {\em et al} \cite{henny}, the experiment is in
essence a demonstration that in Fermi systems the incident carrier
stream is noiseless. The pioneering character of the experiment by
Henny {\em et al} \cite{henny} and an experiment by Oliver {\em et al}
\cite{oliver} which we discuss below lies in the demonstration of the
possibility of measuring current-current correlation in electrical
conductors \cite{buscience}. Henny {\em et al} \cite{henny} measured
not only the shot noise but used the four-terminal geometry of
Fig.~\ref{edge1} to provide an elegant and interesting demonstration
of the fluctuation-dissipation relation, Eq. (\ref{FDT1}).  
\begin{figure}
\epsfxsize=6.cm{\centerline{\epsfbox{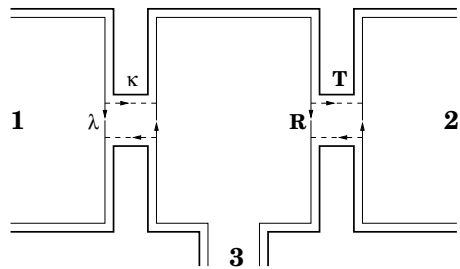}}}
\vspace{0.3cm}
\caption{Three-probe geometry illustrating the experiment by Henny
{\em et al} \protect\cite{henny}.}
\label{oberholzer}
\end{figure}

Now it is interesting to ask, what happens if this complete
population of the available states is destroyed (the incident carrier
stream is {\em not} noiseless any more). This can be achieved by
inserting an additional quantum point contact in the path of the
incident carrier beam (see Fig. \ref{oberholzer}). We denote the
transmission and reflection probability of this first quantum point
contact by $\kappa$ and $\lambda =1 -\kappa$, and the transmission and
reflection probability of the second quantum point contact by $T$ and
$R$ as above. The scattering matrix of this system has the form 
\begin{eqnarray} \label{edgescat2}
s = \left( \matrix{ -i \lambda^{1/2} & 0 & \kappa^{1/2} \cr 
(\kappa T)^{1/2} & -i R^{1/2} & -i (\lambda T)^{1/2} \cr 
-i (\kappa R)^{1/2}&  T^{1/2}  &   - (\lambda R)^{1/2} } 
\right).
\end{eqnarray}
The phases in this experiment play no role and here have been chosen 
to ensure the unitarity of the scattering matrix. In the
zero-temperature limit with a voltage difference $V$ between contact
$1$ and contacts $2$ and $3$ (which are at the same potential) the
noise power spectra are  
\begin{eqnarray} \label{edgescat3}
S = \frac{e^2 \vert V \vert}{\pi\hbar} \left( \matrix{ \kappa \lambda
& -\kappa \lambda T&  -\kappa \lambda R \cr  
-\kappa \lambda T &\kappa T (1-\kappa T)  &  -\kappa^{2} TR \cr 
-\kappa \lambda R&   - \kappa^{2}TR & \kappa R (1-\kappa R) } 
\right).
\end{eqnarray}
The correlation function between transmitted and reflected beams 
$S_{23} = S_{32} =  - \kappa^{2}RT$ is proportional to 
the square of the transmission probability in the first 
quantum point contact. For $\kappa = 1$ the incident beam is
completely filled, and the results of Henny {\em et al} \cite{henny}
are recovered. In the opposite limit, as $\kappa$ tends to zero,
almost all states in the incident carrier stream are empty, and 
the anti-correlation between transmitted and reflected beams also
tends to zero.  

\subsubsection{Three-terminal structures in zero magnetic field}
\label{scat0268} 

A current-current correlation was also measured in an experiment 
by Oliver {\em et al} \cite{oliver} in a three-probe structure in zero
magnetic field. This experiment follows more closely the suggestion 
of Martin and Landauer \cite{ML92} to consider the 
current-current correlations in a Y-structure. Ref. \cite{ML92}, like 
Ref. {\cite{Buttiker90}, analyzes the noise power spectrum in the
zero frequency limit. Early experiments on a three-probe structure by
Kurdak {\em et al} \cite{Kurdak} were dominated by $1/f$--noise and
did not show any effect. 

Here the following remark is appropriate. Strictly speaking, the
Hanbury Brown -- Twiss (HBT) effect is a coincidence measurement. In
the optical experiment the intensity fluctuation $dI_{\alpha}(t)$ is
measured and correlated with the intensity  fluctuation
$dI_{\beta}(t+\tau)$, where $\tau$ is a short time smaller than the
response time $\tilde\tau$ of the detector. The coincidence rate
$C_{\alpha \beta}$ is thus    
\begin{equation} \label{coin1}
C_{\alpha \beta} = (1/2\tilde\tau) \int_{0}^{\tilde\tau} d\tau \langle
d{\hat I}_{\alpha}(t) d{\hat I}_{\beta}(t+\tau) + 
d{\hat  I}_{\beta}(t+\tau) d{\hat I}_{\alpha}(t) \rangle
\end{equation}
The coincidence rate is related to the frequency dependent noise power
spectrum by  
\begin{equation} \label{coin2}
C_{\alpha \beta} = (1/\tilde\tau) \int_{0}^{\tilde\tau} d\tau \int
d\omega e^{i \omega \tau} S_{\alpha\beta}(\omega)
\end{equation}
In Section \ref{freq}, we discuss the frequency dependence of the noise
power spectrum in more detail. Typically its lowest characteristic
frequencies  are given by $RC$-times. In principle, such a measurement
should, therefore, be able to give information on the frequency
dependence of the noise power spectrum. In the experiment of Oliver
{\em et al} the resolution time $\tilde\tau$ is probably long compared
to such intrinsic time scales, and thus the experiment is effectively
determined by the white noise limit of the power spectrum.  

Let us now briefly consider a Y-shaped conductor \cite{ML92} and
discuss  its correlations in the white noise limit. We assume that the
same voltage $V$ is applied between the terminals $1$ and $2$, and $1$
and $3$: $\mu_1 = \mu_2 + eV = \mu_3 + eV$. For zero temperature, the
general formula (\ref{noiset0}) yields the following expression for
the cross-correlations of currents in leads $2$ and $3$,  
\begin{equation} \label{Ynoise}
S_{23} = -\frac{e^3 \vert V \vert}{\pi\hbar} {\rm Tr} \left[
s^{\dagger}_{21} s_{21} s^{\dagger}_{31} s_{31} \right],
\end{equation}
which is negative in accordance with the general considerations. 
Note the formal similarity of this result to the
shot-noise formula in the two terminal geometry given by 
Eq. (\ref{shottr}). In the single channel limit, if we assume 
that there is no reflection back into contact $1$,  
Eq. (\ref{Ynoise}) becomes $S_{23} = -(e^3 \vert V
\vert/\pi\hbar)T(1-T)$, where $T$ is the transmission probability from
$1$ to $2$. This simple result underlines  (the formal) equivalence of
scattering at a QPC with separation of transmitted  and reflected
streams and scattering at a reflectionless Y-structure. The
experiments by Oliver {\em et al} \cite{oliver} confirm these
theoretical predictions.  
 
The experiments by Henny {\em et al} \cite{henny} and Oliver {\em et
al} \cite{oliver} test the partitioning of a current stream. If the
incident carrier stream is noiseless, the resulting current
correlation is negative already due to current conservation
alone. Therefore, experiments are desirable, which test electron
statistical effects (and the sign of correlations) in situations where
current conservation plays a much less stringent role.  

\subsubsection{Exchange Hanbury Brown -- Twiss effects}
\label{scat0269} 

\begin{figure}
\epsfxsize=6.cm{\centerline{\epsfbox{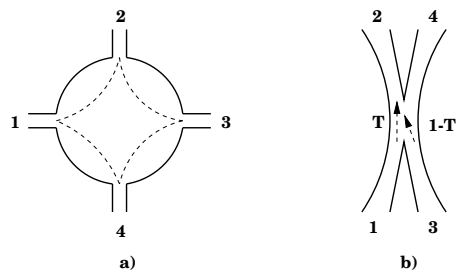}}}
\vspace{0.3cm}
\caption{Four-terminal conductors for Hanbury Brown -- Twiss
exchange effects (a), quantum
point contact geometry (b). Terminals are numbered by
digits. The dashed line in (a) indicates a phase-sensitive trajectory
which contributes to exchange terms in Eq. (\ref{HBTgeneral}). Arrows
in (b) indicate non-zero transmission probabilities.}
\label{HBTfig}
\end{figure}

Another HBT experiment was proposed in Ref. \cite{Buttiker91} (see
also Ref. \cite{Buttiker92}). It is based on the comparison of the
noise generated in the presence of {\em two} incident currents with
the noise generated by one source only. We first present a general
discussion and later consider a number of applications. Consider the
four-terminal structure of Fig.~\ref{HBTfig}a. We will be interested
in cross-correlations of currents at the contacts $2$ and $4$,  
\begin{displaymath}
S \equiv -S_{24}.
\end{displaymath} 
The quantity $S$ defined in this way is always {\em positive}. Now we
discuss three different ways of applying voltage. The first one
(to be referred to as experiment A) is to apply a voltage to the
reservoir $1$, $\mu_1 - eV = \mu_2 = \mu_3 = \mu_4$. In the next one
(experiment B) the voltage is applied to $3$, $\mu_1 = \mu_2 = \mu_3 -
eV = \mu_4$. Finally, in the experiment C the identical\footnote{We
note in passing that if {\em different} voltages are applied to $1$
and $3$, the correlation functions (\ref{HBTgeneral}) imply 
that $S_{23}$ cannot be an analytic function of the two
voltages. Thus, our four-terminal conductor is a non-linear circuit
element due to exchange effects.} voltages are applied to $1$ and $3$,
$\mu_1 - eV = \mu_2 = \mu_3 - eV = \mu_4$. Results for zero
temperature are readily derived from Eq. (\ref{noiset0}) and read  
\begin{eqnarray} \label{HBTgeneral}
S_A & = & \frac{e^3 \vert V \vert}{\pi\hbar} \Xi_1, \ \ \ S_B =
\frac{e^3 \vert V \vert}{\pi\hbar} \Xi_2, \nonumber \\
S_C & = & \frac{e^3 \vert V \vert}{\pi\hbar} \left( \Xi_1 + \Xi_2 +
\Xi_3 + \Xi_4 \right),  
\end{eqnarray}
where 
\begin{eqnarray} \label{Xis}
\Xi_1 & = & {\rm Tr} \left[ s^{\dagger}_{21} s_{21} s^{\dagger}_{41}
s_{41} \right], \nonumber \\
\Xi_2 & = & {\rm Tr} \left[ s^{\dagger}_{23} s_{23} s^{\dagger}_{43}
s_{43} \right], \nonumber \\ 
\Xi_3 & = & {\rm Tr} \left[ s^{\dagger}_{21} s_{23} s^{\dagger}_{43}
s_{41} \right] \nonumber \\
\Xi_4 & = & {\rm Tr} \left[ s^{\dagger}_{23} s_{21}
s^{\dagger}_{41} s_{43} \right]. 
\end{eqnarray}
The quantities $S_A$ and $S_B$ are determined by transmission
probabilities from $2$ and $4$ to $1$ and $3$, respectively, and are
not especially interesting. New information is contained in
$S_C$. In systems obeying classical statistics, the experiment C would
be just a direct superposition of the experiments A and B, $S_C = S_A
+ S_B$. The additional terms $\Xi_3$ and $\Xi_4$ in the rhs of
Eq. (\ref{HBTgeneral}) are due to quantum (Fermi) statistics of the
electrons. These terms now invoke products of scattering matrices 
which are in general not real valued. These terms are not products
of two pairs of scattering matrices as in Eq. (\ref{shottr}) or 
Eq. (\ref{Ynoise}) but contain four scattering matrices 
in such a way that we are not able to distinguish from which of the 
two current carrying contacts a carrier was incident. 
For future convenience, we define the quantity $\Delta S =
S_C - S_A - S_B$, which indicates the fermionic analog of the HBT
effect. One can show \cite{Buttiker92} that for finite temperatures
the corresponding correction for bosons is of the same form but has
the opposite sign, hence it will be called ``exchange
contribution''. One more remarkable feature of the result
(\ref{HBTgeneral}) is that the exchange correction $\Delta S$ is {\em
phase sensitive}. Indeed, it represents the contribution of
trajectories indicated by the dashed line in Fig.~\ref{HBTfig}a
(traversed in both directions), and thus is proportional to $\exp(\pm
i\phi)$, with $\phi$ being the phase accumulated during the motion
along the trajectories. For this reason, one cannot generally predict
the sign of $\Delta S$: the only restrictions are that all the
quantities $S_A$, $S_B$, and $S_C$ need to be positive.  

Gramespacher and one of the
authors~\cite{Gramespa98,Gramespa99,ButtikerQH92}  
considered a particular geometry where the leads $2$ and $4$ are
tunneling contacts locally coupled to the sample ({\em e.g.} scanning
tunneling microscope tips). In this case, the exchange contribution
can directly be expressed in terms of the wave functions (scattering
states) 
\begin{eqnarray} \label{gram69}
\Delta S & = & \frac{1}{\pi^2\hbar^2} \sum_{mn} \frac{1}{v_{1m}v_{3n}}
\nonumber \\
& \times & {\rm Re} \ \left\{ \psi_{1m} (\bbox{r}) \psi_{1m}^*
(\bbox{r}') \psi_{3n}^* (\bbox{r}) \psi_{3n} (\bbox{r}') \right\},  
\end{eqnarray}
where the sum is over all transverse channels $m$ in the lead $1$ and
$n$ in the lead $3$; $\bbox{r}$ and $\bbox{r}'$ are the points
to which the contacts $2$ and $4$ couple, respectively, and
$\psi_{\alpha k}$ is the wave function of the corresponding scattering
state. Thus, the exchange contribution explicitly depends on phases of
the wave functions.    
 
We investigate now the general expression for the four-terminal
phase-sensitive HBT effect (\ref{HBTgeneral}) for various systems. Our
concern will be the sign and relative magnitude of the exchange
contribution $\Delta S = S_C - S_A - S_B$.  

{\bf Disordered systems}. Naively, one might assume that in a
disordered medium the phase accumulated along the trajectory,
indicated by the dashed line in Fig.~\ref{HBTfig}a, is random. Then
the phase-sensitive exchange contribution would be zero after being
averaged over disorder. Thus, this view implies $S_C = S_A + S_B$.  

A quantitative analysis of these questions was provided by the authors 
of this review in Ref.~\cite{Blanter97}. In this work
the scattering matrices in Eq. (\ref{Xis}) are expressed through 
Green's functions to which disorder averaging was applied using
the diagram technique. The key result found in
Ref.~\cite{Blanter97} is that the naive picture mentioned above,
according to which one might expect no exchange effects after disorder
averaging, is completely wrong.  {\em Exchange effects survive
disorder averaging}.  The reason can be understood if the principal
diagrams (which contain four diffusion propagators) are translated
back into the language of electron trajectories. One sees then that
the typical trajectory does not look like the dashed line in
Fig.~\ref{HBTfig}b. Instead, it looks like a collection of dashed
lines shown in Fig.~\ref{HBTdiffus}a: the electron diffuses from 
contact $1$ to some intermediate point $5$ in the bulk of the sample
(eventually, the result is integrated over the coordinate of point
$5$), then it diffuses from $5$ to $2$ and back from $2$ to $5$ {\em
precisely} along the same trajectory, and so on, until it returns from
$5$ to $1$ along the same diffusive trajectory as it started. Thus,
there is no phase enclosed by the trajectory. This explains why the
exchange contribution survives averaging over disorder; apparently,
there is a classical contribution to the exchange correlations which
requires knowledge only of Fermi statistics, but no information about
phases of scattering matrices. Indeed, a classical theory of
ensemble averaged exchange effects was subsequently proposed by
Sukhorukov and Loss \cite{SL98,SLlong}.   
\begin{figure}
\epsfxsize=8.cm{\centerline{\epsfbox{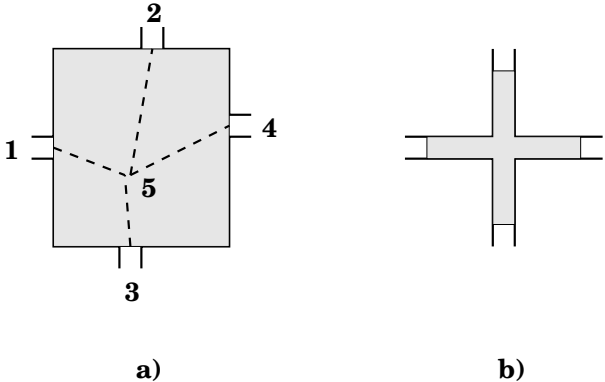}}}
\vspace{0.3cm}
\caption{Examples of four-terminal disordered conductors. Disordered
area is shaded. Dashed line denotes diffusive motion between its
ends.}  
\label{HBTdiffus}
\end{figure}

Once we determined that exchange correction $\Delta S$ exists in
diffusive conductors, we must evaluate its sign and relative
magnitude. We only describe the results qualitatively; details can be
found in Refs. \cite{Blanter97,SLlong}. Two specific geometries have
been investigated: the disordered box (Fig.~\ref{HBTdiffus}a) and
the disordered cross (Fig.~\ref{HBTdiffus}b). For the box, the
exchange correction $\Delta S$ is negative, {\em i.e.} exchange
suppresses noise ($S_C < S_A + S_B$). The effect is quite
considerable: The correction is of the same order of magnitude as the
classical contributions in $S_A$ and $S_B$, and is suppressed only by a
numerical factor. For the cross, the exchange contribution is positive
-- exchange enhances noise -- but the magnitude is by powers of $l/L$
smaller than $S_A$ and $S_B$. Here $l$ and $L$ are the mean free path
and the length of the disordered arms, respectively. Thus, neither
sign nor magnitude of the exchange effects is predetermined in
diffusive systems: they are geometry and disorder dependent, and the
only limitation is $S_C > 0$.  

Gramespacher and one of the authors \cite{Gramespa98,Gramespa99}
considered a geometry of a disordered wire (along the axis $z$)
between the contacts $1$ ($z=0$) and $3$ ($z=L$), coupled locally at
the points $z$ and $z'$ to the contacts $2$ and $4$, respectively, via
high tunnel barriers (these latter can be viewed as scanning tunneling
microscope tips)\footnote{Ref. \cite{Gramespa99} also considers a
three-terminal structure (a (disordered) wire with a single STM tip
attached to it). The fluctuations of the current through the tip are
in this case proportional to the local distribution function of
electrons at the coupling point, see Eq. (\ref{distrib671}) below.}
and evaluated Eq. (\ref{gram69}). It was found that the exchange
effect is {\em positive} in the case, {\em i.e.} it enhances noise,
irrespectively of the position of the contacts $2$ and $4$. For the
particular case when both tunnel contacts are situated symmetrically
around the center of the wire at a distance $d$, $z = (L-d)/2$ and $z'
= (L+d)/2$, the relative strength of the exchange term is  
\begin{displaymath}
\frac{\Delta S}{S_C} = \frac{1}{3} \left[ 2 + \frac{d}{L} - 2 \left(
\frac{d}{L} \right)^2 \right], 
\end{displaymath} 
and reaches its maximum for $d = L/4$. We see that the exchange
effect in this case generally has the same order of magnitude as
the classical terms $S_A$ and $S_B$. 

{\bf Chaotic cavities}. A similar problem in chaotic cavities was
addressed in Ref. \cite{Langen97} (see also Ref. \cite{BlanterUFN}). 
Similarly to disordered systems, it was discovered that exchange
effects survive on average. An additional feature is however that the
exchange effects in chaotic cavities are {\em universal}. 
That of course is a consequence of the assumption that the cavity 
can be described by using Dyson's circular ensemble. For open
cavities, one finds  
\begin{equation} \label{langen}
\Xi_1 = \Xi_2 = -3\Xi_3 = -3\Xi_4 = \frac{3}{4}
\frac{N_{\perp}^3}{16N_{\perp}^2 - 1},
\end{equation}
where we assumed that all leads are identical and support $N_{\perp}$
transverse channels. This implies $S_A = S_B$, $S_C = 4S_A/3$, or
$\Delta S = -2S_A/3$. Thus, exchange effects suppress noise in open
chaotic cavities.  

The situation changes if the cavity is separated from the leads by
tunnel barriers. Assuming that the transmission coefficients of all
barriers in all transverse channels are identical and equal $T$,
$N_{\perp}T \gg 1$, Ref. \cite{Langen97} finds 
\begin{eqnarray*} 
\left\{ \begin{array}{c} \Xi_1 = \Xi_2 \\ \Xi_3 = \Xi_4 \end{array}
\right\} = \frac{N_{\perp}T}{64} \left\{ \begin{array}{c} T + 2 \\
-3T + 2 \end{array} \right\}.
\end{eqnarray*}
Thus, for $T= 2/3$ the exchange effect changes sign: If the barriers
separating the cavity from the reservoirs are high enough, the
exchange enhances the correlations. In the limit of very opaque 
barriers, $T \to 0$, we have $S_C = 2(S_A + S_B)$: the exchange
correction is the same as the classical contributions $S_A$ and $S_B$.

{\bf Edge channels in the quantum Hall effect regime}. Interesting
tests of exchange (interference) effects can also be obtained in high
magnetic fields using edge channels. In fact this leads to a simple
example where the phase dependence of the exchange effect is indeed
essential \cite{Buttiker92}. Imagine that the system is placed into
a strong magnetic field, and the transport is only due to edge
channels. We assume that there is only one edge state flowing from $1$
through $2$ and $3$ to $4$ and back to $1$ (Fig.~\ref{HBTQH}). 
Furthermore, for simplicity we assume that all the
leads are identical, and the transmission probability to enter from
the lead to the edge state is $T$. A direct calculation gives
\cite{Buttiker92}  
\begin{eqnarray*}
\Xi_1 & = & \Xi_2 = \frac{T^4 (1-T)^2}{[1 + (1-T)^4 - 2(1-T)^2\cos
\phi]^2}, \nonumber \\ 
\Xi_3 & = & \Xi_4^* = \frac{T^4 (1-T)^2 \exp(i\phi)}{[1 + (1-T)^4 -
2(1-T)^2\cos \phi]^2}, 
\end{eqnarray*}
where $\phi$ is the phase accumulated along the whole trajectory, and
the phase dependence in the denominator appears due to the possibility
of multiple traversals of the full circle. We have
\begin{equation} \label{phasex}
S_C = \frac{2e^3 \vert V \vert}{\pi\hbar}\frac{T^4 (1-T)^2}{[1 +
(1-T)^4 - 2(1-T)^2\cos \phi]^2} \left(1 - \cos{\phi} \right).  
\end{equation}

\begin{figure}
\epsfxsize=6.cm{\centerline{\epsfbox{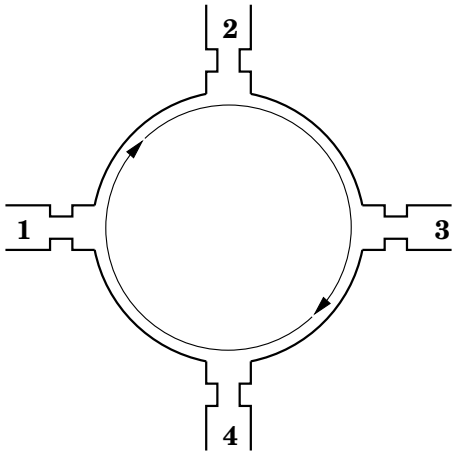}}}
\vspace{0.3cm}
\caption{Hanbury Brown -- Twiss effect with edge states.}  
\label{HBTQH}
\end{figure}

Here the term with $1$ represents the ``classical'' contributions $S_A
+ S_B$, while that with $\cos \phi$ is accountable for the exchange
effect $\Delta S$. We see that, depending on $\phi$, exchange effects
may either suppress (down to zero, for $\phi = 0$) or enhance (up to
$2(S_A + S_B)$, for $\phi = \pi$) total noise. This is an example
illustrating the maximal phase sensitivity which the exchange effect
can exhibit. 

This simple example gives also some insight on how exchange effects 
survive ensemble averaging. Since the phase $\phi$ occurs not only
in the numerator but also in the denominator, the average of
Eq. (\ref{phasex}) over an ensemble of cavities with the phase $\phi$
uniformly distributed in the interval from $0$ to $2\pi$ is non-zero
and given by  
\begin{equation} \label{phaseav}
\langle S_C \rangle = 
\frac{2e^3 \vert V \vert}{\pi\hbar}\frac{T^4 (1-T)^2}{[1 -(1-T)^2]^3
[1 + (1-T)^2]} .   
\end{equation}
The ensemble averaged exchange contribution vanishes both in the limit
$T = 0$ and in the limit $T =1$. 

Note that if another order of contacts is chosen, $1 \to 3 \to 2 \to
4$, the whole situation changes: the exchange term is now phase
insensitive and has a definite sign (negative, {\em i.e.} exchange
suppresses noise) \cite{Buttiker92}. This is because the trajectories
responsible for exchange terms do not form closed loops in this case.  

{\bf Experiments}. The phase-sensitive Hanbury Brown--Twiss effect
discussed above has not so far been probed in experiments. However, a
related experiment was carried out by Liu {\em et al}
\cite{Liupre98,Liu98}, who measured the mean squared fluctuations
$S_{33}$ of the current in the lead $3$, of a four probe structure,
applying voltages in the same {\em three-fold} ways that we have
discussed.    

Prior to the description of experimental results, we discuss briefly 
a measurement of $S_{33}$ on the quantum Hall conductor of
Fig. \ref{edge1}. If current is incident from contact $4$ (experiment
A), or contact $2$ (experiment B) alone we have for the current
fluctuations at contact $3$ $S_{33}^A = S^B_{33} = (e^3 \vert V
\vert/\pi\hbar) T(1-T)$. On the other hand if currents are incident
both from contact $4$ (experiment C) and contact $2$, all states are
now completely filled and thus in the zero-temperature limit $S^C_{33}
=0$. Thus, in comparison to experiments A or B there is a complete
reduction of the shot noise at contact $3$ in experiment C: The
spectral density $S^C_{33}$ is {\em suppressed down to zero}. 
\begin{figure}
\epsfxsize=7.cm{\centerline{\epsfbox{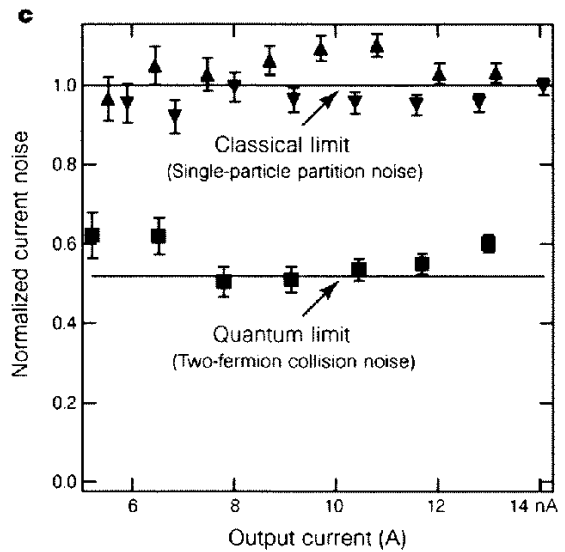}}}
\vspace{0.3cm}
\caption{Experimental results of Liu {\em et al}
\protect\cite{Liu98}. Upward and downward triangles correspond to the
situations when only one of the two input contacts is open (values
$S_A$ and $S_B$, respectively); squares indicate the case when both
input contacts are effective ($S_C$).}
\label{HBTLiu}
\end{figure}

In the experiment of Liu {\em et al}~\cite{Liupre98,Liu98}
the mean square current fluctuations are measured in zero magnetic
field in a conductor in which a left input ($1$) and output contact
($3$) are separated by a thin barrier from a right input contact ($2$)
and output contact ($4$) (see Fig. \ref{HBTLiu}). The input contacts
form QPC's and are adjusted to provide transmission close to 1. The
output contacts support a number of channels. In this experiment it
thus not possible to fill all outgoing states in contact $3$
completely and there is thus only a limited reduction of noise in
experiment C compared to experiments $A$ and $B$. The experimentally
observed ratio $S_{C}/(S_A + S_B) = 0.56$. 

It is also useful to compare the experiment of Liu {\em et al}
\cite{Liupre98,Liu98} simply with a chaotic cavity connected to point
contacts which are fully transparent $T=1$ \cite{Langen97}. Then using
Eq. (\ref{langen}) one finds a ratio $S_{C}/(S_A + S_B) = 2/3$ which
is surprisingly close to what was observed in the experiment.  

\subsubsection{Aharonov -- Bohm effect} \label{scat0260}

The Aharonov -- Bohm (AB) effect tests the sensitivity to a magnetic
flux $\Phi$ of electrons on a trajectory which enclose this flux. In
the pure AB-effect the electron does not experience the magnetic
field, the electron trajectory is entirely in a field free region. It
is a genuine quantum effect, which is a direct consequence of the
gauge invariance of the velocity and the wave nature of electrons. The
simplest geometry demonstrating the AB effect in electric transport is
a ring coupled to two reservoirs and threaded by a magnetic flux, as
shown in Fig.~\ref{ABring}a. Then, the AB effect is manifest in a
periodic flux dependence of all the transport properties.

Qualitatively different phenomena arise in weak and strong magnetic
fields, and these two cases need to be considered separately. 

\begin{figure}
\epsfxsize=8.5cm{\centerline{\epsfbox{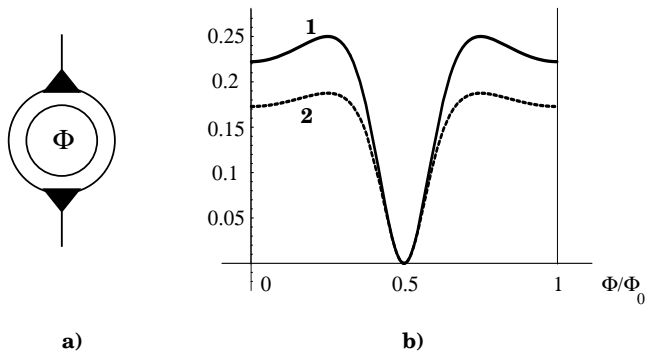}}}
\vspace{0.3cm}
\caption{(a) Geometry of a ring threaded by a flux $\Phi$,
demonstrating the Aharonov--Bohm effect in a weak magnetic field. (b)
Conductance in units of $e^2/2\pi\hbar$ (1) and shot noise power in
units of $e^3V/\pi\hbar$ (2) as a function of the AB flux
$\Phi/\Phi_0$ for a particular value of the phase $\phi = \pi/2$.} 
\label{ABring}
\end{figure}

{\bf Weak magnetic fields}. In this regime, the transmission
coefficient(s) (and, subsequently, conductance) of the two-terminal
structure shown in Fig.~\ref{ABring}a, is a periodic function of
the external flux, with the period $\Phi_0 = 2\pi\hbar c/e$. The
resulting conductance is sample-specific, and, in particular, it is
very sensitive to the phase of the trajectory enclosing the flux. In
phase coherent many-channel conductors the AB-oscillations in the
conductance represent a small correction to a flux insensitive
(classical) background conductance\footnote{In disordered systems, the
ensemble averaged conductance exhibits AB oscillations with the period
of $\Phi_0/2$. These oscillations are, like the weak localization
correction, associated with the interference of two electron
trajectories running in opposite directions \cite{AAS81}. Weak
localization effects and shot noise have not so far been
investigated.}. Thus the AB effect is most dramatic in single channel
rings~\cite{Gefen84,Buttiker84} where the flux induced modulations 
of the conductance are of the order of the conductance itself. 
Shot noise in such a structure was studied by Davidovich and Anda
\cite{Davidovich} using the nonequilibrium Green's functions
technique. They considered a one-channel ring and used a tight-binding
description of the ring and leads. Here we will give another
derivation, based on the scattering approach. A related issue was
discussed by Iannaccone, Macucci, and Pellegrini \cite{IMP97AB}, who
studied noise in a multiply-connected geometry using the scattering
approach, and found that if there is no transmission from the left 
part of the ring to the right part and vice versa
(Fig.~\ref{ABring}a), noise of the left and right parts add up
classically. In particular, this means that such a system would not
exhibit an AB effect.  

We follow Refs. \cite{Gefen84,Buttiker84} which study the
transmission coefficient of single-channel rings connected to external
leads (Fig.~\ref{ABring}a). Our purpose here is to illustrate only the
principal effect, and therefore we consider the simple case without
scattering in the arms of the ring. We also assume that the ring is
symmetric. Formulae for shot noise in more complicated situations can
be readily produced from Ref. \cite{Gefen84,Buttiker84}, though, to
the best of our knowledge, they have never been written down
explicitly.   

We describe the ``beam splitters'', separating the leads from the ring
(black triangles in Fig.~\ref{ABring}a) by the scattering
matrix\cite{Buttiker84}  
\begin{eqnarray} \label{beamsplit}
s_b = \left( \matrix{ -(a+b) & \epsilon^{1/2} & \epsilon^{1/2} \cr
\epsilon^{1/2} & a & b \cr \epsilon^{1/2} & b & a } \right), 
\end{eqnarray}
where the parameter $\epsilon$, $0 < \epsilon < 1/2$, is responsible
for the coupling of the ring to the lead, and 
\begin{eqnarray*}
a & = & \frac{1}{\sqrt{2}} \left( \sqrt{1-2\epsilon} - 1 \right),
\nonumber \\ 
b & = & -\frac{1}{\sqrt{2}} \left( \sqrt{1-2\epsilon} + 1 \right).
\end{eqnarray*}
Specializing to the case of the ring which is ideally coupled to the
leads, $\epsilon = 1/2$, we obtain for the transmission coefficient
\cite{Buttiker84} 
\begin{equation} \label{ABweak}
T (\Phi) = \frac{(1 + \cos \theta) \sin^2 \phi}{(1 + \cos \theta -
\cos 2\phi)^2 + (1/2) \sin^2 2\phi}, 
\end{equation} 
where $\theta = 2\pi\Phi/\Phi_0$, and $\phi$ is the phase accumulated
during the motion along a half of the ring (without magnetic
field). Now the conductance (\ref{condbasic}) $G = (e^2/2\pi\hbar)T$
and the shot noise (\ref{shot2term}) $S = (e^3 \vert V
\vert/\pi\hbar)T(1-T)$ are immediately expressed as functions of the
applied magnetic flux. They are strongly dependent on the phase
$\phi$, which is sample-specific. In particular, both the conductance
and the shot noise vanish for $\phi =0$ or $\phi=\pi$. This is a
consequence of the symmetry assumed here: If the leads are attached
asymmetrically to the ring  the transmission coefficient stays finite
for any value of the phase \cite{busquid}. The dependence of
conductance and shot noise on flux $\Phi$ for a particular value $\phi
= \pi/2$ is shown in Fig.~\ref{ABring}b. We reemphasize that the
flux dependence shown depends strongly on the sample specific phase
$\phi$. 

{\bf Strong magnetic fields}. Now we turn to the situation of the
quantum Hall effect, where transport current is carried by the edge
states. A remarkable feature of this regime is that
a two-terminal ring without backscattering does not exhibit the
AB effect. The edge states (Fig.~\ref{ABQH}a) exist in
different regions of space, and thus do not interfere.
Indeed, the absence of backscattering, which precludes the AB-effect, 
is just the condition for conductance quantization \cite{bu88}.
We cannot have both a quantized conductance and an AB-effect. 
\begin{figure}
\epsfxsize=8.5cm{\centerline{\epsfbox{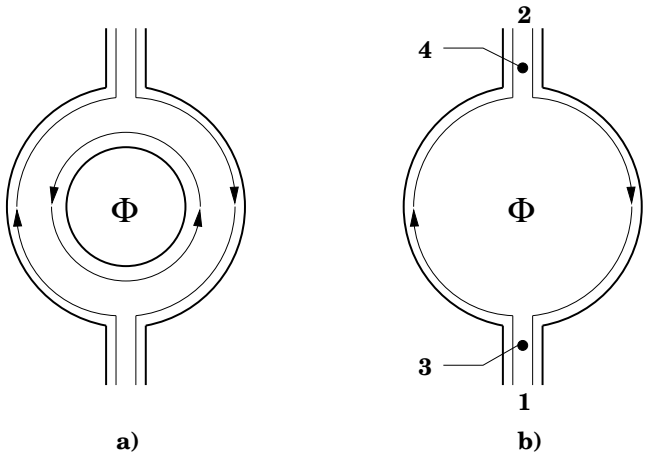}}}
\vspace{0.3cm}
\caption{(a) Geometry of a ring in strong magnetic field. Edge states
are shown. (b) Four-terminal geometry which facilitates separation of
scattering and AB effects.} 
\label{ABQH}
\end{figure}

The question which we want now to address is the following: Can one
observe AB effects in the noise, which the fourth order interference
effect in a situation when they do not exist in the conductance which
is only a second order interference effect? We noticed already that
shot noise is a phase sensitive effect, and contains interference
terms, absent in the conductance. The shot noise contains non-real
terms composed of four scattering matrix elements. This is the case
already in the  two-terminal shot noise formula when it is expressed
in the natural basis. In the two terminal case the appearance of such
products depends, however, on the basis we chose: The shot noise is a
function of transmission probabilities only, if it is evaluated in the
eigen channel basis. However, in a multi-terminal geometry, such
products appear naturally if we consider current-current
cross-correlations. We call these non-real products {\em exchange
interference} terms, since they are a manifestation of interference
effects in multi-particle wave functions (Slater determinants of single
particle wave functions) which result from the indistinguishability of
carriers. In contrast, the effects we have already seen in weak
fields, which contribute to the conductance as well as to the shot
noise are a consequence of second order or {\em direct interference}.
 
The two-terminal geometry of Fig.~\ref{ABQH}a is not appropriate for
the observation of the exchange interference effects, since shot noise
vanishes without scattering between edge channels. If scattering
is introduced, shot noise becomes finite, but at the same time
the conductance becomes sensitive to the flux, due to direct
interference. One can try to separate AB effects in the shot noise due
to direct and exchange interference, but this is awkward. 

A possible way out was proposed in Ref. \cite{ButtikerQH92}, which
suggested four-terminal geometries with two weak coupling contacts. We
follow here a subsequent, clearer discussion given in
Ref. \cite{Gramespa99}. The geometry is shown in
Fig.~\ref{ABQH}b. This is a quantum dot in a strong magnetic field
coupled via two quantum point contacts to reservoirs. The ring
geometry is actually not needed in the experiment and serves only for
conceptual clarity. Current flows between contacts $1$ and $2$, and
the contacts $3$ and $4$ are inserted locally at the quantum point
contact between the edge states in the leads. The magnetic field is
such that the two-probe conductance is quantized, but weak enough such
that at the quantum point contact the left- and right-going wave
functions of the two edge channels overlap. As is well known, the fact
that the wave functions in the quantum point contact overlap, does not
destroy the quantization, as long as the potential of the quantum
point contact is smooth. We take the scattering matrix relating the
amplitudes of carriers in the contact $3$ (or $4$) with those in the
edge channels nearby to be of the same form (\ref{beamsplit}) as for
the ``beam splitter'' discussed above. Here it is now essential to
assume that coupling is weak thus we take $\epsilon \ll 1$. In the
experiment proposed in Refs. \cite{ButtikerQH92,Gramespa99} the same
voltage $V$ is applied simultaneously to the contacts $1$ and
$2$. Then we have   
\begin{eqnarray*}
S_{33} & = & S_{44} = \frac{2e^3 \vert V \vert}{\pi\hbar} \epsilon,
\nonumber \\  
S_{34} & = & -\frac{2e^3 \vert V \vert}{\pi\hbar} \epsilon^2 \left[ 1
+ \cos(\phi + 2\pi\Phi/\Phi_0) \right],
\end{eqnarray*}
where $\phi$ is a certain phase. The relative value of $S_{34}$ as
compared to $S_{33}$ is $\epsilon$. At the same time, the corrections
to the conductance and the shot noise due to direct interference are
proportional to $\epsilon^2$. Thus, in our geometry up to the terms of
$\epsilon^2$ conductance is not renormalized by the AB effect, while
shot noise feels it due to its two-particle nature. This is thus a
geometry where the AB effect manifests itself in the fourth order
interference and modulates the Hanbury Brown -- Twiss effect (the
current-current cross-correlation at contacts $3$ and $4$). 
  
\subsection{Inelastic scattering. Phase breaking} \label{scat027}

Throughout this Section, we treated the mesoscopic systems as
completely phase coherent. In reality, there is always at least some
inelastic or phase breaking scattering present. The scattering
approach as it was used here, is based on the carrier transmission at
a definite energy. In contrast, electron-electron or electron-phonon
interactions can change the energy of a carrier. Thus a scattering
theory of such processes has to be based on a scattering amplitudes
which permit incoming and outgoing particles to have different
energies. To our knowledge, the extension of scattering theory of
electrical transport within such a generalized scattering matrix
approach has not been worked out. It is, however, possible to make
progress even within the scattering approach used so far: To treat
phase breaking theoretically we often proceed by inventing a
Hamiltonian system with many degrees of freedom while we are
interested in the behavior of only a subsystem. Similarly it is
possible to arrive at an approach which describes inelastic
transitions and phase breaking by first considering a completely 
phase-coherent conductor with one or a continuum of additional
voltage probes which are purely fictitious
\cite{Buttiker86,ButtikerIBM}. The additional fictitious voltage
probes act as dephasers on the actual conductor of interest. This
approach has been widely used to investigate the effect of dephasing
on conductance. We refer the reader here only to a few early works
\cite{Buttiker86,ButtikerIBM,amato,gagel}. In this subsection we
illustrate the application of these ideas to noise. Other approaches,
based on Green's function techniques, have also been invoked to derive
results for strongly correlated systems (see Section \ref{strcorr}). 
Furthermore, on the purely classical level, it proved to be
rather simple to extend the fluctuating Boltzmann equation approach to
include interactions. For the results on interaction and noise in
double barrier resonant tunneling structures and metallic diffusive
conductors the reader is addressed to Sections \ref{Langevin} and
\ref{BoltzmannL}, respectively. The approach which uses voltage probes
as dephasers is interesting because of its conceptual clarity and
because of its close relation to experiments: The effect of additional
voltage probes can easily be tested experimentally with the help of 
gates which permit to switch off or on a connection to a voltage probe
(see {\em e.g.} Ref. \cite{expdeph}).  

{\bf Voltage probes as dephasers.} Consider a mesoscopic conductor
connected to $N$ (real) contacts. To introduce inelastic scattering,
we attach a number $M$ of purely fictitious voltage probes to this
conductor. The entire conductor with its $N+M$ contacts is phase
coherent and exhibits the noise of a purely phase coherent
conductor. However, elimination of the $M$ fictitious voltage probes
leads to an effective conduction problem for which the conductance and
the noise depend on inelastic scattering processes
\cite{Buttiker91,Been92,Liu941,Liu942,Lithuania}. 
Depending on the properties of the fictitious voltage probes, three
different types of inelastic scattering can be realized, which de Jong
and Beenakker \cite{Jong96} classify as ``quasi-elastic scattering''
(phase breaking), ``electron heating'', and ``inelastic
scattering''. Now we describe these types of probes separately. This
division corresponds to the distinction of $\tau_{\phi}$, $\tau_{ee}$,
and $\tau_{in}$. We emphasize that only a microscopic theory can give
explicit expressions for these times. What the approach based on
fictitious voltage probes can do is to find the functional dependence of
the conductance or the noise on these times. 

The results for interaction effects in double-barrier structures seem
to be well established by now. In contrast, for diffusive metallic
wires with interactions the situation is less clear. For discussion,
the reader is addressed to Section \ref{BoltzmannL}.  

In this subsection, we assume that the system is {\em charge neutral},
{\em i.e.} there is no pile-up of charge. This charge neutrality is
normally provided by Coulomb interactions, which thus play an
important role. If this is not the case, one can get different
results, like for resonant tunneling quantum wells with charging
(Section \ref{Langevin}) or quantum dots in the Coulomb blockade
regime (Section \ref{strcorr}). 

{\bf Inelastic scattering}. We begin the discussion 
with the strongest scattering processes which lead to carrier energy
relaxation and consequently also energy dissipation. 
Physically, this may correspond to electron-phonon scattering. To
simulate this process, we consider a two-terminal structure in the
conceptually simple case where we add only one fictitious voltage
probe\footnote{There are a number of works treating the effect
of distributed dephasing (many voltage probes) on conduction processes
(see for instance Ref. \cite{Brouwer97}). Results on inelastic
scattering on shot noise with several probes can be found in
Refs. \cite{Been92,Jong92,Jong96}.} (marked as $3$, see
Fig.~\ref{dephasfig}). As in our treatment of noise in multi-probe
conductors  we assume that all reservoirs (also the voltage probe
reservoir $3$) are characterized by Fermi distribution functions (for
simplicity, we only consider zero-temperature case). We take $\mu_2 =
0$ and $\mu_1 = eV$; the chemical potential $\mu_3$ is found from the
condition that at a voltage probe the current $I_3$ vanishes {\em
at any moment of time}. Then an electron which has left the conductor
and escaped into the reservoir of the voltage probe must immediately
be replaced by another electron that is reinjected from the voltage
probe into the conductor with an energy and phase which are
uncorrelated with that of the escaping reservoir. This approach to
inelastic scattering was applied to noise in interacting mesoscopic
systems in Ref. \cite{Buttiker91} (where only the average current was
taken to vanish), and the analysis for an instantaneously vanishing
current was presented in Ref. \cite{Been92}.  

\begin{figure}
\epsfxsize=7.cm{\centerline{\epsfbox{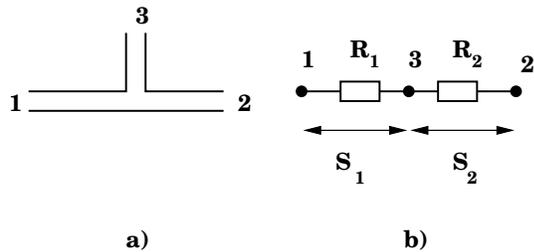}}}
\vspace{0.3cm}
\caption{(a) Setup with intermediate electrode. (b) Equivalent circuit
for the case when there is no direct transmission from $1$ to $2$.}
\protect\label{dephasfig}
\end{figure}
To proceed, we introduce transmission probabilities from the lead
$\alpha$ to the lead $\beta$, 
\begin{displaymath}
T_{\alpha\beta} = \mbox{\rm Tr}\ s^{\dagger}_{\alpha\beta}
s_{\alpha\beta}, \ \ \ \sum_{\beta} T_{\alpha\beta} = \sum_{\beta}
T_{\beta\alpha} = N_{\alpha},
\end{displaymath}
$N_{\alpha}$ being the number of transverse channels in the lead
$\alpha$. Currents can be then written as
\begin{eqnarray*}
I_1 & = & \frac{e^2}{2\pi\hbar} \left[ T_{12} eV + T_{13} (eV - \mu_3)
\right] + \delta I_1, \nonumber \\
I_2 & = & \frac{e^2}{2\pi\hbar} \left[ -T_{21} eV - T_{23}
\mu_3\right] + \delta I_2, \nonumber \\
I_3 & = & \frac{e^2}{2\pi\hbar} \left[ -T_{31} (eV - \mu_3) + T_{32}
\mu_3 \right] + \delta I_3, 
\end{eqnarray*}
where $\delta I_{\alpha}$ are the fluctuating parts of the currents
(each of them is zero on average), correlated according to Eq.
(\ref{noiset0}) with $\mu_3$ replaced by the
average $\langle \mu_3 \rangle$. The current conservation implies
$\delta I_1 + \delta I_2 + \delta I_3 = 0$. The requirement $I_3 = 0$
implies now that the electrochemical potential 
at the voltage probe is a fluctuating function of time $\mu_3 (t)$
given by 
\begin{equation} \label{dephasvolt}
\mu_3 = \frac{T_{31}}{T_{31}+T_{32}} eV +  \frac{2\pi\hbar}{e^2}
\frac{\delta I_1 + \delta I_2}{T_{31} + T_{32}}.
\end{equation}
The first term on the right hand side 
represents the average chemical potential $\langle
\mu_3 \rangle$, while the second one is a fluctuating correction.
The two-terminal conductance (defined according to $\langle I_2
\rangle = G V$) depends only on the average potential $\langle
\mu_3 \rangle$ and is given by \cite{Buttiker86,ButtikerIBM} 
\begin{equation} \label{dephascond}
G = \frac{e^2}{2\pi\hbar} \left[ T_{12} + \frac{T_{13}T_{23}}{T_{13}
+ T_{23}} \right].
\end{equation}
In Eq. (\ref{dephascond}) the probability $T_{12}$
describes coherent transmission, 
whereas the second term is the incoherent contribution.

Due to the fluctuations of the chemical potential $\mu_3$, the random
part of the current $I_1$ is now
\begin{displaymath}
\Delta I_1 = \delta I_1 - \frac{T_{13}}{T_{31} + T_{32}} \left(
\delta I_1 + \delta I_2 \right),
\end{displaymath}
and due to the current conservation the random part of the current
$I_2$ is the same with the opposite sign. Now the expression for the
shot noise power $S_{11}$ can be easily obtained from
Eq. (\ref{noiset0}), but it is rather cumbersome in the general
case. In the following, we consider only the {\em fully incoherent
case} $T_{12} = 0$, when there is no direct transmission from $1$ to
$2$: Every carrier on its way from contact $1$ to $2$ enters the
electrode $3$ with the probability one. This condition also implies
$T_{\alpha 3} = T_{3\alpha}$. Essentially, the fully incoherent case
means that the two parts of the system, from $1$ to $3$, and from $3$
to $1$, are resistors which add classically. In particular, the
conductance Eq. (\ref{dephascond}) contains now only the second term,
which now just states that two consecutive incoherent scatterers
exhibit a resistance which is equal to the series resistance. For the
current correlations, we obtain $S_{11} = S_{22} = -S_{12} = -S_{21}$
with      
\begin{equation} \label{dephasnois1}
S_{11} = \frac{e^3 \vert V \vert}{\pi\hbar} \frac{T_{23}^2 \mbox{\rm
Tr} \left[ s_{11} s^{\dagger}_{11} s_{13} s^{\dagger}_{13} \right] +
T_{13}^2 \mbox{\rm Tr} \left[ s_{22} s^{\dagger}_{22} s_{23}
s^{\dagger}_{23} \right]}{\left( T_{13} + T_{23} \right)^3}. 
\end{equation}

This expression can be re-written in the following transparent manner:
First, we define the resistances of the parts of the system between
$1$ and $3$, $R_{1} = 2\pi\hbar/(e^2 T_{13})$, and between $2$ and
$3$, $R_{2} = 2\pi\hbar/(e^2 T_{23})$. The total resistance between
$1$ and $2$ is given by $R = R_{1} + R_{2}$. Now the voltage drop
between $1$ and $3$ is $eVR_1/R$ and between $3$ and $2$ is
$eVR_2/R$. Taking this into account, the noise power measured between
$1$ and $3$ is (Fig.~\ref{dephasfig}),  
\begin{displaymath}
S_1 = \frac{e^3 \vert V \vert}{\pi\hbar} \frac{R_1}{R} \mbox{\rm Tr}
\left[ s_{13} 
s^{\dagger}_{13} \left( 1 - s_{13} s^{\dagger}_{13} \right) \right],  
\end{displaymath}
and noise power measured between $2$ and $3$ is,  
\begin{displaymath}
S_2 = \frac{e^3 \vert V \vert }{\pi\hbar} \frac{R_2}{R} \mbox{\rm Tr}
\left[ s_{23} s^{\dagger}_{23} \left( 1 - s_{23} s^{\dagger}_{23}
\right) \right].  
\end{displaymath}
Now we can 
write Eq. (\ref{dephasnois1}) as \cite{Been92} ($S \equiv S_{11}$) 
\begin{equation} \label{dephasvoltfl}
R^2 S = R_1^2 S_1 + R_2^2 S_2.
\end{equation}
The meaning of Eq. (\ref{dephasvoltfl}) is obvious if we realize that
$R^2S$ is the voltage fluctuations (for an infinite external impedance
circuit) across the whole conductor. The right hand side is just a sum
of voltage fluctuations from $1$ and $3$, and from $3$ to $2$. Thus,
Eq. (\ref{dephasvoltfl}) states nothing but that the voltage
fluctuations are additive.  

Another form of Eq. (\ref{dephasvoltfl}) is useful \cite{Shimizu92}. 
We introduce the noise suppression factors $F_1 = S_1R/(2e \vert V
\vert)$ and $F_2 = S_2R/(2e \vert V \vert)$ in the first and second
resistor. For the Fano factor of the whole system we obtain
\begin{equation} \label{Fanodephas}
F = \frac{R_1^2 F_1 + R_2^2 F_2}{(R_1 + R_2)^2}. 
\end{equation}

First of all, we now evaluate Eq. (\ref{Fanodephas}) for the case of
the double-barrier structure, where the intermediate electrode is
placed between the barriers. Physically, this would correspond to 
strong inelastic scattering inside the quantum well --- in contrast to
the quantum-mechanical discussion of the previous subsection, which
implicitly requires full phase coherence. Taking into account that for
high barriers $F_1 = F_2 = 1$, and that the resistances $R_1$ and
$R_2$ are inversely proportional to the tunneling rates $\Gamma_L$ and
$\Gamma_R$, respectively, we immediately arrive at
Eq. (\ref{Fanodbt}), {\em i.e.} the result for the fully coherent
case\footnote{Lund B\o \ and Galperin \protect\cite{Bo1} performed
microscopic calculation of noise in quantum wells in transverse
magnetic field with the account of electron-phonon scattering
(phonon-assisted tunneling). They found that the Fano factor is
suppressed by inelastic scattering, as compared with non-interacting
value. This result clearly contradicts to the conclusions of this
subsection; we presently do not understand the reasons for this
discrepancy.}. We thus see that even though inelastic scattering
modifies both the conductance and shot noise of the resonant tunneling
structure, it leaves the Fano factor unchanged. This statement, due to
Chen and Ting \cite{Chen92} and Davies {\em et al} \cite{Davies92},
will be again demonstrated in Section \ref{Langevin}, where the
derivation of Eq. (\ref{Fanodbt}) based on a classical Langevin
approach (which corresponds to the absence of quantum coherence) is
presented.   

Next we consider a quasi-one-dimensional geometry and assume for a
moment that the lead $3$ divides the wire into two identical
parts. Then in Eq. (\ref{Fanodephas}) $R_1 = R_2$, $F_1 = F_2$, and we
obtain $F = F_1/2$. Thus, the Fano factor of the whole wire is one
half of the noise measured in each segment.  

This result, which describes {\em local} inelastic scattering in the
middle of the wire, can be generalized to {\em uniform} inelastic
scattering. For this purpose we introduce a certain length $L_i$
associated with inelastic scattering; we assume that $L_i$ is much
shorter than the total length of the wire $L$. One must then consider
initially a conductor with $N_i = L/L_i$ additional fictitious voltage
probes separated by distances $L_i$ along the conductor. In the fully
incoherent case this picture is equivalent to $N_i$ identical
classical resistors connected in series. The Fano factor of this
system is then the Fano factor of the phase-coherent segment divided
by $N_i$. In particular, if the wire is diffusive, the suppression
factor\footnote{Of course, not only the average of the noise vanishes
in macroscopic system, but also fluctuations. De Jong and Beenakker
found $\mbox{\rm r.m.s.} S \propto N_i^{-5/2}$. The weak localization
correction to shot noise decreases as $N_i^{-2}.$} is
\cite{Been92,Jong92} $F = (3N_i)^{-1}$. Thus, the conclusion is the
following: Inelastic scattering suppresses shot noise. A {\em
macroscopic} system (large compared to an inelastic scattering length,
$N_i \gg 1$) exhibits no shot noise. This is a well known fact, the
absence  of shot noise of macroscopic conductors is used to stabilize
lasers.  

Shimizu and Ueda \cite{Shimizu92} and Liu and Yamamoto
\cite{Liu941,Liu942} provided a similar discussion of noise
suppression in the crossover regime between mesoscopic behavior and 
classical circuit theory (macroscopic behavior). Liu, Eastman, and
Yamamoto \cite{Eastman} performed Monte Carlo simulations of shot
noise and included explicitly electron-phonon scattering. 

{\bf Quasi-elastic scattering}. In contrast to inelastic scattering,
dephasing processes leave the energy essentially invariant. To
simulate a scattering process which destroys phase but leaves the
energy invariant we now have to consider a special voltage probe. We
require that the additional electrode conserves not only the total
current, but also the current {\em in each small energy interval}
\cite{Jong96}. Such a voltage probe will not give rise to energy
relaxation and dissipates no energy.  

From the condition that the current in each energy interval vanishes 
we find that the distribution function in the reservoir of the voltage
probe is given by 
\begin{equation} \label{dephasfun}
f_3(E) = \frac{G_1 f_1 (E) + G_2 f_2(E)}{G_1 + G_2},
\end{equation}
where $f_1$ and $f_2$ are the Fermi functions at the reservoirs $1$
and $2$, and the conductances are $G_1 = R_1^{-1}$, $G_2 =
R_2^{-1}$. We have assumed again that there is no direct transmission
between $1$ and $2$. Straightforward calculation gives for the Fano
factor 
\cite{Jong96}  
\begin{equation} \label{Fanodephas1}
F = \frac{R_1^3 F_1 + R_2^3 F_2 + R_1^2 R_2 + R_1 R_2^2}{(R_1 +
R_2)^3}. 
\end{equation} 
We analyze now this result for various situations. First, we see that
for a ballistic wire divided by a dephasing electrode into two parts,
$F_1 = F_2 = 0$, shot noise does not vanish (unlike
Eq. (\ref{Fanodephas1})). We obtain $F = R_1 R_2 (R_1 +
R_2)^{-2}$. Thus for a ballistic system, which is ideally noiseless, 
dephasing leads to the appearance of shot noise. 

For the strongly biased resonant double-barrier structure, we have
$F_1 = F_2 = 1$, and obtain again the result (\ref{Fanodbt}), which is
thus insensitive to dephasing. 

For metallic diffusive wires, $F_1 = F_2 = 1/3$,
Eq. (\ref{Fanodephas1}) yields for the ensemble averaged Fano factor
$F=1/3$ independent on the location of the dephasing voltage probe,
i. e. independent of the ratio of $R_1$ and $R_2$. Thus, our
consideration indicates that the noise suppression factor for metallic
diffusive wires is also insensitive to dephasing, at least when the
dephasing is local (in any point of the sample). This result hints 
that the Fano factor of an ensemble of metallic diffusive wires is not
sensitive to dephasing even if the latter is uniformly distributed. 
Indeed, de Jong and Beenakker \cite{Jong92} checked this by coupling
locally a dephasing reservoir to each point of the sample. Already at
the intermediate stage their formulae coincide with those obtained
classically by Nagaev \cite{Nagaev92}. This  proves that
introducing dephasing with fictitious, energy conserving voltage probes
is in the limit of complete dephasing equivalent to the
Boltzmann-Langevin approach. That inelastic scattering, and not
dephasing, is responsible for the crossover to the macroscopic regime,
has been recognized by Shimizu and Ueda \cite{Shimizu92}.  

The effect of phase breaking on the shot noise in chaotic cavities 
was investigated by van Langen and one of the authors \cite{Langen97}.
For a chaotic cavity connected to reservoirs via quantum point
contacts with $N_L$ and $N_R$ open quantum channels, the Fano factors 
vanish $F_1 = F_2 = 0$, and since  $R_{1} = \pi\hbar/e^{2}N_L$ and
$R_{2} = \pi\hbar/e^{2}N_R$, the resulting Fano factor is given by
Eq. (\ref{Fanochaot}), {\em i.e.} it is identical with the result that
is obtained from a completely phase coherent, quantum mechanical
calculation. Thus for chaotic cavities, like for metallic diffusive
wires, phase breaking has no effect on the ensemble averaged noise
power.   

{\bf Electron heating}. This is the third kind of inelastic
scattering, which implies that energy can be exchanged between
electrons. Only the total energy of the electron subsystem is
conserved. Physically, this corresponds to electron-electron
scattering. Within the voltage probe approach, it is taken into
account by including the reservoir $3$, with chemical potential
$\mu_3$ determined to obtain zero (instantaneous) electrical current
and a temperature $T_3$, which is generally different from the lattice
temperature (or the temperature of the reservoirs $1$ and $2$), to
obtain zero (instantaneous) energy flux. For a detailed discussion we
refer the reader to the paper by de Jong and Beenakker \cite{Jong92},
here we only mention the result for two identical diffusive conductors
at zero temperature\footnote{The temperature of the reservoirs $1$ and
$2$ is zero. The temperature of the intermediate reservoir is in this
case \cite{Jong92} $k_B T_3 = (\sqrt{3}/2\pi)e \vert V \vert$.}. The
Fano factor is in this case $F \approx 0.38$, which is {\em higher}
than the $1/3$-suppression for the non-interacting case. We will see in
Section \ref{BoltzmannL} that the classical theory also predicts shot
noise enhancement for the case of electron heating.    

{\bf Intermediate summary}. Here are the conclusions one can draw from
the simple consideration we presented above.

\begin{itemize}

\item Dephasing processes do not renormalize the 
ensemble averaged shot noise power (apart from weak localization
corrections, which are destroyed by dephasing). In particular, this
statement applies to metallic diffusive wires, chaotic cavities, 
and resonant double barrier structures.  

\item Inelastic scattering renormalizes even the ensemble 
averaged shot noise power: A macroscopic sample exhibits no shot
noise. An exception is the resonant double barrier structure, subject
to a bias large compared to the resonant level width. Under this
condition neither the conductance nor the shot noise of a double
barrier are affected.   

\item As demonstrated for metallic diffusive wires 
electron heating enhances noise. 

\end{itemize}

The last statement implies the following scenario for noise in
metallic diffusive wires \cite{Steinbach}. There exist three inelastic
lengths, responsible for dephasing ($L_1$), electron heating ($L_2$)
and inelastic scattering ($L_3$). We expect $L_1 < L_2 < L_3$. Indeed,
requirements for dephasing (inelastic scattering) are stronger
(weaker) than those for electron heating. Then for the wires with
length $L \ll L_2$ the Fano factor equals $1/3$ and is not affected by
inelastic processes; for $L_2 \ll L \ll L_3$ it is {\em above} $1/3$,
and for $L \gg L_3$ it goes down and disappears as $L \to \infty$. 

In Section \ref{BoltzmannL} we will reconnect to the results
presented here within the classical Boltzmann-Langevin approach.

\section{Scattering theory of frequency dependent noise spectra}
\label{freq} 

\subsection{Introduction. Current conservation} \label{freq031}

The investigation of frequency dependent transport, in particular,
noise, is important, since it can reveal information about internal
energy scales of mesoscopic systems, not available from dc
transport. On the other hand, the investigation of the dynamic noise
is a  more difficult task than the investigation of quasi-static
noise. This is true experimentally, since frequency dependent
measurements require a particularly careful control of the measurement
apparatus (one wants to see the capacitance of the sample and not that
of the coaxial cable connecting to the measurement apparatus), and it
is also true theoretically. Addressing specifically mesoscopic
systems, the conceptual difficulty is that generally it is meaningless
to consider the dynamic response of non-interacting electrons. Since
this point remains largely unappreciated in the literature, we give
here a brief explanation of this statement. Consider the following
system of equations of classical electrodynamics\footnote{For
simplicity, we assume the lattice dielectric constant to be uniform
and equal to one.},  
\begin{eqnarray}
& & \bbox{j} = \bbox{j}_p + \frac{1}{4\pi} \frac{\partial
\bbox{E}}{\partial t}; \ \ \ \bbox{E} = -\nabla \varphi;
\label{curdefin} \\
& & \nabla^2 \varphi = -4\pi \rho; \label{Poissoneq} \\
& & \mbox{\rm div} \ \bbox{j}_p + \frac{\partial \rho}{\partial t}
= 0. \label{contineq} 
\end{eqnarray}
Here $\bbox{j}$, $\varphi$ and $\rho$ are the density of the
electric current (particle current), the electric potential, and the
charge density, respectively; $\bbox{E}$ is the electric
field. Equation (\ref{curdefin}) states that the {\em total} current
$\bbox{j}$ is a sum of the particle current $\bbox{j}_p$ and
the displacement current, represented by the second term on the
rhs. Equations (\ref{Poissoneq}) and (\ref{contineq}) are the Poisson
equation and the continuity equation, respectively; they must be
supplemented by appropriate boundary conditions. We make the following
observations. 

(i) Equations (\ref{curdefin}), (\ref{Poissoneq}), and
(\ref{contineq}), taken together, yield $\mbox{\rm div} \ \bbox{j}
= 0$: the total current density has neither sources nor sinks. This is
a general statement, which follows entirely from the basic equations
of electrodynamics, and has to be fulfilled in any system. Theories
which fail to yield a source and sink free {\em total} current density
cannot be considered as correct. The equation $\mbox{div} \ \bbox{j}
= 0$ is a necessary condition for the current conservation, as it was
defined above (Section \ref{scat}).   

(ii) The particle current $\bbox{j}_p$ is generally not
divergenceless, in accordance with the continuity equation
(\ref{contineq}), and thus, is not necessarily conserved. To avoid 
a possible misunderstanding, we emphasize that it is not a mere
difference in definitions: The experimentally measurable quantity is
the {\em total} current, and not the {\em particle} current. Thus,
experimentally, the fact  that the particle current is not conserved
is irrelevant.  

(iii) The Poisson equation (\ref{Poissoneq}), representing
electron-electron interactions, is crucial to ensure the conservation
of the total current. This means that the latter cannot be generally
achieved in the free electron model, where the self-consistent
potential $\varphi$ is replaced by the external electric potential.

(iv) In the static case the displacement current is zero, and the
particle current alone is conserved. In this case the self-consistent
potential distribution $\varphi (\bbox{r})$ is generally also
different from the external electrostatic potential, but to linear
order in the applied voltage the conductance is determined only by the
total potential difference. As a consequence of the Einstein relation
the detailed spatial variation of the potential in the interior of the
sample is irrelevant and has no effect on the total current.    

In mesoscopic physics the problem is complicated since a sample is
always a part of a larger system. It interacts with the nearby gates
(used to define the geometry of the system and to control the number
of charges in the system). Thus a complete solution of the above
system of equations is usually a hopeless task without some serious
approximations. The theoretical task is to choose idealizations and
approximations which are compatible with the basic conservation laws
expressed by the above equations. For instance, we might want to
describe interactions in terms of an effective (screened) interaction
instead of the full long range Coulomb interaction. It is then
necessary to ensure that such an effective interaction indeed leads to
the conservation of current.  

Three frequency dependent types of noise spectra should be
distinguished: (i) Finite-frequency noise at equilibrium or in the
presence of dc voltage. (ii) Zero-frequency noise in the presence of
an ac voltage; the resulting spectrum depends on the frequency of the
ac-voltage. (iii) Finite-frequency noise in the presence of an ac
voltage; this quantity depends on two frequencies. Here we are
interested mostly in the first type of noise spectra; the second one
is only addressed in subsection \ref{freq033}. We re-iterate that,
generally, one can {\em not} find the  ac conductance and the current
fluctuations from a non-interacting model. Even the finite frequency
current-current correlations (noise) at equilibrium or in the presence
of a dc voltage source, which are of primary interest in this Section,
can {\em not} be treated without taking account interactions. A simple
way to see this is to note that due to the fluctuation-dissipation
theorem, the equilibrium correlation of currents in the leads $\alpha$
and $\beta$ at finite frequency, $S_{\alpha\beta} (\omega)$, is
related to the corresponding element of the conductance matrix,
$S_{\alpha\beta} (\omega) = 2k_B T [G_{\alpha\beta} (\omega) +
G^*_{\beta\alpha} (\omega)]$. The latter is a response to average
current in the lead $\alpha$ to the ac voltage applied to the lead
$\beta$, and is generally interaction-sensitive. Thus, calculation of
the quantity $S_{\alpha\beta} (\omega)$ also requires a treatment of
interactions to ensure current conservation.     

We can now be more specific and make the same point by looking at
Eq. (\ref{noisefr}) which represents the fluctuations of the {\em
particle current} at finite frequency. Indeed, for $\omega = 0$ the
current conservation $\sum_{\alpha} S_{\alpha\beta} = 0$ is guaranteed
by the unitarity of the scattering matrix: The matrix $A (\alpha,
E,E)$ (Eq. (\ref{curmatr3})) contains a product of two scattering
matrices taken {\em at the same energy}, and therefore it obeys the
property $\sum_{\alpha} A(\alpha, E, E) = 0$. On the other hand, for
finite frequency the same matrix $A$ should be evaluated at two
different energies $E$ and $E+\hbar\omega$, and contains now a product
of two scattering matrices taken {\em at different energies}. These
scattering matrices generally do not obey the property $\sum_{\alpha}
s^{\dagger}_{\alpha\beta} (E) s_{\alpha\gamma}(E+\hbar\omega) =
\delta_{\beta\gamma}$, and the current conservation is not fulfilled:
$\sum_{\alpha} S_{\alpha\beta} (\omega) \ne 0$.  

Physically, this lack of conservation means that there is charge
pile-up inside the sample, which gives rise to displacement
currents. These displacement currents restore current conservation,
and thus need to be taken into account. It is exactly at this stage
that a treatment of interactions is required. Some  progress in this
direction is reviewed in this Section later on.

It is sometimes thought that there are situations when displacement
currents are not important. Indeed, the argument goes, there is always
a certain energy scale $\hbar\omega_c$, which determines the energy
dependence of the scattering matrices. This energy scale is set by 
the level width (tunneling rate) $\Gamma$ for resonant tunnel
barriers, the Thouless energy $E_c = \hbar D/L^2$ for metallic
diffusive wires ($D$ and $L$ are the diffusion coefficient and length
of the wire, respectively), and the inverse Ehrenfest time ({\em i.e.}
the time for which an electron loses memory about its initial position
in phase space) for chaotic cavities. The scattering matrices may be
thought as energy independent for energies below $\hbar\omega_c$. Then
for frequencies below $\omega_c$ we have $\sum_{\alpha}
s_{\alpha\beta}^{\dagger} (E) s_{\alpha\gamma}(E+\hbar\omega) \sim
\delta_{\beta\gamma}$, and the unitarity of the scattering matrix
assures current conservation, $\sum_{\alpha} S_{\alpha\beta} =
0$. However, there are time-scales which are not set by the carrier
kinetics, like $RC$-times which reflect a collective charge response
of the system. In fact, from the few examples for which the ac
conductance has been examined, we know that it is the collective times
which matter\footnote{Exceptions are frequency dependent weak
localization corrections \cite{allar} which depend in addition to the
RC-time also on the dwell time \cite{brbu}, and perfect ballistic
wires which have a charge neutral mode determined by the transit time
as the lowest collective mode \cite{bhb}.}. Displacement currents can
be neglected only, if we can assure that these collective times are
much shorter than any of the kinetic time-scales discussed above. 
Furthermore, there are problems which can only be treated by taking 
interactions into account, even at arbitrarily low frequency: Later on
in this Section we will discuss the RC-times of mesoscopic  
conductors capacitively coupled to a gate. The noise induced into a
nearby capacitor is proportional to the square of the frequency. Naive
discussions which do not consider the energy dependence of the
scattering matrix cannot predict such currents.  

Let us at first consider the range of frequencies that are much
smaller than any inverse kinetic time scale and smaller than any
inverse collective response time. This case is in some sense trivial,
since the energy dependence of the scattering matrices is neglected, 
and the system is now not probed on the scale $\omega_c$. As a
consequence there is no novel information on the system compared to a
zero frequency noise measurement. In this case, as we will see, the
entire frequency dependence of the noise is due to the frequency
dependence of the Fermi functions. However, it is the low-frequency
measurements which are more easily carried out, and therefore there is
some justification to discuss noise spectra in this frequency
interval.  

The rest of the Section is organized as follows. Subsection
\ref{freq032} treats fluctuations of the particle current of
independent electrons, either at equilibrium or in the presence of a
dc bias, in the regime when the scattering matrices can be assumed to
be frequency independent. Subsection \ref{freq033} generalizes the
same notions for noise caused by an ac bias. Afterwards, we relax the
approximation of the energy independence of scattering matrices, and
in subsections \ref{freq034} and \ref{freq035} consider two simple
examples. In both of them Coulomb interactions prove to be
important. Though the theory of ac noise is far from being completed,
we hope that these examples, representing the results available by now
in the literature, can stimulate further research in this direction.

In this Section, we only review the quantum-mechanical description of
frequency dependent noise, based on the scattering approach. 
Alternatively, the frequency dependence of shot noise in diffusive
conductors may be studied based on the classical Boltzmann-Langevin 
approach
\cite{Naveh97,Nagaev98,Naveh99f,Nagaev981,Naveh981,Naveh982}. These
developments are described in Section \ref{BoltzmannL}.   

\subsection{Low-frequency noise for independent electrons: at
equilibrium and in the presence of dc transport} \label{freq032}

{\bf General consideration.} This subsection is devoted to
low-frequency noise, in a regime where the scattering matrices are
energy independent. We take the frequency, temperature and the voltage
all below $\omega_c$, and below any frequencies associated with the
collective response of the structure. For simplicity we only consider
the two-terminal case, $\mu_L = eV$, $\mu_R = 0$, $V \ge 0$. We
emphasize again that in this approach the internal energy scales of
mesoscopic conductors cannot be probed, nor is there a manifestation
of the collective modes. The frequency dependence of noise is entirely
due to the Fermi functions.    

Our starting point is Eq. (\ref{noisefr}), which in this form is given
in Ref. \cite{Buttikerfreq92}. Taking the scattering matrices to be
energy independent, we write
\begin{eqnarray} \label{noisfr2}
S(\omega) & \equiv & S_{LL} (\omega) = \frac{e^2}{2\pi\hbar} \left\{
\sum_n T_n^2 \int dE \right. \nonumber \\
& \times & \left. \left[ f_{LL} (E, \omega) + f_{RR} (E, \omega)
\right]  + \sum_n T_n \left( 1 - T_n \right) \right. \nonumber \\
& \times & \left. \int dE \left[ f_{LR} (E, \omega) + f_{RL} (E,
\omega) \right] \right\},   
\end{eqnarray}
with the abbreviation 
\begin{eqnarray} \label{fermicom}
f_{\alpha\beta} (E, \omega) & = & f_{\alpha} (E) \left[ 1 - f_{\beta}
(E + \hbar\omega) \right] \nonumber \\
& + & \left[ 1 - f_{\alpha} (E) \right] f_{\beta} (E + \hbar\omega). 
\end{eqnarray}
Here the $T_n$'s are, as before, (energy independent) transmission
coefficients. Performing the integration, we obtain
\begin{eqnarray} \label{noisfr3}
S(\omega) & = & \frac{e^2}{2\pi\hbar} \left\{ 2\hbar\omega \coth
\left( \frac{\hbar\omega}{2k_BT} \right) \sum_n T_n^2 \right. \nonumber
\\ 
& + & \left. \left[ \left( \hbar\omega + eV \right) \coth \left(
\frac{\hbar\omega + eV}{2k_BT} \right) \right. \right. \\
& + & \left. \left. \left( \hbar\omega - eV \right) \coth \left(
\frac{\hbar\omega - eV}{2k_BT} \right) \right] \sum_n T_n \left( 1 -
T_n \right) \right\}. \nonumber
\end{eqnarray}
This formula expresses the noise spectral power for arbitrary
frequencies, voltages, and temperatures (all of them are assumed to be
below $\omega_c$). The frequency dependent functions in Eq. 
(\ref{noisfr3}) are obtained already in the discussions of noise based
on the tunneling Hamiltonian approach for junctions \cite{Dahm} (see
also Refs. \cite{Rogovin74,BJMS,Schoen}). In this approach one expands
in the tunneling probability, and consequently, to leading order,
terms proportional to $\sum_n T_n^2$ are disregarded. The full
expression Eq. (\ref{noisfr3}) including the terms proportional to
$T_n^{2}$ was derived by Khlus \cite{Khlus} assuming from the outset
that the scattering matrix is diagonal. It is a general result for an
arbitrary scattering matrix, if the $T_{n}$'s are taken to be the the
eigenvalues of $t^{\dagger}t$. Later the result of Khlus was
re-derived by Yang \cite{EYang} in connection with the QPC; the
many-channel case was discussed by Ueda and Shimizu \cite{Ueda93}, Liu
and Yamamoto \cite{Liu942}, and Schoelkopf {\em et al} \cite{Prober1}.  
\begin{figure}
{\epsfxsize=7.cm\centerline{\epsfbox{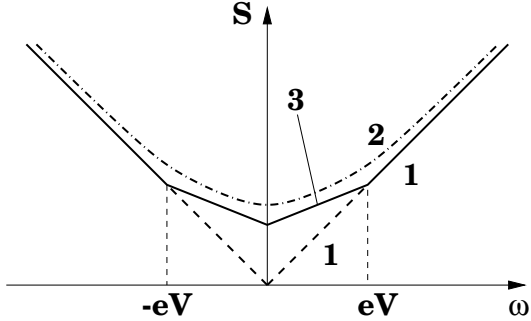}}}
\vspace{0.3cm}
\caption{Frequency dependence of noise for zero temperature,
(Eq. (\ref{noisfr4}), solid line), and finite temperatures
(dash-dotted line, $2$). Line $1$ shows equilibrium noise, $S = e^2
\vert \omega \vert \sum_n T_n/\pi$. Line $3$ corresponds to the
upper line of Eq. (\ref{noisfr4}). In this figure we set $\hbar = 1$.}
\protect\label{freq301}
\end{figure}

We note first that for $\omega = 0$, Eq. (\ref{noisfr3}) reproduces
the results for thermal and shot noise presented in Section \ref{scat}
(for the two-terminal case and energy independent transmission
coefficients). Furthermore, at equilibrium ($V = 0$) it gives 
the Nyquist noise,
\begin{equation} \label{noisfr5}
S_{eq} (\omega) = \frac{e^2\omega}{\pi} \coth \left(
\frac{\hbar\omega}{2k_BT} \right) \sum_n T_n,  
\end{equation}
as implied by the fluctuation-dissipation theorem. For zero
temperature we obtain from Eq. (\ref{noisfr3}) 
\begin{equation} \label{noisfr4}
S(\omega) =  \frac{e^2}{\pi\hbar} 
\end{equation}
\begin{eqnarray*}
\times \left\{ \protect\matrix{ \hbar\vert
\omega \vert \sum_n T_n^2 + eV \sum_n T_n (1-T_n), & \hbar \vert
\omega \vert < eV \nonumber \\ \hbar\vert \omega \vert \sum_n T_n, &
\hbar \vert \omega \vert > eV }. \right. \nonumber 
\end{eqnarray*}
The frequency dependence is given by a set of straight lines. For zero
frequency, the result for shot noise is reproduced, $S =
(e^3V/\pi\hbar) \sum_n T_n (1-T_n)$. At higher frequencies
the spectral density increases, and for any finite $\omega$ it is not
proportional to the voltage any more. Thus, for finite frequency, like
for finite temperature, we do not have pure shot noise. At
$\hbar\omega = \pm eV$ the noise spectral power has a discontinuous
derivative, and for $\vert \omega \vert > eV/\hbar$ the noise spectrum
tends to the equilibrium value Eq. (\ref{noisfr5}) determined by the
zero-point quantum fluctuations, independent of the voltage. Finite
temperature smears the singularities since now the Fermi functions 
are continuous. The noise spectrum is shown in Fig.~\ref{freq301}. It
can also be represented differently, if we define excess noise (for
$k_BT = 0$) as the difference between the full noise power
Eq. (\ref{noisfr4}) and equilibrium noise, $S_{ex} (\omega) =
S(\omega) - e^2 \vert \omega \vert \sum_n T_n/\pi$. The excess noise
is given by  
\begin{displaymath}
S_{ex} (\omega) = \frac{e^2}{\pi\hbar} \sum_n T_n \left( 1 - T_n
\right) \left( eV - \hbar \vert \omega \vert \right), \ \ \ \hbar
\vert \omega \vert \le eV, 
\end{displaymath}  
and zero otherwise. 

Eq. (\ref{noisfr3}) is general and valid for all systems under the
conditions it was derived. Instead of discussing it for all the
examples mentioned in Section \ref{scat}, we consider only the
application to a metallic diffusive wire.  

{\bf Metallic diffusive wires.} 
Performing the disorder averages of transmission coefficients with the
distribution function (\ref{Tdistrib}), we find the result obtained
earlier by Altshuler, Levitov, and Yakovets \cite{Yakovets}, 
\begin{eqnarray} \label{noisfr6}
S(\omega) & = & \frac{1}{3} G \left\{ 4
\hbar\omega \coth \left( \frac{\hbar\omega}{2k_BT} \right)
\right. \nonumber \\
& + & \left. \left( \hbar\omega + eV \right) \coth \left(
\frac{\hbar\omega + eV}{2k_BT} \right) \right. \nonumber \\
 & + & \left. \left( \hbar\omega - eV \right) \coth \left(
\frac{\hbar\omega - eV}{2k_BT} \right) \right\},  
\end{eqnarray}
where $G = e^2N_{\perp}l/2\pi\hbar L$ is the conductance of a wire
with mean free path $l$ and length $L$. 
\begin{figure}
{\epsfxsize=6.cm\centerline{\epsfbox{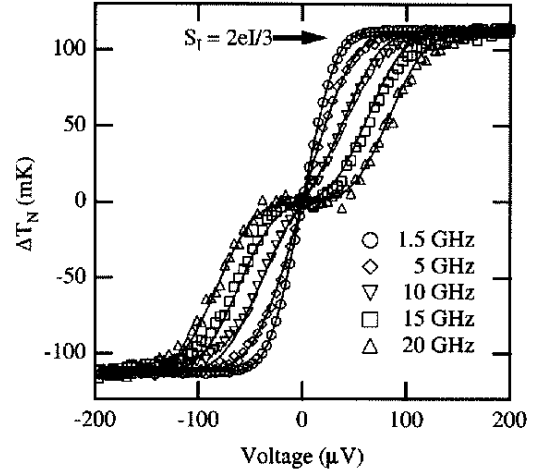}}}
\vspace{0.3cm}
\caption{Experimental results of Schoelkopf {\em et al}
\protect\cite{Prober1} for the frequency dependence of noise in
metallic diffusive wires. Solid lines for each frequency indicate the
theoretical result (\protect\ref{noisfr6}). The quantity shown on the
vertical axis is essentially $\partial S/\partial V$.} 
\label{diffus33}
\end{figure}
An experimental investigation of the frequency dependent noise in
diffusive gold wires is presented by Schoelkopf {\em et al}
\cite{Prober1}; this paper reports one of the only two presently
available measurements of ac noise. They find a good agreement with
Eq. (\ref{noisfr6}), where the electron temperature $T$ was used as a
fitting parameter\footnote{Due to effects of electron heating; see
Section \ref{BoltzmannL}.} (the same for all frequencies). The results
of Ref. \cite{Prober1} are plotted in Fig.~\ref{diffus33} as a
function of voltage for different frequencies; theoretical curves are
shown as solid lines. In the experiment, the highest frequency
corresponded approximately to the Thouless energy, and in this regime
Eq. (\ref{noisfr6}) is well justified.       

We have emphasized earlier that in this frequency regime no internal
dynamics of the system is probed. The fact that Eq. (\ref{noisfr6})
agrees with experiment is a consequence of the strong screening in
metals. In poor metals, the RC-times might become long. A finite value
of the screening length may permit charge fluctuations and
consequently modify the noise behavior even at relative low
frequencies. We address this issue in Section \ref{BoltzmannL}.  

{\bf Inelastic scattering.} Now one can ask:  What is the effect of
inelastic scattering on frequency dependent noise in the regime, where
the scattering matrices can be taken to be {\em energy
independent}. This problem was studied by Ueda and Shimizu
\cite{Ueda93}, and later by Zheng, Wang, and Liu \cite{Zheng95}, who
included electron-phonon interaction directly into the scattering
approach, and by Liu and Yamamoto \cite{Liu942}, who used an approach
based on dephasing voltage probes. The general conclusion is, that 
like in the case of the zero frequency, inelastic scattering
suppresses noise\footnote{In contrast, the effect of dephasing on the
finite frequency noise seems not to have been investigated.}. 

\subsection{Low-frequency noise for independent electrons:
Photon-assisted transport} \label{freq033}

Now we generalize the results of the preceding subsection to
the case when the applied voltage is time-dependent; the scattering
matrices are still assumed not to depend on energy. The
fluctuations in the presence of a potential generated by an ac
magnetic flux were treated by Lesovik and Levitov \cite{Lesovik94};
the fluctuation spectrum in the presence of oscillating voltages 
applied to the contacts of the sample was obtained in
Ref. \cite{Pedersen98}. The results are essentially the same; below we
follow the derivation of Ref. \cite{Pedersen98}. 

We consider a two-terminal conductor; the chemical potential of the
right reservoir is kept fixed (we assume it to be zero), while the
left reservoir is subject to a constant voltage $\bar V$ plus the
oscillating component $U(t) = V(\Omega) \cos \Omega t$. One cannot
simply use the stationary scattering theory as described in
Section \ref{scat}. Instead, the scattering states in the left lead
are now solutions of the time-dependent Schr\"odinger equation,  
\begin{eqnarray} \label{asympac}
\psi_n(\mbox{\bf r},E,t) & = & \chi_{Ln} \left(\bbox{r}_{\perp}\right)
e^{ik_{Ln}z-iEt/\hbar} \nonumber \\ 
& \times & \sum_{l = -\infty}^{\infty} J_l 
\left( \frac{eV(\Omega)}{\hbar\Omega} \right) e^{-il\Omega t}.
\end{eqnarray} 
Thus, in the presence of an oscillating voltage each state with the
central energy $E$ is split to infinitely many subbands with
energies $E + l\hbar \Omega$, which have smaller spectral weight. 
Following the literature on tunneling, this phenomenon is called {\em
photon-assisted transport}, since electrons with higher energies ($l >
0$) have higher transmission probabilities, and might propagate
through the sample thanks to the additional energy. 

Now we use a formal trick, assuming that the oscillating potential
only exists asymptotically far from the sample (and there
Eq. (\ref{asympac}) is valid), and decays slowly towards the
sample. Thus, there is a certain portion of the left lead, where there
is no oscillating potential, but still no scattering. The annihilation
operators in this part of the left lead have thus the form
\begin{displaymath}
\hat a_{Ln} (E) = \sum_l J_l \left( \frac{eV(\Omega)}{\hbar\Omega}
\right) \hat a'_{Ln} (E - l\hbar\Omega),
\end{displaymath}
where the operators $\hat a'$ describe the states of the left
reservoir. Instead of Eq. (\ref{cur2}), we obtain
\begin{eqnarray} \label{curac2}
\hat I_L (t) & = & \frac{e}{2\pi\hbar} \sum_{\alpha\beta} \sum_{mn}
\int dE dE' e^{i(E-E')t/\hbar} \nonumber \\
& \times & \sum_{lk} J_l \left(
\frac{eV_{\alpha}}{\hbar\Omega} \right) J_k \left(
\frac{eV_{\beta}}{\hbar\Omega} \right) \\
& \times & \hat a'^{\dagger}_{\alpha m} (E - l\hbar\Omega)
A_{\alpha\beta}^{mn} (L; E,E') \hat a'_{\beta n} (E' - k\hbar\Omega),
\nonumber 
\end{eqnarray}
and we set $V_L = V(\Omega)$, $V_R = 0$. Finally, we assume that the
frequency is not too high, so that the left reservoir can be
considered to be at (dynamic) equilibrium at any instant of time. Then
the averages of the operators $\hat a'$ are essentially equilibrium
averages, expressed through the Fermi functions, $f_L  = f_F (E -
e\bar V)$ and $f_R = f_F (E)$. 

In the presence of a time-dependent voltage, the correlation function
$S_{LL}$ (\ref{noisedef}) depends not only on the time difference
$t-t'$, but also on the absolute time $\tau = (t+t')/2$. In the
following, we are interested in the noise spectra on a time scale long
compared to $\Omega^{-1}$. Then the noise power can be averaged over
$\tau$,
\begin{displaymath}
S(t - t') = \frac{1}{\tilde\tau} \int_0^{\tilde\tau} d\tau S(t-t',
\tau), \ \ \ \tilde\tau = \frac{2\pi}{\Omega}.
\end{displaymath}
Leaving more general cases aside, we only give an expression for
the zero-frequency component of $S_{LL}$
\cite{Lesovik94,Pedersen98}, 
\begin{eqnarray} \label{curac3}
S_{LL} (\omega = 0, \Omega) & = & \frac{e^2}{2\pi\hbar} 
\sum_{\alpha\beta} \int dE \sum_{l} J_l^2 \left( \frac{e(V_{\alpha} -
V_{\beta})}{\hbar\Omega} \right) \nonumber \\ 
& \times & \mbox{\rm Tr} \ \left[ A_{\alpha\beta} (L) A_{\beta\alpha}
(L) \right] f_{\alpha\beta} (E , l \hbar \Omega),  
\end{eqnarray} 
where the matrices $A$ are explicitly assumed to be energy
independent. For zero external frequency, $\Omega = 0$, only the
Bessel function with $l = 0$ survives, and should be taken equal to
one for any $\alpha$ and $\beta$; then we reproduce the zero-frequency
expression (\ref{noise0}). Performing the energy integration and
introducing the transmission coefficients, we obtain \cite{Prober2} 
\begin{eqnarray} \label{patnoisefin}
S(\Omega) & = & \frac{e^2}{2\pi\hbar} \left\{ 4 k_BT \sum_n T_n^2
\right. \nonumber \\
& = & \left. \sum_{l = -\infty}^{\infty} J_l^2 \left( 
\frac{eV(\Omega)}{\hbar\Omega} \right) \sum_n T_n
\left( 1 - T_n \right) \right. \\
& \times & \left. \left[ \left( l \hbar \Omega +
e\bar V \right) \coth \left( \frac{l \hbar \Omega + e \bar V}{2k_BT}
\right) \right. \right. \nonumber \\
& + & \left. \left. \left( l \hbar \Omega - e\bar V \right) \coth  
\left( \frac{l \hbar \Omega - e\bar V}{2k_BT} \right) \right] \right\}
. \nonumber 
\end{eqnarray} 
For zero temperature, Eq. (\ref{patnoisefin}) exhibits singularities
at voltages $\bar V = l \hbar \Omega/e$: The derivative $\partial
S/\partial \bar V$ is a set of steps. The height of each step depends
on the ac voltage due to the Bessel function in
Eq. (\ref{patnoisefin}).  

Lesovik and Levitov \cite{Lesovik94} considered a geometry of (an
almost closed) one-channel loop of length L connected to two
reservoirs. The loop contains a scatterer, and is pierced by the
time-dependent magnetic flux $\Phi(t) = \Phi_a \sin \Omega t$. The
time dependent flux generates an electric field and thus an internal
voltage $U(t) = U_a \cos \Omega t$ with $eU_a = 2\pi (\Phi_{a}
/\Phi_{0}) (L/2\pi R) \hbar \Omega $, where $L$ is the length of the
segment on the circle with radius $R$. In addition, a constant voltage
$\bar U$ is applied. The magnetic flux can be
incorporated in the phase of the scattering matrix, and the previous
analysis is easily generalized for this case. Ref. \cite{Lesovik94}
found that for zero temperature $\partial S/\partial \bar U$ is
again a step function of voltage. Steps occur at $\bar U = l \hbar
\Omega/e$, and the height of each step is $\lambda_l = T (1-T) J_l^2
(eU_a/\hbar \Omega)$, where $T$ is the transmission coefficient. In
the Ref. \cite{Lesovik94} the argument of the Bessel function is
written in terms of the ratio of fluxes $2\pi\Phi_{a} /\Phi_{0}$ and
the effect is called a {\em non-stationary Aharonov-Bohm
effect}. However, we emphasize that what is investigated is the
response to the external electric field generated by the oscillating
flux. This is a classical response, unrelated to any Aharonov-Bohm 
type effect.  

It is remarkable that in the case of energy independent 
transmission probabilities the response to the electric field 
considered in Ref. \cite{Lesovik94} is the same as that of an 
oscillating voltage $V \equiv U_a $ applied to a contact. 

Levinson and W\"olfle \cite{Levinson99} considered a related problem:
the noise for the transmission through a barrier with an oscillating
random profile (originating, for instance, from the external
irradiation). The latter is represented by a one-dimensional potential
\begin{displaymath} 
U(x,t) = U_0(x) + \delta U(x,t).
\end{displaymath}
The random component $\delta U$ is assumed to be zero on average, and
its second moment is a function of $t-t'$. In this case,
current-current correlations,  $\langle \delta I(t) \delta I(t')
\rangle$, for each particular realization of the random potential
depend on both times $t$ and $t'$. However, after averaging over
disorder realizations, the resulting noise only depends on $t-t'$ and
can be Fourier transformed. The scattering matrices are energy
independent for frequencies below the inverse time of flight through
the barrier, $\omega \ll v_F /L$, with $L$ being the length of the
barrier.  

A remarkable feature of this model is that if the barrier and the
irradiation are symmetric, $U_0 (x) = U_0(-x)$, and $\delta U(x,t) =
\delta U(-x, t)$, and no voltage is applied between the reservoirs,
there is no current generated by the irradiation. On the other hand,
a {\em non-equilibrium} contribution to noise exists. In particular,
when the second moment of the random potential is $\overline{\delta
U(x,t) \delta U(x', t')} = V_0^2 \delta(x - x')$, it happens to have
the same frequency structure as the equilibrium one (\ref{noisfr5}),
but with the coefficient proportional to $V_0^2$. The voltage applied
to the reservoirs is, as usual, one more source of non-equilibrium
noise. For  further details, we refer the reader to
Ref. \cite{Levinson99}.   
\begin{figure}
{\epsfxsize=7.cm\centerline{\epsfbox{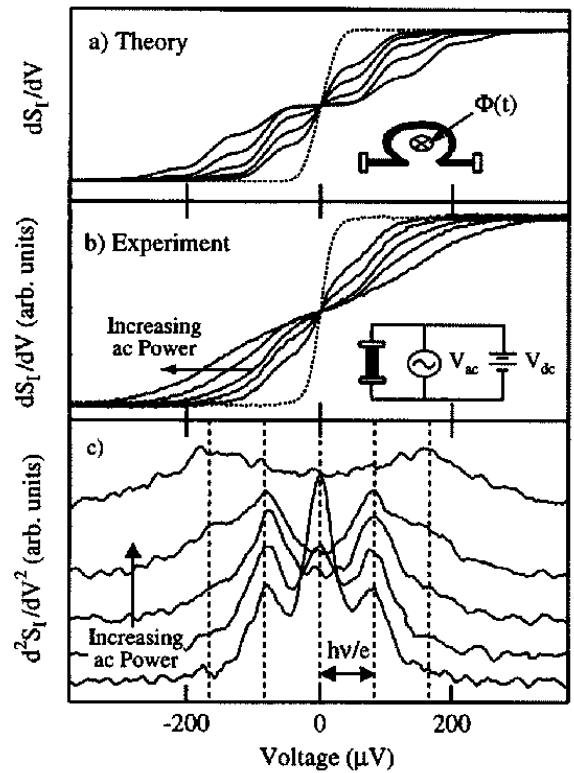}}}
\vspace{3.5cm}
\caption{(a) Theoretical results (\protect\ref{patnoisefin}) for
various amplitudes of the ac voltage; dotted line shows the dc
results; (b) Experimental results of Schoelkopf {\em et al}
\protect\cite{Prober2} for the same parameters; (c)
Experimental results plotted as $\partial^2 S/\partial \bar V^2$ for
different ac voltage amplitudes. The frequency $\Omega$ is fixed.} 
\label{diffus34}
\end{figure}

Experimentally, noise in response to a simultaneous dc
voltage and ac voltage {\it applied to the contacts of the sample} was
studied by Schoelkopf {\em et al} \cite{Prober2} in phase coherent
metallic diffusive wires\footnote{As usual, for metallic diffusive
wires $\sum_n T_n^2$ and $\sum_n T_n (1 - T_n)$ must be replaced by
$2lN_{\perp}/3L$ and $lN_{\perp}/3L$, respectively.}. They measured
zero-frequency noise as a function of voltage $\bar V$ in the GHz
range. The results are presented in Fig.~\ref{diffus34} as $\partial
S/\partial \bar V$ and $\partial^2 S/\partial \bar V^2$. The latter
quantity is expected to have sharp peaks at the resonant voltages
$\bar V = l \hbar \Omega/e$. Indeed, three peaks, corresponding to $l
= 0, \pm 1$ are clearly seen; others are smeared by temperature and
not so well pronounced.    

Ref. \cite{Pedersen98} emphasized the need for a self-consistent
calculation of photon-assisted transport processes even in the case
that the only quantity of interest are the currents or noise-spectra
measured at zero frequency. The true electric field in the interior of
the conductor is not the external field. The fact that the
experimental results agree rather well with the simple results
presented here (which do not invoke any self-consistency) is probably
a consequence of the effective screening of the metallic
conductor. The true potential is simply linear in the range of
frequencies investigated experimentally. A self-consistent spectrum
for photon-assisted noise spectra can probably be developed along the
lines of Ref. \cite{Pedersen98}. 

\subsection{Noise of a capacitor} \label{freq034}

Now we turn to the problems where the energy dependence of the
scattering matrices is essential. Rather than trying to give the
general solution (which is only available for the case when the
potential inside the system is spatially uniform
\cite{ButtikerHandel,BTP93}), we provide a number of examples which
could serve as a basis for further investigations of finite frequency
noise.  

The simple case of shot noise in a ballistic wire was studied by Kuhn,
Reggiani, and Varani \cite{Kuhn921,Kuhn922}. They found signatures of
the inverse flight time $v_F/L$. However, the interactions are taken
into account only implicitly via what the authors call ``quantum
generalization of the Ramo-Shockley theorem''. We do not know in 
which situations this approach is correct, and it certainly cannot be
correct universally. Thus, even this simple case cannot be considered
as solved and needs further consideration. 

\begin{figure}
{\epsfxsize=6.cm\centerline{\epsfbox{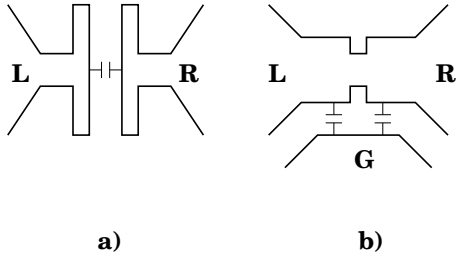}}}
\vspace{0.3cm}
\caption{A mesoscopic capacitor (a); a mesoscopic conductor vis-a-vis
a gate (b).}
\protect\label{freq303}
\end{figure}
We start from the simplest system --- a mesoscopic capacitor
(Fig. \ref{freq303}a), which is connected via two leads to
equilibrium reservoirs. Instead of the full Poisson equation 
interactions are described with the help of a geometrical capacitance
$C$. There is no transmission from the left to the right plate, and
therefore there is no dc current from one reservoir to the other. 
Moreover, this system {\em does not exhibit any noise} even at finite 
frequency if the scattering matrix is energy independent. Indeed, if
only the matrices $s_{LL}$ and $s_{RR}$ are nonzero, we obtain from
Eq. (\ref{noisefr})    
\begin{eqnarray} \label{capac0}
S^{(0)}_{\alpha\beta} (\omega) & = & \frac{e^2}{2\pi\hbar}
\delta_{\alpha\beta} \int dE \ \mbox{\rm Tr}\ \left\{ \left[ 1 -
s^{\dagger}_{\alpha\alpha} (E) s_{\alpha\alpha} (E+\hbar\omega)
\right] \right. \nonumber \\
& \times & \left. \left[ 1 - s^{\dagger}_{\alpha\alpha} (E +
\hbar\omega) s_{\alpha\alpha} (E) \right] \right\} \nonumber \\
& \times & f_{\alpha}(E) (1-f_{\alpha}(E+ \hbar \omega)).  
\end{eqnarray}
Here we used the superscript $(0)$ to indicate that the fluctuations
of the particle current, and not the total current, are discussed. 
If the scattering matrices are energy independent, Eq. (\ref{capac0})
is identically zero due to unitarity. Another way to make the same
point is to note that since $S_{LR}=0$, the only way to conserve
current would be $S_{LL} = 0$. 

Before proceeding to solve this problem, we remark that
Eq. (\ref{capac0}) describes an equilibrium fluctuation spectrum and
via the fluctuation dissipation theorem $S^{(0)}_{\alpha\alpha}
(\omega) = 2\hbar\omega g_{\alpha\alpha} (\omega)  \coth
({\hbar\omega}/{2k_BT})$  is related to the real part of a conductance
$g_{\alpha \alpha} (\omega)$ given by   
\begin{eqnarray} \label{capac1}
g_{\alpha\alpha} (\omega) 
& = & \frac{e^{2}}{\hbar} 
\int dE \ \mbox{\rm Tr}\left\{ \left[ 1 -
s^{\dagger}_{\alpha\alpha} (E) s_{\alpha\alpha} (E+\hbar\omega)
\right] \right\} \nonumber \\
& \times & \frac{f_{\alpha}(E) - f_{\alpha}(E+ \hbar \omega)}{\hbar
\omega } .   
\end{eqnarray}
In what follows, 
the fluctuation dissipation theorem for the non-interacting system 
also ensures this theorem for the interacting system. 

Current conservation is restored only if interactions are taken into
account. Below we assume that charging effects are the only
manifestation of interactions \cite{BTP93,Lithuania}. We first present
the general result which is, within the limitations stated above,
valid for arbitrary frequencies. We then consider in detail the low
frequency expansion of this result which can be expressed in
physically appealing quantities: an {\em electrochemical} capacitance
and a {\em charge relaxation} resistance.  

{\bf General result}. Our starting point are the particle current
operators in the left and right lead (\ref{cur3}), $\hat I^{(0)}_L
(t)$ and $\hat I^{(0)}_R (t)$.  Their fluctuation spectra are
determined by  Eq. (\ref{capac0}). Now we must take into account that
the total currents are in fact not just the particle currents but
contain an additional contribution generated by the fluctuating
electrostatic potential on the capacitor plates. We introduce the {\em
operators} of the potential on the left $\hat u_L (t)$ and right $\hat
u_R (t)$ plate. The fluctuation of the total current through the lead
$\alpha$ can be written in operator form as follows, 
\begin{equation} \label{capacfl1}
\Delta \hat I_{\alpha} (t) = \delta \hat I_{\alpha} (t) + \int dt'
\chi_{\alpha} (t - t') \delta \hat u_{\alpha} (t'), \ \ \ \alpha =
L,R. 
\end{equation}  
Here we introduced $\delta \hat I_{\alpha} (t) = \hat
I_{\alpha}^{(0)} (t)  - \langle I^{(0)}_{\alpha} \rangle$ and $\delta
\hat u_{\alpha} (t) = \hat u_{\alpha} (t)  - \langle u_{\alpha}
\rangle$. Furthermore, $\chi_{\alpha}$ is the response function 
which determines the current generated at contact $\alpha$ in response
to an oscillating potential on the capacitor plate. For the simple
case considered here, it can be shown \cite{BTP93} that this response
function is directly related to the the ac conductance,
Eq. (\ref{capac1}), for {\em non-interacting electrons}, which gives
the current through the lead $\alpha$ in response to a voltage applied
to the same lead, $\chi_{\alpha} (\omega) = - g_{\alpha\alpha}
(\omega)$. The minus sign is explained by noting that the current
fluctuation is the response to $\mu_{\alpha} - u_{\alpha}$ rather than
$u_{\alpha}$. The operators $\Delta \hat I$, and not $\delta \hat I$,
determine the experimentally measured quantities.    

The total current in this system is the displacement current. In the
capacitance model the charge of the capacitor $\hat Q$ is given by
$\hat Q = C (\hat u_L - \hat u_R)$. Note that this is just the Poisson
equation expressed with the help of a geometrical capacitance. To the
extent that the potential on the capacitor plate can be described by a
uniform potential this equation is valid for all frequencies. Then the
fluctuations of the current through the left and right leads are  
\begin{eqnarray} \label{capacfl2}
\Delta \hat I_L (t) = \frac{\partial}{\partial t} \delta \hat Q
(t) = C \frac{\partial}{\partial t} \left[ \delta \hat u_L (t) -
\delta \hat u_R (t) \right],  
\end{eqnarray}
and $\Delta \hat I_L (t) = - \Delta \hat I_R (t)$. Thus, the
conservation of the total current is assured. In contrast to the
non-interacting problem, the currents to the left and right are now
completely correlated. Equations (\ref{capacfl1}) and (\ref{capacfl2})
can be used to eliminate the voltage fluctuations. The result is
conveniently expressed in the frequency representation,
\begin{eqnarray*}
\Delta \hat I_L  =  - \Delta \hat I_R & = & \frac{i\omega C}{g_{LL}
g_{RR} - i\omega C (g_{LL} + g_{RR} )} \nonumber \\
& \times & \left[ g_{LL} \delta \hat I_R - g_{RR} \delta \hat I_L
\right]. 
\end{eqnarray*}
The noise power $S \equiv S_{LL}$ becomes
\begin{eqnarray} \label{capacfl3}
S (\omega) & = & \frac{\omega^2 C^2}{\vert g_{LL} g_{RR} - i\omega C
(g_{LL} + g_{RR} ) \vert^2} \nonumber \\
& \times & \left[ \vert g_{RR} \vert^2 S_{LL}^{(0)}
+ \vert g_{LL} \vert^2 S_{RR}^{(0)} \right]. 
\end{eqnarray} 
Eq. (\ref{capacfl3}) expresses the frequency dependent noise spectrum
of the interacting system in terms of the conductances
$g_{LL}(\omega), g_{RR}(\omega)$ and the noise spectra $S_{LL}^{(0)},
S_{RR}^{(0)}$ of the problem without interactions. Together with
Eq. (\ref{capac0}), this is the result of the first step of our
calculation. Note that the noise of the capacitor depends only
implicitly, through the scattering matrices and the Fermi functions,
on the stationary (dc) voltage difference across the capacitor. 
Eq. (\ref{capacfl3}) is thus valid independently on whether the
potentials in the two reservoirs are the same or not.  

So far, our consideration is valid in the entire frequency range up to
the frequencies at which the concept of a single potential no longer
holds: $\omega \ll e^2/\epsilon d$, where $\epsilon$ and $d$ are the
static susceptibility and the distance between the capacitor plates,
respectively. 

{\bf Low-frequency expansion}. Now we turn to the low-frequency
expansion of Eq. (\ref{capacfl3}), leaving only the leading term. The
expansion of $g_{\alpha\alpha}$ can be easily obtained \cite{BTP93},
\begin{eqnarray} \label{expansion1}
g_{\alpha\alpha} (\omega) & = & -i\omega e^2 \nu_{\alpha} {\cal
A}_{\alpha} + O (\omega^2), \nonumber \\ 
\nu_{\alpha} {\cal A}_{\alpha} & = & -\frac{1}{2\pi i} \int dE
\frac{\partial f_{\alpha}}{\partial E} \mbox{\rm Tr} \ \left(
s^{\dagger}_{\alpha\alpha} (E) \frac{\partial s_{\alpha\alpha}
(E)}{\partial E} \right), 
\end{eqnarray} 
where $\nu_{\alpha}$ is the density of states per unit area and
${\cal A}_{\alpha}$ is the area of the cross-section of the plate
$\alpha$. The fact that the density of states can be expressed in
terms of the scattering matrix is well known \cite{dashen,stafford}. 
Expanding Eq. (\ref{capac0}), we write\footnote{For zero temperature,
the expansion of Eq. (\ref{capac0}) starts with a term proportional to
$\vert \omega \vert^3$ rather than $\omega^2$. As a result, $4k_BT$ is
replaced by $2\hbar \vert \omega \vert$ in the final expression
(\protect\ref{capacfin}) for noise $S$.} 
\begin{eqnarray*}
S_{\alpha\alpha}^{(0)} & = & \frac{e^{2} \hbar \omega^2 k_B T}{\pi}
\int dE \left( - \frac{\partial f_{\alpha}}{\partial E} \right)
\mbox{\rm Tr} \ \left( s^{\dagger}_{\alpha\alpha} (E) \frac{\partial
s_{\alpha\alpha}  (E)}{\partial E} \right)^2 \\
& + & O (\vert \omega \vert^3).  
\end{eqnarray*}
Now we introduce the {\em electrochemical capacitance},
\begin{equation} \label{elchemcap}
C_{\mu}^{-1} \equiv C^{-1} + (e^2 \nu_L {\cal A}_L)^{-1} + (e^2 \nu_R
{\cal A}_R)^{-1},    
\end{equation}
and the {\em charge relaxation resistances}, 
\begin{eqnarray} \label{Rq}
R_{q\alpha} & = & \frac{\hbar}{4\pi e^2 \nu_{\alpha}^2 {\cal
A}_{\alpha}^2} \int dE \left( -\frac{\partial f_{\alpha}}{\partial
E}\right) \nonumber \\ 
& \times & \mbox{\rm Tr} \ \left( s^{\dagger}_{\alpha\alpha} (E)
\frac{\partial s_{\alpha\alpha} (E)}{\partial E} \right)^2 . 
\end{eqnarray}
These two quantities, which determine the RC-time of the mesoscopic
structure, now completely specify the low frequency noise of the
capacitor. A little algebra gives  \cite{BTP93,Lithuania}
\begin{equation} \label{capacfin}
S = 4k_B T\omega^2 C_{\mu}^2 \left( R_{qL} + R_{qR} \right) + O
(\omega^3).  
\end{equation}
We remark that the charge relaxation resistance is determined by 
{\it half} the resistance quantum ${\pi\hbar}/{e^{2}}$ and not
${2\pi\hbar}/e^2$, reflecting the fact that each plate of the
capacitor is coupled to one reservoir only. 

We note now that the low-frequency expansion of the ac conductance
(admittance) of the same capacitor \cite{BTP93} has a form
\begin{equation} \label{admitt}
G(\omega) = -i\omega C_{\mu} + \omega^2 C_{\mu}^2 \left( R_{qL} +
R_{qR} \right) + O (\omega^3).
\end{equation}
Thus, we see that the fluctuation-dissipation theorem is also obeyed
for the interacting system. Again, the noise spectrum of this
equilibrium system contains the same information as the admittance. We
emphasize once more that it is not required that the electrochemical
potential of the two plates of the capacitor are identical. The
results given above also hold if there is a large dc voltage applied
across the capacitor.  

The electrochemical capacitance and the charge relaxation resistance
have been evaluated for a number of examples. In the limit of one
quantum channel only, the charge relaxation resistance is universal,
independent of the properties of the scattering matrix, and given by
$R_{q} ={\pi\hbar}/{e^2}$. This is astonishing in view of the fact
that if a tunnel barrier is inserted in the channel connecting the
capacitor plate to the reservoir one would expect a charge relaxation
resistance that diverges as the tunnel barrier becomes more and more
opaque. For a chaotic cavity connected via a perfect single-channel 
lead to a reservoir and coupled capacitively to a macroscopic gate,
the distribution of the electrochemical capacitance has been given in
Ref. \cite{goparb}. In this case the charge relaxation resistance, as
mentioned above, is universal and given by $R_{q} ={\pi\hbar}/{e^2}$. 
For a chaotic cavity coupled via an $N$-channel quantum point contact
to a reservoir and capacitively coupled to a macroscopic gate, the
capacitance and charge relaxation resistance can be obtained from the
results of Brouwer and one of the authors \cite{brbu}. For large $N$,
for an ensemble of chaotic cavities, the capacitance fluctuations are
very small, and the averaged charge relaxation resistance is given by
$R_q = {2\pi\hbar}/{e^{2}N}$. If a tunnel barrier is inserted into the
contact, the ensemble averaged resistance is \cite{bee99} $R_q =
{2\pi\hbar}/{e^{2}TN}$ for a barrier which couples each state inside
the cavity with transmission probability $T$ to the reservoir. In
accordance with our expectation the charge relaxation resistance is
determined by the two-terminal tunnel barrier resistance $R_t =
{2\pi\hbar}/{e^{2}TN}$. For additional examples we refer the reader to
Ref. \cite{bukor} which presents an overview of the known charge
relaxation resistances $R_q$ for mesoscopic conductors. 

\subsection{Shot noise of a conductor observed at a gate}
\label{freq035} 

It is interesting to ask what would be measured at a gate that couples
capacitively to a conductor which is in a transport state. In such a
situation, in the zero-temperature limit, the low frequency noise in
the conductor is the shot noise discussed in this Review. Thus we can
ask, what are the current fluctuations capacitively induced into a
gate due to the shot noise in the nearby conductor?  To answer this
question we consider a mesoscopic conductor vis-a-vis a macroscopic
gate \cite{PvLB}. The whole system is considered as a three-terminal
structure, with $L$ and $R$ labeling the contacts of the conductor,
and $G$ denoting the gate (Fig.~\ref{freq303}b). The gate and the
conductor are coupled capacitively with a geometrical capacitance
$C$. For a macroscopic gate the fluctuations of the potential within
the gate are small and can be neglected. Finally, the most crucial
assumption is that the potential inside the mesoscopic conductor is
uniform and may be described by a single (fluctuating) value $u$. This
assumption is often made in the discussion of the Coulomb blockade
effect, but it is in reality almost never satisfied\footnote{There are
no quantum-mechanical calculations of frequency dependent noise with
the potential profile taken into account available in the
literature. However, when the conductor is a perfect wire, with  a
nearby gate, the existing calculation of ac conductance
\protect\cite{BHB} can be generalized to calculate noise. One has to
start from the field operators, write a density operator, and solve
the Poisson equation as an operator equation for the field operator of
the (electro-chemical) potential. This lengthy calculation leads to an
obvious result: there is no non-equilibrium noise in the absence of
backscattering. Though this outcome is trivial, we hope that the same
approach may serve as a starting point to solve other problems, like a
wire with backscattering. A related discussion was developed
{\em classically} for the frequency dependent noise in diffusive
conductors
\cite{Naveh97,Nagaev98,Naveh99f,Nagaev981,Naveh981,Naveh982},  
see Section \ref{BoltzmannL}.}.     

We provide a solution to this problem by extending the discussion 
of the mesoscopic capacitor. We start from the operators of the
particle currents, $\hat I_{\alpha}^{(0)}$. We have $\hat I_G^{(0)} =
0$, since without interactions there is no current through a
macroscopic gate, and correlations of other current operators are
given by Eq. (\ref{noisefr}). We introduce the operator of potential
fluctuation $\delta \hat u(t)$ in the conductor; the charge
fluctuation is $\delta \hat Q(t) = C\delta \hat u(t)$. Since there is
transmission from the left to the right, we write for the fluctuations
of the {\em total} currents, 
\begin{eqnarray} \label{gatefl1}
\Delta \hat I_{\alpha} (t) & = & \delta \hat I_{\alpha} (t) +
\int_{-\infty}^{t} dt^{\prime} \chi_{\alpha} (t - t') \delta \hat
u(t^{\prime} ), \ 
\ \ \alpha = L,R \nonumber \\
\Delta \hat I_G (t) & = & - C \frac{\partial}{\partial t} \delta \hat
u(t). 
\end{eqnarray}
with the condition $\Delta \hat I_{L}^{(0)} + \Delta \hat I_{R}^{(0)}
= C \partial \delta \hat u(t)/\partial t$, which ensures the current
conservation. It is important that the quantities $\chi_{\alpha}$,
which determine the response of currents at the terminals to the
potential {\em inside} the sample, must be evaluated {\em at
equilibrium}, and, since this potential is time dependent, the ac
current should be taken. The entire dependence on $u$ of the average
current is due to scattering matrices. In the semi-classical
approximation they depend on the combination $E - eu$, and the
derivative with respect to the internal potential is essentially the
derivative with respect to energy. Retaining only the leading order in
frequency, we obtain (see {\em e.g.} \cite{Buttiker93}) 
\begin{eqnarray} \label{expansion2}
\chi_{\alpha} (\omega) & = & i\omega e^2
N_{\alpha} + O(\omega^2), \nonumber \\
N_{\alpha} & = & -\frac{1}{4\pi i} \int dE \frac{\partial f}{\partial
E} \sum_{\beta} \nonumber \\
& \times & \mbox{\rm Tr} \left( s^{\dagger}_{\alpha\beta}
\frac{\partial s_{\alpha\beta}}{\partial E} - \frac{\partial
s^{\dagger}_{\alpha\beta}}{\partial E} s_{\alpha\beta} (E) \right), 
\end{eqnarray}
which is the analog of Eq. (\ref{expansion1})
for a multi-probe conductor. The quantities
$N_{\alpha}$, called {\em emittances} in Ref. \cite{Buttiker93}, obey
the rule $N_L + N_R = \nu {\cal A}$ (with $\nu$ and ${\cal A}$ being
the density of states and the area/volume of the conductor,
respectively). They have the meaning of a density of the scattering
states which describes the electrons exiting eventually through the
contact $\alpha$, irrespectively of the contact they entered through.

Combining Eqs. (\ref{gatefl1}) and ({\ref{expansion2}), we obtain 
\begin{eqnarray*}
\Delta \hat I_L & = & \left[ 1 - e^2 N_L K \right] \delta \hat I_L -
e^2 N_L K \delta \hat I_L, \nonumber \\
\Delta \hat I_R & = & -e^2 N_R K \delta \hat I_L + \left[ 1 - e^2 N_R
K \right] \delta \hat I_R, \nonumber \\
\Delta \hat I_G & = & -CK \left[ \delta \hat I_L + \delta \hat I_R
\right],  
\end{eqnarray*} 
where $K \equiv (C + e^2 N_L + e^2 N_R)^{-1}$ plays the role of an
effective interaction which determines the change in the electrostatic
potential inside the conductor in response to a variation of the
charge inside the conductor. In the following, we are only interested
in the fluctuations of the current through the gate. For this
quantity, which vanishes for zero frequency, we obtain 
\begin{equation} \label{gatefl2}
S_{GG} = C^2 K^2 \left[ S_{LL}^{(0)} + S_{LR}^{(0)} + S_{RL}^{(0)} +
S_{RR}^{(0)} \right].  
\end{equation}
We note that the sum of all the current fluctuation spectra in
Eq. (\ref{gatefl2}) is just the fluctuation spectrum of the total
charge on the conductor: The continuity equation gives \cite{PvLB}
$\sum_{\alpha} \hat I_{\alpha}(\omega) =  
i\omega e \hat{\cal N} (\omega)$ where $\hat{\cal N}$ is the operator 
of the charge in the mesoscopic conductor. From the current operator
Eq. (\ref{cur3}) we obtain 
\begin{eqnarray} \label{qop}
\hat{\cal N}(t) & = & \frac{e}{2\pi\hbar} \sum_{\alpha\beta\gamma}
\sum_{mn} \int dE dE' e^{i(E-E')t/\hbar} \nonumber \\
& \times & \hat a^{\dagger}_{\beta m}
(E) {\cal N}_{\beta\gamma}^{mn} (E,E') \hat a_{\gamma n} (E'),
\end{eqnarray}
with the non-diagonal density of states elements
${\cal N}_{\beta\gamma}^{mn}$ 
\begin{eqnarray} \label{qop1}
{\cal N}_{\beta\gamma}^{mn} (E, E') & = & \frac{1}{2\pi i(E-E')}
\sum_{\alpha}  \left[\delta_{mn} \delta_{\alpha\beta}
\delta_{\alpha\gamma} \right. \nonumber \\
& - & \left. \sum_{k} s^{\dagger}_{\alpha\beta; mk}(E)
s_{\alpha\gamma; kn} (E') \right],
\end{eqnarray}
or in matrix notation ${\cal N}_{\beta\gamma} =  ({1}/{2\pi i\omega})
\sum_{\alpha} A_{\beta\gamma} (\alpha, E, E+ \hbar \omega )$ with the
current matrix $A_{\beta\gamma} (\alpha, E, E')$ given by
Eq. (\ref{curmatr3}). Thus, instead of Eq. (\ref{gatefl2}) we can also
express the fluctuation spectrum of the current at the gate in terms
of the charge fluctuation spectrum  
\begin{equation} \label{gatefl3}
S_{GG} = e^2 C^2 K^2 \omega^{2} S_{NN}^{(0)} ,  
\end{equation}
with 
\begin{eqnarray} \label{gatefl4}
S^{(0)}_{NN} (\omega)  & = & \frac{e^2}{2\pi \hbar}
\sum_{\gamma\delta} \sum_{mn} \int dE {\cal N}_{\gamma\delta}^{mn} 
(E, E + \hbar\omega) \nonumber \\
& \times & {\cal N}_{\delta\gamma}^{nm} (E + \hbar\omega, E)
\left\{ f_{\gamma} (E) \left[ 1 - f_{\delta} (E +
\hbar\omega) \right] \right. \nonumber \\
& + & \left. \left[ 1 - f_{\gamma} (E) \right] f_{\delta}
(E + \hbar\omega) \right\}.   
\end{eqnarray}
Eq. (\ref{gatefl4}) is, in the absence of 
interactions, the general fluctuation spectrum of the charge 
on a mesoscopic conductor.

In the zero-temperature limit, we obtain \cite{PvLB}
\begin{eqnarray} \label{fateflfin}
& & S_{GG} \\
& = & 2C_{\mu}^2 \omega^2 \left\{ \matrix{ R_q \hbar \vert \omega
\vert + R_V (eV - \hbar \vert \omega \vert), & \hbar \vert \omega
\vert < eV, \cr
R_q \hbar \vert \omega \vert, & \hbar \vert \omega \vert > eV, }
\right.  \nonumber
\end{eqnarray}
where $V>0$ is voltage applied between left and right reservoirs.
Here the electro-chemical capacitance is
\begin{equation} \label{gatefl33}
C_{\mu}^{-1} = C^{-1} + \left[ e^2 \left( N_L + N_R \right)
\right]^{-1}, 
\end{equation}
the charge relaxation resistance\footnote{In terms of the previous
subsection, this is the charge relaxation resistance of the
conductor. The charge relaxation resistance of the macroscopic gate is
zero.} reads 
\begin{equation} \label{gatefl5}
R_q = \frac{\pi\hbar}{e^2} \sum_{\gamma\delta = L,R} \mbox{\rm Tr} \
\left( \mbox{$\cal{N}$}_{\gamma\delta}
\mbox{$\cal{N}$}^{\dagger}_{\gamma\delta} \right) \left[ \sum_{\gamma}
\mbox{\rm Tr} \ \mbox{$\cal{N}$}_{\gamma\gamma} \right]^{-2},
\end{equation}
and the non-equilibrium resistance is
\begin{equation} \label{gatefl6}
R_V = \frac{2\pi\hbar}{e^2} \mbox{\rm Tr} \ \left(
\mbox{$\cal{N}$}_{LR} \mbox{$\cal{N}$}^{\dagger}_{LR} \right) \left[
\sum_{\gamma} \mbox{\rm Tr} \ \mbox{$\cal{N}$}_{\gamma\gamma}
\right]^{-2}, 
\end{equation}
with the notation 
\begin{displaymath}
\mbox{$\cal{N}$}_{\gamma\delta} = \frac{1}{2\pi i} \sum_{\alpha}
s^{\dagger}_{\alpha\gamma} \frac{\partial s_{\alpha\delta}}{\partial
E}. 
\end{displaymath}
It can be checked easily that equilibrium noise, $S^{eq}_{GG} =
2C_{\mu}^2 R_q \hbar \vert \omega \vert^3$, satisfies the
fluctuation-dissipation theorem. The resistance $R_q$ can be extracted
from the ac conductance as well. However, non-equilibrium noise is
described by another resistance, $R_V$, which is a new quantity. It
probes directly the {\em non-diagonal density of states} elements 
$\mbox{$\cal{N}$}_{LR}$ of the charge operator. The non-diagonal
density of states elements which describe the charge fluctuations 
in a conductor in the presence of shot noise can be viewed as the
density of states that is associated with a simultaneous current
amplitude at contact $\gamma$ and contact $\delta$, regardless through
which contact the carriers leave the sample. These density of states 
can be also viewed as blocks of the Wigner-Smith time delay matrix
$(2\pi i)^{-1} s^{\dagger} ds/dE$.   

For a saddle-point quantum point contact the resistances $R_q$  and
$R_V$ are evaluated in Ref. \cite{PvLB}. In the presence of a
magnetic field $R_V$ has been calculated for a saddle point model by
one of the authors and Martin \cite{buma}, and is shown in
Fig. \ref{freq305}. For a chaotic cavity connected to two single
channel leads both resistances are random quantities, for which the
whole distribution function is known \cite{PvLB}. Thus, the resistance
$R_q$ (in units of $2\pi\hbar/e^2$) assumes values between $1/4$ and
$1/2$, with the average of $3/8$ (orthogonal symmetry) or $5/14$
(unitary symmetry). The resistance $R_V$ lies in the interval between
$0$ and $1/4$, and is on average $1/12$ and $1/14$ for orthogonal and
unitary symmetry, respectively.  
\begin{figure}
{\epsfxsize=8.cm\centerline{\epsfbox{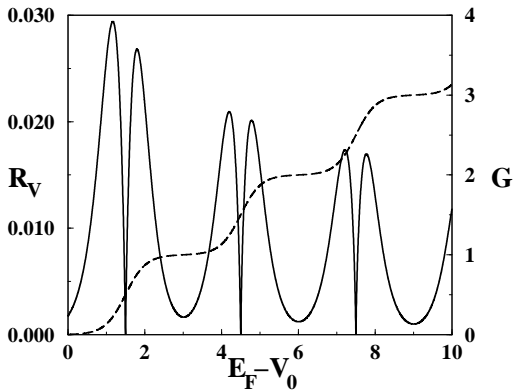}}}
\vspace{0.3cm}
\caption{$R_V$ (solid line, in units of $2\pi\hbar/e^2$) and the
conductance $G$ (dashed line, in units of $e^2/2\pi\hbar$) as a
function of $E_F - V_0$ for a saddle point QPC with $\omega_x/\omega_y
= 1$ and $\omega_c/\omega_x = 4$, $\omega_c$ is the cyclotron
frequency. After Ref. \protect\cite{buma}.} 
\protect\label{freq305}
\end{figure}

\section{Shot noise in hybrid normal and superconducting structures}
\label{hybr}

The dissipationless current (supercurrent) in superconductors is a
property of a ground state, and therefore is noiseless -- it is not
accompanied by any fluctuations. However, noise appears if 
the superconductor is in contact with piece(s) of normal metals. This
Section is devoted to the description of shot noise in these {\em
hybrid structures}, which exhibit a variety of 
interesting phenomena. Following the point of view of the previous 
Sections, we present here a description which is based on an extension
of the scattering approach to hybrid structures. 

\subsection{Shot noise of normal-superconductor interfaces}
\label{hybr41} 

{\bf Simple NS interface, scattering theory and general expressions.}
We consider first an interface of normal metal and superconductor
(NS).  If the applied voltage is below the superconducting gap
$\Delta$, the only mechanism of charge transport is Andreev reflection
at the NS interface: an electron with energy $E$ approaching the
interface from the normal side is converted into a hole with energy
$-E$. The velocity of the hole is directed back from the interface to
a normal metal. The missing charge $2e$ on the normal side 
appears as a new Cooper pair on the superconducting side. There is,
of course, also a reverse process, when a Cooper pair recombines
with a hole in the normal conductor, and creates an electron. 
At equilibrium, both processes have the same probability, and there is
thus no net current.  However, if a voltage is applied, a finite
current flows across the NS interface.  

The scattering theory which we described in Section \ref{scat} 
has to be extended to take into account the Andreev scattering
processes. We give here only a sketch of the derivation; a more
detailed description, as well as a comprehensive list of references,
may be found in Refs. \cite{BeenRMP,Datta}. To set up a scattering
problem, we consider the following geometry (Fig. \ref{hybr401}): The
boundary between normal and superconducting parts is assumed to be
sharp,  and elastic scattering happens inside the normal metal (shaded
region) only. The scattering region is separated from the NS interface
by an ideal region $2$, which is much longer than the wavelength, and
thus we may there use asymptotic expressions for the
wave functions. This spatial separation of scattering from the
interface is artificial. It is not really necessary; our consideration
leading to Eq. (\ref{hybrcond1}) and (\ref{hybrnoise1}) does not
rely on it. Furthermore, for simplicity, we assume that the number of
transverse channels in the normal lead $1$ and in the intermediate
normal portion $2$ is the same.
\begin{figure}
{\epsfxsize=6.cm\centerline{\epsfbox{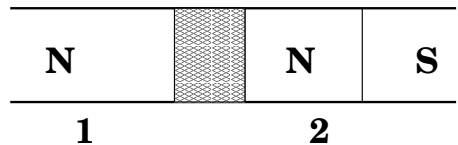}}}
\vspace{0.3cm}
\caption{A simplified model of an NS interface adopted for the
scattering description. Scattering is assumed to happen in the shaded
area inside the normal metal, which separates the ideal normal parts
$1$ and $2$. Andreev reflection is happening strictly at the interface
separating $2$ and the superconductor.}
\protect\label{hybr401}
\end{figure}

We proceed in much the same way as in Section \ref{scat} and define
the annihilation operators in the region $1$ asymptotically far from
the scattering area, $\hat a_{1en}$, which annihilate {\em electrons}
incoming on the sample. These electrons are described by wave functions
$\chi_{1n} (\bbox{r}_{\perp}) \exp(ik_Fz)$ with unit incident
amplitude, where the coordinate $z$ is directed towards the
superconductor (from the left to the right in Fig.~\ref{hybr401}), 
the index $n$ labels the transverse channels. Here we neglected the
energy dependence of the wave vector, anticipating the fact that only
energies close to the Fermi surface will play a role in transport. 
Similarly, the operator $\hat b_{1en}$ annihilates {\em electrons} in
the outgoing states in the region $1$, $\chi_{1n} (\bbox{r}_{\perp})
\exp(-ik_Fz)$.  

For {\em holes}, we define an annihilation operator in the incoming
states in the region $1$ as $\hat a_{1hn}$, and the corresponding
wave function is $\chi_{1n} (\bbox{r}_{\perp}) \exp(-ik_Fz)$. Note
that though this wave function is identical to that for outgoing
electrons, it corresponds to the incoming state with energy $-E$. The
velocity of these holes is directed towards the interface. The
annihilation operator for holes in the outgoing states, $\hat
b_{1hn}$, is associated with the wave function $\chi_{1n}
(\bbox{r}_{\perp}) \exp(ik_Fz)$. Creation operators for electrons and
holes are defined in the same way. Thus, the difference with the
scattering theory for normal conductors is that we now have an extra
index, which assumes values $e,h$ and discriminates between electrons
and holes. 

The electron and hole operators for the outgoing states 
are related to the electron and hole operators of the incoming 
states via the scattering matrix, 
\begin{eqnarray} \label{scathybr2}
\left( \matrix{ \hat b_{1e} \cr \hat b_{1h} } \right) = s \left(
\matrix{ \hat a_{1e} \cr \hat a_{1h} } \right) \equiv \left(
\matrix{ s_{ee} & s_{eh} \cr s_{he} & s_{hh} } \right)
\left( \matrix{ \hat a_{1e} \cr \hat a_{1h} } \right), 
\end{eqnarray}
where the element $s_{ee}$ gives the outgoing electron current
amplitude in response to an incoming electron current amplitude,
$s_{he}$ gives the outgoing hole  current amplitude in response to an
incoming electron current amplitude, {\em etc}. The generalized
current operator (\ref{cur1}) for electrons and holes in region $1$ is
\begin{eqnarray} \label{hybrop1}
\hat I_L (t) & = & \frac{e}{2\pi\hbar} \int_0^{\infty} dE dE'
e^{i(E-E')t/\hbar} \nonumber \\
& \times & \mbox{\rm Tr} \ \left[ \hat a^{\dagger}_{1e} (E)
\hat a_{1e} (E') -  \hat a^{\dagger}_{1h} (E) \hat a_{1h}(E') \right.
\nonumber \\
& - & \left. \hat b^{\dagger}_{1e} (E) \hat b_{1e}(E') + \hat
b^{\dagger}_{1h} (E) \hat b_{1h}(E') \right],
\end{eqnarray}
or, equivalently,
\begin{eqnarray} \label{hybrop2}
\hat I_L (t) & = & \frac{e}{2\pi\hbar} \sum_{\alpha\beta}
\int_0^{\infty} dE dE' e^{i(E-E')t/\hbar} \nonumber \\
& \times & \mbox{\rm Tr} \left[ \hat
a^{\dagger}_{1\alpha} (E) A_{\alpha\beta} (E, E') \hat a_{1\beta}
(E') \right], 
\end{eqnarray}
where we have again introduced electron-hole indices $\alpha$ and
$\beta$, and the trace is taken over channel indices. The matrix $A$
is given by 
\begin{eqnarray*}
A (E, E') = \Lambda - s^{\dagger} (E) \Lambda s (-E'), \ \ \ \Lambda =
\left( \matrix{ 1 & 0 \cr 0 & -1 }\right),
\end{eqnarray*}
with the matrix $\Lambda$ discriminating between electron and holes. 
Introducing the distribution functions for electrons $f_e (E) = [
\exp[(E - eV)/k_BT] + 1]^{-1}$ and holes $f_h(E) = [\exp[(E+eV)/k_BT]
+ 1]^{-1}$, and acting in a similar way as in Section \ref{scat} for
normal systems, we obtain from the current operator and the usual
quantum statistical assumptions for the averages and correlations of
the electron and hole operators in the normal reservoir the
zero-temperature conductance   
\begin{equation} \label{hybrcond1}
G = \frac{e^2}{\pi\hbar} \mbox{\rm Tr} \ \left[ s^{\dagger}_{he}
s_{he} \right]
\end{equation}
and the shot noise power 
\begin{equation}  \label{hybrnoise1}
S = \frac{4e^3 \vert V \vert }{\pi\hbar} \mbox{\rm Tr} \ \left[
s^{\dagger}_{he} s_{he} \left( 1 - s^{\dagger}_{he} s_{he} \right)
\right]  
\end{equation}
in the zero-temperature limit up to linear order in the applied
voltage. Here we have made use of the unitarity of the scattering
matrix, in particular, $s^{\dagger}_{ee}s_{ee} +
s^{\dagger}_{he}s_{he} = 1$, and of the particle-hole symmetry. 
As a consequence, both the conductance and the noise can be expressed
in terms of $s_{he}$ only. We emphasize that Eqs. (\ref{hybrcond1})
and  (\ref{hybrnoise1}) are completely general: In particular, they do
not require a clean NS interface and the spatial separation of the
scattering region of the normal conductor from the interface. However,
without such additional assumptions the evaluation of the scattering
matrix can be very difficult.   

To gain more insight we now follow Beenakker \cite{BeenRMP} and
assume, as shown in Fig.~\ref{hybr401}, that a perfect region of
normal conductor is inserted between the disordered part of the
conductor and the NS-interface. In region $2$, incoming states for
electrons and outgoing states for holes have wave functions
proportional to $\exp(-ik_Fz)$, while outgoing states for electrons
and incoming states for holes contain the factor $\exp(ik_Fz)$. We
also define annihilation operators $\hat a_{2en}$, $\hat a_{2hn}$,
$\hat b_{2en}$, $\hat b_{2hn}$, and creation operators for this
region.  

The scattering inside the normal lead is described by a $4N_{\perp}
\times 4N_{\perp}$ scattering matrix $s_N$ ($N_{\perp}$ being the
number of transverse channels),
\begin{eqnarray} \label{scathybr1}
\left( \matrix{ \hat b_{1e} \cr \hat b_{2e} \cr \hat b_{1h} \cr \hat
b_{2h} } \right) =  s_N \left( \matrix{ \hat a_{1e} \cr \hat a_{2e}
\cr \hat a_{1h} \cr \hat a_{2h} } \right),
\end{eqnarray}
where operators like $\hat b_{1e}$ are vectors, each component
denoting an individual transverse channel. The elastic scattering in
the normal region does not mix electrons and holes, and therefore in
the electron-hole decomposition the matrix $s_N$ is diagonal,
\begin{displaymath}
s_N = \left( \matrix{ s_0(E) & 0 \cr 0 & s_0^*(-E) } \right), \ \ \
s_0 (E) = \left( \matrix{ r_{11} & t_{12} \cr t_{21} & r_{22} }
\right),
\end{displaymath}
Here $s_0(E)$ is the usual $2N_{\perp} \times 2N_{\perp}$ scattering
matrix for electrons, which contains reflection and transmission
blocks.

To leading order in $\Delta/E_F$ (if both the normal conductors and
the superconductor have identical Fermi energies) Andreev reflection
at a clean interface is described by a $2N \times 2N$ scattering
matrix, which is off-diagonal in the electron-hole decomposition, and
is given by 
\begin{eqnarray} \label{scatand1}
\left( \matrix{ \hat a_{2e} \cr \hat a_{2h} } \right) = \gamma \left(
\matrix{ 0 & \exp(i\phi) \cr \exp(-i\phi) & 0 } \right)
\left( \matrix{ \hat b_{2e} \cr \hat b_{2h} } \right),
\end{eqnarray}
with $\gamma \equiv \exp [ -i\arccos (E/\Delta)]$. With some algebra
we can now find expressions for the scattering matrices $s_{ee},
s_{eh}, \dots $ in terms of the Andreev reflection amplitude and the
scattering matrix of the normal region \cite{BeenRMP}, 
\begin{eqnarray} \label{scathybr3}
s_{ee} (E) & = & r_{11} (E) + \gamma^2 t_{12} (E) r^*_{22} (-E)
\nonumber \\
& \times & \left[ 1 - \gamma^2 r_{22} (E) r^*_{22} (-E) \right]^{-1}
t_{21} (E), \nonumber \\
s_{eh} (E) & = & \gamma \exp(i\phi) t_{12} (E) \nonumber \\
& \times & \left[ 1 - \gamma^2 r^*_{22} (-E) r_{22} (E) \right]^{-1}
t^*_{21} (-E), \nonumber \\ 
s_{he} (E) & = & \gamma \exp(-i\phi) t^*_{12} (-E) \nonumber \\
& \times & \left[ 1 - \gamma^2 r_{22} (E) r^*_{22} (-E) \right]^{-1}
t_{21} (E), \\ 
s_{hh} (E) & = & r^*_{11} (-E) + \gamma^2 t^*_{12} (-E) r_{22} (E)
\nonumber \\
& \times & \left[ 1 - \gamma^2 r^*_{22} (-E) r_{22} (E) \right]^{-1}
t^*_{21} (-E). \nonumber
\end{eqnarray}
These matrices express amplitudes for an electron (hole) incoming from
the left to be eventually reflected as an electron (hole). The
corresponding probability is given by the squared absolute value of
the matrix element. In the following, we only consider the case when
no magnetic field is applied to the structure. Then the matrix $s_0
(E)$ is symmetric. For $e \vert V \vert \ll \Delta$ one has $\gamma =
-i$. Taking again the particle-hole symmetry into account, we obtain
with the help of Eqs. (\ref{hybrcond1}), (\ref{scatand1}) the
conductance  
\begin{equation} \label{hybrcond2}
G = \frac{e^2}{\pi\hbar} \sum_n \frac{T_n^2}{(2-
T_n)^2},
\end{equation}
and using Eq. (\ref{hybrnoise1}), the shot noise
\begin{eqnarray}  \label{hybrnoise2}
S & = & \frac{4e^3 \vert V \vert}{\pi\hbar} \mbox{\rm Tr} \ \left[
s^{\dagger}_{he} s_{he} \left( 1 - s^{\dagger}_{he} s_{he} \right)
\right] \nonumber \\ 
& = & \frac{16e^3 \vert V \vert }{\pi\hbar} \sum_n \frac{T_n^2 (1 -
T_n)}{(2 - T_n)^4},  
\end{eqnarray}
where $T_n$ are eigenvalues of the matrix $t_{12}^{\dagger}t_{12}$,
{\em i.e.} transmission eigenvalues of the {\em normal region}
(evaluated at the Fermi surface). As in normal conductors, 
channels with $T_n = 0$ and $T_n = 1$ do not contribute to the
noise. Note that it is the fact that we have chosen to express the
conductance and the noise in terms of the eigenvalues of the normal
region which gives rise to the non-linear eigenvalue expressions given
by Eqs. (\ref{hybrcond2}), (\ref{hybrnoise2}). In terms of the
eigen channels of $s_{he}$ the resulting expression would be formally
identical to the conductance and the noise of a normal conductor.  

The expression (\ref{hybrnoise2}) was obtained by Khlus \cite{Khlus}
using a Keldysh approach for the case when the normal metal and
the superconductor are separated by a tunnel barrier. He also
investigated the finite temperature case and derived the Nyquist
noise. The results were re-derived within the scattering approach by
Muzykantskii and Khmelnitskii \cite{MKsup}. The general case was
studied by de Jong and Beenakker \cite{Jong94} in the framework of the
scattering approach; we followed their work in the course of the above
derivation. Martin \cite{Martin96} obtains the same results using
statistical particle counting arguments and investigates the crossover
between shot and thermal noise (see below).  

{\bf Applications}. If the normal and superconducting electrodes are
separated by a {\em tunnel barrier}, all the transmission coefficients
$T_n$ can be taken to be the same, $T_n = T$ (not to be confused with
temperature). Then we obtain  
\begin{eqnarray*}
G & = & \frac{e^2 N_{\perp}}{\pi\hbar} \frac{T^2}{(2-T)^2}, \nonumber
\\
S & = & \frac{16e^3 \vert V \vert N_{\perp}}{\pi\hbar}
\frac{T^2(1-T)}{(2-T)^4},  
\end{eqnarray*}
with $N_{\perp}$ the number of transverse channels. For the Fano
factor this yields \cite{Khlus,Jong94}
\begin{equation} \label{hybrfano}
F = \frac{S}{2eGV} = \frac{8(1-T)}{(2-T)^2}.
\end{equation}
For low transparency $T \ll 1$ the Fano factor tends to the value of
$2$. This corresponds to the notion that shot noise in NS junctions is
essentially the result of uncorrelated transfer of particles with 
charge $2e$. The shot noise is super-Poissonian ($F > 1$) for $T <
2(\sqrt{2} - 1) \approx 0.83$. For open barriers ($T = 1$) the shot
noise vanishes.   

Refs. \cite{Khlus,Martin96} have also shown that the crossover between
shot and Nyquist noise happens at the temperature $k_B T = 2e\vert V
\vert$, which is one more manifestation of the doubling of the
effective charge. 

For the case of a {\em disordered} normal metal,
Eqs. (\ref{hybrcond2}) and (\ref{hybrnoise2}) have to be averaged over
impurity configurations. Using the distribution function of the
transmission coefficients in the disordered region (\ref{Tdistrib}),
we obtain for the Fano factor \cite{Jong94} $F = 2/3$. This is twice
as high as for a normal disordered wire. Mac\^edo \cite{Macedo2}
obtains the weak localization correction and mesoscopic fluctuations
of the shot noise power. In particular, he finds that the mean square
of the shot noise power scales as the shot noise power itself, and in
this sense the fluctuations are universal (as for normal diffusive
conductors).   

De Jong and Beenakker \cite{Jong94} analyze the case when both a
disordered normal metal and tunnel barrier are present, and describe
the crossover between the two limiting regimes which are obtained in
the absence of a disordered region or in the absence of tunnel
barrier.  

Naidenov and Khlus \cite{Naidenov} and Fauch\`ere, Lesovik, and 
Blatter \cite{Fauchere} analyze the situation when the normal and
the superconducting electrodes are separated by a resonant double
barrier (in particular, this may correspond to the situation of
resonant impurities in the insulating layer separating the two
electrodes). Ref. \cite{Naidenov} considers the one-channel 
sample specific case (no averaging) and discussed the resonant
structure of the conductance and the noise. Using the distribution
function of transmission eigenvalues and assuming that the barrier is
symmetric, Ref. \cite{Fauchere} finds for the ensemble average a Fano
factor $F = 3/4$. This should be contrasted with the result $F = 1/2$
for the corresponding normal symmetric resonant double barrier. 

Schep and Bauer \cite{Schep} investigate the effect of a disordered
interface separating the normal metal and the superconductor. The Fano
factor also, of course, equals $F = 3/4$, which is higher than the
value $2/3$ discussed above.   

{\bf Non-linear regime}. Khlus \cite{Khlus}, and subsequently
Anantram and Datta \cite{Anantram} (who used the scattering
approach), considered noise in the non-linear regime. Without giving 
details, we only mention the case of an ideal NS interface coupled to
a perfect wire for which all the transmission coefficients $T_n$ are
equal to one. This ideal contact does not exhibit shot noise for
voltages below $\Delta/e$, as is seen from Eq. (\ref{hybrnoise2}). The
physical reason is that in this case the scattering process is not
random: an electron approaching the interface is converted into a hole
with probability one and sent back. However, as the voltage increases
above $\Delta/e$, quasiparticle states in the superconductor become
available, and electrons can now tunnel into the superconductor
without being reflected as holes ({\em imperfect Andreev
reflections}). This induces noise even for an ideal interface. For
still higher voltages, an even broader range of energies is
involved. However, for energies $E \gg \Delta$ (almost) all electrons
tunnel into the superconductor without being Andreev reflected. Since
the interface is ideal, this process is also noiseless. Thus, noise is
produced only by electrons with energies higher than $\Delta$, but
with energies which are in the vicinity of $\Delta$. This implies
\cite{Khlus,Anantram} that the shot noise is zero for $e \vert V \vert
< \Delta$, then starts to grow rapidly, and saturates when the voltage
becomes of the order of several $\Delta/e$. The saturation value is
found \cite{Khlus} to be  
\begin{equation} \label{saturNS}
S_{\max} = \frac{4e^2N_{\perp}\Delta}{15\pi\hbar}.
\end{equation}

For a non-ideal conductor ($0 < T_n < 1$ for at least one channel) the
same mechanism leads to the crossover from Eq. (\ref{hybrnoise2}) at
low voltages to Eq. (\ref{shot2term}) for high voltages.  

For barriers of low transparency $T \ll 1$ (for instance, when there
is an insulating layer at some distance from the NS interface) another
mechanism for non-linear noise takes place, as discussed by
Fauch\`ere, Lesovik and Blatter \cite{Fauchere}. In such a geometry
the phases of the scattering matrix are energy sensitive. For $e \vert
V \vert \ll \Delta$ we obtain, similarly to Eqs. (\ref{hybrcond2}) and
(\ref{hybrnoise2}), formulae for non-linear transport,
\begin{equation} \label{hybrcond21}
I = \frac{e}{\pi\hbar} \sum_n \int_0^{eV} \frac{T_n^2}{T_n^2 + 2 (1 -
T_n)(1 - \cos(\alpha_n(E))} dE, 
\end{equation} 
and noise,
\begin{equation} \label{hybrnoise21}
S = \frac{8e^2}{\pi\hbar} \sum_n \int_0^{e \vert V \vert} \frac{T_n^2
(1-T_n) (1 - \cos(\alpha_n(E))}{[T_n^2 + 2 (1 - T_n)(1 -
\cos(\alpha_n(E))]^2} dE. 
\end{equation}
Here we assumed the transmission probabilities $T_n$ to be energy
independent. The phase $\alpha_n$ is
\begin{eqnarray*}
\alpha_n(E) & = & \phi_n(E) - \phi_n(-E) - 2\arccos (E/\Delta) \\
& = & 4Ed/\hbar v_n - 2\arccos (E/\Delta),
\end{eqnarray*}
where $d$ is the distance between the insulating layer and the NS
interface, and $v_n$ is the velocity in the channel $n$. $\phi_n (E)$
is the phase that an electron with energy $E$ acquires during a
round-trip between the NS interface and the insulating layer. For $E =
0$ we have $\alpha_n = \pi$, and thus in the linear regime
Eqs. (\ref{hybrcond21}) and (\ref{hybrnoise21}) are reduced to
Eqs. (\ref{hybrcond2}) and (\ref{hybrnoise2}), respectively.      

These expressions can be interpreted as follows. The part of the
normal metal between the NS interface and the tunnel barrier serves as
an {\em Andreev resonant double barrier}. The electron entering this
region travels to the NS interface, is converted into a hole, then
this hole makes a round-trip, and is converted to an electron, which
returns to the barrier. The total phase gain during this trip is
$\alpha_n (E)$. The ``transmission probability'' of this process (the
integrand in Eq. (\ref{hybrcond21})) shows a pronounced resonance
structure near the energies where the phase $\alpha_n (E)$ equals
$2\pi m$ with integer $m$. Explicitly, for each channel $n$, we have a
set of resonances ({\em Andreev -- Kulik bound states}
\cite{Fauchere}) $E_m = (\pi \hbar v_n/4d)(2m + 1)$. Thus, the
behavior of the transmission probability is similar to that describing
resonant tunneling in the double barrier normal system.   

Specializing further to the case of one channel with velocity $v_F$
and transmission coefficient $T \ll 1$, we write the analog of the 
Breit-Wigner formula 
\begin{equation} \label{hybrcond3}
I = \frac{eT^2}{\pi\hbar} \sum_m \int_0^{eV} \frac{dE}{T^2 + (4d/\hbar
v_F)^2 (E - E_m)^2},
\end{equation}
and
\begin{eqnarray} \label{hybrnoise3}
S & = & \frac{4e^2T^2}{\pi\hbar} \left( \frac{4d}{\hbar v_F} \right)^2
\sum_m \nonumber \\
& \times & \int_0^{e \vert V \vert} \frac{(E-E_m)^2 dE}{[T^2 +
(4d/\hbar v_F)^2 (E - E_m)^2]^2}.
\end{eqnarray}
We see that both the current and the noise power show plateaus as a 
function of applied voltage; sharp transitions between the plateaus
take place at resonances, when $e \vert V \vert = E_m$. In particular,
when the voltage $e V $ lies between the resonances (plateau regime),
$E_M < e \vert V \vert < E_{M+1}$, we have $I = eTv_F/(4d)$, and $S =
2eI$. Thus, already after the first resonance, the Fano factor assumes
the value $F = 1$, the same as for the normal structure. The
explanation is that the transport through Andreev bound states, which
dominates in this regime, is not accompanied with the formation of
Cooper pairs, and thus the usual classical Schottky value is
restored. These considerations should be supplemented by an analysis
of the charge and its fluctuations and the role of screening (see
Section \ref{Langevin}). 

{\bf ND interfaces}. Zhu and Ting \cite{Zhu99} considered shot noise
of the interface between a normal metal and a superconductor with a
$d$-wave symmetry. For this purpose, they generalized the scattering
approach for this situation and subsequently performed numerical
studies. Now the shot noise depends on the orientation of the
superconducting order parameter at the interface. Zhu and Ting
\cite{Zhu99} investigated only one particular orientation, when the
gaps felt by electrons and holes are of the same magnitude but of
different signs. 

The results they found are drastically different from those for
$s$-wave superconductors. In the tunneling regime, the Fano factor is
zero (rather than $2$) for low voltages. It grows with voltage and
saturates at $F = 1$ for $e \vert V \vert \gg \Delta$. Thus, shot
noise is below the Poisson value, defined with respect to the normal
metal, for all voltages. In contrast, in the ballistic limit the Fano
factor is enhanced as compared with $s$-wave superconductors. 

To our knowledge, Ref. \cite{Zhu99} is the only paper addressing shot
noise in hybrid structures with non-trivial symmetry of the order
parameter.    

{\bf Frequency dependence}. The frequency dependence of the noise of
NS interfaces is easy to find in the situation when the scattering
matrices of the normal region may be assumed to be energy independent. 
Lesovik, Martin and Torr\`es \cite{Torres} have investigated this case.
Generalizing Eq. (\ref{noisfr2}) to the case of NS interface, we
write  
\begin{eqnarray} \label{nsfr1}
S(\omega) & = & \frac{e^2}{\pi\hbar} \left\{
\sum_n D_n^2 \int dE \left[ f_{ee} (E, \omega) + f_{hh} (E, \omega)
\right] \right. \\
& + & \left. \sum_n D_n \left( 1 - D_n \right) \int dE \left[ f_{eh}
(E, \omega) + f_{he} (E, \omega) \right] \right\},  \nonumber
\end{eqnarray}
with $D_n \equiv T_n^2(2-T_n)^{-2}$. Performing the integration, we
obtain 
\begin{eqnarray} \label{nsfr2}
& & S(\omega)  =  \frac{e^2}{\pi\hbar} \left\{ 2\hbar\omega \coth
\left( \frac{\hbar\omega}{2k_BT} \right) \sum_n D_n^2
\right. \\
& + & \left. \left[ \left( \hbar\omega + 2eV \right) \coth \left(
\frac{\hbar\omega + 2eV}{2k_BT} \right) \right. \right. \nonumber \\
& + & \left. \left. \left( \hbar\omega - 2eV \right) \coth \left(
\frac{\hbar\omega - 2eV}{2k_BT} \right) \right] \sum_n D_n \left( 1 -
D_n \right) \right\}, \nonumber
\end{eqnarray}
which gives the noise frequency spectrum for arbitrary frequencies,
voltages, and temperatures (provided all of them are much below
$\Delta$). For $V = 0$ Eq. (\ref{nsfr2}) agrees with the
fluctuation-dissipation theorem. At zero temperature, we obtain
\cite{Torres} 
\begin{equation} \label{nsfr3}
S(\omega) = \frac{2e^2}{\pi\hbar} 
\end{equation}
\begin{eqnarray*}
\times \left\{ \matrix{ \hbar\vert
\omega \vert \sum_n D_n^2 + 2e \vert V \vert \sum_n D_n (1-D_n), &
\hbar \vert \omega \vert < 2e \vert V \vert, \nonumber \\ \hbar\vert
\omega \vert \sum_n D_n, & \hbar \vert \omega \vert > 2e \vert V \vert
. } \right. \nonumber
\end{eqnarray*}
This expression is quite similar to Eq. (\ref{noisfr4}) which
describes zero temperature noise frequency spectrum in the normal
contact. One evident difference is that the transmission coefficients
$T_n$ are replaced by $D_n$, due to the modification of scattering
by Andreev reflections. Another observation is that the electron
charge is now doubled. Thus, instead of the singularity at the
frequency $\hbar\omega = \pm eV$ in a normal metal we have now the
singularity at $\hbar\omega = \pm 2eV$. This is yet one more
manifestation of the fact that transport in NS structures is related
to the transmission of Cooper pairs.  

Even more interesting effects are expected when the frequency becomes
of order $\Delta$. In this case the total scattering matrix, however,
can by no means assumed to be energy independent, and the
self-consistent treatment of interactions\footnote{By this now we mean
self-consistency in both the charge {\em and} the superconducting
order parameter $\Delta$.} is needed, as we discussed in Section
\ref{freq}. A step in this direction has been done in
Ref. \cite{MGB1}, which analyzes noise of a NS interface measured at a
capacitively coupled gate and only considers the charge
self-consistency. In accordance with the general conclusions
of Section \ref{freq}, the leading order in frequency for this noise
is given by $S_{GG} = 2C_{\mu}^2 \omega^2 R_V e \vert V \vert$, where
$R_V$ is determined by the properties of the interface. In particular,
when the normal part contains a quantum point contact and a new
channel opens, $R_V = 9\pi\hbar/2e^2$, whereas for a normal quantum
point contact $R_V = 0$ in this situation. 

{\bf Multi-terminal devices}. Consider now a multi-probe hybrid
structure, which contains a number of normal and a number of
superconducting leads (the superconducting leads are taken at the same
chemical potential). The current operator Eq. (\ref{hybrop1}) can be
written for each lead of a multi-terminal structure connected to a
superconductor. This leads to a second quantization formulation of the
current-current correlations put forth by Anantram and Datta
\cite{Anantram}. At each normal contact, labeled $\alpha$, the current
is the sum of an electron current $I_\alpha^e$ and a hole current
$I_\alpha^h$. In terms of the scattering matrix the resulting current
correlations are
\begin{eqnarray} \label{gencorfluc.5}
\langle \Delta I_\alpha^\mu\Delta I_\beta^\nu\rangle & =& \frac{q^\mu
q^\nu}{\pi\hbar} \sum_{\gamma\lambda\atop\delta\kappa} \int_0^ \infty
dE\ {\rm Tr}\left[ 
A_{\gamma\lambda,\delta\kappa}(\alpha\mu)
A_{\delta\kappa,\gamma\lambda}(\beta\nu)\right] \nonumber \\
& \times & f_\gamma^\lambda(E) \left[ 1-f_\delta^\kappa(E) \right],  
\end{eqnarray}
where $q^e = -e$ and $q^h = e$. Here the indices $\alpha$, $\beta$,
$\gamma$, $\delta$ label the terminals, and $\kappa$, $\lambda$,
$\mu$, $\nu$ describe the electron-hole decomposition and may assume
values $e$ and $h$. Basing on Eq. (\ref{gencorfluc.5}), Anantram and
Datta predict that, though correlations {\em at the same contact}
are always positive, like in the case of normal structures, those {\em
at different contacts} also may in certain situations become
positive. (We remind the reader that cross-correlations are quite
generally negative in normal devices, as discussed in
Section~\ref{scat}). Similar results have been obtained by Martin
\cite{Martin96} using statistical particle counting arguments. 

Qualitatively, this conclusion may be understood in the following
way. There are two types of processes contributing to noise. First,
an electron or hole can be simply reflected from the interface without
being Andreev reflected. In accordance with the general
considerations, this reflection tends to make the cross-correlation
negative. On the other hand, processes involving Andreev reflection
(an electron is converted into a hole or vice versa) provide transport
of particles {\em with opposite charges}. Due to
Eq. (\ref{gencorfluc.5}), these processes are expected to push the
cross-correlations towards positive values. This {\em interplay 
between normal scattering and Andreev reflections} determines the
total sign of the cross-correlations. This is, indeed, seen from
the expressions of Anantram and Datta \cite{Anantram}, who decompose
current correlations at different contacts into a sum of positive
definite and negative definite contributions. 

The interpretation which may be found in the literature, that positive
cross-correlations in hybrid structures are due to the bosonic nature
of Cooper pairs, does not seem to be plausible. Indeed, the
microscopic theory of superconductivity never uses explicitly the Bose
statistics of Cooper pairs. In particular, Eq. (\ref{gencorfluc.5})
only contains the (Fermi) distribution functions of electrons and
holes, but not the distribution function of Cooper pairs. 

Quantitative analysis of this effect would also require the next step,
which is to express the scattering matrix $s$ through the scattering
matrices of the normal part of the device (multi-terminal analog of
Eq. (\ref{scathybr3})). The multi-terminal correlations could then be 
studied for various systems, similarly to the discussion for the
normal case (see Section \ref{scat}). Analytical results are currently
only available for systems with an ideal NS interface, where the
matrix $s$ is fully determined by Andreev reflection. 

Anantram and Datta \cite{Anantram} consider a three-terminal device
with two normal contacts and a contact to the superconductor. The
superconductor connects to the normal system via two NS-interfaces
forming a loop which permits the application of an Aharonov-Bohm
flux. The conductor is a perfect ballistic structure and the NS
interfaces are also taken to be ideal. In this system, shot noise is
present for arbitrary voltages, since the electron emitted from the
normal contact $1$, after (several) Andreev reflections may exit
through the normal contact $1$ or $2$, as an electron or as a
hole. Specifically, Ref. \cite{Anantram} studies cross-correlations of
current at the two normal contacts, and finds that they may be both
positive and negative, depending on the phases of Andreev reflection
which in their geometry can be tuned with the help of an Aharonov-Bohm
flux. 

Another three-terminal geometry, a wave splitter connected to a
superconductor, is discussed by Martin \cite{Martin96} and Torr\`es
and Martin \cite{Torres2}. The cross-correlation in the normal leads
depends on the parameter $\epsilon$ which describes the coupling to
the superconducting lead (see Eq. (\ref{beamsplit})), $0 < \epsilon <
1/2$ \cite{Gefen84,Buttiker84}. For an ideal NS interface,
Ref. \cite{Torres2} finds that the cross-correlations are positive for
$0 < \epsilon < \sqrt{2} - 1$ (weak coupling) and negative for
$\sqrt{2} - 1 < \epsilon < 1/2$. Torr\'es and Martin \cite{Torres2}
also report numerical results for disordered NS interfaces, showing
that disorder enhances positive cross-correlations. 

A particularly instructive example has been analyzed in
Ref. \cite{thesis} by Gramespacher and one of the authors of this
Review. They investigate the shot noise measurement with a
tunneling contact (STM tip) which couples very weakly to a normal
conductor which is in turn coupled to a superconductor. If both the
normal reservoir and the superconductor are taken at the same
potential $\mu_0$, and the tunneling tip at potential $\mu$, they find
from Eq. (\ref{gencorfluc.5}) the following correlations,
\begin{eqnarray}
\langle\Delta I_1^e\Delta I_{tip}^e\rangle & = & \alpha
4\pi^2\nu_{tip}|t|^2\nu(x_e,1_e)\, , \\ \langle\Delta I_1^h\Delta
I_{tip}^e\rangle & = & -\alpha 4\pi^2\nu_{tip}|t|^2\nu(x_e,1_h)\,
,
\\ \langle\Delta I_1^e\Delta I_{tip}^h\rangle & = & \langle\Delta
I_1^h\Delta I_{tip}^h\rangle = 0\, ,
\end{eqnarray}
with $\alpha=-(e^2/\pi\hbar)\Delta\mu$, $\Delta\mu = \mu - \mu_0$, and
$\vert t \vert^2$ the coupling energy of the tip to the sample. Here
$\nu(x_e,1_e)$ is the electron density  generated at the coupling
point $x$ due to injected electrons and $\nu(x_e,1_h)$ is the electron
density at the coupling point due to holes injected by the normal
reservoir. The total correlation of the currents at contact 1 and 2 is
the sum of all four terms. In the absence of a magnetic field, the
correlations are proportional to the injected {\it net} charge density
$q(x)=\nu(x_e,1_e)-\nu(x_h,1_e)$, and given by 
\begin{eqnarray}
\langle\Delta I_1\Delta I_{tip}\rangle & = &
-\frac{e^2}{\pi\hbar}\Delta\mu 4\pi^2\nu_{tip}|t|^2 q(x) \nonumber \\
& = & -2G_0\Delta\mu\frac{q(x)}{p(x)}
\end{eqnarray}
where $p(x) = \nu(x_e,1_e) + \nu(x_h,1_e)$ is the total particle
density of states and $G_0=(e^2/2\pi\hbar)4\pi^2\nu_{tip}|t|^2p(x)$ is
the tip to sample conductance. This result states, that if at the
point $x$ the electrons injected from contact $1$ generate a hole
density at $x$ which is larger than the electron density at $x$, the
injected charge becomes negative and the corresponding correlation
becomes positive. A more detailed analysis suggest that this effect is
of order $1/N$, where $N$ is the number of channels. 

Up to now positive correlations in hybrid structures have been
theoretically demonstrated only for single channel conductors. This
leaves open the question, on whether or not, ensemble averaged shot
noise spectra can in fact have a positive sign in hybrid structures.

{\bf Experiments}. The only experiment on shot noise in NS structures
currently available was performed by Vystavkin and Tarasov
\cite{Vystavkin} long before the current interest on shot noise in
mesoscopic systems started. For this reason, they did not study noise
systematically, and only concluded that in certain samples it was
suppressed below the value $2e\langle I \rangle$. 

Recently, Jehl {\em et al} \cite{Jehl} experimented with an Nb/Al/Nb
structure at temperatures above the critical temperature for Al, but
below that for Nb. They estimate that the length of the Al region was
longer than the thermal length, which means that the multiple Andreev
reflection processes (see below) are suppressed. Thus, qualitatively
their SNS structure acts just as two incoherent NS interfaces, and the
expected effective charge is $2e$ (the Fano factor for the diffusive
system is $2/3$). Indeed, the measurements show that the Fano factor
for high temperatures is $1/3$ for all voltages (in accordance with
the result for a metallic diffusive wire), while for lower
temperatures it grows. The low-temperature behavior is found
to be in better agreement with the value $F = 2/3$ for an NS
interface, than with the $F=1/3$ prediction for normal systems, though
the agreement is far from perfect. The experimental results are shown
in Fig. \ref{hybr403}. 
\begin{figure}
{\epsfxsize=8.5cm\centerline{\epsfbox{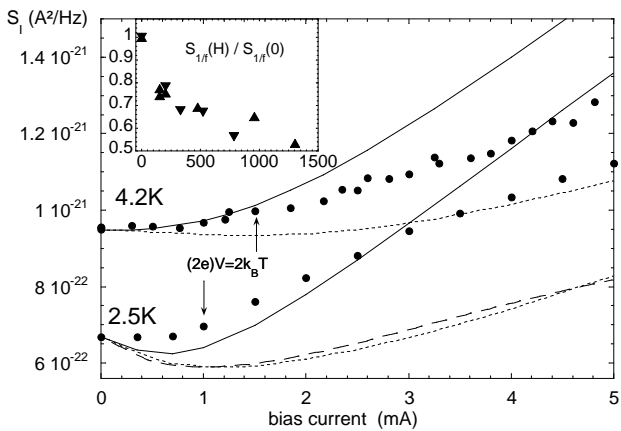}}}
\vspace{0.3cm}
\caption{Experimental results of Jehl {\em et al}
\protect\cite{Jehl}. Solid and dotted lines are theoretical curves
corresponding to the effective charges $2e$ and $e$, respectively.}  
\label{hybr403}
\end{figure}

A clear experimental demonstration of the shot noise doubling with
clean NS interfaces remains to be performed.   

\subsection{Noise of Josephson junctions} \label{hybr42}

Josephson junctions are contacts which separate two superconducting
bulk electrodes by an insulating barrier. We briefly describe here
noise properties for the case when the transmission of this barrier is
quite low; other, more interesting, cases are addressed in the next
subsection.  

The transport properties of Josephson junctions can be summarized as
follows. First, at zero voltage a {\em Josephson current} may flow
across the junction, $I = I_0 \sin \phi$, where $\phi$ is the
difference of the phases of the superconducting order parameter
between the two electrodes. In addition, for finite voltage tunneling
of quasiparticles between the electrodes is possible. For zero
temperature this {\em quasiparticle current} only exists when the
voltage exceeds $2\Delta/e$; for finite temperature an (exponentially
small) quasiparticle current flows at any voltage.  

The Josephson current is a property of the ground state of the
junction, and therefore it does not fluctuate. Hence, shot noise in
Josephson junctions is due to the quasiparticle current\footnote{As
stated by Likharev in his 1979 review \cite{Likharev79}, {\em ``the
most important results of all the theories of fluctuations in the
Josephson effect is that the only intrinsic source of fluctuations is
the normal current rather than the supercurrent of the junction''}.},
and basically coincides with the corresponding shot noise properties
of normal tunnel barriers. For zero temperature, there is no shot
noise for voltages below $2\Delta/e$. Thermal and shot noise in
Josephson junctions are analyzed in detail by Rogovin and Scalapino
\cite{Rogovin74}, and have been measured by Kanter and Vernon, Jr. 
\cite{Kanter70,Kanter701}.    

For completeness, we mention that if a voltage $V(t)$ is applied
across the junction, as a consequence of gauge invariance, the
Josephson current becomes time-dependent, 
\begin{displaymath}
I(\phi) = I_0 \sin \left[ \phi + \frac{2e}{\hbar} \int_0^t V(t) dt
\right].  
\end{displaymath}
Then, due to any fluctuations of the voltage $V(t)$ (like thermal and
shot noise) the Josephson junction starts to radiate.
If the voltage $V$ is time-independent on average,
the spectral density of this radiation is centered around the resonant
frequencies $\omega = 2enV/\hbar$, and fluctuations determine the
width of the maxima, {\em the linewidth of the Josephson
radiation}. This effect, analyzed by Stephen
\cite{Stephen68,Stephen69}, turned out to be an effective experimental
tool for detecting voltage fluctuations in Josephson junctions. It is
the subject of many theoretical
\cite{LO68,Dahm,Likharev72,Rogovin74,Aronov75,Landa77,Zorin81,Lesovik942}
and experimental \cite{Parker67,Dahm,Vernet,Kirschman,Tarasov98} papers.   

\subsection{Noise of SNS hybrid structures} \label{hybr43}

Now we address the limit in which the two superconducting
electrodes are separated by a region in which the motion is ballistic,
or, at least, the transmission probability is not too small; for
definiteness we will consider the quantum point contact connected to
superconducting banks. To make a distinction between the tunnel
Josephson junction described in the previous subsection and the case
of interest here, we refer to these systems as SNS structures; for
convenience, we only describe the one-channel case, and first consider
the perfect contact for which the transmission probability is equal to
one. Furthermore, we consider the case of a constriction with a length
(distance between superconducting electrodes) small compared to the
superconducting coherence length. 

{\bf Equilibrium noise}. In many respects, SNS structures are
different from the tunnel Josephson junctions. A phase-coherent SNS
structure supports a discrete set of {\em subgap} states. Carriers are
trapped between two NS interfaces which act as Andreev mirrors. An
electron reaching one of the interfaces is reflected as a hole and
travels back to the other interface where it is reflected as an
electron. The subgap states are known as Andreev-Kulik states
\cite{Kulik}. In the particular case of interest here, there are two
subgap states \cite{Tsukada} with energies $\epsilon_{\pm}(\phi) = \pm
\Delta \cos (\phi/2)$, which carry Josephson current. These states
have a width $\gamma(\phi) \equiv \gamma[\epsilon(\phi)]$,
which can appear, for instance, due to electron-phonon interactions
(see {\em e.g.} Ref. \cite{AVZ}). As a consequence the SNS system can
undergo fluctuations between a ground state and an excited state. The
situation encountered here exhibits a close analogy to the low-lying
excitations in a normal metal ring penetrated by an Aharonov-Bohm
loop. If the ring is closed, the excitation away from the ground state
has to be described in a canonical ensemble. If the ring is coupled
via a side branch to an electron reservoir, carrier exchange is
permitted, and the discussion has, as in the problem at hand, to be
carried out in the grand-canonical ensemble
\cite{Buttiker85a,Buttiker85b}. At equilibrium the occupation
probability of the two states is $f_{+} = f(\epsilon_{+}(\phi))$ and
$f_{-} = f(\epsilon_{-}(\phi))$, where $f$ is the Fermi distribution
with energy measured away from the center of the gap of the
superconductor. Note that $f_{-} = 1 - f_{+}$. To investigate the
dynamics of this system, in the presence of a bath permitting
inelastic transitions, we investigate the relaxation of the
non-equilibrium distribution $\rho_{\pm}$ towards the instantaneous
equilibrium distribution function with the help of the
Debye-Boltzmann-like equations  
\begin{equation}\label{debye}
d\rho_{\pm}/dt = -\gamma (\rho_{\pm} -f_{_{\pm}}(t)). 
\end{equation}
If the system is driven out of equilibrium, the instantaneous 
distribution function is time-dependent. Eq. (\ref{debye}) states that
the non-equilibrium distribution tries to follow the instantaneous
distribution but can do that at best with a time lag determined by
$\gamma^{-1}$. The time-dependent readjustment of the distribution
$\rho$ is achieved with inelastic processes and is thus
dissipative. To find the resulting noise we investigate the response
of the current to a small oscillating phase $\delta \phi (\omega)
e^{i\omega t}$ superimposed on the dc phase. The current is $I =
-(e/\hbar)[(d\epsilon_{+}/d\phi) \rho_{+} + (d\epsilon_{-}/d\phi)
\rho_{-}]$. The Josephson relation, $d\phi /dt = - (2e/\hbar)V$, leads
to a conductance which in the zero frequency limit is given by
\cite{Buttiker85b}  
\begin{equation} \label{Nyquistloop}
G (\phi) = \left. \left(\frac{2e^{2}}{\gamma\hbar^{2}}\right) 
\left(\frac{d\epsilon}{d\phi}\right)^{2} 
\left(-\frac{df}{d\epsilon}\right) \right\vert_{\epsilon =
\epsilon_{+}}.  
\end{equation}
The resulting thermal noise of the Josephson current follows from the
fluctuation dissipation theorem $S = 4k_BT G$ and, as found by Averin
and Imam \cite{Imam} and Mart\'in-Rodero, Levy Yeyati, and
Garc\'ia-Vidal \cite{Rodero96}, is given by 
\begin{equation} \label{NyquistSNS}
S = \frac{2}{\gamma(\phi)} \left( \frac{e\Delta}{\hbar}
\frac{\sin(\phi/2)}{\cosh(\epsilon_0/2k_BT)} \right)^2.
\end{equation}
The peculiar feature of both the conductance (\ref{Nyquistloop})
and the noise (\ref{NyquistSNS}) is their divergence as the 
damping $\gamma$ tends to zero. Furthermore, since $(-df/d\epsilon)$
in Eq. (\ref{Nyquistloop}) is proportional to $1/k_BT$, the Nyquist
noise given by Eq. (\ref{NyquistSNS}) is not proportional to $k_BT$,
as in open systems. Since in the zero temperature limit $\gamma$ can
be expected to tend rapidly to zero, the Nyquist noise may actually
grow as the temperature is lowered.   

Instead of a small amplitude ac oscillation of the phase, 
we can also consider a phase that linearly increases with time, $\phi
= 2eVt/\hbar$. In a junction without dissipation, we now have an
ac Josephson effect. In a system which permits inelastic transitions,
the ac Josephson current will be accompanied by a dissipative
current. If the induced voltage is small, we can to linear order in
$V$ again determine the conductance $G_{J}$, where the index $J$ is to
remind us that this conductance occurs in parallel with the ac
Josephson effect. Note that the two energy bands $\epsilon_{\pm}$
cross at $\phi = \pm \pi$. To describe this crossing, we extend the
range of $\phi$ from $-2\pi$ to $2\pi$. Ref.~\cite{Buttiker85b} finds
that $G_J$ and $G$, as given by Eq. (\ref{Nyquistloop}), are related,
\begin{equation} \label{gvv}
G_J = \frac{1}{4\pi} \int_{-2\pi}^{2\pi} d\phi G(\phi) = 
\frac{e^{2}}{2 \pi \gamma\hbar^{2}} \int_{-2\pi}^{2\pi}
\left(\frac{d^{2}\epsilon_+}{d\phi^{2}}\right) f_{+}(\phi) d\phi.
\end{equation}
Thus, $G_{J}$ is inversely proportional to the ``effective mass''
(weighted by the equilibrium distribution function). We could of
course derive this result directly from Eq. (\ref{debye}). In the zero
temperature limit the effective mass is $1/m^{*} = 2\Delta/\pi$, and
the conductance $G_{J}$ is finite and given by $G_{J} =2e^2\Delta/\pi
\gamma \hbar^2$. From the fluctuation dissipation theorem, we obtain
an equilibrium noise 
\begin{equation} \label{nnn41}
S = \frac{4e^2 \Delta k_BT}{\pi\hbar^2 \gamma},
\end{equation} 
which (unlike Eq. (\ref{NyquistSNS})) is proportional to the
temperature. Eq. (\ref{nnn41}) was obtained by Averin and Imam
\cite{Imam} and Cuevas, Mart\'in-Rodero, and Levy Yeyati
\cite{Cuevas}. 
  
{\bf Non-equilibrium noise}. For larger voltages but still $eV \ll
\Delta$ ($V > 0$), the average dc current becomes a non-linear
function of voltage. The current peaks for $eV \sim \hbar\gamma$ when
dissipation due to the mechanism described above is maximal. We can no
longer invoke the fluctuation-dissipation theorem to find the
noise. Instead, a direct calculation of the current-current
correlation function $\langle \hat I(t_1) \hat I(t_2) + \hat I(t_2)
\hat I(t_1) \rangle$ is needed. Since the current oscillates with
frequency $2eV/\hbar$ and its harmonics, the correlation function
depends not only on the time difference $t-t'$, as for noise away from
stationary states, but also periodically on the total time $t = (t_1 +
t_2)/2$. Averin and Imam \cite{Imam}, and Cuevas, Mart\'in-Rodero, and
Levy Yeyati \cite{Cuevas} used the Green's function technique to
obtain results for the noise power $S(\omega)$ which is this
correlation function averaged over $t$ and Fourier-transformed with
respect to $t_1 - t_2$. For the discussion of this far from
equilibrium noise, we refer the reader to
Refs. \cite{Imam,Cuevas}. For low voltages $eV \ll \hbar \gamma$, and
for an energy independent damping constant $\gamma$, the calculation
reproduces Eq. (\ref{nnn41}), whereas for  $\hbar\gamma \ll eV \ll
\Delta$ the noise actually {\em decreases} with voltage\footnote{This
expression is not explicit in Ref. \protect\cite{Imam}, but can be
easily derived in the limit of strictly zero temperature.}, 
\begin{equation} \label{SNSshot}
S = \left( 3 - \frac{8}{\pi} \right)
\frac{\gamma\Delta^2}{\hbar V^2}.
\end{equation}  
The voltage dependence of the noise is quite unusual in this case, and
exhibits a peak for $eV \sim \hbar\gamma$. 

The result (\ref{SNSshot}) has the following interpretation. The
mechanism of charge transport in SNS structures for $eV \ll \Delta$ is
{\em multiple Andreev reflections} (MARs) \cite{Klapwijk82}. Imagine
an electron with energy $E$, $\Delta < E < \Delta + eV$ (measured from
the chemical potential of the right contact), emanating from
the right contact. During the motion in the normal region it loses the
energy $eV$, and thus when it arrives at the left superconducting bank
it has an energy below $\Delta$. This electron may only be Andreev
reflected and converted into a hole, which (due to the opposite sign
of the charge) loses the energy $eV$ again. The hole is again Andreev
reflected at the right interface, and this process goes
on\footnote{MAR is a fully coherent process. It cannot take place, 
for instance, if the length of the junction is longer than the phase
breaking length. In this limit the systems acts rather as two
independent NS interfaces. Another limitation is $eV \gg
\hbar\gamma$.}, until the energy of the initial electron falls below
$-\Delta$. The number of these MARs is equal to $2\Delta/eV$. In each
individual Andreev reflection  the charge $2e$ is transferred to or
from the condensate, and to avoid double counting, we must only take
the reflections happening at the same interface. Therefore the whole
MAR process is accompanied by a transfer of charge $2\Delta/V$ (for
$eV \ll \Delta$). In view of this, the noise (\ref{SNSshot}) may be
interpreted \cite{Imam} as ``shot noise'' of the $2\Delta/eV \gg 1$
charge quanta. In this sense this noise is giant: it greatly exceeds
the Poisson value $2eI$. The general expression for noise can be
written explicitly as a sum of contributions of Andreev reflections
of different orders \cite{Imam,Cuevas}. 

Now we briefly discuss the case of a non-ideal contact, {\em i.e.}
when the two electrodes are separated by a barrier of arbitrary
transparency. First, for a finite but small reflection coefficient
$1-T$ an additional source of noise is given by Landau-Zener
transitions between the two subgap states, as discussed by Averin
\cite{Averin96}. The probability of these transitions, which exist
even at zero temperature, is $\nu = \exp(-\pi (1-T) \Delta/eV)$, and
noise is caused by the randomness of these transitions. Naveh and
Averin \cite{Naveh99} obtained the following result for the noise due
to this mechanism, 
\begin{displaymath}
S = \frac{8e\Delta^2}{\pi\hbar V} \nu (1 - \nu).
\end{displaymath}
They also considered the generalization to the multi-channel case and
analyzed a structure with a normal diffusive conductor between the two
superconductors. Taking into account the distribution of transmission
eigenvalues (\ref{Tdistrib}) of a normal conductor yields
\cite{Naveh99} 
\begin{displaymath}
S = \frac{(2\Delta)^{3/2}}{(eV)^{1/2}} G (\sqrt{2} - 1),
\end{displaymath}
where $G$ is the Drude conductance of the normal region. The noise
diverges for low voltages as $V^{-1/2}$. 

If the transparency of the barrier is low, we return to the case of 
a classical Josephson junction. The amplitude of a MAR process
containing $n$ Andreev reflections is proportional to $T^n$, where $T$
is the transmission probability of the junction. Thus, for the
classical case MAR's are totally suppressed. The case of arbitrary
transparency was investigated by Cuevas, Mart\'in-Rodero and Levy
Yeyati \cite{Cuevas}, who described the crossover between these two
regimes.  

Bezuglyi {\em et al} \cite{Bezuglyi} considered a tunnel barrier (an
insulating layer) inserted in the middle of a long SNS
constriction. This geometry is different from the standard Josephson
junction problem. Instead, bound Andreev states (similar to what has
been discussed before for the NS interface with a barrier) are formed
in both parts of the normal region, separated by the insulating
layer. An electron in the left part, before being converted to a hole
at the left NS interface, is oscillating many times before it tunnels
(as an electron or a hole) through the barrier, and starts oscillating
again. This picture resembles \cite{Bezuglyi} transport in a diffusive
metallic wire, which gives us a hint that shot noise may be suppressed
in comparison with its ``giant Poisson'' value, {\em i.e.} the value
corresponding to the effective charge $2\Delta/V$. Indeed,
Ref. \cite{Bezuglyi} finds that in the limit of low voltages the
effective charge is $2\Delta/3V$, which surprisingly reminds us of the
$1/3$--suppression of shot noise in metallic diffusive wires. 

For high voltages $eV \gg \Delta$ imperfect Andreev reflections lead
to the saturation of noise, similar to NS structures. For an ideal
junction the saturation value $S = 8e^2 \Delta/(15\pi\hbar)$, found by
Hessling {\em et al} \cite{Hessling96}, is two times as large as for an
ideal NS interface (\ref{saturNS}). For a non-ideal contact, shot noise
in this regime equals its normal state value $S = 2e\langle I
\rangle$; there is also a voltage independent contribution ({\em excess
noise}) \cite{Cuevas,Naveh99}. The origin of this excess noise are MAR
processes, and the physics is similar to that encountered in the
discussion of excess current (see {\em e.g.} Ref. \cite{Klapwijk82}).

{\bf Experiments}. Recently a number of efforts have been made to
observe the giant shot noise, caused by multiple Andreev
reflections. Experiments by Misaki, Saito, and Hamasaki \cite{Misaki},
and Misaki {\em et al} \cite{Misaki97} used a sandwich of Nb and NbN
superconducting films, separated by a point contact. However, the
giant shot noise was not observed in these experiments, possibly
because of additional scattering inside the point contact.

An experimental observation of MAR-enhanced shot noise is reported 
by Dieleman {\em et al} \cite{Dieleman}. They investigate noise in a 
NbN/MgO/NbN structure, where the two superconducting layers (NbN) are
separated by an insulator. The main feature observed in this
experiment was a decrease of the Fano factor (which is theoretically
predicted to be $2\Delta/eV$ for $eV \ll \Delta$) for voltages up to
$2\Delta$. For voltages $eV \sim 2\Delta/n$, $n \in {\cal Z}$, a
step-like structure is observed (Fig.~\ref{hybr402}).  

Hoss {\em et al} \cite{Hoss} carried out measurements on Nb/Au/Nb,
Al/Au/Al, and Al/Cu/Al junctions, where the Au and Cu layers were
essentially diffusive conductors. They observe a well pronounced peak
in the voltage dependence of the shot noise for low voltages (much
less than $\Delta/e$). In addition, they also plot the Fano factor,
which turns out to be linear in $V^{-1}$ in the whole range of
voltages, but with the coefficient {\em higher} than $2\Delta/e$. This
discrepancy with theory is not understood. 
\begin{figure}
{\epsfxsize=8.cm\centerline{\epsfbox{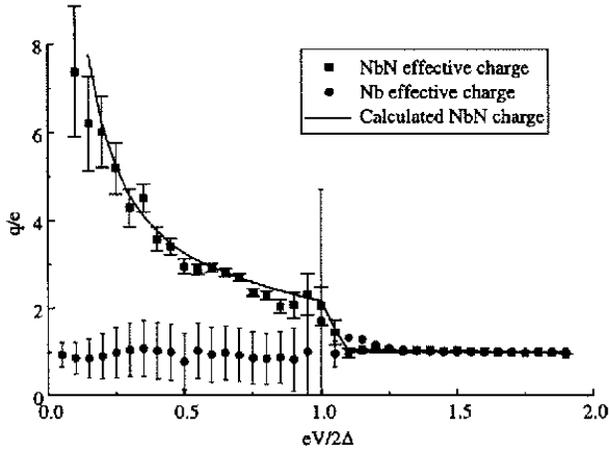}}}
\vspace{0.3cm}
\caption{Experimental results of Dieleman {\em et al}
\protect\cite{Dieleman}. The Fano factor (plotted here as an effective
charge, black squares) is compared with the theoretical prediction
(solid line).}  
\label{hybr402}
\end{figure}

In the experiments by Jehl {\em et al} \cite{Jehl} on long Nb/Al/Nb
SNS contacts, the two NS interfaces act effectively independently, and
MAR processes are suppressed. Thus, the physics of this experiment
resembles more that of a single NS interface, as we have discussed
above.

\section{Langevin and master equation approach to noise:
Double-barrier structures} \label{Langevin}

\subsection{Quantum-mechanical versus classical theories of shot
noise} \label{lang51}

In this and the next Section, we consider classical theories of shot
noise in various systems. By doing this, we leave the main road that
started from basic quantum mechanics and, through a number of exact
transformations and well-justified approximations, lead us to the
final results for shot noise. In contrast, classical theories are
mostly based on the Langevin approach, which has a conceptually much
weaker and less transparent foundation. Indeed, the Langevin equation
is equivalent to the Fokker-Planck equation under the condition that
the random Langevin sources are Gaussian distributed\footnote{Given
the results for the distribution of transmitted charge (Appendix
\ref{con082}), it is apparent that the Langevin sources are {\em not}
Gaussian distributed. Possibly, this does not affect the shot noise,
which is related to the second cumulant of the Langevin sources.} (see
{\em e.g.} Ref. \cite{Kampen}). In turn, the Fokker-Planck equation is
derived from the master equation in the diffusion approximation, and
this procedure determines the pair correlation function of Langevin
sources. In practice, however, such a basic derivation is usually not
presented. The correlation function is written based on some {\em ad
hoc} considerations rather than derived rigorously. For double-barrier 
structures, which are considered in this Section, many results have
been derived directly from the master equation, and thus are far
better justified than many discussions for other structures. The next
Section is devoted to the Boltzmann-Langevin approach in disordered
conductors, and, to our knowledge, no attempt to obtain the final
results from the master equation, or to justify microscopically the
starting Boltzmann equation with Langevin sources, has ever been
performed\footnote{Quite recently Nagaev \cite{Nagaev991} has shown
that the zero frequency results for shot noise in metallic diffusive
wires which are obtained in a quantum-mechanical Green's functions
technique, are equivalent to those available from the
Boltzmann-Langevin approach, even if the interactions are taken into
account. This is a considerable step forward, but it still does not
explain {\em why} the Boltzmann-Langevin approach works.}. 

The reality, though it may be surprising to some readers, is that in
{\em all} available cases when the results of classical calculations
of shot noise can be compared to exact quantum results {\it averaged
over an ensemble}, based on the scattering or Green's functions
approaches, they turn out to be identical. It is the fact that for
many systems the ensemble averaged quantities are classical which
makes classical Boltzmann-Langevin theories of shot noise in mesoscopic
conductors credible even in those situations, where quantum results
are not available. The fact that the ensemble averaged quantum result
and the classical result agree is best illustrated by considering for
a moment a metallic diffusive wire. A calculation of the conductance
can be performed purely quantum-mechanically by finding the scattering
matrix computationally or using random matrix theory. After ensemble
averaging the leading order result for the conductance is just the
Drude result for the conductance of the wire which we can find by
solving a diffusion equation. This situation persists if we consider
the noise: the leading order of the noise, the 1/3-suppression of shot
noise in a metallic diffusive wire can be found by ensemble averaging
a quantum-mechanical calculation \cite{Been92} and/or from a classical
consideration \cite{Nagaev92}. If the two procedures would not agree
to leading order, it would imply a gigantic quantum effect. Of course,
effects which are genuinely quantum, like the Aharonov-Bohm effect,
weak localization, or the quantum Hall effect, cannot be described
classically. This consideration also indicates the situations where we
can expect differences between a quantum approach and a classical
approach: Whenever the leading order effect is of the same order as
the quantum corrections we can obviously not find a meaningful
classical description.  

The Langevin approach essentially takes the Poissonian incoming stream
of particles and represents it as a random fluctuating force acting
even inside the system. In the language of the scattering approach,
this would mean that the Poissonian noise of the input stream is
converted into the partition noise of the output stream. Another
possible classical approach to the shot noise would be to take the
Poissonian input stream as a sequence of random events, and to obtain
the distribution of the carriers in the output stream after the
scattering events took place. To our knowledge, this approach has not
been realized precisely in this form. Landauer \cite{Landauer93}
discusses noise in diffusive metallic conductors from a similar point
of view, but does not calculate the distribution of
outgoing particles. Chen {\em et al} \cite{Chen911} attempt to
modify the distribution function in ballistic systems to take into
account the Pauli principle, assuming that the time the particle
spends inside the system is finite. Barkai, Eisenberg, and Schuss
\cite{Barkai96} and van Kampen \cite{Kampen97} consider the case when
the electrons may arrive from two reservoirs, and are transmitted or
reflected with certain probabilities. The Pauli principle forbids two
electrons to be in the channel simultaneously. 

Though there is no doubt that this approach, if realized, would yield
the same value of the shot noise as more powerful methods, it would
still help to visualize the results and it might allow a simple
generalizations to the interacting systems. Raikh \cite{Raikh91} and
Imamo\=glu and Yamamoto \cite{Imamoglu93} have suggested a
generalization to the Coulomb blockade regime. Raikh \cite{Raikh91}
shows how the noise in the Coulomb blockade regime may be
expressed if the transformation from the Poisson input stream to the
correlated output stream is known for {\em non-interacting}
electrons. Imamo\=glu and Yamamoto \cite{Imamoglu93} assume that the
Poisson distribution is modified in some particular way by the finite
charging energy, and are able to obtain sub-Poissonian shot noise
suppression. We treat shot noise in the Coulomb blockade regime later
on (Section \ref{strcorr}) by more elaborate methods.  

\subsection{Suppression of shot noise in double-barrier structures}
\label{lang52} 

We consider now transport through quantum wells, which were
described quantum-mechanically in Section \ref{scat}. The tunneling
rates through the left $\Gamma_L$ and the right $\Gamma_R$ barrier are
assumed to be much lower than all other characteristic energies,
including temperature\footnote{In terms of the quantum-mechanical
derivation, this would mean that the transmission coefficient
(\ref{breitw}) is replaced by $T(E) = 2\pi\Gamma_{Ln}\Gamma_{Rn}
\Gamma_n^{-1} \delta(E - E_n^r)$.}. Introducing the distribution
function in the well $f_w (E)$, we write the charge of the well $Q_w$
in the form  
\begin{equation} \label{charge501}
Q_w = e\nu_2 {\cal{A}} \sum_{n} \int_0^{\infty} dE_z dE_{\perp} f_w
(E_z + E_{\perp}) \delta (E_z - E_n^r), 
\end{equation}
where the energy $E_z$ in the well is counted from the band bottom
$eU$ in the well\footnote{Note that the notations here and below
differ from those introduced in Section \protect\ref{scat}: All the
energies are measured from the corresponding band bottoms.}
(Fig.~\ref{rtw1}), and the sum is taken over all the resonant levels.

Now we introduce the charges $Q_L (t)$ and $Q_R (t)$ which have passed
through the left and right barriers, respectively, from the time
$t=-\infty$ until the time $t$. At any instant of time the charge of
the well is $Q_w(t) = Q_L(t) - Q_R(t)$. The time evolution of the
charge $Q_L$ is determined by the rate equation, 
\begin{equation} \label{rate502}
\dot Q_L = e \left( \gamma_{L\rightarrow} - \gamma_{L\leftarrow}
\right), 
\end{equation}
where $\gamma_{L\rightarrow}$ and $\gamma_{L\leftarrow}$ are
transition rates through the left barrier, from the reservoir to the
well and from the well to the reservoir, respectively. We have
\begin{eqnarray} \label{ratestun1}
& & \gamma_{L\rightarrow} = \frac{\nu_2 {\cal{A}}}{\hbar} \sum_{n}
\int_{e(V-U)}^{\infty} dE_z \int_0^{\infty} dE_{\perp} \Gamma_{Ln}
\delta (E_z - E_n^r) \nonumber \\
& \times & f_L (E_z + E_{\perp} + eU - eV) \left[ 1 - f_w (E_z +
E_{\perp}) \right], \nonumber \\  
& & \gamma_{L\leftarrow} = \frac{\nu_2 {\cal{A}}}{\hbar} \sum_{n}
\int_{e(V-U)}^{\infty} dE_z \int_0^{\infty} dE_{\perp} \Gamma_{Ln}
\delta (E_z - E_n^r)\nonumber \\ 
& \times &  \left[ 1 - f_L (E_z + E_{\perp} + eU - eV) \right] 
f_w (E_z + E_{\perp}). 
\end{eqnarray}
Here we wrote the distribution function of the left (right) reservoir
in such a way that the energy is counted from the chemical potential
of the left (right) reservoir. Similarly, $\dot Q_R =
e(\gamma_{R\rightarrow} - \gamma_{R\leftarrow})$ is expressed in terms
of the transition rates through the right barrier, which are written
analogously to Eq. (\ref{ratestun1}). 

Further progress is easy in two cases: either the tunneling
rates $\Gamma_{L,Rn}$ do not depend on $n$ (and equal $\Gamma_{L,R}$),
or there is only one resonant level of the longitudinal motion in the
relevant range of energies. Taking into account Eq. (\ref{charge501}),
we obtain 
\begin{eqnarray} \label{rateeq1}
\dot Q_L = I_L - \hbar^{-1} \Gamma_L (Q_L - Q_R), \nonumber \\
\dot Q_R = I_R + \hbar^{-1} \Gamma_R (Q_L - Q_R),
\end{eqnarray}
where we have introduced 
\begin{eqnarray} \label{currents501}
I_L & = & \frac{e\nu_2 {\cal{A}}}{\hbar} \Gamma_L \sum_{n}
\int_{e(V-U)}^{\infty} dE_z \nonumber \\
& \times & \int_0^{\infty} dE_{\perp} \delta (E_z -
E_n^r) f_L (E_z + E_{\perp} + eU - eV) \nonumber, \\ 
I_R & = & -\frac{e\nu_2 {\cal{A}}}{\hbar} \Gamma_L \sum_{n}
\int_{0}^{\infty} dE_z \nonumber\\  
& \times & \int_0^{\infty} dE_{\perp} \delta
(E_z - E_n^r) f_R (E_z + E_{\perp} + eU).
\end{eqnarray}
Note that we derived the rate equations (\ref{rateeq1}) without
specifying the distribution function $f_w$. Thus, within the
approximations used here, the rate equations are the same,
independently of the relaxation rate in the well (which determines the
distribution $f_w$).  

The current, for instance, through the left barrier, is given by the
sum of the particle current $I_{Lp} = \dot Q_L$ and the displacement
current $I_{Ld} = c_L \dot Q_w/(c_L + c_R)$, where we have introduced
the capacitances $c_L$ and $c_R$ of the left and the right barrier,
respectively. Using this gives for the total current
\begin{equation} \label{Shockley}
I = \frac{c_L \dot Q_R + c_R \dot Q_L}{c_L + c_R},
\end{equation}
which is often cited as Ramo--Shockley theorem\footnote{Many papers in
the field are flawed since they use $I = (\dot Q_L + \dot Q_R)/2$ and
subsequently claim the validity of the results for arbitrary
barriers. Similarly, evaluating Eq. (\ref{Shockley}) with the help of
the free electron results for $\dot Q_L$ and $\dot Q_R$ leads to a
current conserving answer but not to a self-consistent result. Compare
Eqs. (\ref{dbtfreq}) and (\ref{noisnlin512}).}. The calculation of the
current through the right barrier yields the same result,
demonstrating that the total current is conserved.   

Up to now, we have only discussed average quantities. The idea of the
Langevin approach is that the (current) fluctuations can be calculated
from the same rate equations (\ref{rateeq1}), if random currents ({\em
Langevin sources}) are added to their right-hand sides. We write
\begin{eqnarray} \label{rateeq2}
\dot Q_L = I_L - \hbar^{-1} \Gamma_L (Q_L - Q_R) + \xi_L(t), \nonumber
\\ 
\dot Q_R = I_R + \hbar^{-1} \Gamma_R (Q_L - Q_R) + \xi_R (t),
\end{eqnarray}
where the Langevin sources $\xi_{L,R} (t)$ have the following
properties. They are zero on average, $\langle \xi_{\alpha} (t)
\rangle = 0$, $\alpha = L,R$. Furthermore, they are correlated only
for the same barrier, and the correlation function describes 
Poissonian shot noise at each barrier, 
\begin{equation} \label{sourcecorr}
\langle \xi_{\alpha} (t) \xi_{\beta} (t') \rangle = e \langle I
\rangle \delta(t - t') \delta_{\alpha\beta},
\end{equation}
where $\langle I \rangle$ is the average current. To find the noise
power, we do not need to specify higher cumulants of the Langevin
sources. As we mentioned above, the definition (\ref{sourcecorr}) is
intuitive rather than the result of a formal derivation. However, the
results we obtain in this way coincide with those found by ensemble
averaging the quantum-mechanical results.  

The equations (\ref{rateeq2}) are linear and can be easily solved. The
average current is
\begin{equation} \label{curlang1}
\langle I \rangle = \frac{\Gamma_R I_L + \Gamma_L I_R}{\Gamma}, 
\end{equation}
where again $\Gamma = \Gamma_L + \Gamma_R$. The finite frequency shot
noise power is found to be
\begin{eqnarray} \label{dbtfreq}
S(\omega) & = & 2e \langle I \rangle \left[ \frac{c_L^2 + c_R^2}{c^2}
\right. \nonumber \\
& + & \left. \frac{2}{\Gamma^2 + (\hbar\omega)^2} \left( \Gamma^2
\frac{c_L c_R}{c^2} - \Gamma_L \Gamma_R \right) \right],
\end{eqnarray}
with the definition $c \equiv c_L + c_R$. 

For zero frequency Eq. (\ref{dbtfreq}) gives the result
(\ref{Fanodbt}), $F = (\Gamma_L^2 + \Gamma_R^2)/\Gamma^2$. For high
frequencies $\hbar\omega \gg \Gamma$ we have $S(\omega) =
2e\langle I \rangle (c_L^2 + c_R^2)/c^2$. This expression can be
obtained from Eq. (\ref{Shockley}) based on the assumption that the
particle currents $\dot Q_L$ and $\dot Q_R$ fluctuate independently
and according to the Poisson shot noise. The crossover frequency
between these regimes is $\omega \sim \Gamma/\hbar$: In accordance
with general expectations, the frequency dependence of the shot noise 
is governed by time-scales inherent to the conductor. We
note finally that for a symmetric double barrier, $c_L = c_R$ and
$\Gamma_L = \Gamma_R$, the noise power (\ref{dbtfreq}) is frequency
independent and equal to $S(\omega) = e\langle I \rangle$.  

Note also that Eq. (\ref{dbtfreq}) does not contain the high-frequency
Nyquist noise (\ref{noisfr4}), which is proportional to $\hbar\vert
\omega \vert$: Zero-point noise is quantum-mechanical and cannot be
reproduced by a classical discussion.    

The classical derivation of Eq. (\ref{dbtfreq}), based on the master
equation approach, was given independently by Chen and Ting
\cite{Chen92} ($c_L = c_R$, zero frequency limit) and by Davies {\em
et al} \cite{Davies92} (general case), and later by Chen \cite{Chen93}
($c_L = c_R$, arbitrary frequency) and M\"uller {\em et al}
\cite{Muller97} ($\omega = 0$). Sun and Milburn
\cite{Milburn98,Milburn99} and Milburn \cite{Milburn99a} developed a
quantum master equation approach and also derived the same result
(\ref{dbtfreq}). The Langevin approach was applied to 
double-barrier structures in Ref. \cite{BlanterMoriond}; here we have
given an extended version of the derivation. We also remark that the
frequency dependence of the noise of the double-barrier structure
(\ref{dbtfreq}) was obtained by Runge \cite{Runge} ($c_L = c_R$)
and by Lund B\o\ and Galperin \cite{Bo} (general case) using the
non-equilibrium Green's functions method.    

Beenakker and de Jong \cite{Jong95,Jong96} consider the two-barrier
suppression using the conceptually similar Boltzmann-Langevin
approach, described in the Section \ref{BoltzmannL}. They also
investigate the case of $n$ identical barriers in series and obtain
for the Fano factor 
\begin{displaymath}
F = \frac{1}{3} \left[ 1 + \frac{n(1-T)^2(2+T) - T^3}{[T + n(1-T)]^3}
\right],  
\end{displaymath}
where $T$ is the transmission coefficient of a barrier. This
expression gives $F = 1 - T$ for $n=1$, $F = 1/3$ for $n \to
\infty$ (which mimics a diffusive wire), and reduces to $F = 1/2$ for
$T \ll 1$ for the two-barrier case, $n = 2$. Thus, the crossover of
the ensemble averaged shot noise between a two-barrier and
many-barrier (diffusive) system can be described classically.   

The theory we have presented above has a number of drawbacks, which we
discuss now. First, the charges $Q_L$ and $Q_R$ are assumed to be
continuous, and thus Coulomb blockade effects cannot be treated in
this way (see Section \ref{strcorr}). Even the charging effects which
exist if charge quantization can be neglected are not properly taken
into account. The discussion thus far has neglected to include the
response to the fluctuating electric potential in the well. Thus the
Ramo-Shockley theorem, as it has been used here, and is applied in
much of the literature, leads to a current conserving, but, as we
discuss below, not to a self-consistent result for the frequency
dependence of the shot noise. In the next subsection we show that
charging effects can lead to the enhancement of shot noise above the
Poisson value.  

Furthermore, our consideration is limited to zero
temperature\footnote{To use the kinetic equation formalism, we had to
assume $k_BT \gg \Gamma$; on the other hand Eq. (\ref{sourcecorr})
states $k_BT \ll eV$.}, and it is not immediately clear how the
Langevin approach should be modified in this case to reproduce correct
expressions for the Nyquist noise. Another limitation is that our
derivation assumes that the tunneling rates through each of the
resonant levels are the same. If this is not the case, the rate
equations do not have the simple form (\ref{rateeq1}), but instead
start to depend explicitly on the distribution function $f_w$. Whereas
the above derivation does not require any assumptions on the
distribution of the electrons in the well $f_w$ ({\em i.e.} any
information on the inelastic processes inside the well), generally
this information is required, and it is not {\em a priori} clear
whether the result on noise suppression depends on the details of the
inelastic scattering.  

The last two issues are relatively easily dealt with in the more
general master equation approach. Chen and Ting \cite{Chen92}, Chen
\cite{ChenMPL93}, and independently Davies {\em et al} \cite{Davies92}
solved the master equation in the {\em sequential tunneling limit},
when the electrons, due to very strong inelastic scattering inside the
well, relax to the equilibrium state. They found the results to be
identical to those obtained by quantum-mechanical methods (which
require quantum coherence, {\em i.e.} absence of inelastic
scattering), and concluded that inelastic processes do not affect shot
noise suppression in quantum wells. Later, Iannaccone, Macucci, and
Pellegrini \cite{IMP97} solved the master equation allowing explicitly
for arbitrary inelastic scattering, and found that the noise
suppression factor is given by Eq. (\ref{Fanodbt}) at zero temperature
for arbitrary inelastic scattering provided the reservoirs are
ideal. They also studied other cases and temperature effects. This
picture seems to be consistent with the results obtained
quantum-mechanically by attaching voltage probes to the sample
(subsection \ref{scat027}).  

We note here, however, that there is no consensus in the literature
concerning this issue. First, we discuss the results obtained
quantum-mechanically by Davies, Carlos Egues, and Wilkins
\cite{Davies95}. They start from the exact expression (\ref{rtwtc1})
and average it over the phase $\phi$, allowing for inelastic
scattering (dephasing). Instead of assuming that the phase is a
uniformly distributed random variable, they postulated $\langle \exp
(i\phi_1 + \phi_2) \rangle = \langle \exp (i\phi_1) \rangle\langle
\exp (i\phi_2) \rangle$. For the Fano factor (at zero frequency) they
obtain in this way  
\begin{displaymath} 
F = 1 - \frac{2\Gamma_L\Gamma_R}{\Gamma^2} \left( \frac{\Gamma}{\Gamma
+ \Gamma_{in}/2} \right),
\end{displaymath} 
where $\Gamma_{in}$ is the rate of inelastic scattering, proportional
to $1 - \langle \exp (i \phi) \rangle$. Thus, in their model inelastic
scattering {\em enhances} the Fano factor, driving noise towards the
Poisson value. While it is clear that for certain models of inelastic
scattering shot noise is affected by interactions, we do not see a
direct relation of the model of Ref. \cite{Davies95} to the voltage
probe models which we consider in subsection \ref{scat027} and which
yield the result that the Fano factor is interaction insensitive. 

Furthermore, Isawa, Matsubara, and Ohuti \cite{Isawa98}, using the
Green's functions approach, find that inelastic processes leading to 
sequential tunneling affect the Fano factor. Their theory, however, is
explicitly not current conserving. We mentioned already the result by
Lund B\o \ and Galperin \cite{Bo1}, who report {\em suppression} of
the Fano factor by electron-phonon interactions. Their results clearly
contradict the conclusions based on the voltage probe models. 

A related issue is investigated by Sun and Milburn
\cite{Milburn98,Milburn99}, who apply the quantum master equation
to the analysis of noise in a double-well (triple-barrier)
structure. With this approach, they are able to study the case when
the two wells are coupled coherently. Their results show an abundance
of regimes depending on the relation between the coupling rates to the
reservoirs, elastic rates, and the coupling between the two wells.   

To conclude this subsection, we address here one more problem. The
result (\ref{Fanodbt}) predicts that the noise suppression factor may
assume values between $1/2$ and $1$, depending on the asymmetry of the
double barrier. The question is whether interaction effects, under
some circumstances, may lead to Fano factors above $1$
(super-Poissonian noise) or below $1/2$. Relegating the problem of
super-Poissonian noise to the next subsection, we only discuss here
the possible suppression of shot noise below $1/2$. Experimentally,
noise suppression below $1/2$ was observed in early experiments by
Brown \cite{Brown92}, and recently by Przadka {\em et al}
\cite{Przadka} and by Kuznetsov {\em et al} \cite{Mendez}. 

Early papers on the subject (Han and Barnes \cite{Han91}, Alam and
Khondker \cite{Alam92}, Sheng and Chua \cite{Chua94}, Jahan and
Anwar \cite{Jahan95} ) predict shot noise suppression down to zero
either with frequency or even at zero-frequency. Following
Ref. \cite{Han91}, these works treat current fluctuations as a
superposition of density and velocity fluctuations, with a 
self-consistent treatment of interaction effects. However, apparently
they did not include the partition noise ($T(1-T)$) in their
consideration. Since it is precisely the partition noise which
produces the minimal suppression of $1/2$, it is not quite surprising
that their theory predicts lower suppression factors.  

Sugimura \cite{Sugimura} and, independently, Carlos Egues, Hershfield,
and Wilkins \cite{Egues94} propose a model in which the states in the
well are inelastically coupled to the degrees of freedom of the
reservoirs. This model, indeed, yields noise suppression below $1/2$
in a limited parameter range. The minimal suppression factor given in
Ref. \cite{Egues94} is $0.45$, which is still above the experimental
data. This direction of research looks promising, but certainly
requires more efforts.  

\subsection{Interaction effects and super-Poissonian noise
enhancement} \label{lang53}

Prior to the discussion of charging effects, we briefly comment on how
quantum wells operate in the strongly non-linear regime ($eV \sim
E_F$), provided charging effects are not important. For simplicity, we
assume that there is only one resonant level of the longitudinal
motion $E_0 \equiv E_0^r$ in the relevant range of energies, and all
others levels lie too high to be of importance, $E_n^r \gg eV, E_F$
for $n > 0$. We assume also $E_0 > E_F$, and $V > 0$, then electrons
from the right reservoir cannot enter the well.  

Modifying Eq. (\ref{qwc00}) to take into account that the band bottoms
$eV$ in the left reservoir and $U$ in the quantum well are
now finite, and substituting the transmission coefficient, 
$T(E_z) = 2\pi \Gamma_L \Gamma_R \Gamma^{-1} \delta(E_z - E_0)$, we
obtain for the average current
\begin{equation} 
\langle I \rangle = \frac{e \nu_2 {\cal{A}}}{\hbar}
\frac{\Gamma_L\Gamma_R}{\Gamma} \left( eV - eU + E_F - E_0 \right),
\end{equation}
provided $E_0 + eU - E_F < eV < E_0 + eU$, and zero otherwise. This
dependence, which, of course, could also be obtained classically
from the rate equations, is shown in Fig. \ref{fig501}a, solid
line. The current drops abruptly to zero for $eV = E_0 + eU$, which
corresponds to the passage of the band bottom of the left reservoir
through the resonant level in the well. When the smearing of the
resonance due to finite tunneling rates is taken into account (the
transmission coefficient is not replaced by a delta-function), the
$I$--$V$ curve becomes smeared (dashed line in Fig. \ref{fig501}a),
and the region of {\em negative differential resistance} develops
around $eV \sim E_0 + eU$. This was noted by Tsu and Esaki in their
early paper \cite{TsuEsaki}, and subsequently observed experimentally
in Ref. \cite{Esaki1}.  

Now we consider charging effects. The new ingredient is now that the
electrostatic potential in the well $U$ is not an independent
parameter any more, but is a function of voltage, which must be
calculated self-consistently.  Moreover, it has its own dynamics and
may fluctuate; we are going to show that the fluctuations of $U$ may
considerably enhance noise. Theoretical papers emphasizing the
necessity of charging effects for the $I$--$V$ curve are too numerous
to be cited here. For noise, the necessity of a self-consistent
treatment in the negative differential resistance region was
illustrated in the Green's function approach by Levy Yeyati, Flores,
and Anda \cite{Flores93}, who, however, did not take into account the
fluctuations of $U$. Iannaccone {\em et al} \cite{Iann98} suggested
that these fluctuations may lead to the enhancement of noise above the
Poisson value, and provided numerical results supporting this
statement. A self-consistent analytical theory of noise in quantum
wells including the fluctuations of the band bottom 
was developed by the authors \cite{BB99} in the framework of the
scattering approach\footnote{The quantity $U$ serves then as an
operator which obeys an operator Poisson equation.}. The
classical theory yields the same results \cite{BlanterMoriond};
here we give the classical derivation.   
\begin{figure}
{\epsfxsize=8.5cm\centerline{\epsfbox{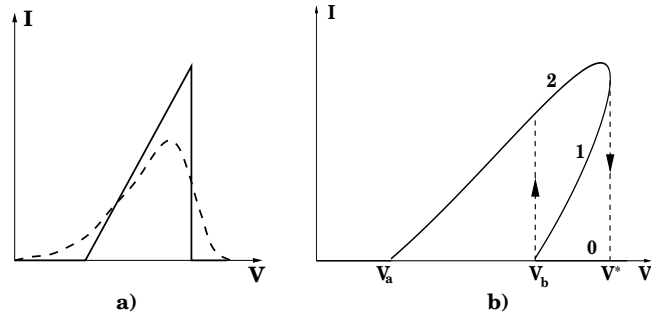}}}
\vspace{0.3cm}
\caption{$I$--$V$ characteristics of the quantum well: (a) charging
effects are neglected; (b) charging effects are taken into
account. The values of parameters for (b) are $a_L = a_R$, $c_L =
c_R$, $E_F = 2E_0/3$, $e^2\nu_2{\cal{A}} = 10c_L$. For this case $eV^*
= 2.70 E_0$. }
\label{fig501}
\end{figure}

In the strongly non-linear regime which we discuss here, the energy
dependence of the tunneling rates becomes important. To take this into
account, we use a simple model and treat each barrier as
rectangular. The transmission probability through a rectangular
barrier determines the partial decay width of the resonant level, 
\begin{eqnarray} \label{tunrates511}
\Gamma_L (E_z) & = & a_L E_z^{1/2} (E_z + eU - eV)^{1/2} \nonumber \\
& \times & \theta(E_z) \theta(E_z + eU - eV); \nonumber \\
\Gamma_R (E_z) & = & a_R E_z^{1/2} (E_z + eU)^{1/2} \theta(E_z)
\theta(E_z + eU),
\end{eqnarray}
where $a_L$ and $a_R$ are dimensionless constants (the case of a
symmetric well corresponds to $a_L = a_R$, {\em not} to $\Gamma_L =
\Gamma_R$). We emphasize that 
the partial decay widths  
$\Gamma_{L,R}$ are now functions of $V$ and $U$.  

Consider first the average, stationary quantities. Equations 
(\ref{rateeq1}) are still valid for our case (now $I_R = 0$, since 
$E_0 > E_F$). However, $U$ is no longer an independent variable, but
is related to the charge in the well $Q_L - Q_R$. Assuming that the
interaction effects can be described by a charging energy only, we
write this relation in the form 
\begin{equation} \label{charging51}
(Q_L - Q_R) = c_L (U-V) + c_R U.
\end{equation}
This equation just states that the total charge of the capacitor
equals the sum of charges at the left and the right plates. Equations
(\ref{rateeq1}) and (\ref{charging51}) must now be solved together,
using the expressions for the partial decay widths
(\ref{tunrates511}). Combining them, we obtain a closed equation for
$U$,  
\begin{eqnarray} \label{bbottom51}
\hbar c \dot U & = & e \nu_2 {\cal{A}} \Gamma_L
\left( eV - eU + E_F - E_0 \right) \nonumber \\
& \times & \theta (eV - eU + E_F - E_0) \theta (E_0 + eU - eV)
\nonumber \\
& - & \left( \Gamma_L + \Gamma_R \right) \left[ c_L (U - V) + c_R U
\right],  
\end{eqnarray}
where $\Gamma_{L,R} \equiv \Gamma_{L,R} (E_0)$.

We analyze now the stationary solutions\footnote{Analytic expressions
for the stationary solutions may be obtained, since they
obey a cubic equation. However, these expressions are to cumbersome
and not very transparent. Instead, we have chosen to give a
qualitative discussion, illustrating it by numerical results.} of
Eq. (\ref{bbottom51}). For $0 < eV < eV_a \equiv c(E_0 - E_F)/c_R$ the
only solution is $U = U_0 \equiv c_LV/c$, which corresponds to the
charge neutral well (see Eq. (\ref{charging51})): The resonant level
is pushed up too high to allow any charge in the well. At $V = V_a$
the resonant level passes through the Fermi level of the left
reservoir, and with a further increase of voltage, the well is charged,
$U > U_0$. For $eV_a < eV < eV_b \equiv c E_0/c_R$, 
Eq. (\ref{bbottom51}) has only one stationary solution; however, for
$V > V_b$ three solutions develop. One of them is $U = U_0$ and
corresponds to the charge neutral well; two other solutions $U_1 <
U_2$ describe the charged well. It can be seen from
Eq. (\ref{bbottom51}) that the solution $U_1$ is unstable, while $U_0$
and $U_2$ are stable. As the voltage $V$ grows, the solutions $U_1$
and $U_2$ move towards each other, and at the voltage $V^*$ (which is
referred below as the {\em instability threshold}) merge. For higher
voltages $V > V^*$, the only stationary solution is $U_0$: The well is
charge neutral again, since the resonant level lies too low.    

Thus, the new feature due to charging effects is the multi-stability of
the system in the range of voltages between $V_b$ and $V^*$. The
$I$--$V$ characteristics of the quantum well with charging are shown
in Fig. \ref{fig501}b; in the multi-stability range the marks $0$, $1$,
and $2$ refer to the solutions $U_0$, $U_1$, and $U_2$, respectively. 
The current is only non-zero if the well is charged. The instability
is manifest in the hysteretic behavior shown by dashed lines in
Fig. \ref{fig501}b: when the voltage is increasing, the well stays
charged (solution $U_2$) until $V^*$, and then jumps to the
zero-current state ($U_0$); if the voltage is decreasing, the current
is zero until $V_b$, and then the jump to the charged state ($U_2$)
happens. This hysteresis was apparently observed experimentally
\cite{GoldmanTsui}. A finite value of the tunneling rates smears all
these features, causing a finite current for all values of
voltage. Furthermore, the instability range shrinks, and a region of
negative differential resistance appears close to the instability
threshold $V^*$. For $\Gamma \sim (V^* - V_b)$ the instability
disappears, and the $I$--$V$ characteristics resemble the dashed
curve in Fig. \ref{fig501}a: The multi-stable regime is only
pronounced for wells formed with high tunnel barriers. 

Now we turn to the calculation of noise. The principal difficulty
which we encounter for the charged well is the following. Equations
(\ref{rateeq1}) are now {\em non-linear}, since they depend on the
potential in the well $U$ in a non-linear way, and $U$, in turn,
is related to the charges $Q_L$ and $Q_R$ via
Eq. (\ref{bbottom51}). The Langevin sources, however, can only be
added to {\em linear} equations \cite{Kampen}. Thus, we have to
linearize our set of equations\footnote{The quantum-mechanical theory
\protect\cite{BB99} also has been developed in linear
approximation.}. Restricting ourselves to the voltage range $V_a < V <
V^*$, we write for the potential in the well
\begin{displaymath}
U(t) = U_2 + \Delta U(t), 
\end{displaymath}
where $U_2 (V)$ is a stable stationary solution corresponding to the
charged well, and $\Delta U(t)$ are fluctuations. Expanding in $\Delta
U$ and adding the Langevin sources with the same properties as before
to the resulting {\em linear} equations, we write
\begin{eqnarray} \label{Langevin511} 
\dot Q_L & = & \langle I \rangle +\frac{1}{\hbar c} \left[ \hbar J -
\Gamma_L (c + c_0) \right] \left( Q_L - Q_R \right) + \xi_L (t);
\nonumber \\ 
\dot Q_R & = & \langle I \rangle + \frac{1}{\hbar c} \left[ \hbar J + 
\Gamma_R (c + c_0)\right] \left( Q_L - Q_R \right) + \xi_R (t),
\nonumber \\
\end{eqnarray}
where the $U$-dependent tunneling rates $\Gamma_{L,R}$ are evaluated
for $U = U_2$, and the quantities $c_0 = - \partial \bar Q /\partial
U$ and $J = \partial (\Gamma_R \bar Q)/\hbar \partial U$ (also taken
for $U = U_2$; $\bar Q \equiv e\nu_2 {\cal A} (eV - eU + E_F - E_0)
\Gamma_L/\Gamma$ is the average charge of the well) are the response
of the average charge and current to the increment of the potential in
the well. The average current in Eq. (\ref{Langevin511}) is $\langle I
\rangle = \Gamma_R \bar Q/\hbar$.    

The noise spectrum, which follows from Eqs. (\ref{Langevin511}), is
\begin{eqnarray} \label{noisnlin512}
S(\omega) & = & 2e \langle I \rangle \left\{ \frac{c_L^2 + c_R^2}{c^2}
+ \frac{2}{c^2\hbar^2\omega^2 + \Gamma^2(c+c_0)^2} \right. \\
& \times & \left. \left[ (c+c_0)^2 \left[ -\Gamma_L \Gamma_R +
\frac{c_Lc_R\Gamma^2}{c^2} \right] \right. \right. \nonumber \\
& - & \left. \left. \hbar J (c+c_0)(\Gamma_L -
\Gamma_R) + \hbar^2 J^2 \right]  \right\}. \nonumber  
\end{eqnarray}
The noise power (\ref{noisnlin512}) is strongly voltage dependent
via the quantities $c_0$, $J$, and $\Gamma$. In particular, for $V
\to V^*$ the combination $c + c_0$ tends to zero, as it is seen from
Eq. (\ref{bbottom51}). If the charging effects do not play a role, we
may set $c_0 = J = 0$ and recover the result (\ref{dbtfreq}).  

The frequency structure of Eq. (\ref{noisnlin512}) is identical to
that of Eq. (\ref{dbtfreq}); the crossover frequency is $\hbar\omega
\sim \Gamma (c + c_0)/c$, and drops down to zero for $V = V^*$. For
high frequencies the usual result $S = 2e\langle I \rangle (c_l^2 +
c_R^2)/c^2$ is recovered. For zero frequency, we obtain a voltage
dependent Fano factor, 
\begin{equation} \label{Fano512}
F = \frac{1}{2} + 2\frac{(\Lambda - \Delta \Gamma)^2}{\Gamma^2}, \ \
\ \Delta\Gamma \equiv (\Gamma_L-\Gamma_R)/2.
\end{equation}
We have introduced the ``interaction energy''
\begin{displaymath}
\Lambda \equiv \frac{\hbar J}{c_L + c_R + c_0},
\end{displaymath}
which has the form of a dimensionless conductance $\hbar J/e^2$
multiplied by an effective charging energy of the well $e^2/(c_L + c_R
+ c_0)$. This quantity contains the relevant information about the
charging effects of the well. Eq. (\ref{noisnlin512}) is a
self-consistent result in contrast to Eq. (\ref{dbtfreq}) found by
inserting the free electron currents into the Ramo-Shockley
formula. We reproduce this result for $\Lambda = 0$. We re-emphasize
that a calculation of ac conductance or noise using the free-electron
results for the currents and the Ramo-Shockley formula is not in
general a sound procedure: It assumes that the  self-consistent
contribution to the currents arising from the internal potential
oscillations or fluctuations can be neglected. 
\begin{figure}
{\epsfxsize=6.cm\centerline{\epsfbox{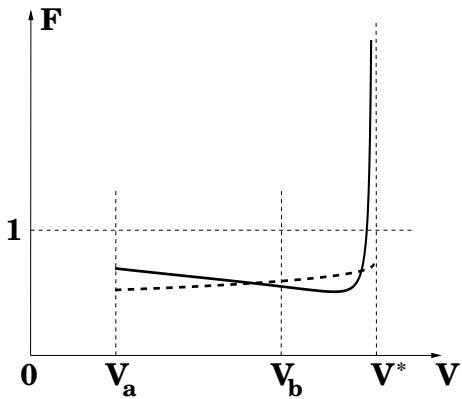}}}
\vspace{0.3cm}
\caption{Voltage dependence of the Fano factor (\protect\ref{Fano512})
for the same set of parameters as Fig. \protect\ref{fig501} (solid
line); Fano factor (\protect\ref{Fanodbt}) for a charge neutral
quantum well (dashed line). }
\label{fig502}
\end{figure}

For $V \to V^*$ the denominator $c + c_0$ of the interaction energy
$\Lambda (V)$ quite generally diverges as $(V^* - V)^{-1/2}$, while
the numerator $J$ stays finite. Thus, the Fano factor diverges
according to $(V^* - V)^{-1}$. In particular, close enough to the
instability threshold $V^*$ the Fano factor increases above one: The
noise becomes {\em super-Poissonian}. At the onset of current, for $V
= V_a$ the Fano factor can be calculated in closed form, and one has
$1/2 < F(V_a) < 1$. Thus, we describe the transition from
sub-Poissonian to super-Poissonian noise. The minimal possible value
of the Fano factor in this theory is $F = 1/2$. It can be shown that
the voltage dependence of the Fano factor is peculiar, and, depending
on the relative value of the charging effects $e^2 \nu_2 {\cal{A}}/c$,
noise may be {\it either suppressed as compared with the
non-interacting value} (\ref{Fanodbt})  for low voltages and enhanced
for high voltages (``weak interaction scenario''),{\it  or enhanced},
then suppressed, and only after that enhanced again (``strong
interaction scenario''). For details, see Ref. \cite{BB99}. The
voltage dependence of the Fano factor is illustrated in
Fig. \ref{fig502}. The divergence of noise is clearly seen.  

The finite value of the partial decay widths (tunneling rates) smears
the singularities, and, in particular, induces a finite value of the
Fano factor for $V = V^*$. Another source of deviations from the
experimental results is our linearization procedure. For voltages
close to the instability threshold the linear approximation is clearly
insufficient, and large fluctuations (transitions between the state
$U_2$ and the state $U_0$) must be taken into account. These
fluctuations also would induce a finite value of noise for $V =
V^*$. We do not see any reason, however, why these processes should
suppress noise below the Poisson value in the whole range of voltages.  

Now we briefly review various predictions of the possibility of
super-Poissonian noise enhancement in quantum wells. Brown
\cite{Brown92} theoretically predicted that noise can be enhanced
above the Poissonian value because the energy dependence of
transmission probabilities will be affected by the applied voltage in
a non-linear way. He, however, did not include partition noise in his
consideration, so that for the energy independent transmission
probabilities noise is fully Poissonian. Jahan and Anwar
\cite{Jahan95}, who also found super-Poissonian noise enhancement,
included self-consistent effect at the level of the stationary
transmission probabilities, but also did not take partition noise into
account. As we already mentioned, an explanation of the
super-Poissonian noise in terms of the potential fluctuations was
given by Iannaccone {\em et al} \cite{Iann98}, and the analytical
theory of this enhancement, identifying the relevant energy scales,
was proposed in Refs. \cite{BB99,BlanterMoriond}.   

Experimentally, enhancement of noise in quantum wells, as the voltage
approaches the range of negative differential resistance, was observed
already in the early experiments by Li {\em et al} \cite{LiRTQW} and
by Brown \cite{Brown92}. The super-Poissonian shot noise in the
negative differential resistance range was observed by Iannaccone {\em
et al} \cite{Iann98}. Kuznetsov {\em et al} \cite{Mendez} have
presented a detailed investigation of the noise oscillations from
sub-Poissonian to super-Poissonian values of a resonant quantum well
in the presence of a parallel magnetic field. The magnetic field leads
to multiple voltage ranges of negative differential resistance and
permits a clear demonstration of the effect. Their results are shown
in Fig. \ref{Mendezres}.  

To conclude this Section, we discuss the following issue. To obtain
the super-Poissonian noise enhancement, we needed multi-stable behavior
of the $I$--$V$ curve; in turn, the multi-stability in quantum
wells is induced by charging effects. It is easy to see, however, that
the charging (or, generally, interaction) effects are not required to
cause the multi-stability. Thus, if instead of a voltage controlled
experiment, we discuss a current controlled experiment, the $I$--$V$
characteristics for the uncharged quantum well (Fig.~\ref{fig501}) are
multi-stable for {\em any} external current. For the case of an
arbitrary load line there typically exists a finite range of external
parameters where multi-stable behavior is developed. Furthermore,
the quantum wells are not the only systems with multi-stability; as
one well-known example we mention Esaki diodes, where the
multi-stability is caused by the structure of the energy bands
\cite{PHHT}. 
\begin{figure}
{\epsfxsize=7.cm\centerline{\epsfbox{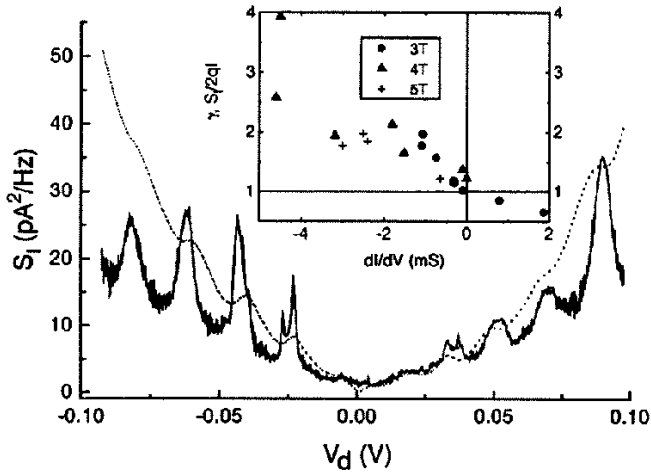}}}
\vspace{0.3cm}
\caption{Experimental results of Kuznetsov {\em et al}
\protect\cite{Mendez} which show noise in resonant quantum wells in
parallel magnetic field. A dotted line represents the Poisson
value. It is clearly seen that for certain values of applied bias
voltage the noise is super-Poissonian.}  
\label{Mendezres}
\end{figure}
 
Usually such bistable systems are discussed from the point of view of
telegraph noise, which is due to spontaneous random transitions
between the two states. This is a consideration complementary to the
one we developed above. Indeed, in the linear approximation the system
does not know that it is multi-stable. The shot noise grows
indefinitely at the instability threshold only because the state
around which we have linearized the system becomes unstable rather
than metastable. This is a general feature of linear fluctuation
theory. Clearly the divergence of shot noise in the linear
approximation must be a general feature of all the systems with
multi-stable behavior. Interactions are not the necessary ingredient
for this shot noise enhancement. On the other hand, as we have
discussed, the transitions between different states, neglected in the
linear approximation, will certainly soften the singularity and drive
noise to a finite value at the instability threshold. To describe in
this way the interplay between shot noise and random telegraph noise
remains an open problem.

\section{Boltzmann-Langevin approach to noise: Disordered systems}
\label{BoltzmannL} 

\subsection{Fluctuations and the Boltzmann equation} \label{bol61}

In this Section we describe the generalization of the Langevin method
to disordered systems. As is well known, the evolution of the
(average) distribution function $\bar f(\bbox{r}, \bbox{p},
t)$ is generally described by the Boltzmann equation,  
\begin{equation} \label{Boltzmann1}
\left( \partial_t + \bbox{v} \nabla + e\bbox{E}
\partial_{\bbox{p}} \right) \bar f(\bbox{r}, \bbox{p}, t)
= I[\bar f] + I_{in} [\bar f]. 
\end{equation}
Here $\bbox{E}$ is the local electric field, $I_{in}[\bar f]$ is
the inelastic collision integral, due to the electron-electron and
electron-phonon scattering (we do not have to specify this integral
explicitly at this stage), and $I[\bar f]$ is the electron-impurity
collision integral. For a $d$-dimensional disordered system of volume
$\Omega$ it is written as    
\begin{eqnarray} \label{colint}
I[ \bar f(\bbox{r}, \bbox{p}, t)] & = & \Omega \int
\frac{d\bbox{p}'}{(2\pi\hbar)^d} \left[ \bar J (\bbox{p}', \bbox{p}, 
\bbox{r}, t) - \bar J (\bbox{p}, \bbox{p}', \bbox{r}, 
t) \right], \nonumber 
\\ 
\bar J (\bbox{p}, \bbox{p}', \bbox{r}, t) & \equiv &  \tilde W
(\bbox{p}, \bbox{p}', \bbox{r}) \bar f(\bbox{r},
\bbox{p}, t) \left[ 1 - \bar f(\bbox{r}, \bbox{p}', t)
\right], \nonumber \\
\end{eqnarray}
where we have introduced the probability $\tilde W$ of scattering per
unit time from the state $\bbox{p}$ to the state $\bbox{p}'$
due to the impurity potential $U$, 
\begin{displaymath}
\tilde W (\bbox{p}, \bbox{p}', \bbox{r}) =
\frac{2\pi}{\hbar} \vert U_{\bbox{p} \bbox{p}'} \vert^2 \delta
\left[ \epsilon(\bbox{p}) - \epsilon(\bbox{p}') \right].
\end{displaymath}
Thus, the impurity collision integral can be considered as the sum of
particle currents $J$ to/from the state $\bbox{p}$ from/to all the
possible final states $\bbox{p}'$, taken with appropriate signs.  

The fluctuations are taken into account via the {\em
Boltzmann-Langevin} approach, introduced in condensed matter physics
by Kogan and Shul'man \cite{KoganShulman}. This approach assumes that
the particle currents between the states $\bbox{p}$ and
$\bbox{p}'$ fluctuate due to the randomness of the scattering
process and partial occupation of the electron states. We write  
\begin{equation} \label{Lang601}
J (\bbox{p}, \bbox{p}', \bbox{r}, t) =\bar J (\bbox{p}, \bbox{p}',
\bbox{r}, t) + \delta J (\bbox{p}, 
\bbox{p}', \bbox{r}, t), 
\end{equation}
where the average current $\bar J$ is given by Eq. (\ref{colint}),
and $\delta J$ represents the fluctuations. Then the actual
distribution function $f(\bbox{r}, \bbox{p}, t)$, which is the
sum of the average distribution $\bar f$ and fluctuating  part of the
distribution $\delta f$,  
\begin{equation} \label{distfun1}
f(\bbox{r}, \bbox{p}, t) = \bar f(\bbox{r}, \bbox{p},
t) + \delta f(\bbox{r}, \bbox{p}, t), 
\end{equation}
obeys a Boltzmann equation which contains now in addition a
fluctuating Langevin source $\xi$ on the right hand side,
\begin{eqnarray} \label{Bolzmann2}
& & \left( \partial_t + \bbox{v} \nabla + e\bbox{E}
\partial_{\bbox{p}} \right) f(\bbox{r}, \bbox{p}, t)
= I[f] + I_{in} [f] + \xi (\bbox{r}, \bbox{p}, t), \nonumber
\\ 
& & \xi (\bbox{r}, \bbox{p}, t) = \Omega \int
\frac{d\bbox{p}'}{(2\pi\hbar)^d} \left[ \delta J (\bbox{p}', \bbox{p}, 
\bbox{r}, t) - \delta J (\bbox{p}, \bbox{p}', \bbox{r}, t) \right].
\nonumber \\
\end{eqnarray}

These Langevin sources are zero on average, $\langle \xi \rangle =
0$. To specify the fluctuations, Kogan and Shul'man
\cite{KoganShulman} assumed that the currents $J (\bbox{p}, \bbox{p}',
\bbox{r}, t)$ are {\em independent elementary 
processes}. This means that these currents are correlated only when
they describe the same process (identical initial and final states,
space point, and time moment); for the same process, the correlations
are taken to be those of a Poisson process. Explicitly, we have 
\begin{eqnarray} \label{fluct601}
& & \left\langle \delta J (\bbox{p}_1, \bbox{p}_2, \bbox{r},
t) \delta J (\bbox{p}_1', \bbox{p}_2', \bbox{r}', t')
\right\rangle \nonumber \\
& = & \frac{(2\pi\hbar)^{2d}}{\Omega} \delta \left(
\bbox{p}_1 - \bbox{p}_1' \right) \delta \left( \bbox{p}_2 - 
\bbox{p}_2' \right) \nonumber \\
& \times & \delta(\bbox{r} - \bbox{r}')  \delta(t - t') \bar J
(\bbox{p}_1, \bbox{p}_2, \bbox{r}, t).    
\end{eqnarray}
Eq. (\ref{fluct601}) then implies the following correlations between
the Langevin sources,
\begin{eqnarray} \label{fluct602}
& & \left\langle \xi (\bbox{r}, \bbox{p}, t) \xi (\bbox{r}',
\bbox{p}', t') \right\rangle = \delta(\bbox{r} - \bbox{r}')
\delta(t-t') G(\bbox{p}, \bbox{p}', \bbox{r}, t), \nonumber \\
\end{eqnarray} 
where 
\begin{eqnarray} \label{fluct603}
G(\bbox{p}, \bbox{p}', \bbox{r}, t) & = & \Omega \left\{
\delta(\bbox{p} - \bbox{p}') \int d\bbox{p}'' \right. \nonumber \\
& \times & \left. \left[ \bar
J(\bbox{p}'', \bbox{p}, \bbox{r}, t) + \bar J(\bbox{p}, \bbox{p}'',
\bbox{r}, t) \right] \right. \nonumber \\ 
& - & \left. \left[ \bar J(\bbox{p}, \bbox{p}', \bbox{r},
t) + \bar J(\bbox{p}', \bbox{p}, \bbox{r}, t) \right]
\right\}.  
\end{eqnarray}
Note that the sum rule
\begin{equation} \label{sumrule601}
\int d\bbox{p} G(\bbox{p}, \bbox{p}', \bbox{r}, t) =
\int d\bbox{p}' G(\bbox{p}, \bbox{p}', \bbox{r}, t) =
0,
\end{equation}
is fulfilled. This sum rule states that the fluctuations only
redistribute the electrons over different states, but do not change
the total number of particles. 

These equations can be further simplified in the important case (which
is the main interest in this Section) when all the quantities are
sharply peaked around the Fermi energy. Instead of the momentum
$\bbox{p}$, we introduce then the energy $E$ and the direction of
the momentum $\bbox{n} = \bbox{p}/p$. The velocity and the
density of states are assumed to be constant and equal to $v_F$ and
$\nu_F$, respectively. We write 
\begin{displaymath} 
\Omega \tilde W(\bbox{p}, \bbox{p}', \bbox{r}) = \nu_F^{-1}
\delta(E - E') W(\bbox{n}, \bbox{n}', \bbox{r}), 
\end{displaymath} 
where $W$ is the probability of scattering from the state $\bbox{n}$
to the state $\bbox{n}'$ per unit time at the space point 
$\bbox{r}$. Furthermore, we will be interested only in the
stationary regime, {\em i.e.} the averages $\bar f$ and $\bar J$ ({\em
not} the fluctuating parts) are assumed to be time
independent. Eliminating the electric field $\bbox{E}$ by the
substitution $E \to E - e\varphi (\bbox{r})$, with $\varphi$ being
the potential, we write the Boltzmann-Langevin equation in the form   
\begin{equation} \label{Boltzmann2}
\left( \partial_t + v_F\bbox{n} \nabla \right) f(\bbox{r},
\bbox{n}, E, t) = I[f] + I_{in} [f] + \xi(\bbox{r}, \bbox{n}, E, t),   
\end{equation}
where the Langevin sources $\xi$ are zero on average and are
correlated as follows, 
\begin{eqnarray} \label{fluct604}
& & \left\langle \xi(\bbox{r}, \bbox{n}, E, t) \xi(\bbox{r}',
\bbox{n}', E', t') \right\rangle = \frac{1}{\nu_F}
\delta(\bbox{r} - \bbox{r}') \delta(t-t') \nonumber \\
& \times & \delta(E - E') G(\bbox{n}, \bbox{n}', \bbox{r},
E), 
\end{eqnarray}
\begin{eqnarray} \label{fluct605} 
& & G(\bbox{n}, \bbox{n}', \bbox{r}, E) \equiv \int
d\bbox{n}'' \left[ \delta (\bbox{n} - \bbox{n}' ) - \delta
(\bbox{n}' - \bbox{n}'' ) \right] \nonumber \\
& \times & \left[ W(\bbox{n},
\bbox{n}'', \bbox{r}) \bar f (1 - \bar f'') + W(\bbox{n}'', \bbox{n},
\bbox{r}) \bar f'' (1 - \bar f) \right]. \nonumber \\   
\end{eqnarray}
We used the notations $\bar f \equiv \bar f(\bbox{r}, \bbox{n}, E)$
and $\bar f'' \equiv \bar f(\bbox{r}, \bbox{n}'', 
E)$. The expression for the current density,
\begin{equation} \label{cur601}
\bbox{j} (\bbox{r}, t) = ev_F \int d\bbox{n} \ dE \
\bbox{n} f(\bbox{r}, \bbox{n}, E, t)  
\end{equation}
(with the normalization $\int d\bbox{n} = 1$), completes the
system of equations used in the Boltzmann-Langevin method. We remark
that in this formulation the local electric potential does not appear
explicitly: for systems with charged carriers such as electric 
conductors the electric field is coupled to the (fluctuating) charge
density via the Poisson equation. We will return to this point when we
discuss situations in which a treatment of this coupling is essential.  

\subsection{Metallic diffusive systems: Classical theory of 1/3 --
noise suppression and multi-probe generalization} \label{bol62}

Equations (\ref{Boltzmann2}), (\ref{fluct604}), and (\ref{fluct605})
are very general and may be applied to a large variety of systems. 
Now we turn to the case of metallic diffusive systems, where these
equations may be simplified even further and eventually can be
solved. A classical theory of noise in metallic diffusive wires was
proposed by Nagaev \cite{Nagaev92}, and subsequently by de Jong and
Beenakker \cite{Jong95,Jong96}. Sukhorukov and Loss \cite{SL98,SLlong}
gave another derivation of the shot noise suppression for the
two-terminal wire and, more importantly, generalized it to treat
conductors of arbitrary geometry and with an arbitrary number of
contacts. Below we give a sketch of the derivation, following
Ref. \cite{SLlong}, to which the reader is addressed for further
details.      

The distribution function in metallic diffusive systems is almost
isotropic. We then separate it into symmetric and asymmetric parts,
\begin{equation} \label{split601}
f(\bbox{r}, \bbox{n}, E, t) = f_0(\bbox{r}, E, t) +
\bbox{n} \bbox{f}_1(\bbox{r}, E, t).
\end{equation}
For simplicity, we consider here the relaxation time approximation for
the electron-impurity collision integral,  
\begin{displaymath} 
I [f] = -\frac{\bbox{n}\bbox{f}_1}{\tau (\bbox{r})},
\ \ \frac{n_{\alpha}}{\tau (\bbox{r})} = \int d\bbox{n}' \
W(\bbox{n}, \bbox{n}', \mbox{r}) \left[ n_{\alpha} -
n'_{\alpha} \right],
\end{displaymath}
where $\tau$ is the average time a carrier travels between collisions
with impurities. This approximation is valid when the scattering is
isotropic ($W$ depends only on the difference $\vert \bbox{n} -
\bbox{n}' \vert$). The full case is analyzed in
Ref. \cite{SLlong}. Integrating Eq. (\ref{Boltzmann2}) first with
$\int d\bbox{n}$, and then with $\int \bbox{n} \ d\bbox{n}$, we obtain
\begin{eqnarray} \label{interm601}
D\nabla \cdot \bbox{f}_1 & = & l \bar I_{in}[f], \ \ \ \bar I_{in}
[f] \equiv \int d\bbox{n} \ I_{in} [f], \nonumber \\
\bbox{f}_1 & = & -l \nabla f_0 + \tau d \int  \bbox{n} \ \xi
d\bbox{n}, 
\end{eqnarray}
where we have introduced the mean free path\footnote{This is the
transport mean free path; this definition differs by a numerical
factor from that used in Section \protect\ref{scat}. Following the
tradition, we are keeping two different definitions for the scattering
approach and the kinetic equation approach.} $l = v_F \tau$ and the
diffusion coefficient $D = v_F l/d$. We assume the system to be {\em
locally} charge neutral. Integrating Eq. (\ref{interm601}) with
respect to energy, we obtain for the local fluctuation of the current
\begin{eqnarray} \label{curfl601}
& & \delta \bbox{j} + \sigma \nabla \delta \varphi = \delta
\bbox{j}^s, \ \ \ \nabla \cdot \delta \bbox{j} = 0, \nonumber \\
& & \delta \bbox{j}^s (\bbox{r}, t) = el\nu_F \int \bbox{n} \
\xi(\bbox{r}, \bbox{n}, E, t)\ d\bbox{n}\ dE.   
\end{eqnarray}
Here $\sigma = e^2 \nu D$ is the conductivity, and $\varphi
(\bbox{r})$ is the electrostatic potential. The currents $\bbox{j}^s$
are correlated as follows, 
\begin{equation} \label{fluct611}
\left\langle \delta j^s_l (\bbox{r}, t) \delta j^s_m
(\bbox{r}', t') \right\rangle = 2\sigma \delta_{lm}
\delta(\bbox{r} - \bbox{r}') \delta(t - t') \Pi(\bbox{r}),
\end{equation}
where the quantity $\Pi$ is expressed through the isotropic part of
the average distribution function $\bar f$,
\begin{equation}\label{loctemp611}
\Pi(\bbox{r}) = \int dE \bar f_0 (\bbox{r}, E) \left[ 1 - \bar
f_0 (\bbox{r}, E) \right].  
\end{equation}
The distribution $\bar f_0$ obeys the equation
\begin{equation} \label{avfun611}
D\nabla^2 \bar f_0 (\bbox{r}, E) + \bar I_{in}[\bar f_0 - l
\bbox{n} \cdot \nabla \bar f_0] = 0. 
\end{equation}
The standard boundary conditions for the distribution function are the
following. Let $L_n$ denote the area of contact $n$ ($1 \le n \le N$),
and $\Omega$ the rest of the surface of the sample. At contact $n$,
the non-equilibrium distribution function $\bar f_0$ is determined by
the equilibrium Fermi distribution function in the reservoir $n$,
$\bar f_0\vert_{L_n} = f_n (E)$, whereas away from the contacts the
current perpendicular to the surface must vanish and thus $\bbox{N}
\cdot \nabla \bar f_0 \vert_{\Omega} = 0$, where $\bbox{N}$ is the
outward normal to the surface. 

Eqs. (\ref{curfl601}) and (\ref{fluct611}) can be used to find the
current-current fluctuations if the non-equilibrium carrier
distribution is known. Thus we proceed first to find the
non-equilibrium distribution function, solving Eq. (\ref{avfun611}). 
We follow then Ref. \protect\cite{SL98} and find the {\em 
characteristic potentials}\footnote{For an arbitrary conductor the 
electrostatic potential can be expanded as $\varphi(\bbox{r}) =
\sum \phi_n (\bbox{r}) V_n$; Ref. \protect\cite{Buttiker93} calls the
coefficients $\phi_n$ {\em characteristic potentials}. We remark that
in the absence of inelastic processes, the 
average distribution function can be written as a linear combination
of the equilibrium reservoir functions $f_n$, $\bar f (\bbox{r}) =
\nu_F^{-1} \sum \nu_n (\bbox{r}) f_n$. Refs. 
\cite{Gramespa98,Gramespa99} call the coefficients $\nu_n$
{\em injectivities}. In the diffusive metallic conductor of interest
here the characteristic potentials and injectivities are the same
functions up to a factor given by the local density of states
$\nu_F$. Such an equivalence does not hold, for instance, in systems
composed of different metallic diffusive conductors, and in general the
characteristic potentials and injectivities may have a quite different
functional form. Here the use of the characteristic potentials has the
advantage that it takes effectively the local charge neutrality into
account.}  $\phi_n$, which on the ensemble average obey the Poisson
equation $\nabla^2 \phi_n = 0$, with the boundary conditions 
\begin{displaymath} 
\phi_n\vert_{L_m} = \delta_{mn}; \ \ \ \bbox{N} \cdot \nabla
\phi_n \vert_{\Omega} = 0. 
\end{displaymath}
In terms of the characteristic potentials the electrostatic potential
is \cite{Buttiker93}
\begin{displaymath}
\varphi(\bbox{r}) = \sum_n \phi_n (\bbox{r}) V_n,
\end{displaymath} 
where $V_n$ is the voltage applied to the reservoir $n$. 
Note that $\sum_{n}\phi_n (\bbox{r}) =1$ at every space point 
as a consequence of the invariance of the electrical properties 
of the conductor under an arbitrary overall voltage shift.
With the help of the characteristic potentials, the conductance matrix
(which we, as before, define as $I_m = G_{mn} V_n$, $I_m$ being the
current through $L_m$ directed into the sample), we obtain 
\begin{equation} \label{cond611}
G_{mn} = \sigma \int d\bbox{r} \nabla \phi_m \nabla \phi_n
\end{equation} 
(the integration is carried out over the whole sample). 
The conductances are independent of the electrical (non-equilibrium)
potential inside the conductor. To see this one can re-write
Eq. (\ref{cond611}) in terms of a surface integral.   

Multiplying Eq. (\ref{curfl601}) by $\nabla \phi_n$ and
integrating over the whole volume we obtain the fluctuation of the
current through the contact $n$. The potential fluctuations
$\delta\varphi (\bbox{r})$ actually play no role and are
eliminated due to the boundary condition that they vanish at the
contacts. At zero temperature, and to linear order in the applied
voltage, this is exact: Internal potential fluctuations play a role
only in the non-linear voltage dependence of the shot noise and in its
temperature dependence. Taking into account the form of the
correlation function (\ref{fluct611}), we find the noise power,
\begin{equation} \label{nois611}
S_{mn} = 4 \sigma \int d\bbox{r} \nabla \phi_n (\bbox{r})
\nabla \phi_m (\bbox{r}) \Pi(\bbox{r}).   
\end{equation}
Eq. (\ref{nois611}), together with the equation (\ref{avfun611}) for
the distribution function $\bar f_0$, is the general result for the
multi-terminal noise power within the classical approach. At
equilibrium $\Pi(\bbox{r}) = k_BT$, and Eq. (\ref{nois611})
reproduces the fluctuation-dissipation theorem. 

We next apply Eq. (\ref{nois611}) to calculate noise suppression in
metallic diffusive wires, for the case when the inelastic processes are
negligible, $\bar I_{in} = 0$. We consider a wire of length $L$
and width\footnote{We use two-dimensional notations, $d = 2$.} $W
\ll L$, situated along the axis $x$ between the point $x=0$
(reservoir L) and $x=L$ (R). The voltage $V$ is applied to the left
reservoir. There are only two characteristic potentials, 
\begin{equation} \label{charpot612}
\phi_L = 1 - \phi_R = 1 - x/L,
\end{equation}
which obey the diffusion equation and do not depend on the transverse
coordinate. The average distribution function is found as $f_0 (x) =
\phi_L (x) f_L + \phi_R (x) f_R$, and thus the quantity $\Pi$ for zero
temperature is expressed as
\begin{displaymath} 
\Pi(x) = eV \phi_L (x) [1 - \phi_L (x)].
\end{displaymath}
Subsequently, we find the conductance $G_{LL} = \sigma W/L$, and the
shot noise 
\begin{equation} \label{nois612} 
S_{LL} = \frac{4e\sigma WV}{L^2} \int_0^L dx \frac{x}{L} \left( 1 -
\frac{x}{L} \right) = \frac{2e\langle I \rangle}{3}.  
\end{equation}
As we mentioned earlier, this expression is due to Nagaev
\cite{Nagaev92}. The Fano factor is $1/3$, in accordance with
the results found using the scattering approach (Section
\ref{scat}). 

For purely elastic scattering the distribution function $\bar f_0$ in
an arbitrary geometry quite generally can be written as 
\begin{equation} \label{distfun613}
\bar f_0 (\bbox{r}, E) = \sum_n \phi_n (\bbox{r}) f_n (E). 
\end{equation}
This facilitates the progress for multi-probe geometries. Sukhorukov
and Loss \cite{SL98,SLlong} obtain general expressions for the
multi-terminal noise power and use them to study the Hanbury Brown --
Twiss effect in metallic diffusive conductors. The quantum-mechanical
theory of the same effect can be found in Ref. \cite{Blanter97}. 

\subsection{Interaction effects} \label{bol63}

Interaction effects are relatively easy to deal with in the
Boltzmann-Langevin approach, in contrast to the difficulties
encountered by the scattering theory.  

{\bf Electron-electron interactions}. An important feature of
electron-electron interactions is that they do not change the total
momentum of the electron system\footnote{This is only correct if 
Umklapp processes can be neglected. The influence of Umklapp processes
on shot noise in mesoscopic systems has not been
investigated.}. Generally, therefore, electron-electron scattering
alone cannot cause transport, and in particular it cannot cause
noise. Technically, this is manifested in the fact that the form of
current-current fluctuations is given by the same expression
(\ref{nois611}), as in the non-interacting case. However,
electron-electron scattering processes alter the distribution function
$\bar f_0$, and thus the value of the shot noise. 

The inelastic collision integral for electron-electron 
interactions $\bar I_{in} [f]$ (which we denote $\bar I_{ee} [f]$)
generally has the form 
\begin{eqnarray} \label{colint621}
\bar I_{ee} [f(E)] & = & \int dE' d\omega K(\omega) \left\{ f(E) f(E')
\right. \nonumber \\
& \times & \left. \left[ 1 - f(E - \omega) \right] \left[ 1 - f(E' +
\omega) \right] \right. \\
& - & \left. f(E - \omega) f(E' + \omega) \left[ 1 - f(E) \right]
\left[ 1 - f(E') \right] \right\}, \nonumber 
\end{eqnarray}
where the kernel $K(\omega)$ for disordered systems must be found from
a microscopic theory. For three-dimensional metallic diffusive systems
it was obtained by Schmid \cite{Schmid74}; for two- and
one-dimensional systems the zero-temperature result may be found in
Ref. \cite{AAr}. For finite temperatures, a self-consistent treatment
is needed \cite{Blanter96}. This kernel turns out to be a complicated
function of disorder and temperature. In particular, in 1D for zero
temperature it diverges as $K(\omega) \propto \omega^{-3/2}$ for
$\omega \to 0$. The strength of the interaction is characterized by a
time $\tau_{ee}$, which we call the {\em electron-electron scattering
time}.   

Thus, one needs now to solve Eq. (\ref{avfun611}) for $\bar f_0$,
calculate the quantity $\Pi(\bbox{r})$, and substitute it into the
expression (\ref{nois611}) to obtain the noise. Up to now, only two
limiting cases have been discussed analytically. First, for weak
interactions $D/L^2 \gg \tau^{-1}_{ee}$ (with $L$ being the typical
size of the system) the collision integral in Eq. (\ref{avfun611}) may
be treated perturbatively. Since electron-electron interactions are of
minor importance, the distribution function $\bar f_0$ in this case is
still given by Eq. (\ref{distfun613}). Nagaev \cite{Nagaev95},
assuming a specific form of $K(\omega)$, found that the shot noise
power is {\em enhanced} due to the electron-electron interactions
compared to the non-interacting result.  

Another regime where progress is possible is that of strong
scattering, $D/L^2 \ll \tau^{-1}_{ee}$. In this situation, electrons
undergo many scattering events before leaving the system, and one can
expect that the distribution function at every point is close to the
equilibrium distribution. Indeed, in the leading order, the diffusion
term in Eq. (\ref{avfun611}) can be neglected, and the distribution
function must then be one for which the collision integral is
zero. The collision integral (\ref{colint621}) vanishes identically
for the Fermi distribution, and thus the distribution function assumes
the form of a local equilibrium Fermi function with the potential
$\varphi(\bbox{r})$ and a local effective temperature $T(\bbox{r})$, 
\begin{equation} \label{hotelec1}
\bar f_0 (\bbox{r}, E) = \left[ \exp \left( \frac{E -
e\varphi(\bbox{r})}{k_BT(\bbox{r})} \right) + 1 \right]^{-1} .
\end{equation}   
In the literature this distribution function is also referred to as a
{\em hot electron distribution}. Eq. (\ref{avfun611}) is now used to
find the effective temperature profile $T(\bbox{r})$. Consider the
quantity 
\begin{equation} \label{endens621}
w(\bbox{r}) = \int dE \ E \left[ \bar f_0 (E) - \theta(E -
e\varphi(\bbox{r})) \right],
\end{equation} 
which up to a coefficient and an additive constant is the total energy
of the system. The substitution of Eq. (\ref{hotelec1}) gives
$w(\bbox{r}) = \pi^2 T^2(\bbox{r})/6$. Then the application of
Eq. (\ref{avfun611}) yields   
\begin{equation} \label{temp622}
\frac{\pi^2}{6} \nabla^2 \left[ T^2 (\bbox{r}) \right]  = -
\frac{e^2}{2} \nabla^2 \varphi^2 (\bbox{r}) = -e^2E^2(\bbox{r}), 
\end{equation}
where we have taken into account $\nabla^2 \varphi = 0$ to obtain the
final result. Actually, the term on the right-hand side is
proportional to the Joule heating $\bbox{jE}$, and the equation
states that this energy losses are spent to heat the electron gas. 

In the following, we again specialize to the case of a
quasi-one-dimensional metallic diffusive wire between $x = 0$ and $x =
L$. In this case $E = V/L$, and Eq. (\ref{temp622}) must be solved
with the boundary conditions $T(0) = T(L) = T_0$, with $T_0$ being the
bath temperature. For the temperature profile we then obtain
\begin{equation} \label{temp623}
T(x) = \left[ T_0^2 + \frac{3}{\pi^2} \left( \frac{eV}{L} \right)^2 x
(L - x) \right]^{1/2}.
\end{equation}
Substituting this into Eq. (\ref{nois611}), we find the shot noise for
hot electrons. In particular, when the bath temperature $T_0$ equals
zero, we find for the Fano factor
\begin{equation} \label{Fano621}
F = \frac{\sqrt{3}}{4} \approx 0.43.
\end{equation}
The result (\ref{temp623}) is due to Kozub and Rudin
\cite{Kozub95,Kozub951} and Nagaev \cite{Nagaev95}. The multi-terminal
generalization is given by Sukhorukov and Loss
\cite{SL98,SLlong}. Eq. (\ref{temp623}) states that shot noise for hot
electrons is actually {\em higher} than for non-interacting
electrons. This agrees with the notion which we obtained considering
the scattering approach (Section \ref{scat}): Electron heating
enhances the shot noise. Indeed, enhanced (as compared to the
$1/3$--suppression) shot noise was observed experimentally by
Steinbach, Martinis, and Devoret \cite{Steinbach}. Their experimental
data are shown in Fig. \ref{devoretfig} for a particular
sample. Shorter samples in the same experiment \cite{Steinbach}
exhibit Fano factors which are between $1/3$ and $\sqrt{3}/4$. The
$1/3$--suppression of shot noise and crossover from the diffusive to
the hot-electron regime was very carefully studied by Henny {\em et
al} \cite{Henny}, see subsection \ref{scat0264}. 

The hot-electron result (\ref{Fano621}) is actually independent of the
details of electron-electron interaction (independent of the kernel
$K(\omega)$ in Eq. (\ref{colint621})). The crossover between $F = 1/3$
and $F = \sqrt{3}/4$ {\em does} depend on this kernel. Nagaev
\cite{Nagaev981} and Naveh \cite{Naveh981} studied this crossover
numerically for a particular form of $K(\omega)$ which assumes that
there is no interference between elastic and electron-electron
scattering. They suggested that information on the strength of the
electron-electron scattering may be extracted from the zero-frequency
noise measurements.  
\begin{figure}
{\epsfxsize=8.cm\centerline{\epsfbox{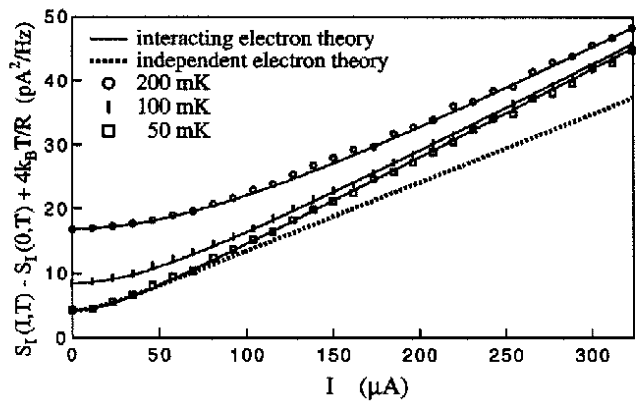}}}
\vspace{0.3cm}
\caption{Shot noise observed at the same sample for three different
temperatures by Steinbach, Martinis, and Devoret
\protect\cite{Steinbach}. Dashed and solid lines indicate the
$1/3$--suppression and the hot electron result $F=\sqrt{3}/4$. The
temperature is the lowest for the lowest curve.}  
\label{devoretfig}
\end{figure}

At this point, we summarize the information we obtained from the
scattering and Boltzmann-Langevin approaches about the effects of the
electron-electron interactions on noise. There are two characteristic
times, one responsible for dephasing processes, $\tau_{\phi}$, and
another one due to inelastic scattering (electron heating),
$\tau_{ee}$. We expect $\tau_{\phi} \ll \tau_{ee}$. Dephasing does not
have an effect on noise, and thus for short enough samples, $D/L^2 \gg
\tau_{ee}^{-1}$, the Fano factor is $1/3$ (irrespectively of the
relation between $D/L^2$ and $\tau_{\phi}$). For long wires, $D/L^2
\ll \tau_{ee}^{-1}$, the Fano factor equals $\sqrt{3}/4$ due to
electron heating. For even longer wires the electron-phonon
interactions become important (see below), and shot noise is
suppressed down to zero.  

The microscopic theory, however, predicts {\em three} characteristic
times responsible for electron-electron scattering in disordered
systems (for a nice qualitative explanation, see Ref. \cite{AAK}). One
is the dephasing time $\tau_{\phi}$, another one is the energy
relaxation time, $\tau_E$, and the third one, for which we keep the
notation $\tau_{ee}$, has the meaning of the average time between
electron-electron collisions. While the dephasing time is
quantum-mechanical and cannot be accounted for in the classical
theory, the information about $\tau_{ee}$ and $\tau_E$ is contained in
the collision integral (\ref{colint621}). For three-dimensional
disordered systems, all the three times coincide: The presence of
dephasing always means the presence of inelastic scattering and
electron heating. The situation is, however, different in two- and
one-dimensional systems, where the three times differ parametrically,
with the relation \cite{AAK,Blanter96} $\tau_{ee} \ll \tau_{\phi} \ll
\tau_E$. This is in apparent contrast with the intuitive predictions
of the scattering approach, and opens a number of questions. First, it
is not clear whether $\tau_E$ or $\tau_{ee}$ is responsible for the
$1/3$ -- $\sqrt{3}/4$ crossover in the classical theory. Then, the
role of dephasing, which is not taken into account in the
Boltzmann-Langevin approach, may need to be revisited. These questions
may only be answered on the basis of a microscopic theory.

We must point out here that interactions, besides altering the
amplitude of shot noise, also create an additional source of noise in
diffusive conductors, since the moving electrons produce fluctuating
electromagnetic field, as discovered by von Oppen and Stern
\cite{Stern97}. Referring the reader to Ref. \cite{Stern97} for
details, we mention that this noise is proportional to $V^2$ for low
voltages, and to $\vert \omega \vert^{2/(4-d)}$ for $D/L^2 \ll \hbar
\vert \omega \vert \ll k_BT$. As a function of frequency, this noise
saturates for low frequencies, and vanishes for $\hbar \vert \omega
\vert \gg k_B T$.  

{\bf Electron-phonon interactions}. In contrast to electron-electron
scattering, electron-phonon interactions {\em do} change the total
momentum of electrons and cause a finite resistance and noise by
themselves. Noise in ballistic, one-channel quantum wires due to
electron-phonon interactions within the Boltzmann-Langevin approach
was studied by Gurevich and Rudin \cite{Gurevich95,Gurevich951}, who
start directly from Eq. (\ref{Boltzmann2}) with the electron-phonon
collision integral on the right-hand side (no impurity scattering). 
They consider only the situation of weak interactions, when the
electron distribution function is not modified by inelastic
scattering. They discover that, for this case, the main effect is
the absence of noise when the Fermi energy lies below the threshold
energy $E_{th} = 2p_Fs$, where $s$ is the sound velocity. The
appearance of this threshold is due to the fact that the maximal wave
vector of acoustic phonons which interact with electrons is
$2p_F/\hbar$. For $E_F > E_{th}$ shot noise grows. The suppression of
shot noise in long wires, which is a consequence of the equilibration
of the electron distribution function due to strong interactions, was
beyond the scope of Refs. \cite{Gurevich95,Gurevich951}.    

In disordered systems, one has to take into account three
effects. First, elastic scattering modifies the electron-phonon
collision integral \cite{Schmid73}, which assumes different forms 
depending on the spatial dimension and degree of disorder. Then, it
affects the distribution function of electrons. Finally, interactions
modify the resistance of the sample, and the expression for the shot
noise does not have the simple form of Eq. (\ref{nois611}). The
distribution function of phonons, which enters the electron-phonon
collision integral, must be, in principle, found from the Boltzmann
equation for phonons, which couples with that for electrons. 

The standard approximations used to overcome these difficulties and to
get reasonable analytical results are as follows. First, for low
temperatures, the contribution to the resistance due to
electron-phonon collisions is much smaller than that of
electron-impurity scattering. Thus, one assumes that
Eq. (\ref{nois611}) still holds, and electron-phonon collisions only
modify the distribution function for electrons and the form of the
collision integral. Furthermore, phonons are assumed to be in
equilibrium at the lattice temperature. This approach was taken by
Nagaev, who calculates the effects of electron-phonon scattering on
the shot noise of metallic diffusive wires for the case when
electron-electron scattering is negligible \cite{Nagaev92}, and
subsequently for the case of hot electrons \cite{Nagaev95}. In
addition, careful numerical studies of the role of the electron-phonon
interaction in noise in metallic diffusive conductors (starting from
the Boltzmann-Langevin approach) are performed by Naveh, Averin, and
Likharev \cite{Naveh98}, and Naveh \cite{Naveh982}. The results depend
on the relation between temperature, applied voltage, and
electron-phonon interaction constant, and we refer the reader to
Refs. \cite{Nagaev92,Nagaev95,Naveh98,Naveh982} for details. The only
feature we want to mention here is that for constant voltage $eV \gg
k_B T$ and strong electron-phonon scattering, the Fano factor
decreases with the length of the wire. For both non-interacting and
hot electrons the specific form of the collision integral which was
assumed leads to the dependence $F \propto L^{-2/5}$, which is
different from the prediction $F \propto L^{-1}$ of the simple model
of voltage probes (Section \ref{scat}). The exponent $2/5$ is,
however, strongly model dependent, and should not be taken very
seriously.  

A simple way to see that shot noise is suppressed by the
electron-phonon scattering is \cite{Kozub95,Jong96} to assume that, in
the limit of strong interactions, the distribution function of
electrons is the Fermi function with the temperature equal to the bath
temperature. Eq. (\ref{nois611}) gives then a noise power 
which is just the Nyquist value, {\em i.e.} in this case the shot
noise is completely suppressed. 

Experimentally, for low temperatures, electron-phonon scattering is
less effective than electron-electron collisions, and therefore one
expects that, with the increase of the length of the wire, one first
goes from the non-interacting regime $F = 1/3$ to the hot-electron
regime, $F = \sqrt{3}/4$. For even longer wires electron-phonon
collisions play a role, and the Fano factor decreases down to
zero. 

\subsection{Frequency dependence of shot noise} \label{bol64}

While discussing the classical approach to the shot noise suppression,
we explicitly assumed that the sample is {\em locally charge neutral}:
Charge pile-up is not allowed in any volume of any size. As a result,
we obtained white (frequency independent) shot noise. 

In reality, however, there is always a finite (though small) screening
radius, which in the case when the system is locally three-dimensional
has the form $\lambda_0 = (4\pi e^2 \nu)^{-1/2}$. As Naveh, Averin,
and Likharev \cite{Naveh97} and Turlakov \cite{Turlakov99} point out,
if one of the dimensions of a disordered sample becomes comparable
with $\lambda_0$, the pile-up of the charge may modify the frequency
dependence of noise, though it leaves the zero-frequency noise power
unchanged. If all the dimensions of the sample exceed the screening
radius (which is typically the case for metallic, and often also for
semiconducting) mesoscopic systems, the charge pile-up inside the
sample is negligible, and noise stays frequency independent until at
least the plasma frequency, which in three-dimensional structures is
very high. 

The situation is different if the sample is capacitively connected to
an external gate. As we have seen in the framework of the scattering
approach (Section \ref{freq}), the fact that the sample is now
charged, leaves the zero-frequency noise unchanged, but strongly
affects the frequency dependence of the shot noise. An advantage of
the Boltzmann-Langevin approach is that it can treat these effects
analytically, calculating the potential distribution inside the
sample and making use of it to treat the current fluctuations. The
general program is as follows. Instead of the charge neutrality
condition, one uses the full Poisson equation, relating potential and
density fluctuations inside the sample. In their turn, density
fluctuations are related to the current fluctuations via the
continuity equation. Finally, one expresses the current fluctuations
via those of the potential and the Langevin sources. Thus, the
system of coupled partial differential equations with appropriate
boundary conditions needs to be solved. The solution is
strongly geometry dependent and has not been written down for an
arbitrary geometry. Particular cases, with a simple geometry, were
considered by Naveh, Averin, and Likharev \cite{Naveh97,Naveh99f},
Nagaev \cite{Nagaev98,Nagaev981}, and Naveh
\cite{Naveh981,Naveh982}. Without even attempting to give derivations,
we describe here the main results, referring the reader to these
papers for more details.  

A conductor in proximity to a gate can be charged vis-a-vis the gate.
We can view the conductor and the gate as the two plates of a
capacitor. In the limit where the screening length is much larger than
the wire radius, only a one-dimensional theory is needed. It is this
atypical situation which is considered here. For a wire of
cross-section ${\cal A}$ and a geometrical capacitance $c$ per unit
length its low frequency dynamics is characterized by the
electrochemical capacitance\footnote{For a discussion of
electrochemical capacitance and ac conductance see Refs. \cite{BTP93}
and \cite{BHB}; see also Section \protect\ref{freq}.} $c_{\mu}^{-1}
= c^{-1} + (e^{2} {\cal A}\nu)^{-1}$ 
which is the parallel addition of the geometrical capacitance and the
quantum capacitance $(e^{2} {\cal A} \nu)^{-1}$. Here we have assumed
that the potential is uniform both along the wire and more importantly
also in the transverse direction of the wire. Any charge accumulated
in the wire can dynamically relax via the reservoirs connected to the
wire and via the external circuit which connects the wire and the
gate. For a zero-external impedance circuit this relaxation generates
a charge relaxation resistance $R_{q}$ (see the discussion in
Section \ref{freq}) which for a metallic diffusive conductor is of the
order of the sample resistance $R = L (\sigma{\cal A})^{-1}$. With
these specifications we expect that a metallic diffusive wire in
proximity of a gate is characterized by a frequency $\omega_{RC} =
1/R_qC_{\mu}$ which is given by $\omega_{RC} = \sigma{\cal A}/(cL^2) +
\sigma/(e^{2} {\nu} L^{2})$. (Refs. \cite{Naveh97,Naveh99f} express
$\omega_{RC}$ in terms of a generalized diffusion constant $D' = D +
\sigma {\cal A}/c$ using the Einstein relation $\sigma = e^{2} \nu D$,
such that $\hbar \omega_{RC} = \hbar D'/L^2$ has the form of a
Thouless energy). For $\omega \ll \omega_{RC}$, noise measured at the
contacts to the wire is dominated by the white-noise zero-frequency
contribution (the Fano factor equals $1/3$ for independent electrons
or $\sqrt{3}/4$ for hot electrons). For frequencies higher than
$\omega_{RC}$  the spectrum measured at the contacts of the wire
starts to depend on the details of the system, and for infinite
frequency the Fano factor tends to a constant value, which may lie
above as well as below the non-interacting value. This is because the
zero-temperature quantum noise $S \propto \hbar \vert \omega \vert$
cannot be obtained by classical means: Thus, all the results of this
subsections are applicable only outside of the regime when this source
of noise is important. In particular, for zero temperature this means
$\omega < e\vert V \vert$. The crossover to the quantum noise was
recently treated by Nagaev \cite{Nagaev991} using the Green's
functions technique.  

It is also assumed that the frequency is much below the inverse
elastic scattering time, $\omega \ll \tau^{-1}$; outside this regime,
the diffusion approximation is not valid. As emphasized in work on
chaotic cavities \cite{PvLB} experiments which measure the noise at
the gate can also be envisioned: This has the advantage that even for
frequencies much smaller than $\omega_{RC}$ the noise is frequency
dependent and in fact can for metallic systems also be expected to be
determined by $R_q$ and $C_{\mu}$. 

Nagaev \cite{Nagaev98,Nagaev981} considers a circular conductor of
length $L$ and radius $R$, surrounded by a circular gate. For $L \ll
R$ (short wire) he finds that the noise spectrum measured at a contact
is frequency independent. However, generally the frequency
dependence is quite pronounced. Thus, for the case when the
distribution function is described by Eq. (\ref{distfun613}), and
wires are long, $L \gg R$, the correlation of currents taken at the
same contact, $S_{LL}$, for $\omega \gg \omega_c$ tends to the Poisson
value $2e\langle I \rangle$, while the current correlation at
different contacts, $S_{LR}$, rapidly falls off with frequency, and is
exponentially small for $\omega \gg \omega_c$. Note that the current
is not conserved in this system, since an ac current is also
generated at the gate. For hot electrons, the high-frequency noise is
given by \cite{Nagaev981} $S_{LL} \sim e \langle I \rangle
\alpha^{1/6}$, where $\alpha$ is a dimensionless parameter
proportional to the interaction strength. This result seems
paradoxical, since for the hot electrons the distribution function is
independent of the form of the collision integral, and thus the
results for noise at any frequency are not expected to depend on this
form. The resolution of this paradox, as pointed out by Nagaev
\cite{Nagaev981}, is that close to the contacts for hot electrons the
term $\nabla^2 f$ diverges, and in some sense the electrons close to
the contacts are never hot.  

Naveh, Averin, and Likharev \cite{Naveh99f} consider analytically
a geometry in which a planar contact is situated above a gate. 
Instead of solving the Poisson equation, they assumed that the
potential profile $\varphi(x)$ is proportional to the charge density
$\rho(x)$ in the insulating layer which separates the sample from the
gate. Thus, the screening radius is set to be zero, and the charge
pile-up in the sample is effectively forbidden\footnote{Numerical
results for the same model with charge pile-up (only non-thermalized
electrons) were previously provided by Nagaev \cite{Nagaev98}.}. They
studied the current correlation at the same cross-section, $S(x)$. For
non-thermalized electrons, the high-frequency noise ($\omega \gg
\omega_c$) {\em inside} the conductor grows even for zero
temperature, $S(x) \propto (\omega/\omega_c)^{1/2} [x(L-x)]^{1/2}$. At
the contacts this noise power turns to zero, and one obtains $S_{LL}
\equiv S(x = 0) = e \langle I \rangle$. Note that this value is the
same as for a double-barrier structure with symmetric
capacitances. The difference with the result $S_{LL} = 2e \langle I
\rangle$ originates from the fact that in this mode the charge pile-up
in the contact is forbidden. In addition, the frequency dependent
noise in the same model is studied numerically by Naveh for
electron-electron \cite{Naveh981} and electron-phonon \cite{Naveh982}
interactions.    

\subsection{Shot noise in non-degenerate conductors} \label{bol65}

{\bf Diffusive conductors.} Recently, Gonz\'alez {\em et al}
\cite{Gonzalez98}, motivated by the diversity of proofs for the $1/3$
noise suppression in metallic diffusive wires, performed numerical
simulations of shot noise in {\em non-degenerate} diffusive
conductors. They have taken the interaction effects into account via
the Poisson equation. For the disorder a simple model with an energy
independent scattering time was assumed. They found that the noise
suppression factor for this system, within the error bars, equals
$1/3$ and $1/2$ in three- and two-dimensional systems, respectively. A
subsequent work \cite{Gonzalez981} gives the suppression factor $0.7$
for $d=1$. 

An analytic theory of shot noise in these conductors was proposed by
Beenakker \cite{Been99nd}, and subsequently by Nagaev \cite{Nagaevnd}
and Schomerus, Mishchenko, and Beenakker
\cite{Schomerus99,Schomerus991}. The general conclusion is as
follows. Shot noise in non-degenerate diffusive conductors is {\em
non-universal} in the sense that it depends on the details of the
disorder (the energy dependence of the elastic scattering time) and
the geometry of the sample. In the particular case, when the elastic
relaxation time is energy independent (corresponding to the
simulations by Gonz\'alez {\em et al} \cite{Gonzalez98,Gonzalez981}),
the suppression factors are close to $1/3$, $1/2$, and $0.7$, but not
precisely equal to these values. Below we give a brief sketch of the
derivation, following Ref. \cite{Schomerus99}.    

We start from the Boltzmann equation (\ref{Boltzmann1}) and add
Langevin sources on the right-hand side. Three points now require
special attention: (i) for a degenerate gas $\bar f \ll 1$, and thus
the factors $(1 - \bar f)$ in the collision integral (\ref{colint})
can be taken equal to 1; (ii) since carriers are now distributed over
a wide energy range, all the quantities must be taken 
energy-dependent rather than restricted to the Fermi surface. In
particular, the velocity is $v(E) = (2E/m)^{1/2}$, and the energy
dependence of the mean free path matters, (iii) the screening length
is energy dependent, and the system generally may not be regarded as
charge neutral. We employ again the $\tau$-approximation for the
electron-impurity collision integral. For a while, we assume the
relaxation time $\tau$ to be energy independent. We also neglect all
kinds of inelastic scattering. Separating the distribution function
into symmetric and asymmetric parts (\ref{split601}), and introducing
the (energy resolved) charge and current densities, 
\begin{eqnarray*}
\rho(\bbox{r}, E, t) = e \nu(E) f_0 (\bbox{r}, E, t); \\
\bbox{j} (\bbox{r}, E, t) = \frac{1}{d} e v(E) \nu(E)
\bbox{f}_1 (\bbox{r}, E, t), 
\end{eqnarray*}
we obtain two equations, analogous to
Eq. (\ref{interm601}),
\begin{equation} \label{nden631}
\nabla \cdot \bbox{j} + e \bbox{E} (\bbox{r}, t)
\partial_E \bbox{j} = 0
\end{equation}
and
\begin{equation} \label{nden632}
\bbox{j} = -D(E) \nabla \rho - \sigma (E) \bbox{E} (\bbox{r}, t)
\partial_E f_0 + \delta \bbox{j} (\bbox{r}, E, t) = 0.  
\end{equation}
Here $D(E) = 2E\tau/md$ and $\sigma(E) = e^2 \nu(E) D(E)$ are the
energy dependent diffusion coefficient and conductivity,
respectively. The electric field $\bbox{E}$ is coupled to the
charge density via the Poisson equation,
\begin{equation} \label{nden633}
\nabla \cdot \bbox{E} (\bbox{r}, t) = \int dE \rho(\bbox{r}, E, t), 
\end{equation}
and the energy resolved Langevin currents $\delta \bbox{j}$ are
correlated in the following way (cf. Eq. (\ref{fluct611})),
\begin{eqnarray} \label{fluct631}
& & \left\langle \delta j_l (\bbox{r}, E, t) \delta j_m 
(\bbox{r}', E', t') \right\rangle = 2\sigma(E) \delta_{lm}
\delta(\bbox{r} - \bbox{r}') \nonumber \\
& \times & \delta(E - E') \delta(t - t') \bar f_0 (\bbox{r}, E,
t). 
\end{eqnarray}

Equations (\ref{nden631}), (\ref{nden632}), and (\ref{nden633})
describe the response of the system to the fluctuations
(\ref{fluct631}) of the Langevin currents. The main complication as
compared to the degenerate case is that the fluctuations of the
electric field now play an essential role, and cannot be
neglected. For this reason, results for the shot noise 
depend dramatically on the geometry. 

Further progress can be achieved for the quasi-one-dimensional
geometry of a slab $0 < x < L$, located between two reservoirs 
of the same cross-section. One more approximation is the regime of 
space-charge limiting conduction, corresponding to the boundary 
condition $\bbox{E} (x=0,t) = 0$. This condition means that 
the charge in the contacts is well screened, $L \gg L_s \gg L_c$, 
with $L_c$ and $L_s$ being the screening length in the contacts 
and the sample, respectively. Equations (\ref{nden631}),
(\ref{nden632}), and (\ref{nden633}) are supplemented by the absorbing
boundary condition at another contact, $\rho(x=L, t) = 0$.   

Now one can calculate the potential profile inside the sample and
subsequently the shot noise power. In this way, within the
approximation of energy independent relaxation times, Schomerus,
Mishchenko, and Beenakker \cite{Schomerus99} obtain the following
results, 
\begin{eqnarray} \label{ndenres1} 
F = \left\{ \matrix{ 0.69, \ \ \ \ d=1 \cr 0.44, \ \ \ \ d=2 \cr 0.31,
\ \ \ \ d=3 
} \right. .
\end{eqnarray}
The numerical values (\ref{ndenres1}) are, indeed, close to $0.7$,
$1/2$  and $1/3$, respectively, in accordance with the numerical
results by  Gonz\'alez {\em et al} \cite{Gonzalez98,Gonzalez981}, but
not precisely equal to them. 

It is now worthwhile to mention that the results (\ref{ndenres1}), in 
contrast to the $1/3$--suppression of shot noise in the degenerate
diffusive conductors, are {\em not universal}. Whereas the $1/3$
suppression is independent of the geometry of the sample,
degree of the disorder, or local dimensionality, the values
(\ref{ndenres1}), being geometry independent, {\em do} depend on the
dimensionality of the sample. Furthermore, they {\em do} depend on the
disorder, and this dependence enters through their sensitivity to the
energy dependence of the relaxation time\footnote{In the case of
metallic diffusive wires, this dependence is irrelevant since the
relaxation time is evaluated at the Fermi surface.}, as noticed by
Nagaev \cite{Nagaevnd}. Schomerus, Mishchenko, and Beenakker
\cite{Schomerus991} investigated the case $\tau(E) \propto
E^{\alpha}$, $-1/2 \le \alpha \le 1$, which are the only values of
$\alpha$ compatible with the regime of space-charge limited
conduction. In particular, $\alpha = -1/2$ corresponds to scattering
on short-ranged impurities. They found that the Fano factor in $d=3$
crosses over monotonically from $F = 0.38$ ($\alpha = -1/2$) to $F =
0$ ($\alpha = 1$). There is no shot noise in this model for $\alpha >
1$.   

This example clearly demonstrates that Fermi statistics are not
necessary to suppress shot noise, in accordance with general
expectations.   

{\bf Ballistic non-degenerate conductors}. Bulashenko, Rub\'i, and
Kochelap \cite{Kochelap99} address the noise in charge limited 
ballistic conductors. They consider a two-terminal semiconductor
sample with heavily doped contacts. Carriers in the semiconductor
exist only due to injection from the contacts which thus determine the
potential distribution inside the sample. The self-consistent field
determines a barrier at which carriers are either completely reflected
or completely transmitted (no tunneling). This system is thus a close
analog of the charge limited shot noise in vacuum tubes.   

It is easy to adapt the Boltzmann-Langevin formulation to this
problem: Since carrier motion inside the conductor is determined by
the Vlasov equation (collisionless Boltzmann equation), 
\begin{equation} \label{balnden1}
\left( \partial_t + v_x \partial_x + e E_x \partial_{p_x} \right) f(x,
\bbox{p}, t) = 0,
\end{equation}
the distribution function $f(x, \bbox{p}, t)$ at any point inside
the sample is determined by the distribution function at the surface
of the sample. The only source of noise arises from the random
injection of carriers at the contacts. Thus the boundary conditions
are   
\begin{eqnarray} \label{balnden2}
f(0,\bbox{p},t)\vert_{v_x > 0} & = & f_L + \delta f_L
(\bbox{p},t),  \nonumber \\ 
f(L,\bbox{p},t)\vert_{v_x < 0} & = & f_R + \delta f_R
(\bbox{p},t). 
\end{eqnarray} 
The stochastic forces $\delta f_{L,R}$ are zero on average, and their
correlation is  
\begin{eqnarray} \label{corrnden2}
& & \langle \delta f_{\alpha} (\bbox{p},t) \delta f_{\beta}  
(\bbox{p}', t') \rangle \propto \delta_{\alpha\beta}
\delta (p_x - p'_x) \delta (t - t') f_{\alpha}(p_x), \nonumber \\
\end{eqnarray}
where $f_{\alpha}(p_x)$ is the Maxwell distribution function
restricted to $p_x > 0$ ($\alpha = L$) or $p_x < 0$ ($\alpha =
R$). Note that in the degenerate case we would have to write an
extra factor $1 - f_{\alpha}$ on the right-hand side of
Eq. (\ref{corrnden2}), thus ensuring that there is no noise at zero
temperature. We have checked already in Section \ref{scat}
that this statement is correct. On the contrary, in the non-degenerate
case $f_{\alpha} \ll 1$, and thus $(1 - f_{\alpha}) \sim 1$. The
results for this regime do not of course allow an extrapolation to
$k_BT = 0$. The crossover between the shot noise behavior in
degenerate and non-degenerate conductors was investigated by
Gonz\`alez {\em et al} \cite{Gonzalez99c} using Monte Carlo
simulations. 

Eq. (\ref{balnden1}) is coupled to the Poisson equation, 
\begin{equation} \label{balnden3}
-dE_{x}/dx = 4\pi \sum_{\bbox{p}} f(x, \bbox{p}, t).
\end{equation}
Solving the resulting equations, Bulashenko, Rub\'i, and
Kochelap \cite{Kochelap99} found that interactions, at least in a
certain parameter range, suppress shot noise below the Poisson
value. The suppression may be arbitrarily strong in long (but still
ballistic) samples. The results are in good agreement with previous
numerical studies of shot noise in the same system by Gonz\'alez {\em
et al} \cite{Gonzalez97,Gonzalez971} and Bulashenko {\em et al}
\cite{Bulashenko98}.   

The crossover between the ballistic and diffusive behavior of the
non-degenerate Fermi gas was numerically studied by Gonz\'alez {\em et
al} \cite{Gonzalez971,Gonzalez981}. Analytical results on this
crossover are presently unavailable.  

A similar problem, a ballistic {\em degenerate} conductor in the
presence of a nearby gate, was posed by Naveh, Korotkov, and Likharev
\cite{NavehKorotkov}.  
  
We conclude this subsection by recalling that the effect of
interactions on noise arises not only in long ballistic structures but
already in samples which are effectively zero dimensional, like
resonant tunneling diodes \cite{Blanter97} or in quantum point
contacts \cite{PvLB}. The added complication in extended structures
arises from the long range nature of the Coulomb interaction.   
  
\subsection{Boltzmann-Langevin method for shot noise suppression in
chaotic cavities with diffusive boundary scattering} \label{bol66}

Now we turn to the classical derivation of the $1/4$--shot noise 
suppression in chaotic cavities. In standard cavities, which are
regular objects, the chaotic dynamics arises due to the complicated
shape of a surface. Thus scattering at the surface is {\em
deterministic} and in an individual ensemble member scattering along
the surface of the cavity is noiseless. Thus it is not obvious how to
apply the Boltzmann-Langevin equation.   
\begin{figure}
{\epsfxsize=6.cm\centerline{\epsfbox{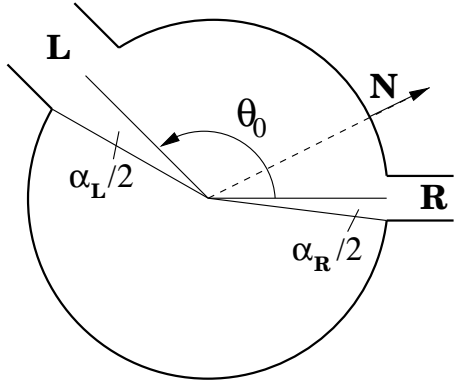}}}
\vspace{0.3cm}
\caption{Geometry of the chaotic cavity with diffusive boundary
scattering.}   
\label{fig602}
\end{figure}
  
However, recently a model of a {\em random} billiard --- a circular
billiard with diffusive boundary scattering --- was proposed
\cite{Tripathi,BMM98} to emulate the behavior of chaotic 
cavities\footnote{Earlier, a similar model was numerically implemented
to study spectral statistics in closed square billiards
\cite{Cuevas96,Cuevas97}.}. It turned out that the model can be
relatively easily dealt with, and Refs. \cite{Tripathi,BMM98} used it
to study spectral and eigenfunction properties of {\em closed}
systems.  Ref. \cite{Blanter99} suggests that the same model may be
used to study the transport properties of the open chaotic cavities
and presents the theory of shot noise based on the Boltzmann-Langevin
approach. 

We consider a circular cavity of radius $R$ connected to the two
reservoirs via ideal leads; the angular positions of the leads are
$\theta_0 - \alpha_L/2 < \theta < \theta_0 + \alpha_L/2$ (left) and
$-\alpha_R/2 < \theta < \alpha_R/2$ (right), see Fig.~\ref{fig602};
$\theta$ is the polar angle. The contacts are assumed to be narrow,
$\alpha_{L,R} \ll 1$, though the numbers of the transverse channels,
$N_{L,R} = p_F R \alpha_{L,R}/\pi\hbar$, are still assumed 
to be large compared to $1$. Inside the
cavity, motion is ballistic, and the average distribution function
$\bar f(\bbox{r}, \bbox{n})$ obeys the equation  
\begin{equation} \label{cavfun01}
\bbox{n} \nabla \bar f(\bbox{r}, \bbox{n}) = 0.
\end{equation}  
At the surface (denoted by $\Omega$) we can choose a diffusive
boundary condition: the distribution function of the particles
backscattered from the surface is constant (independent of $\bbox{n}$)
and fixed by the condition of current conservation\footnote{This is
the simplest possible boundary condition of this kind. For a review,
see Ref. \cite{Okulov}. We expect similar results for any other
diffusive boundary condition.},  
\begin{equation} \label{difcond651}
\bar f(\bbox{r}, \bbox{n}) = \pi\int_{(\bbox{N}\bbox{n}' > 0)}
(\bbox{N} \bbox{n}') \bar f(\bbox{r}, \bbox{n}') d\bbox{n}', \ \ \
\bbox{N} \bbox{n}' < 0,  
\end{equation}
where $\bbox{r} \in \Omega$, $\bbox{N}$ is the outward normal
to the surface, and $\int d\bbox{n} = 1$. Furthermore, we assume
that the electrons coming from the leads are described by the
equilibrium distribution functions, and are emitted uniformly into all
directions. Explicitly, denoting the cross-sections of the left and
the right leads by $\Omega_{L}$ and $\Omega_{R}$, we have 
\begin{equation} \label{cavfun02}
\bar f(\bbox{r}, \bbox{n}) = f_{L,R}, \ \ \ \bbox{r} \in
\Omega_{L,R}; \ \ \bbox{N} \bbox{n}  < 0. 
\end{equation}

Now we can find the average distribution function. Since motion 
away from the boundary is ballistic, the value of the distribution
function Eq. (\ref{cavfun01}) at a point away from the boundary, is
determined by the distribution function at the surface associated with
the trajectory that reaches this point after a scattering event at the
surface. With the boundary conditions (\ref{difcond651}) and
(\ref{cavfun02}), we can then derive an integral equation for $\bar
f(\theta)$,   
\begin{equation} \label{cavfun03}
\left. \bar f(\theta) \right\vert_{\Omega} = \frac{1}{4} \int_{\Omega
+ \Omega_L + \Omega_R}  
\bar f(\theta') \left\vert \sin \frac{\theta - \theta'}{2} \right\vert
d\theta',   
\end{equation}  
subject to the additional
conditions $\bar f(\theta)\vert_{\Omega_{L,R}} =
f_{L,R}$. This exact equation may be considerably simplified in the
limit of narrow leads, $\alpha_{L,R} \ll 1$. Integrals of the 
type $\int_{\Omega_R} F(\theta) d\theta$ can now be replaced
by $\alpha_R F(0)$. This gives for the distribution function   
\begin{eqnarray} \label{cavfun04}
\bar f(\theta) & = & \frac{\alpha_L f_L + \alpha_R f_R}{\alpha_L +
\alpha_R} \nonumber \\
& + &  \frac{g(0) - g(\theta_0)}{4\pi} \frac{\alpha_L
\alpha_R (\alpha_L - \alpha_R)}{(\alpha_L + \alpha_R)^2} \left( f_L -
f_R \right) \nonumber \\  
& + & \frac{g(\theta) - g(\theta-\theta_0)}{4\pi} \frac{\alpha_L
\alpha_R}{(\alpha_L + \alpha_R)} \left( f_L - f_R \right),  
\end{eqnarray}
with the notation
\begin{displaymath}
g(\theta) = \sum_{l=1}^{\infty} \frac{\cos l\theta}{l^2} =
\frac{1}{12} (3\theta^2 - 6\pi\theta + 2\pi^2 ), \ \ \ 0 \le \theta
\le 2\pi. 
\end{displaymath}
The first part of the distribution function (\ref{cavfun04}) does not
depend on energy and corresponds to the random matrix theory (RMT)
results for the transport properties. The second two terms on the
right hand side are not universal and generate sample-specific
corrections to RMT \cite{Blanter99}, which we do not discuss here.   

The conductance is easily found to be 
\begin{displaymath}
G = \frac{e^2}{2\pi\hbar} \frac{N_L N_R}{N_L + N_R},
\end{displaymath}
which is identical to the RMT result. 

The main problem we encounter in attempting to calculate noise via the
Boltzmann-Langevin method is that the system is not described by a
collision integral of the type (\ref{colint}). Instead, the impurity
scattering is hidden inside the boundary condition
(\ref{difcond651}), which, in principle, itself must be derived from
the collision integral. This difficulty can, however, be avoided,
since we can calculate the probability $W (\bbox{n}, \bbox{n}',
\bbox{r})$ of scattering per unit time from the state 
$\bbox{n}$ to the state $\bbox{n}'$ at the point $\bbox{r}$ (which
is, of course, expected to be non-zero only at the 
diffusive boundary). Indeed, this probability is only finite for
$\bbox{N} \bbox{n} > 0$ and $\bbox{N} \bbox{n}' <
0$. Under these conditions it does not depend on $\bbox{n}'$, and
thus equals\footnote{The coefficient $2$, instead of $\pi^{-1}$, is
due to the normalization.} $W (\bbox{n}, \bbox{n}', \bbox{r}) = 2 W'
(\bbox{n}, \bbox{r})$, with $W'$ being the 
probability per unit time to scatter out of the state $\bbox{n}$
at the space point $\bbox{r}$. Imagine now the (short) time
interval $\Delta t$. During this time, the particles which are closer
to the surface than $v_F \bbox{n} \bbox{N} \Delta t$ are 
scattered with probability one, and others are not scattered at
all. Taking the limit $\Delta t \to 0$, we obtain 
\begin{equation} \label{prob65}
W (\bbox{n}, \bbox{n}', \bbox{r}) = \left\{ 
\matrix{ v_F \bbox{n} \bbox{N} \ \delta(R - r), &  
\bbox{n}\bbox{N} >0, \bbox{n}'\bbox{N} < 0 \cr
0, & \mbox{otherwise}
} \right. .
\end{equation}

Imagine now that we have a collision integral, which is
characterized by the scattering probabilities (\ref{prob65}). Then,
the fluctuation part of the distribution function $\delta f$ obeys the
equation\footnote{We assume that there is no inelastic scattering,
$I_{in} \equiv 0$.} (\ref{Boltzmann2}) with the Langevin sources
correlated according to Eq. (\ref{fluct604}). A convenient way to
proceed was proposed by de Jong and Beenakker \cite{Jong95,Jong96},
who showed that quite generally the expression for the noise can be
brought into the form 
\begin{eqnarray} \label{noise661}
S & = & 2e^2\nu_F \int dE \int d\bbox{n} d\bbox{n}' d\bbox{r}
T_R (\bbox{n}, \bbox{r}) T_R (\bbox{n}', \bbox{r}) \nonumber \\
& \times & G(\bbox{n}, \bbox{n}', \bbox{r}, E),  
\end{eqnarray} 
where the function $G$ is given by Eq. (\ref{fluct605}), and $T_R$ is
the probability that the particle at $(\bbox{n}, \bbox{r})$
will eventually exit through the right lead. This probability obeys
$\bbox{n} \nabla T_R = 0$ with the diffusive boundary conditions
at the surface; furthermore, it equals $0$ and $1$ provided the
particle is headed to $\Omega_L$ and $\Omega_R$, respectively. In the
leading order $T_R = \alpha_R/(\alpha_L + \alpha_R)$; however, this
order does not contribute to the noise due to the sum rule
(\ref{sumrule601}). The subleading order is that for $\bbox{n}$
pointing out of the left (right) contact, $T_R = 0 (1)$. Substituting
the distribution function $\bar f = (\alpha_L f_L + \alpha_R
f_R)/(\alpha_L + \alpha_R)$ into Eq. (\ref{noise661}), we obtain for
the Fano factor 
\begin{displaymath}
F = \frac{N_L N_R}{(N_L + N_R)^2}. 
\end{displaymath} 
We have thus presented a purely classical derivation of the Fano
factor of a chaotic cavity (\ref{Fanochaot}). This result was
previously derived with the help of the scattering approach and
RMT theory (subsection \ref{scat0265}).   

To what extent can the billiard with diffusive boundary scattering 
also describe cavities which exhibit deterministic surface scattering?
As we have stated above, if we consider an ensemble member of a cavity
with specular scattering at the surface, such scattering is
deterministic and noiseless. Thus we can definitely not expect the
model with a diffusive boundary layer to describe an ensemble
member. However, to the extent that we are interested in the
description only of ensemble averaged quantities (which we are if we
invoke a Boltzmann-Langevin equation) the diffusive boundary layer
model can also describe the ensemble averaged behavior of cavities
which are purely deterministic. While in an individual cavity, a
particle with an incident direction and velocity generates a definite
reflected trajectory, we can, if we consider the ensemble average,
associate with each incident trajectory, a reflected trajectory of
arbitrary direction. In the ensemble, scattering can be considered
probabilistic, and the diffusive boundary model can thus also be used
to describe cavities with completely deterministic scattering at the
surface. This argument is correct, if we can commute ensemble and
statistical averages. To investigate this further, we present below
another discussion of the deterministic cavity.  

\subsection{Minimal correlation approach to shot noise in
deterministic chaotic cavities} \label{bol67}

In cavities of sufficiently complicated shape, deterministic chaos
appears due to specular scattering at the surface. To provide a
classical description of shot noise in this type of structures,
Ref. \cite{Blanter99} designed an approach which it called a ``minimal
correlation'' approach. It is not clear whether it can be applied to
a broad class of systems, and therefore, we decided to put it in this
Section rather than to provide a separate Section. 

In a mesoscopic conductor, in the presence of elastic scattering only,
the distribution function is quite generally given by
\cite{Gramespa98,Gramespa99}  
\begin{equation} \label{distrib671}
f_{0} (\bbox{r})= \sum_n \frac{\nu_{n}( \bbox{r})}{\nu (\bbox{r})}
f_n,  
\end{equation}
where $\nu_{n}(\bbox{r})$ is the {\em injectivity} of contact $n$ (the
contribution to the local density of states $\nu (\bbox{r})$ of contact
$n$.) For the ensemble averaged distribution which we seek, we can
replace the actual injectivities and the actual local density of
states by their ensemble average. For a cavity with classical contacts
($N_{n} \gg 1$ open quantum channels), the ensemble averaged
injectivities are \cite{brbu} $\langle \nu_{n} (\bbox{r}) \rangle =
\nu_F N_{n}/\sum_{n} N_{n}$, where $\nu_F$ is the ensemble averaged
local density of states. This just states that the contribution of the
$n$-th contact to the local density of states is proportional to its
width (number of quantum channels). Thus the ensemble averaged
distribution function, which we denote by $f_C$, is
\begin{equation} \label{distribav}
f_C = \sum_n \beta_n f_n, \ \ \ \beta_n \equiv N_n/\sum_n N_n.
\end{equation} 

To derive this distribution function classically, we first assert that
the ensemble averaged distribution inside the cavity, called $f_C$, is
a spatially independent constant. This is a consequence of the fact
that after ensemble averaging at any given point within the cavity
there are no preferred directions within the cavity if carriers
conserve their energy (no inelastic scattering). On the ensemble
average the interior of the cavity can be treated as an additional
dephasing voltage probe (see subsection \ref{scat027}). The interior
acts as a dephasing probe since we assume that there is no inelastic
scattering at the probe. Consequently at such a probe the current in
each energy interval is conserved \cite{Jong96}. Thus we must consider
the  energy resolved current.   

Let us denote the current at contact $n$ in an energy interval $dE$
by $J_n(E)$. The total current at the contact is $I_n = \int J_n
dE$. In terms of the distribution function of the reservoirs and the
cavity the energy-resolved current is 
\begin{equation}\label{curen}
J_n (E) = e^{-1} G_n \left( f_n - f_C \right),
\end{equation}
where  $G_n = e^2 N_n/(2\pi\hbar)$ is the (Sharvin) conductance of the
$n$th contact, and $N_n = p_FW_n/\pi\hbar$ is the number of transverse
channels. For energy conserving carrier motion the sum of all currents 
in each energy interval must vanish (subsection \ref{scat027}). This
requirement immediately gives Eq. (\ref{distribav}).  

Using the distribution function Eq. (\ref{distribav}) gives for the
conductance matrix  
\begin{equation} \label{condcla}
G_{mn} = \left(\delta_{mn} - \beta_m \right) G_n.
\end{equation}
This conductance matrix is symmetric, and for the two-terminal case
becomes $G_{LL} = (e^2/2\pi\hbar)(N_LN_R/(N_L + N_R))$, as expected. 

Now we turn to the shot noise. The fluctuation of the current through
the contact $n$ is written as 
\begin{equation} \label{cur673}
\delta I_n  =  -\frac{ep_F}{2\pi\hbar^2} \int_{\Omega_n} d\bbox{n}
d\bbox{r} dE\ (\bbox{nN}_n) \delta f(\bbox{r}, \bbox{n}, E, t),  
\end{equation}
where $\Omega_n$ and $\bbox{N}_n$ denote the surface of the
contact $n$ and the outward normal to this contact. Here $\delta f$ is
the fluctuating part of the distribution function, and for further
progress we must specify how these fluctuations are correlated. 

The terms with $\bbox{nN}_n < 0$ describe fluctuations of the
distribution functions of the {\em equilibrium} reservoirs, $f_n
(E)$. These functions fluctuate due to partial occupation of states
(equilibrium noise); the fluctuations of course vanish for $k_B T =
0$. The equal time correlator of these equilibrium fluctuations quite
generally is (see {\em e.g.} Ref. \cite{Gantsevich})
\begin{eqnarray} \label{equilib671}
& & \langle \delta f(\bbox{r}, \bbox{n}, E, t) \delta f(\bbox{r}', 
\bbox{n}', E', t) \rangle \nonumber \\
& = & \nu_F^{-1} \delta (\bbox{r} -
\bbox{r}') \delta(\bbox{n} - \bbox{n}') \delta(E - E') 
\nonumber \\
& \times &  \bar f(\bbox{r}, \bbox{n}, E, t) \left[ 1 - \bar
f(\bbox{r}, \bbox{n}, E, t) \right],
\end{eqnarray}
where in the reservoirs $\bar f = f_n (E)$. In particular, the cross
correlations are completely suppressed. 

On the other hand, the terms with $\bbox{nN}_n > 0$ describe
fluctuations of the distribution function {\em inside} the cavity. 
These {\em non-equilibrium} fluctuations resemble
Eq. (\ref{equilib671}) very much. Indeed, in the absence of random
scattering the only source of noise are the fluctuations of the
occupation numbers. Furthermore, in the chaotic cavity the cross
correlations should be suppressed because of multiple random
scattering inside the cavity.  Thus, we assume that
Eq. (\ref{equilib671}) is valid for fluctuations of the
non-equilibrium state of the cavity, where the function $f_C (E)$
(\ref{distribav}) plays the role of $\bar f(\bbox{r}, \bbox{n}, E,
t)$. In contrast to the true equilibrium state, these 
fluctuations persist even for zero temperature, since the average
distribution function (\ref{distribav}) differs from both zero and
one.  

Furthermore, for $t \neq t'$ the correlator obeys the kinetic
equation, $(\partial_t + v_F \bbox{n} \nabla) \langle \delta f(t)
\delta f(t') \rangle = 0$ \cite{Gantsevich}. We obtain the following
formula,   
\begin{eqnarray} \label{fluctfun671}
& & \langle \delta f(\bbox{r}, \bbox{n}, E, t) \delta
f(\bbox{r}', \bbox{n}', E', t') \rangle \nonumber \\
& & = \nu_F^{-1} \delta
\left[ \bbox{r} - \bbox{r}' - v_F \bbox{n} (t-t') \right]
\nonumber \\ & & \times \delta (\bbox{n} - \bbox{n}') \delta
(E-E') f_C (1-f_C), 
\end{eqnarray}
which describes strictly ballistic motion and is therefore only valid
at the time scales {\em below} the time of flight. An attempt to use
Eq. (\ref{fluctfun671}) for {\em all times} and insert it to
Eq. (\ref{cur673}) immediately leads to the violation of current
conservation. 

Thus, we must take special care of the fluctuations of the
distribution function inside the cavity for the case when $t-t'$ is
much longer than the dwell time $\tau_d$. In this situation, the
electron becomes uniformly distributed and leaves the cavity through
the $n$th contact with the probability $\beta_n$. For times $t \gg
\tau_d$ (which are of interest here) this can be described by an
instantaneous fluctuation of the isotropic distribution $\delta
f_C(E,t)$, which is not contained in Eq.\ (\ref{fluctfun671}). We
write then
\begin{eqnarray} \label{cur674}
\delta f (\bbox{r}, \bbox{n}, E, t) & = & \delta \tilde f
(\bbox{r}, \bbox{n}, E, t) + \delta f_C (E,t), \nonumber \\
& & \bbox{nN}_n > 0, \ \ \ \bbox{\bf r} \in \Omega_n, 
\end{eqnarray}
where $\delta \tilde f$ obeys Eq. (\ref{fluctfun671}). 

The requirement of the conservation of the number of electrons in the
cavity leads to {\em minimal correlations}\footnote{{\em Minimal}
means here that the correlations are minimally necessary --- the only
non-equilibrium type fluctuations are induced by the current
conservation requirement. This kind of fluctuations was discussed by
Lax \cite{Lax}. The approach similar to what we are discussing was
previously applied in Ref. \cite{Houten91} to the shot noise in
metallic diffusive wires, where it does not work due to the additional
fluctuations induced by the random scattering events.} between $\delta
f_C(E,t)$ and $\delta \tilde f(\bbox{r},\bbox{n}, E, t)$. The
requirement that current is conserved {\em at every instant of time},
$\sum_n \delta I_n = 0$, eliminates fluctuations $\delta f_C$. After
straightforward calculations with the help of Eq.\ (\ref{fluctfun671})
we arrive at the expression \cite{Blanter99}
\begin{equation} \label{noisemult671}
S_{mn} = 2G_{mn} k_B (T + T_C), \ \ \ k_B T_C = \int dE f_C (1 - f_C).
\end{equation}
It is easy to check that Eq. (\ref{noisemult671}) actually reproduces
all the results we have obtained in Section \ref{scat} with the help
of the scattering approach. Explicitly, we obtain 
\begin{displaymath} 
k_B T_C=\frac{e}{2} \sum_{m,n} \beta_m \beta_n (V_n-V_m) \coth
\left( \frac{e(V_n-V_m)}{2k_BT} \right). 
\end{displaymath}
At equilibrium $T_C = T$, and the noise power spectra obey the
fluctuation-dissipation theorem. For zero temperature, in the
two-terminal geometry we reproduce the noise suppression factor
(\ref{Fanochaot}); in the multi-terminal case the Hanbury Brown--Twiss
results \cite{Langen97}, described in Section \ref{scat}, also follow
from Eq. (\ref{noisemult671}). 

The perceptive reader notices the close similarity of the discussion
given above and the derivation of the classical results from the
scattering approach invoking a dephasing voltage probe
\cite{Jong96,Langen97} (subsection \ref{scat027}). The correlations
induced by current conservation are built in in the discussion of the
dephasing voltage probe model of the scattering approach. The
discussion which we have given above, is equivalent to this
approach, with the only difference that fluctuations are at every
stage treated with the help of fluctuating distributions. In contrast,
the scattering approach with a dephasing voltage probe invokes only
stationary, time-averaged distributions.  

The comparison of the Boltzmann-Langevin method with the dephasing
voltage probe approach also serves to indicate the limitations 
of the above discussion. In general, for partial dephasing (modeled by
an additional fictitious lead) inside the cavity connected via leads
with a barrier \cite{Langen97}, even the averaged distribution
function $f_{C}$ can  not be determined simply via Eq. (\ref{curen}):
even on the ensemble average, injectivities are not given simply by
conductance ratios.

\section{Noise in strongly correlated systems} \label{strcorr}

This Section is devoted to the shot noise in strongly correlated
systems. This is a rather wide subject due to the diversity of the
systems considered. There is no unifying approach to treat strongly
correlated systems. Typically, the shot noise in interacting
systems is described by methods more complicated than the scattering
or Langevin approaches. An attempt to present a detailed description
of the results and to explain how they are derived would lead us to
the necessity to write a separate review (if not a book) for each
subject. Our intention is to avoid this, and below we only present
some results for particular systems without an attempt to derive
them. The discussion is qualitative; for a quantitative description,
the reader is referred to the original works. As a consequence, this
Section has the appearance of a collection of independent results.  

\subsection{Coulomb blockade} \label{scorr71}

The term {\em Coulomb blockade} is used to describe phenomena which
show a blockage of transport through a system due to the electrostatic
effects. We recall only some basic facts; the general features of the
Coulomb blockade are summarized in the early review article
\cite{AverinMPS}. The most common technique to describe Coulomb
blockade effects is the master equation approach. 

{\bf Tunnel barriers}. The simplest structure for which one might
think that the Coulomb blockade is significant is a tunnel
junction. The junction is characterized by a capacitance $C$. From the
electrostatic point of view, the system can be regarded as a capacitor
where the tunneling between the electrodes is allowed, {\em i.e.} the
equivalent circuit is the capacitor $C$ connected in parallel with a
resistor $R\propto (\sum_n T_n)^{-1}$. Due to the additional charging
energy $Q^2/2C$, where $Q$ is the charge of the junction, the current
through the junction is blocked ({\em i.e.} exponentially small for
$k_BT \ll e^2/C$) for voltages below $V_c = e/2C$ (for simplicity, we
only consider $V > 0$). The $I$--$V$ curve is essentially $I = R^{-1}
(V - V_c) \theta(V - V_c)$. The singularity at $V = V_c$ is smeared at
elevated temperatures and finite resistances (see below). The shot
noise is Poissonian for low voltages (since the junction is opaque)
and crosses over to the usual $(1-T)$ suppression (for one channel)
for high voltages. We note, however, that this description only
applies to junctions with $T \ll 1$, and, thus, technically shot
noise is Poissonian in the whole range of voltages.   

The Coulomb blockade picture given above holds only provided the
resistance of the sample is much greater than the quantum of the
resistance, $R \gg R_k \equiv 2\pi\hbar/e^2$, and breaks down in the
opposite limit, where the $I$--$V$ characteristics are Ohmic and (for
zero frequency) insensitive to the charging effects. This latter limit
is the subject of work by Ben-Jacob, Mottola and Sch\"{o}n \cite{BJMS}
and Sch\"{o}n \cite{Schoen}.

Lee and Levitov \cite{LeeLev96} investigate the effect of the external
impedance (active resistance $R_{ex}$) on the noise properties of the
tunnel junction. They consider the case $R \gg R_k$ and describe the
crossover between the Ohmic regime ($R_{ex} \ll R$) to the Coulomb
blockade regime ($R_{ex} \gg R$). They also consider finite frequency
effects, but do not take into account the displacement currents. They
arrive at the remarkable conclusion that the current conservation for
noise is violated for high enough frequencies, $(i\omega C)^{-1} \ll
\vert Z(\omega) \vert$, where $Z$ is the external impedance. Thus,
this is an illustration of the statement (Section \ref{freq}) that one
has to take care of the displacement currents, even when the
scattering matrices are not energy dependent, for frequencies higher
than the inverse collective response time, in this particular
situation $(R_{ex}C)^{-1}$.   

{\bf Quantum dots}. An equivalent circuit for a quantum dot with
charging (cited in the Coulomb blockade literature as the {\em
single-electron transistor}, SET) is shown in Fig.~\ref{fig701}a. The
SET is essentially a two-barrier structure with the capacitances
included in parallel to the resistances; in addition, the quantum dot
is capacitively coupled to a gate. We assume $R_{1,2} \gg R_k$. In the
simplest approximation, the role of the gate is simply to fix the
number of electrons which are in the dot in the absence of the driving
voltage $V$, $N_g = CV_g/e$, $C = C_1 + C_2 + C_g$. For a moment we
assume now $V_g = N_g = 0$. Then for low voltages $V < e/2C$ the
tunneling is blocked. The point $V = e/2C$ is degenerate, since the
energies of an empty dot and a dot with one electron are equal. For
$e/2C < V < 3e/2C$ there is one electron in the dot, and so
on. Consequently, the $I$--$V$ curve shows steps (the Coulomb
staircase, Fig.~\ref{fig701}b). The steps are smeared by the
temperature and finite resistance. Actually, the Coulomb staircase is
only well pronounced when the dot is asymmetric, $R_1 \ll R_2$, and in
addition, $C_1 \le C_2$. In the opposite case, there is no current
until $e/2C$, but the $I$--$V$ characteristics for higher voltages are
regular, similarly to the tunnel junction. Finite (fixed) gate voltage
shifts the $I$--$V$ curve; for $N_g = 1/2$ the degeneracy point
happens to be at $V = 0$, {\em i.e.} the system shows Ohmic behavior.

For the case when there is no Coulomb staircase, shot noise was
investigated by Belogolovskii and Levin \cite{Levin91}. Based on the
master equation approach, they report a shot noise
suppression. However the paper is too concise to provide an
understanding of the origin of the noise suppression. Independently,
for this case shot noise was investigated by Korotkov {\em et al}
\cite{Korotkov92} and Korotkov \cite{Korotkov94} using the master
equation approach, and subsequently by Korotkov using the Langevin
approach \cite{Korotkov98}: He adds Langevin sources to the rate
equations, each number of electrons in the dot is described by an
individual fluctuating source. In this case the shot noise is absent
for $V < e/2C$, Poissonian at the onset of current (since the
tunneling rate through one of the barriers dominates, see below), and
for high voltages is described by a Fano factor 
\begin{equation} \label{FanoCB}
F = \frac{R_1^2 + R_2^2}{(R_1 + R_2)^2}. 
\end{equation}
The result (\ref{FanoCB}) is genuinely the
double-barrier Fano factor (\ref{Fanodbt}), since $R_{1,2} \propto
T_{1,2}^{-1}$. This high-voltage behavior was independently obtained
by Hung and Wu \cite{Hung93} using the Green's function technique.  
Korotkov \cite{Korotkov94} also studied the frequency dependence of
shot noise and found that it is a regular function, which for high
frequencies $(RC)^{-1} \ll \omega \ll e^2/\hbar C$, $R^{-1} = R_1^{-1}
+ R_2^{-1}$, saturates at an interaction-dependent value. 

For the general case (Coulomb staircase regime), shot noise was
(independently) analyzed by Hershfield {\em et al} \cite{Davies93}
using a master equation. In the plateau regimes, the transport is via
the only state with a fixed number of electrons, and the shot noise is
Poissonian (up to exponentially small corrections); in particular, for
$V < e/2C$ noise is exponentially small. The situation is, however,
different close to the degenerate points (at the center of the step of
the staircase), since there are now two charge states
available. Hershfield {\em et al} \cite{Davies93} find that the Fano
factor in the vicinity of the degenerate points is $F(V) = (\Gamma_1^2
+ \Gamma_2^2)/(\Gamma_1 + \Gamma_2)^2$. Here $\Gamma_1$ and $\Gamma_2$
are the (tunneling) rates to add an electron through barrier $1$ and
remove an electron through barrier $2$. Though this result looks
similar to the expression for the resonant double-barrier structure
(\ref{Fanodbt}), the important difference is that the rates
$\Gamma_{1,2}$ are now strongly voltage dependent due to the Coulomb
blockade. In particular, at the onset of the step $\Gamma_1 \ll
\Gamma_2$, and $F = 1$. The Fano factor thus shows dips in the region
of the steps. For high voltages the Coulomb blockade is insignificant,
and the Fano factor assumes the double-barrier suppression value
(\ref{FanoCB}). The resulting voltage dependence of the Fano factor is
sketched in Fig.~\ref{fig701}b, upper curve. Hershfield {\em et al}
\cite{Davies93} also perform extensive numerical simulations of shot
noise and show that the structure in the Fano factor disappears in the
symmetric limit $R_1 \sim R_2$.     
\begin{figure}
{\epsfxsize=8.5cm\centerline{\epsfbox{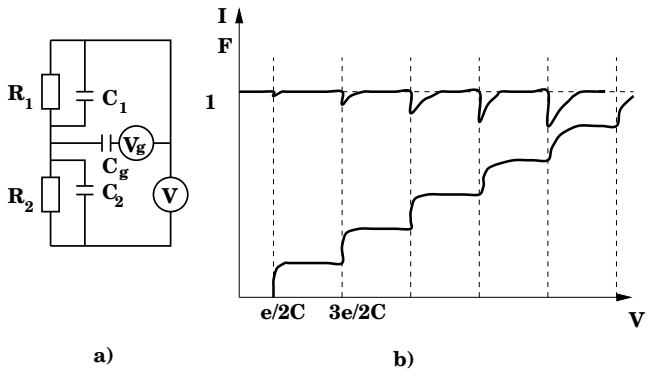}}}
\vspace{0.3cm}
\caption{(a) Equivalent circuit for the single-electron transistor;
(b) A sketch of the $I$--$V$ characteristics (lower curve) and the
Fano factor voltage dependence (upper curve) for the very asymmetric
case $R_1 \ll R_2$, $C_1 \ll C_2$, $V_g = 0$.}   
\label{fig701}
\end{figure}

Similar results were subsequently obtained by Galperin {\em et al}
\cite{GHCZ}, Hanke {\em et al} \cite{Hanke93}, and Hanke, Galperin,
and Chao \cite{Hanke94}, who also investigate the frequency dependence
of shot noise. They point out that the quantity $\partial^2 S/\partial
\omega^2$ is more sensitive to the voltage near the Coulomb blockade
steps than the noise spectral power $S$. Sub-Poissonian shot noise
suppression in the Coulomb blockade regime is numerically confirmed
by Anda and Latg\'e \cite{Latge}, however, at high bias, they find 
that the average current tends to zero. 

Another option is to consider transport properties as a function of
the gate voltage $V_g$. Both current and noise are periodic functions
of $V_g$ with the period of $e/C$ (which corresponds to the period of
$1$ in $N_g$). As pointed out by Hanke, Galperin, and Chao
\cite{Hanke94}, shot noise is then periodically suppressed below the
Poisson value. Wang, Iwanaga, and Miyoshi \cite{Wang98} consider shot
noise in the Coulomb blockade regime in a semiconductor quantum dot,
where the single-particle level spacing is relatively large. In this
situation it is not enough to write the interaction potential in the
dot in the form $Q^2/2C$, and the real space-dependent Coulomb
interaction must be taken instead. The current as a function of the
gate voltage exhibits a number of well-separated peaks, and,
consequently, shot noise is Poissonian everywhere except for the peak
positions, where the Fano factor has dips. Surprisingly enough, the
numerical results of Wang, Iwanaga, and Miyoshi \cite{Wang98} show
that the Fano factor in the dip may be arbitrarily low, certainly
below $1/2$. They explain this as being due to the suppression of the
shot noise by Coulomb interactions.   

We also mention here the papers by Krech, H\"adicke, and M\"uller
\cite{Krech92} and Krech and M\"uller \cite{Krech93}, who, based on a
master equation, conclude that the Fano factor may be suppressed down
to zero in the Coulomb blockade regime; in particular, the Fano factor
tends to zero for high frequencies. Though we cannot point out an
explicit error in these papers, the results seem quite surprising to
us. We believe that in the simple model of the Coulomb 
blockade, when the electrostatic energy is approximated by $Q^2/2C$,
it is quite unlikely that the shot noise is suppressed below $1/2$,
which is the non-interacting suppression factor for the symmetric
double-barrier structure. Clearly, to answer these questions, an
analytic investigation, currently unavailable, is required.   
\begin{figure}
{\epsfxsize=5.5cm\centerline{\epsfbox{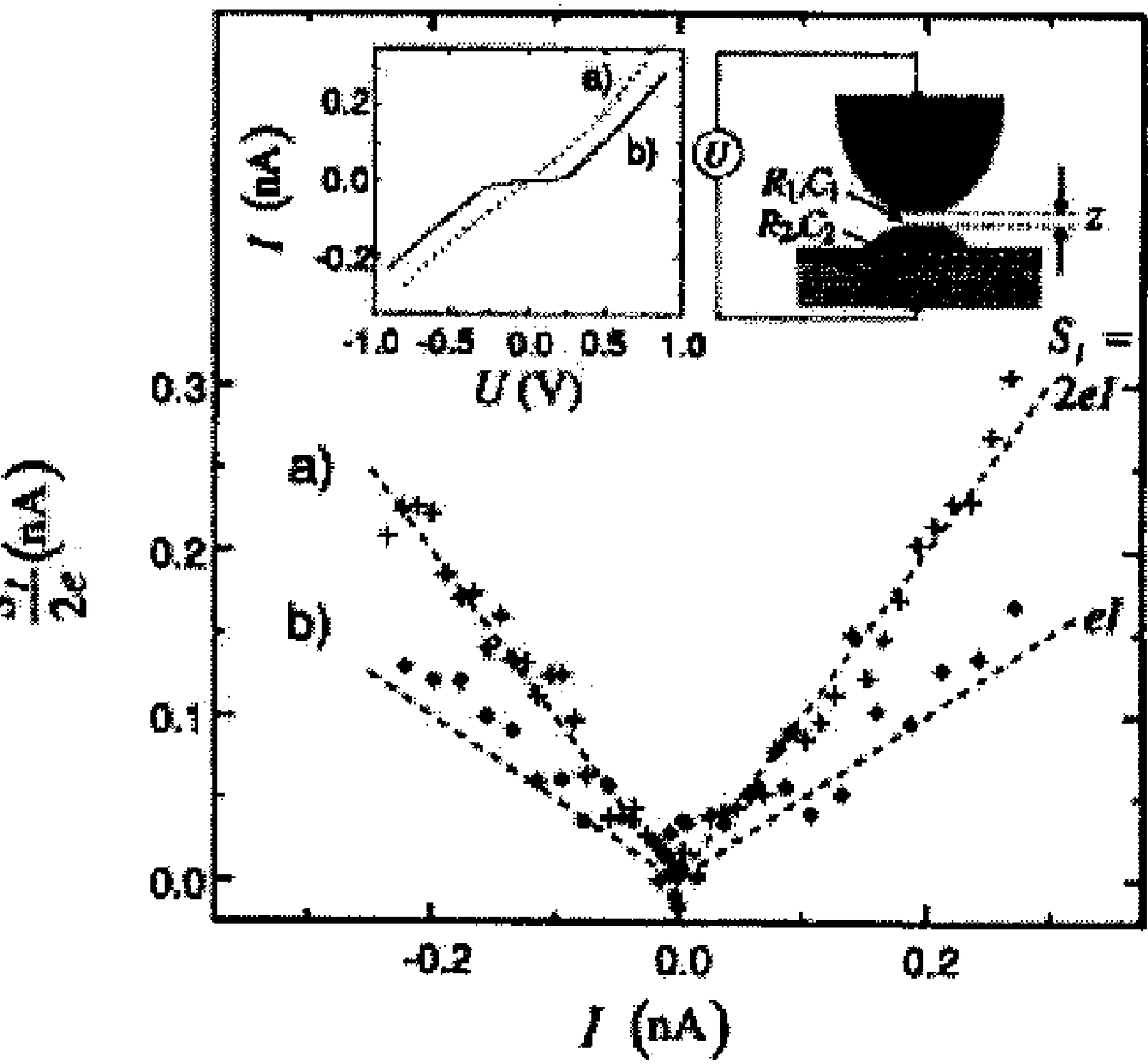}}}
{\epsfxsize=6.cm\centerline{\epsfbox{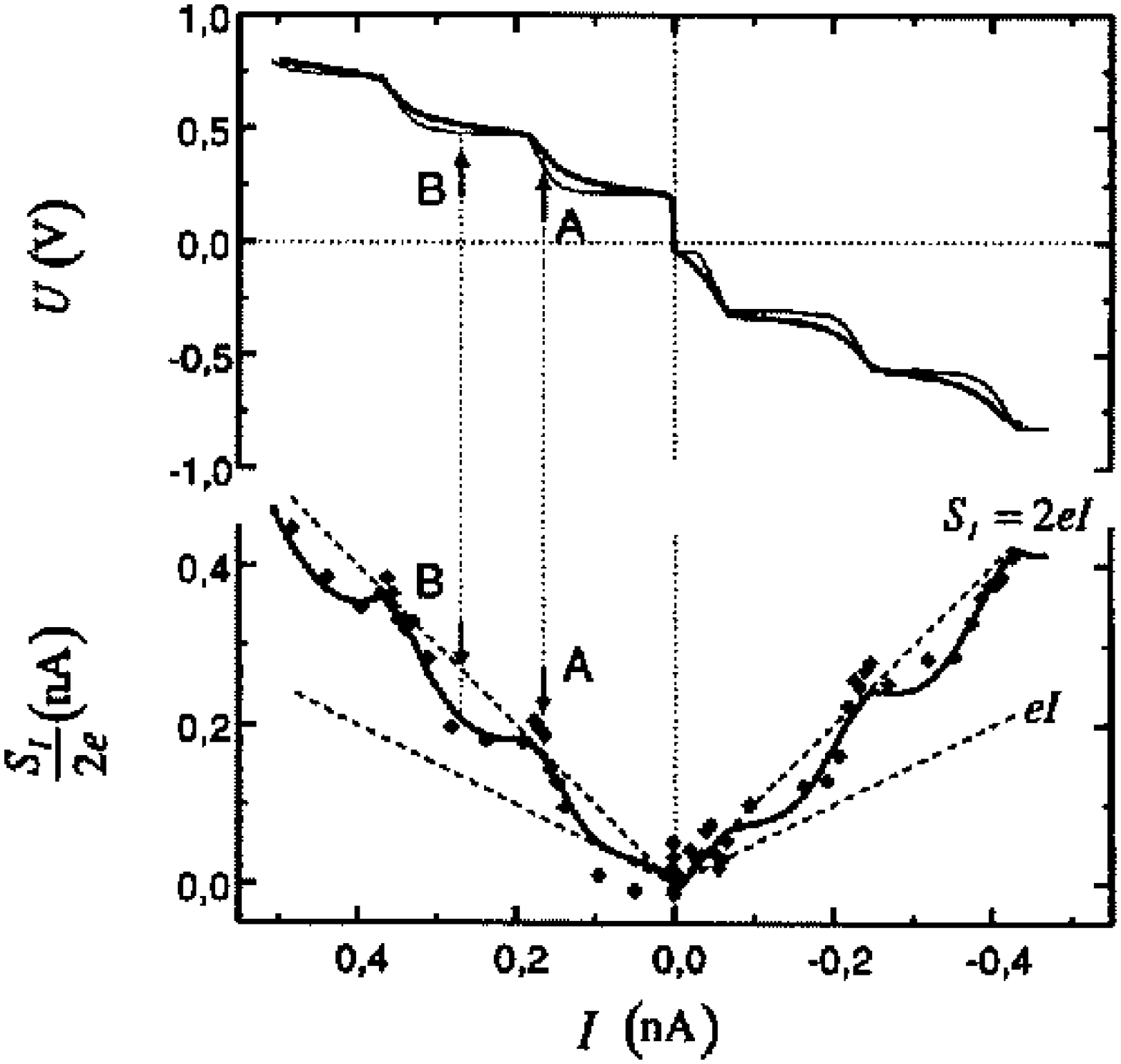}}}
\vspace{0.3cm}
\caption{Experiments by Birk, de Jong, and Sch\"onenberger
\protect\cite{Birk}. Left: shot noise in the quantum dot with $R_2 =
2R_1$ (diamonds); the inset shows the $I$--$V$ characteristics. Right:
$I$--$V$ curve (top) and shot noise (bottom) in a very asymmetric
quantum dot. Solid curves show the theory of Hershfield {\em et al}
\protect\cite{Davies93}. Dashed lines in all cases show the Poisson
value $2e\langle I \rangle$ and the value $e\langle I \rangle$.} 
\label{fig702}
\end{figure}

Experiments on shot noise in the Coulomb blockade regime were
performed by Birk, de Jong, and Sch\"onenberger \cite{Birk}. They put
a nanoparticle\footnote{The difference of this setup with a SET is
that there is no gate.} between the STM tip and metallic surface, and
were able to change the capacitances $C_1$ and $C_2$ and resistances
$R_1$ and $R_2$. For $R_2 = 2R_1$ they found the ``smooth'' Coulomb
blockade $I$-$V$ characteristics (no traces of the Coulomb staircase),
with the noise crossing over from Poissonian to Eq. (\ref{FanoCB})
with the increasing voltage (Fig. \ref{fig702}, left). In
contrast, for the very asymmetric case $R_1 \ll R_2$ Birk, de Jong,
and Sch\"onenberger \cite{Birk} observe the Coulomb staircase and
periodic shot noise suppression below the Poisson value
(Fig. \ref{fig702}, right). They compare their experimental data with
the theory of Hershfield {\em et al} \cite{Davies93} and find
quantitative agreement.  

{\bf Arrays of tunnel barriers}. Consider now a one-dimensional array
of metallic grains separated by tunnel barriers. This array also
shows Coulomb blockade features: The current is blocked below a
certain threshold voltage $V_c$ (typically higher than the threshold
voltage for a single junction), determined by the parameters of the
array. Just above $V_c$, the transport through the array is determined
by a single junction ({\em a bottleneck junction}), and corresponds to
the transfer of charge $e$. Thus, the shot noise close to the
threshold is Poissonian, the Fano factor equals one. With increasing 
voltage more and more junctions are opened for transport, and
eventually a {\em collective state} is established in the array
\cite{AverinMPS}: An addition of an electron to the array causes
the polarization of all the grains, such that the effective charge
transferred throughout the array\footnote{This can be viewed as a
transport of solitons with the fractional charge.} is $e/N$, with $N
\gg 1$ being the number of grains. In this regime, the Fano factor is
$1/N$. The crossover of shot noise from Poissonian to
$1/N$--suppression was obtained by Matsuoka and Likharev
\cite{Matsuoka}, who suggested these qualitative considerations and
backed them by numerical calculations.  

Other interesting opportunities for research open up if the frequency
dependence of shot noise is considered. First, in the collective state
these arrays exhibit {\em single-electron tunneling} oscillations with
the frequency $\omega_s = I/e$, $I$ being the average current. These
oscillations are a consequence of the discreteness of the charge
transfer through the tunnel barrier. The increase of charge on the
capacitor is the result of the development of a dipole across the
barrier. This is a polarization process for which the charge is not
quantized. However, when the polarization charge reaches $e/2$ the
Coulomb blockade is lifted and the capacitance is decharged by the
tunneling transfer of a single electron. Both the continuous
polarization process and the discrete charge transfer process give
rise to currents in the external circuit. It is difference in time
scale for the two processes which causes current oscillations. The
polarization process is much slower than the decharging process. The
discreteness of the charge transfer process is, as we have seen, a key
ingredient needed to generate shot noise. However here, in contrast to
the Schottky problem, the charge transfer occurs at time intervals
which are clearly not given by a Poisson distribution. Instead, here,
ideally the charge transfer process is clocked and there is no shot
noise. However in an array of tunnel junctions, there exists the
possibility to observe both current oscillations and shot noise: In an
array charge transfer occurs via a soliton (a traveling charge
wave). Korotkov, Averin, and Likharev \cite{Korotkov941} and Korotkov
\cite{Korotkov942} find that arrays of tunnel barriers exhibit both a
current that oscillates (on the average) at a frequency $\omega_s =
I/e$, and shot noise which is strongly peaked at the frequency
$\omega_s$.   

Another type of oscillations are {\em Bloch} oscillations, which are
due to the translational symmetry of the array. Their frequency is 
proportional to the voltage drop across each contact; if the array is
long and $V \ll Ne^2/C$, this frequency is approximately $\omega_B =
e^2/4\pi\hbar C$, $\omega_B \gg \omega_s$. The shot noise also peaks
at $\omega_B$, as found by Korotkov, Averin, and Likharev
\cite{Korotkov941}.    

{\bf Hybrid structures}. Qualitatively, the behavior of a
superconducting SET -- the same structure as shown in
Fig.~\ref{fig701}a but with the normal central grain replaced by a
superconducting grain -- is different from the behavior of the normal
SET, since the system is now sensitive to the parity of the number of
electrons on the island. If this number is odd, there is always an
unpaired electron, and thus the energy of these odd states is shifted
up by the superconducting gap $\Delta$ as compared to the even
states. Thus, odd states are unfavorable, and for $e^2/C, eV <
\Delta$, $k_B T \ll \Delta$ cannot be occupied. This {\em parity
effect} \cite{Averin92} is superimposed on the usual Coulomb blockade
behavior, and the general picture can be very complex. In particular, 
in the limiting case $e^2/C \ll \Delta$ and for $N_g = 0$, we
anticipate the following behavior of noise. For $V < e/C$ the current
is Coulomb blocked, and the shot noise is thus Poissonian. To
determine the Fano factor in this case, we must find the processes
which give rise to current slightly above the threshold $V_c =
e/C$. The dominant process is an electron coming from a reservoir
which is Andreev reflected at the interface, and generates a Cooper
pair in the superconducting grain, which is again converted to an
electron and a hole pair at the interface. This process results in the
transfer of an effective charge of $2e$; the Fano factor is thus
$2$. For $e^2/C < eV \ll \Delta$, the Fano factor is expected to be
smaller as the voltage is increased and the Coulomb blockade is
lifted. For $eV \sim \Delta$ single-electron tunneling starts to play
a role; eventually, for $eV \gg \Delta$ shot noise is determined by
the double-barrier value of the Fano factor (\ref{FanoCB}).
 
With the help of the gate, the polarization charge can be adjusted to
be $N_g = 1$. In this case, the system is at the degeneracy point,
since the energies of two ground states with an even number of
carriers are the same. The current is then finite for any bias
\cite{Hekking}, and there is no Coulomb blockade behavior: Cooper pair
tunneling starts from zero voltage.  

These considerations are qualitative, and must be confirmed by
quantitative calculations. Presently, there are not many results
available on noise in interacting hybrid structures. The
investigations of shot noise in SNS and NSN systems in the Coulomb
blockade regime were pioneered by Krech and M\"uller
\cite{Krech94}. They use the same master equation technique as in
their previous works \cite{Krech92,Krech93}, and arrive at the same
unphysical result that the shot noise is always suppressed for high
frequencies down to zero.  

Hanke {\em et al} \cite{Hanke941,Hanke95} perform numerical
calculations of the shot noise in superconducting SETs in the regime
$e^2/C < \Delta$, $N_g = 1$. In this case, there are several threshold
voltages, and the Fano factor is found to peak at each
threshold. Strangely, Refs. \cite{Hanke941,Hanke95} find that the Fano
factor may lie well above 2. We do not currently understand this
result.   

{\bf Ferromagnetic junctions}. Recently, junctions in which one or
several electrodes are ferromagnetic have become of interest in
mesoscopic physics. Many different structures have been proposed and,
in principle, one may discuss shot noise for all these structures;
presently, only one theoretical and one experimental paper exist.  

Bu{\l}ka {\em et al} \cite{Bulka99} consider a structure in which 
two electrodes are separated by a quantum dot in the Coulomb blockade
regime (a ferromagnetic analog of the SET). They assume that there is
no spin-flip scattering in the dot. The difference between the two
spin polarizations in the electrodes is phenomenologically modeled as
a difference in resistances: Thus, if the ferromagnetic reservoir
$1$ is spin-up polarized, the resistance $R_{1\downarrow}$ for
spin-down electrons is much higher than the resistance $R_{1\uparrow}$
for spin-up electrons. Assuming that all the resistances are much
higher than $R_k$, Bu{\l}ka {\em et al} \cite{Bulka99} generalize the
master equation approach to the case of spin-dependent transport, and
present results of a numerical calculation.  

In structures in which both electrodes are ferromagnetic (FNF
junction), the orientation of the magnetic moments of the electrodes
becomes important. To understand what happens, we discuss first the
shot noise suppression in a {\em non-interacting} FNF junction
\cite{blantermagn}. We have already seen that it is relevant at the
steps of the Coulomb staircase. We assume that both electrodes are
made of the same ferromagnetic material, so that the ratio of the
resistances for spin-up and spin-down propagation is the same. The
junction is also assumed to be very asymmetric. Thus, for the parallel
orientation (when the magnetic moments of both reservoirs are
spin-up), we have the resistances  $R_{1\uparrow} = R_0$,
$R_{1\downarrow} = \alpha R_0$, $R_{2\uparrow} = \beta R_0$, and
$R_{2\downarrow} = \alpha \beta R_0$. The constants $\alpha$ and
$\beta$ describe the asymmetry of the spin-up and spin-down
propagation, and the asymmetry of the barrier, respectively. For a
given spin projection, the current is proportional to
$R_1R_2/(R_1+R_2)$, while the shot noise is proportional to $R_1R_2
(R_1^2 + R_2^2)/(R_1+R_2)^3$. Here $R_1$ and $R_2$ must be taken for
each spin projection separately; the total current (shot noise) is
then expressed as the sum of the currents (shot noises) of spin-up and
spin-down electrons. Evaluating in this way the Fano factor, we obtain 
\begin{equation} \label{fer711}
F_{\uparrow\uparrow} = \frac{1 + \beta^2}{(1 + \beta)^2},
\end{equation}
which is precisely the double-barrier Fano factor (\ref{Fanodbt}),
since the Fano factors for spin-up and spin-down electrons are the
same.  

Similarly, for the anti-parallel ($\uparrow - \mbox{\rm N} -
\downarrow$) orientation we write $R_{1\uparrow} = R_0$,
$R_{1\downarrow} = \alpha R_0$, $R_{2\uparrow} = \alpha\beta R_0$, and
$R_{2\downarrow} = \beta R_0$. The Fano factor is
\begin{equation} \label{fer712}
F_{\uparrow\downarrow} = \left( \frac{1+\alpha^2\beta^2}{(1 +
\alpha\beta)^3} + \frac{\alpha^2 + \beta^2}{(\alpha + \beta)^3}
\right) \left( \frac{1}{1 + \alpha\beta} + \frac{1}{\alpha + \beta}
\right)^{-1},
\end{equation}
and depends now on $\alpha$. In the strongly asymmetric junction,
$\alpha, \beta \gg 1$, the Fano factor for the parallel orientation
(\ref{fer711}) becomes $1$. At the same time, for the anti-parallel
orientation Eq. (\ref{fer712}) is entirely determined by
spin-down electrons, and takes the form $F_{\uparrow\downarrow} =
(\alpha^2 + \beta^2)/(\alpha + \beta)^2$. Thus, the shot noise in the
anti-parallel configuration is suppressed as compared to the parallel
case. This situation is, of course, the same in the Coulomb blockade
regime, and this is precisely what Bu{\l}ka {\em et al} \cite{Bulka99}
find numerically. 

In these discussions, it is assumed that both spin channels are
independent. In reality, spin relaxation processes which couple the
two channels might be important. 

It is interesting to consider the structure in which one of the
reservoirs is ferromagnetic and the another one is normal (FNN
junction). Bu{\l}ka {\em et al} \cite{Bulka99} only treat the limiting
case when, say, spin-down electrons cannot propagate in the
ferromagnet at all ($R_{1\downarrow} = \infty$). They find that the
$I$--$V$ curve in this case shows negative differential resistance,
and shot noise may be enhanced above the Poisson value, similarly to
what we have discussed in Section \ref{Langevin} for quantum wells.  

Experimentally, the main problem in ferromagnetic structures is to
separate shot noise and $1/f$--noise. Nowak, Weissman, and Parkin
\cite{Nowak99} measured the low-frequency noise in a tunnel junction
with two ferromagnetic electrodes separated by an insulating
layer. They succeeded in extracting information on shot noise, and
report sub-Poissonian suppression for low voltages, but they did not
perform a systematic study of shot noise, and the situation, both
theoretically and experimentally, is very far from being clear.  

{\bf Concluding remarks on the Coulomb blockade}. A patient reader who
followed this Review from the beginning has noticed that we
have considerably changed the style. Indeed, in the previous Sections
we mostly had to deal with results, which are physically
appealing, well established, cover the field, and in many cases
are already experimentally confirmed. Here, instead, the results are
contradictory, in many cases analytically unavailable, and, what is 
more important, fragmentary --- they do not systematically address the
field. To illustrate this statement, we only give one example; we
could have cited dozens of them. Consider a quantum dot under low
bias; it is typically Coulomb blocked, unless the gate voltage is
tuned to the degenerate state, so that $N_g$ is half-integer. As we
have discussed above, the current dependence on the gate voltage is
essentially a set of peaks, separated by the distance of approximately
$e/C$. Between the peaks (in the {\em valleys}), as we implicitly
assumed, the zero-temperature current is due to quantum tunneling,
and is exponentially small. We concluded therefore, that the Fano
factor is $1$ --- the shot noise is Poissonian. In reality, however,
there are {\em cotunneling} processes --- virtual transitions via the
high-lying state in the dot --- which are not exponentially suppressed
and thus give the main contribution to the current. Cotunneling is a
genuinely quantum phenomenon, and cannot be obtained by means of a 
classical approach. It is clear that the cotunneling processes
may modify the Fano factor in the valleys; moreover, it is a good
opportunity to study quantum effects in the shot noise. This  problem,
among many others, remains unaddressed\footnote{A more complicated
problem -- transport through the double quantum dot in the regime when
the cotuneling dominates -- was investigated by Loss and Sukhorukov
\cite{Loss99} with the emphasis on the possibility of probing
entanglement (see Appendix \ref{con083}). They report that the Fano
factor equals one.}. 

\subsection{Anderson and Kondo impurities} \label{scorr72}

{\bf Anderson impurity model}. In the context of mesoscopic physics,
the Anderson impurity model describes a resonant level with a Hubbard
repulsion. It is sometimes taken as a model of a quantum dot. Commonly
the entire system is described by the tight-binding model with
non-interacting reservoirs and an interaction $U \hat n_{\downarrow}
\hat n_{\uparrow}$ on the site $i=0$ (resonant impurity), with $\hat
n_{\downarrow}$ and $\hat n_{\uparrow}$ being the operators of the
number of electrons on this site with spins down and up,
respectively. Tunneling into the dot is described, as in the
non-interacting case, by partial tunneling widths $\Gamma_L$ and
$\Gamma_R$, which may be assumed to be energy independent. The on-site
repulsion is important (in the linear regime) when $U \gg E_0$, where
$E_0$ is the energy of the resonant impurity relative to the Fermi
level in the reservoirs. In this case, for low temperatures $T < T_K$
the spin of electrons traversing the quantum dot starts to play a
considerable role, and the system shows features essentially similar
to the Kondo effect \cite{Glazman881,NgLee}. Here $T_K$ is a certain
temperature, which is a monotonous function of $U$, and may be
identified with the effective Kondo temperature. For $T < T_K$ the
effective transmission coefficient grows as the temperature is
decreased, and for $T = 0$ reaches the resonant value $T^{\max} =
4\Gamma_L \Gamma_R/(\Gamma_L + \Gamma_R)^2$ (see Eq. (\ref{breitw}))
for {\em all} impurities which are closer to the Fermi surface than
$\Gamma_L + \Gamma_R$. Subsequent averaging over the impurities
\cite{Glazman881} gives rise to the logarithmic singularity in the
conductance for zero temperature. A finite bias voltage $V$ smears the
singularity \cite{Glazman881,Hershfield921}; thus, the differential
conductance $dI/dV$ as a function of bias shows a narrow peak around
$V = 0$ and two broad side peaks for $eV = \pm U$.     

Theoretical results on shot noise are scarce. Hershfield
\cite{Hershfield92} performs perturbative analysis in powers of $U$ of
the shot noise, based on the Green's function approach. His results
are thus relevant for the high-temperature regime, but do not describe
the Kondo physics, which is non-perturbative in $U$. He finds that the
noise, apart from the non-interacting contribution (described by the
Fano factor (\ref{Fanodbt})) contains also an interacting
correction. This correction is a non-trivial function of the applied
bias voltage; it is always positive for a symmetric barrier $\Gamma_L
= \Gamma_R$, but may have either sign for an asymmetric
barrier. This interacting correction is zero for zero bias, and peaks
around a certain energy $E_r$, which is the bare resonant energy
$E_0$, renormalized by interactions. For higher voltages it falls off
with energy, and in the limit $eV \gg U, E_0$ the Fano factor returns
to the non-interacting value (\ref{Fanodbt}). Thus, in this case the
interactions may either enhance or suppress the noise. We also point
out the analogy with the Coulomb blockade results: In the symmetric
case $R_1 = R_2$, the actual Coulomb blockade noise suppression is
always {\em stronger}, than the non-interacting suppression
(\ref{FanoCB}); generally, it may be either stronger or
weaker. Yamaguchi and Kawamura \cite{Yamaguchi94,Yamaguchi96} perform
a complementary analysis by treating the tunneling Hamiltonian
perturbatively. They find that the shot noise is strongly suppressed
as compared with the Poisson value for voltages $eV \sim E_0$ and $eV
\sim E_0 + U$ (the latter resonance corresponds to the addition of the
second electron to the resonant state, which is then shifted upwards
by $U$). 

For the Kondo regime, we expect that, since the effective transmission
coefficient tends to $T^{\max}$ for zero temperature, the shot noise
is a sensitive function of $k_BT$, which for a symmetric barrier
decreases and eventually vanishes as the temperature tends to
zero. This regime is investigated by Ding and Ng \cite{Ding97}, who
complete the Green's functions analysis by numerical simulations. They
only plot the results for the symmetric case $\Gamma_L = \Gamma_R$ and
only for $T = T_K$; the shot noise in this regime is, indeed,
suppressed below the non-interacting value for any applied bias.
Yamaguchi and Kawamura \cite{Yamaguchi96}, treating the tunneling
Hamiltonian perturbatively, report that the Fano factor is suppressed
down to zero at zero bias. Results concerning averages over impurities
are unavailable. 

{\bf Kondo model}. In mesoscopic physics, this is the model of 
two non-interacting reservoirs which couple to the $1/2$-spin in the
quantum dot via exchange interaction. The interacting part of the
Hamiltonian is $\hat H_{int} = J_{\lambda}^{\alpha\beta}
s^{\lambda}_{\alpha\beta} \sigma^{\lambda}$, where $\lambda = x,y,z$  
and $\alpha, \beta = L,R$. Here $s^{\lambda}_{\alpha\beta}$ are the
matrix elements of the electron spin operator in the basis of the
reservoir states, and $\sigma^{\lambda}$ is the spin of the Kondo
impurity. In physical systems the coupling $J_{\lambda}$ is symmetric;
however, to gain some insight and use the exact solutions, other
limits are often considered. 

To our knowledge, the only results on shot noise in the Kondo model
are due to Schiller and Hershfield \cite{Schiller98}, who consider a
particular limiting case ({\em Toulouse limit}), $J_z^{\alpha\beta} =
J_y^{\alpha\beta}$, $J_z^{LR} = J_z^{RL} = 0$, and $J_z^{LL} =
J_z^{RR}$. As a function of the bias voltage, the Fano factor is zero
at zero bias, and grows monotonically. In the high-bias limit the
noise is Poissonian rather than suppressed according to
Eq. (\ref{Fanodbt}). The transport properties of the Kondo model are
strongly affected by an applied magnetic field, which may drive the
Fano factor well above the Poisson value. The frequency dependence of
the shot noise is sensitive to the spectral function of the Kondo
model, and exhibits structure at the inner scales of energy. The
studies \cite{Schiller98}, though quite careful, do not, of course,
exhaust the opportunities to investigate shot noise in strongly
correlated systems, offered by the Kondo model.

\subsection{Tomonaga--Luttinger liquids and fractional quantum Hall
edge states} \label{scorr73}

Many problems concerning (strictly) one-dimensional systems of
interacting electrons may be solved exactly by using specially
designed techniques. As a result, it turns out that in one dimension,
electron-electron interactions are very important. They lead to the
formation of a new correlated state of matter, {\em a
Tomonaga--Luttinger liquid}, which is characterized by the presence of
gapless collective excitations, commonly referred as plasmons. In
particular, the transport properties of the one-dimensional wires are
also quite unusual. We only give here the results which we
subsequently use for the description of noise; a comprehensive review
may be found {\em e.g.} in Ref. \cite{Fisher96}. Throughout the whole
subsection we assume that the interaction is short-ranged and
one-dimensional, $V(x-x') = V_0 \delta(x - x')$, and the voltage $V >
0$.    

For an infinite homogeneous Luttinger liquid the ``conductance'' is 
renormalized by interactions, $G = ge^2/2\pi\hbar$, where the
dimensionless interaction parameter, 
\begin{equation} \label{intparam73}
g = \left( 1 + \frac{V_0}{\pi \hbar v_F} \right)^{-1/2},
\end{equation}
will play an important role in what follows. This parameter equals $1$
for non-interacting electrons, while $g < 1$ for repulsive
interactions. However, if one takes into account the reservoirs, which
corresponds to a proper definition of conductance, the non-interacting
value $e^2/2\pi\hbar$ is restored, and thus the interaction constant
$g$ cannot be probed in this way.  

If one has an infinite system (no reservoirs) with a
barrier\footnote{This barrier is routinely called ``impurity'' in the
literature.}, the situation changes. Even an arbitrarily weak barrier
totally suppresses the transmission in the interacting case $g < 1$,
and for zero temperature there is no linear dc conductance. The two
limiting cases may be treated analytically. For strong barriers (weak
tunneling), when the transmission coefficient is $T \ll 1$, the
$I$--$V$ curve in the leading order in $V$ is  
\begin{equation} \label{condstrong73} 
\langle I \rangle = \frac{e^2}{2\pi\hbar} a T V^{2/g-1},
\end{equation} 
where $a$ is a non-universal (depending on the upper energy cut-off)
constant. In the opposite case of weak reflection, $1 - T \ll 1$, the
interactions renormalize the transmission coefficient, so that the
barrier becomes opaque, and for low voltages we return to the result
(\ref{condstrong73}). On the other hand, for high voltages the
backscattering may be considered as a small correction, and one
obtains  
\begin{equation} \label{condweak73} 
\langle I \rangle = \frac{ge^2}{2\pi\hbar} V - \frac{e^2}{2\pi\hbar} b
T V^{2g-1}, 
\end{equation} 
where $b$ is another non-universal constant. Eq. (\ref{condweak73}) is
only valid when the second term on the right-hand side is small. To
emphasize the difference, we will refer to the weak and strong
tunneling cases (which describe the regimes (\ref{condstrong73}) and
(\ref{condweak73}), respectively), rather than to the cases of
transparent and opaque barriers. In particular, whatever the strength
of the barrier, for low voltages and temperatures the
tunneling is weak\footnote{For the FQHE case (see below) this should
not cause any confusion: {\em Strong tunneling} regime means strong
tunneling through the barrier, which is the same as weak
backscattering, or {\em weak} tunneling of quasiparticles between the
edge states. Conversely, the {\em weak tunneling} regime means that
the edge states are almost not interconnected
(Fig.~\protect\ref{fig703}).}.   

For the transmission through a double-barrier structure resonant
tunneling may take place, but the resonances become infinitely narrow
in the zero-temperature limit.   

What is extremely important for the following is that weak
tunneling is accompanied by a transfer of charge $ge$ between left-
and right-moving particles (we can loosely say that there are
quasiparticles with the charge $ge$ which are scattered back from the
barrier), while in the case of the strong tunneling the charge
transfer across the barrier is $e$ --- there is tunneling of real
electrons.      

Whereas the Luttinger liquid state may, in principle, be observed in
any one-dimensional system, the most convenient opportunity is
offered by the fractional quantum Hall effect (FQHE) edge
states. Indeed, the edge state of a sample in the FQHE regime is a
one-dimensional system, and it may be shown that for the bulk filling
factor $\nu = 1/(2m+1)$, $m \in \cal{Z}$ (Laughlin states), the
interaction parameter $g$ takes the same value $g = 1/(2m+1)$. The
difference with the ordinary Luttinger liquid is that the FQHE edge
states are chiral: the motion along a certain edge is only possible in
one direction. Thus, if we imagine a FQHE strip, the electrons along
the upper edge move, say, to the right, and the electrons along the
lower edge move to the left. For the transport properties we discuss
this plays no role\footnote{A difference for shot noise is that any
experiment with the FQHE edge states is always four-terminal. The
behavior of all correlation functions which we discuss below is the
same.}, and the expressions (\ref{condstrong73}) and
(\ref{condweak73}) remain valid.  
\begin{figure}
{\epsfxsize=8.5cm\centerline{\epsfbox{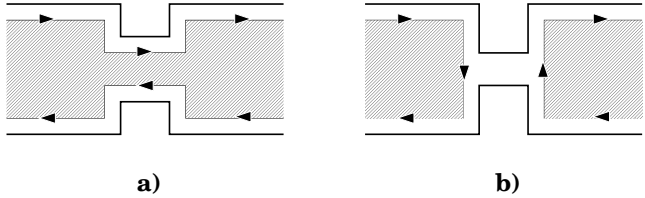}}}
\vspace{0.3cm}
\caption{The tunneling experiment with the FQHE edge states: (a)
strong tunneling; (b) weak tunneling. The shaded areas denote the
location of the FQHE droplet(s).}
\label{fig703}
\end{figure}

In particular, for the FQHE case the quasiparticles with the charge
$\nu e$ may be identified with the Laughlin quasiparticles. The
distinction between the strong and weak tunneling we described above
also gets a clear interpretation (Fig.~\ref{fig703}), which is in this
form due to Chamon, Freed and Wen \cite{Chamon96}. Indeed, consider a
FQHE strip with a barrier. If the tunneling is strong
(Fig.~\ref{fig703}a), the edge states go through the barrier. The
backscattering corresponds then to the charge transfer from the upper
edge state to the lower one, and this happens via tunneling between
the edge states {\em inside} the FQHE strip. Thus, in this case, there
are Laughlin quasiparticles which tunnel. In principle, the electrons
may also be backscattered, but such events have a very low probability
(see below). In the opposite regime of weak tunneling, the strip
splits into two isolated droplets (Fig.~\ref{fig703}b). Now the
tunneling through the barrier is again the tunneling between two edge
states, but it only may happen {\em outside} the FQHE state, where the
quasiparticles do not exist. Thus, in this case, one has tunneling
of real electrons.    

{\bf Theory of dc shot noise}. Shot noise in Luttinger liquids was
investigated by Kane and Fisher \cite{Kane94} using the bosonization
technique\footnote{Ref. \protect\cite{Kane94} also considers the
finite-temperature case and describes the crossover between thermal
and shot noise.}. The conclusion is that for an ideal infinite
one-dimensional system there is no shot noise\footnote{The same is, of
course, also true for a system between the two reservoirs.}. Shot
noise appears once the barrier is inserted. For the strong tunneling
regime the shot noise is   
\begin{equation} \label{noiseweak73}
S = 2ge\left( \frac{ge^2}{2\pi\hbar} V - \langle I \rangle \right),
\end{equation}
where the average current $\langle I \rangle$ is given by
Eq. (\ref{condweak73}) in the regime in which the latter is valid. 
If we introduce the (small) backscattering current $I_b
= (e^2/2\pi\hbar) b T V^{2g-1}$, the shot noise is written as  
\begin{equation} \label{noise731}
S = 2 ge I_b,
\end{equation}
which physically corresponds to the Poisson backscattering stream of
(Laughlin) quasiparticles with the charge $ge$. Eq. (\ref{noise731})
is precisely the analog of the two-terminal expression $S =
(e^3V/\pi\hbar)T(1-T)$ for the non-interacting case, which may be
written as $S = 2e I_b$ for $1 - T \ll 1$ and corresponds to a
Poissonian distributed stream of backscattered electrons. As we have
mentioned already, there is also a contribution to the shot noise
(\ref{noiseweak73}) due to the tunneling of electrons; as explicitly
shown by Auerbach \cite{Auerbach98}, this contribution is
exponentially suppressed. Thus, the shot noise experiments in the
strong tunneling regime open the possibility to measure the fractional
charge. This possibility has been experimentally realized (see
below). 

For the case of weak tunneling, the shot noise is Poissonian with the
charge $e$, $S = 2e\langle I \rangle$: It is determined by the charge
of tunneling electrons. Expressions for noise interpolating between
this regime and Eq. (\ref{noiseweak73}), as well as a numerical
evaluation for $g = 1/3$, are provided by Fendley, Ludwig, and Saleur
\cite{Fendley95}; Fendley and Saleur \cite{Fendley96} and Weiss
\cite{Weiss96} generalize them to finite temperatures.   

For the resonant tunneling process, at resonance, the shot noise is
given by \cite{Kane94}
\begin{equation} \label{noiseweak731}
S = 4ge\left( \frac{ge^2}{2\pi\hbar} V - \langle I \rangle \right) =
4 g e I_b,
\end{equation}
which corresponds to the effective charge $2eg$. This reflects the
fact that in resonance the excitations scatter back in pairs. Safi
\cite{Safi97} argues, however, that the contribution due to the
backscattering of single quasiparticles is of the same order; this
statement may have implications for the Fano factor, which is then
between $g$ and $2g$.  

Sandler, Chamon, and Fradkin \cite{Sandler99} consider a situation
in the strong tunneling regime with a barrier separating the two FQHE
states with different filling factors $\nu_1 = 1/(2m_1+1)$ and $\nu_2
= 1/(2m_2+1)$. One of the states may be in the integer quantum Hall
regime, for instance $m_1 = 0$. In particular, the case of  
$\nu_1 = 1$ and $\nu_2 = 1/3$ may be solved exactly. They conclude
that the noise in this system corresponds to a Poissonian stream
(\ref{noise731}) of backscattered  quasiparticles which are now,
however, not the Laughlin quasiparticles of any of the two FQHE
states. The charge of these excitations, which is measured by the shot
noise, is $\tilde g e$, with $\tilde g^{-1} = (\nu_1^{-1} +
\nu_2^{-1})/2 = m_1 + m_2 + 1$. Thus, for $\nu_1 = 1$ and $\nu_2 =
1/3$ the effective charge is $e/2$. This also implies that the shot
noise experiment cannot distinguish certain combinations of filling
factors: the effective charge of the $1/3$--$1/3$ junction is the same
as that of the $1$--$1/5$ junction. 

We also note that the above results were obtained for infinite
wires (or FQHE edges). Taking into account the electron reservoirs,
as we have mentioned above, changes the conductance of an ideal
wire. However, it is not expected to affect the Fano factor of a wire
with a barrier, which is determined by the scattering processes at the
barrier. On the other hand, Ponomarenko and Nagaosa
\cite{Ponomarenko99} present a calculation which implies that the shot
noise in the wire connected to reservoirs is Poissonian with the
charge $e$. In our opinion, this statement is not physically
appealing, and we doubt that it is correct. Nevertheless, it deserves
a certain attention, and more work is needed in this direction.   

{\bf Frequency-dependent noise}. The frequency dependence of noise was
studied by Chamon, Freed, and Wen \cite{Chamon95,Chamon96}, who first
derived results perturbative in the tunneling strength, and
subsequently were able to find an exact solution for
$g=1/2$. Lesage and Saleur \cite{Lesage97,Lesage971} and Chamon and
Freed \cite{Chamon99} developed non-perturbative techniques valid for
any $g$. We briefly explain the main results, addressing the reader
for more details to Refs. \cite{Chamon96,Lesage97}. The 
frequency dependence of noise is essentially similar to that for
non-interacting electrons in the case when the scattering matrices are
energy independent. There is the $\vert \omega \vert$ singularity for
zero frequency; the singularity itself is not changed by the
interaction, but the coefficient in front of this singularity is
interaction-sensitive\footnote{There is a discrepancy in the
literature concerning this issue: Chamon, Freed, and Wen
\cite{Chamon95} and Chamon and Freed \cite{Chamon99} report that the
coefficient in front of $\vert \omega \vert$ is proportional to
$V^{4(g-1)}$, while Lesage and Saleur find that the singularity is
$V^{4(1-1/g)}$. The reason for this discrepancy is currently 
unclear.}. Furthermore, there is a singularity at the frequency
$\omega = eV/\hbar$, which corresponds to the motion of electrons. One
could expect that the role of interactions consists in the creation of
yet one more singularity at the ``Josephson'' frequency $\omega_J =
geV/\hbar$, reflecting the quasiparticle motion. Indeed, as shown by
Chamon, Freed, and Wen \cite{Chamon96}, shot noise exhibits structure
at this frequency. This structure is, however, smeared by the finite
tunneling probability, and does not represent a true
singularity\footnote{In the perturbative calculation it appears as a
true singularity, though. Chamon, Freed and Wen \cite{Chamon96} call
this a ``fake singularity''.}. Lesage and Saleur \cite{Lesage97} also
predict some structure at the multiple Josephson frequency
$ngeV/\hbar$, which could be a clear signature of interactions. 

{\bf FQHE: Other filling factors}. An interesting question is what
happens for the edge states in the FQHE regime with filling
factors different from $1/(2m+1)$. In this case, the structure of edge
states is more complicated. Although currently there is no general
accord on the precise form of this structure, it is clear that for the
non-Laughlin FQHE states there are two or several edge states,
propagating in the same or opposite directions. Furthermore, it is not
quite clear whether different assumptions for the edge structure would
lead to different predictions for the shot noise. If this is the case,
the shot noise measurements can be used to test the theories of the
structure of the FQHE edge. 

Imura and Nomura \cite{Imura99} apply a Luttinger liquid description
for the investigation of shot noise in the FQHE plateau regime with
the filling factor $\nu = 2/(2m \pm 1)$, $m \in \cal{Z}$. They predict
that, as the gate voltage which forms the barrier changes, the
conductance of the sample crosses from the plateau with the value
$(1/(m\pm 1))(e^2/2\pi\hbar)$ to another plateau $(2/(2m\pm
1))(e^2/2\pi\hbar)$. These expressions apply in the weak scattering
regime, and the full Poissonian noise is determined by the effective
charges $q = e/(m \pm 1)$ and $q = e/(2m \mp 1)$ (rather than $\nu
e$) at the first and second plateaus, respectively. Their results are
in agreement with the experimental data for $\nu = 2/5$, and also with
the picture emerging from the phenomenological consideration of
transport in the composite fermion model (see below). 

{\bf Fractional statistics}. After we have shown that the shot noise
experiments with the FQHE edge states may probe fractional charge of
Laughlin quasiparticles, one may naturally ask the question: Can the
fractional statistics, which the Laughlin quasiparticles are known to
obey, also be probed in the shot noise experiments? This problem was
addressed by Isakov, Martin and Ouvry \cite{Isakov99}, who consider
the two-terminal experiment for the independent charged particles
obeying the exclusion statistics. They obtain that the crossover
between zero-temperature shot noise and Nyquist noise is sensitive to
the statistics of quasi-particles. We believe that the paper
\cite{Isakov99} has a number of serious drawbacks. To start with, the
statistical particle counting arguments which the paper takes as a
departing point, are unable to reproduce the exact results which
follow from the scattering matrix approach for bosons, and thus had to
add {\em ad hoc} certain terms to reproduce these results in the
limiting case. Then, the exclusion statistics apply to an ensemble of
particles and not to single particles; it is not clear
whether the notion of independent particles obeying the exclusion
statistics is meaningful. Finally, in the scattering problem one needs
to introduce the reservoirs, which are not clearly defined in this
case. Having said all this, we acknowledge that the question, which
Isakov, Martin and Ouvry \cite{Isakov99} address, is very
important. Presumably to attack it one must start with the ensemble of
particles; we also note that the effects of statistics are best probed
in the multi-terminal geometry rather that in the two-terminal case. A
demonstration of the HBT-type effect with the FQHE edge states would
clearly indicate the statistics of the quasiparticles.   

{\bf Experiments}. Saminadayar {\em et al} \cite{Glattli97} and,
independently, de-Picciotto {\em et al} \cite{Reznikov97} performed
measurements on a FQHE strip with the filling factor $\nu=1/3$ into
which they inserted a quantum point contact. The transmission
coefficient of the point contact can be modified by changing the gate
voltages. In particular, Saminadayar {\em et al} \cite{Glattli97}
obtained the data showing the crossover from strong tunneling to the
weak tunneling regime. Their results, shown in Fig. \ref{fig704},
clearly demonstrate that in the strong tunneling (weak backscattering)
regime the effective charge of the carriers is $e/3$. As the tunneling
becomes weak, the effective charge crosses over to $e$, as expected
for the strong backscattering regime. They also carefully checked the
crossover between the shot noise and the equilibrium noise, and found
an excellent agreement with the theory.    

Though the results of Refs. \cite{Glattli97} and \cite{Reznikov97}
basically coincide, we must point out an important difference which is
presently not understood. The transmission in the experiment by
de-Picciotto {\em et al} \cite{Reznikov97} is not entirely perfect:
they estimate that for the sets of data they plot the transmission
coefficients are $T = 0.82$ and $T = 0.73$. Taking this into account,
they phenomenologically insert the factor $T$ in the expression for
the shot noise and fit the data to the curve $S = (2e/3)TI_B$. In this
way, they obtain a good agreement between the theory and
experiment. On the other hand, Saminadayar {\em et al}
\cite{Glattli97} fit their data to the curve $S = (2e/3)I_B$ (without
the factor $T$). An attempt to replot the data taking the factor $T$ 
into account leads to an overestimate of the electron charge. The
theory we described above predicts a more complicated dependence than
just the factor $T$; therefore it may be important to clarify this
detail in order to improve our understanding of the theory of FQHE
edge states.  
\begin{figure}
{\epsfxsize=7.cm\centerline{\epsfbox{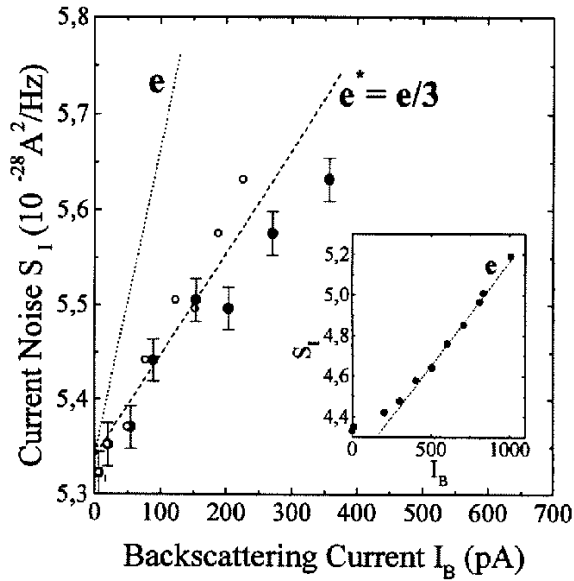}}}
\vspace{0.3cm}
\caption{Experimental results of Saminadayar {\em et al}
\protect\cite{Glattli97} for $\nu = 1/3$ (strong tunneling -- weak
backscattering regime).}   
\label{fig704}
\end{figure}

Reznikov {\em et al} \cite{Reznikov99} performed similar measurements
in the magnetic field corresponding to the filling factor
$\nu=2/5$. Changing the gate voltage (and thus varying the shape of
the quantum point contact) they have observed two plateaus of the
conductance, with the heights $(1/3)(e^2/2\pi\hbar)$ and
$(2/5)(e^2/2\pi\hbar)$, respectively. The noise measurements showed
that the effective charges at these plateaus are $e/3$ and $e/5$,
respectively. These results are in agreement with the subsequent
theory of Imura and Nomura \cite{Imura99}, and also with the 
predictions of the composite fermion model (see below): They assumed
that there are two transmission channels which in turn open with the
gate voltage. This experiment is important since it clearly shows that
what is measured in the FQHE shot noise experiments is not merely a
filling factor (like one could suspect for $\nu=1/3$), but really the
quasiparticle charge.   

\subsection{Composite fermions} \label{scorr74}

An alternative description of the FQHE systems is achieved in terms of
the {\em composite fermions}. Starting from the FQHE state with the
filling factor $\nu = p/(2np\pm 1)$, $n,p \in \cal{Z}$, one can
perform a gauge transformation and attach $2n$ flux quanta to each
electron. The resulting objects (an electron with the flux attached)
still obey the Fermi statistics and hence are called composite
fermions (CF). The initial FQHE state for electrons corresponds in the
mean field approximation to the filling factor $\tilde\nu = p$ for CF,
{\em i.e.} the composite fermions are in the integer quantum Hall
regime with the $p$ Landau levels filled. In particular, the
half-filled Landau level corresponds to the CF in zero magnetic
field. Composite fermions interact electrostatically via their
charges, and also via the gauge fields, which are a measure of the
difference between the actual flux quanta attached and the flux
treated in the mean field approximation. It is important that, at
least in the mean field approximation, the composite Fermions do not
form a strongly interacting system, and therefore may be regarded as
independent particles. One can then proceed by establishing an analogy
with the transport of independent or weakly interacting electrons.  

Von Oppen \cite{Oppen97} considers the shot noise of composite
fermions at the half-filled Landau levels, assuming that the sample is
disordered. He develops a classical theory based on the
Boltzmann-Langevin approach\footnote{To this end, he has to assume
implicitly the existence of two CF reservoirs, described by
equilibrium Fermi distribution functions.} (see Section
\ref{BoltzmannL}), incorporating interactions between them. Though the
fluctuations of both electric and magnetic fields now become
important, in the end he obtains the same result as for normal
diffusive wires: In the regime of negligible interactions between the
CF's, the Fano factor equals $1/3$ and is universal. Likewise, in the
regime when the CF distribution function is in local equilibrium
(analog of the hot electron regime), the suppression factor is
$\sqrt{3}/4$.  

Kirczenow \cite{Kirczenow98} considers current fluctuations in the
FQHE states, appealing to statistical particle counting
arguments. However, he does not take into account any kind of
scattering, and only treats equilibrium (Nyquist) noise for which the
result is known already.    

Shot noise in the FQHE strip ($\nu = 2/(2p+1)$) with a tunnel barrier
was discussed by de Picciotto \cite{Picciotto98} who assumed that the
composite Fermions are transmitted through the quantum point contact
similarly to the non-interacting electrons. Namely, there are $p$
channels corresponding to the $p$ CF Landau levels. Each channel is
characterized by an individual transmission coefficient. As the gate
voltage is changed the channels open (the corresponding transmission
coefficient crosses over from $0$ to $1$), and the conductance
exhibits plateaus with the height $(e^2/2\pi\hbar)l/(2l+1)$, $1 < l <
p$). The shot noise at the plateau $l$ is then given as the Poisson
backscattered current with the effective charge $e/(2l+1)$. Though
this paper is phenomenological and requires further support from
microscopic theory, we note that all the features predicted in
Ref. \cite{Picciotto98} are not only in agreement with the Luttinger
liquid approach by Imura and Nomura \cite{Imura99}, but also were
observed experimentally \cite{Reznikov99} for $\nu = 2/5$ (see
above). In this case there are two channels corresponding to $p=1$ and
$p=2$, which implies conductance plateaus with heights
$(1/3)(e^2/2\pi\hbar)$ and (for higher gate voltage)
$(2/5)(e^2/2\pi\hbar)$; the corresponding charges measured in the shot
noise experiment are $e/3$ and $e/5$, respectively.

\section{Concluding remarks, future prospects, and unsolved problems}
\label{conclus} 

\subsection{General considerations} \label{con081}

In this Section, we try to outline the directions along which the
field of shot noise in mesoscopic systems has been developing, to
point out the unsolved problems which are, in our opinion, important,
and to guess how the field will further develop. A formal summary can
be found at the end of the Section.    

Prior to the development of the theory and the experiments on shot
noise in mesoscopic physics, there already existed a considerable
amount of knowledge in condensed matter physics, electrical
engineering, and especially optics. Both theory and experiments are
available in these fields, and the results are well
established. Similarly in mesoscopic physics, there exists a fruitful
interaction between theory and experiment. However, presently there
are many theoretical predictions concerning shot noise, not only
extensions of the existing theories, but which really address new
sub-fields, which are not yet tested experimentally. Below we give a
short list of these predictions. Like every list, the choice reflects
very much our taste, and we do not imply that the predictions
not included in this list are of minor importance.  
\begin{itemize}

\item $1/4$--suppression of shot noise in chaotic cavities;

\item Multi-terminal effects probing statistics (exchange Hanbury
Brown -- Twiss (HBT) effect; shot noise at tunnel microscope tips; HBT
effect with FQHE edge states); 

\item Frequency dependent noise beyond Nyquist-Johnson (noise
measurements which would reveal the inner energy scales of mesoscopic
systems); current fluctuations induced into gates or other nearby
mesoscopic conductors; 

\item Shot noise of clean NS interfaces; mesoscopic nature of positive
cross-correlations in hybrid structures;

\item Shot noise in high magnetic fields at the half-filled Landau
level;

\item Shot noise in  hybrid magnetic structures. 

\end{itemize}

The theory, in our opinion, is generally well developed for most of
the field and adequately covers it. However, a number of problems
persist: For instance, there is no clear understanding under which
conditions the cross-correlations in multi-terminal hybrid structures
may be positive. Recent work \cite{thesis} suggests that it is only a
mesoscopic quantum contribution which is positive, but that to leading
order the correlations will be negative as in normal
conductors. Considerably more work is required on the frequency
dependence of shot noise and on strongly correlated systems. The
former (Section \ref{freq}) offers the opportunities to probe the
inner energy scales and collective relaxation times of the mesoscopic
systems; only a few results are 
presently available. As for the strongly correlated systems (including
possibly unconventional superconductors), this may become (and is
already becoming) one of the mainstreams of mesoscopic physics; since
even the dc shot noise measurements provide valuable information about
the charge and statistics of quasiparticles, we expect a lot of
theoretical developments in this direction concerning the shot
noise. Some of the unsolved problems in this field may be found
directly in Section \ref{strcorr}, one of the most fascinating being
the possibility of probing the quasiparticle statistics in
multi-terminal noise measurements with FQHE edge states.

One more possible development, which we did not mention in the main
body of this Review, concerns shot noise far from equilibrium under
conditions when the $I$--$V$ characteristics are non-linear. The
situation with non-linear problems resembles very much the frequency
dependent ones: Current conservation and gauge invariance are not
automatically guaranteed, and interactions must be taken into account
to ensure these properties (for a discussion, see {\em e.g.}
Ref. \cite{Pedersen98}). Though in the cases which we cited in the
Review the non-linear results seem to be credible, it is still
desirable to have a gauge-invariant general theory valid for arbitrary
non-linear $I$--$V$ characteristics. It is also desirable to gain
insight and develop estimates of the range of applicability of the
usual theories. Recently Wei {\em et al} \cite{Wei99} derived a gauge
invariant expression for shot noise in the weakly non-linear regime,
expressing it through functional derivatives of the Lindhard function
with respect to local potential fields. They apply the results to the
resonant tunneling diode. Wei {\em et al} \cite{Wei99} also discuss
the limit in which the tunneling rates may be assumed to be energy
independent. Apparently, the theory of Wei {\em et al} does not treat
the effect of fluctuations of the potential inside the sample, which
may be an important source of noise. Furthermore, Green and Das
\cite{GreenDas1,GreenDas2,GreenDas3} proposed a classical theory of
shot noise, based on a direct solution of kinetic equations. They
discuss the possibility to detect interaction effects in the
cross-over region from thermal to shot noise. It is, however, yet to
be shown what results this approach yields in the linear regime and
whether it reproduces, for instance, the $1/3$--suppression of shot
noise in metallic diffusive wires. An application of both of these
approaches to specific systems is highly desirable.    

What we have mentioned above, concerns the development of a field
{\em inside} mesoscopic physics. We expect, however, that
interesting connections will occur across the boundaries of different
fields. An immediate application which can be imagined is the shot
noise of photons and phonons. Actually, noise is much better studied
in optics than in condensed matter physics (see {\em e.g.} the review
article \cite{Kazarinov}), and, as we have just mentioned previously, 
the theory of shot noise in mesoscopic physics borrowed many ideas 
from quantum optics. At the same time, mesoscopic physics gained a
huge experience in dealing, for instance, with disordered and chaotic
systems. Recently a ``back-flow'' of this experience to quantum optics
started. This concerns photonic noise for the transmission through
(disordered) waveguides, or due to the radiation of random lasers or
cavities of chaotic shape. In particular, the waveguides and cavity
may be absorbing or amplifying, which adds new features as compared
with condensed matter physics\footnote{One has to take, of course,
also into account the differences between photon and electron
measurements: Apart from the evident Bose versus Fermi and neutral
versus charged particles, there are many more. For example, the photon
measurement is accompanied by a removal of a photon from the device,
while the total electron number is always conserved.}. For references,
we cite a recent review by Patra and Beenakker \cite{Patra99}. 
Possibly, in the future other textbook problems of mesoscopic physics
will also find their analogies in quantum optics. Phonons are less
easy to manipulate with, but, in principle, one can also imagine the
same class of problems for them. Generally, shot noise accompanies the
propagation of any type of (quasi)-particles; as the last example, we
mention plasma waves.  

\subsection{Summary for a lazy or impatient reader} \label{con085} 

Below is a summary of this Review. Though we encourage the reader to
work through the whole text (and then she or he does not need this
summary), we understand that certain readers are too lazy or too
impatient to do this. For such readers we prepare this summary which
permits to acquire some information on shot noise in a very short
time-span. We only include in this summary the statements which in our
opinion are the most important.  

\begin{itemize}

\item Shot noise occurs in a transport state and is due to
fluctuations in the occupation number of states caused by (i) thermal
random initial fluctuations; (ii) the random nature of
quantum-mechanical transmission/reflection (partition noise), which,
in turn, is a consequence of the discreteness of the charge of the
particles. The actual noise is a combination of both of these
microscopic sources and typically these sources cannot be
separated.

\item  Shot noise provides information about the kinetics of the
transport state: In particular, it can be used to obtain information
on transmission channels beyond that contained in the conductance. In
two-terminal systems, zero-frequency shot noise for non-interacting
electrons is suppressed in comparison with the Poisson value $S = 2 e
\langle I \rangle$.

\item For quantum wells the noise suppression is  $F = S/2e\langle I 
\rangle = (\Gamma_L^2 + \Gamma_R^2)/(\Gamma_L + \Gamma_R)^2$. The
suppression is universal for metallic diffusive wires ($F = 1/3$) and
chaotic cavities ($F = 1/4$).

\item Far from equilibrium, in the vicinity of instability points the
shot noise can exceed the Poisson value.

\item In the limit of low transmission, the shot noise is Poissonian
and measures the charge of transmitting particles. In particular, for
normal metal -- superconductor interfaces this charge equals $2e$,
whereas in SNS systems it is greatly enhanced due to multiple Andreev
reflections. In the limit of low reflection, the shot noise may be
understood as Poissonian noise of reflected particles; in this way,
the charge of quasiparticles in the fractional quantum Hall effect is
measured.

\item For carriers with Fermi statistics, in multi-terminal systems
the zero-frequency correlations of currents at different terminals are
always negative. For Bose statistics, they may under certain
circumstances become positive. These cross-correlations may be used to
probe the statistics of quasiparticles.

\item The ensemble averaged shot noise  may be described both
quantum-mecha\-nically (scattering approach; Green's function
technique) and classically (master equation; Langevin and
Boltzmann-Langevin approach; minimal correlation approach). Where they
can be compared, classical and quantum-mechanical descriptions provide
the same results. Classical methods, of course, fail to describe
genuinely quantum phenomena like {\em e.g.} the quantum Hall effect.

\item As a function of frequency, the noise crosses over from the shot
noise to the equilibrium noise $S \propto \vert \omega \vert$. This is
only valid when the frequency is low as compared with the inner energy
scales of the system and inverse times of the collective
response. For higher frequencies, the noise is sensitive to all these
scales. However, the current conservation for frequency dependent
noise is not automatically provided, and generally is not achieved in
non-interacting systems. 

\item Inelastic scattering may enhance or suppress noise, depending on
its nature. In particular, in macroscopic systems shot noise is always
suppressed down to zero by inelastic (usually, electron-phonon)
scattering. When interactions are strong, shot noise is {\em usually}
Poissonian, like in the Coulomb blockade plateau regime. 

\item We expect that the future development of the field of the shot
noise will be mainly along the following directions: Within the field
of mesoscopic electrical systems: (i) experimental developments; (ii)
frequency dependence; and (iii) shot noise in strongly correlated
systems; and more generally (iv) shot noise in disordered and chaotic
quantum optical systems, shot noise measurements of phonons (and,
possibly, of other quasiparticles). 

\end{itemize}

\section*{Acknowledgements}

We have profited from discussions and a number of specific comments by
Pascal Cedraschi, Thomas Gramespacher, and Andrew M. Martin. 
Some parts of this Review were written at the Aspen Center for Physics
(Y.~M.~B); at the Max-Planck-Institut f\"ur Physik Komplexer Systeme,
Dresden (Y.~M.~B.); at the Centro Stefano Franscini, Ascona
(Y.~M.~B. and M.~B.); and at the Universit\'e de Montpellier II
(Y.~M.~B.). We thank these institutions for hospitality and
support. This work was supported by the Swiss National Science
Foundation via the Nanoscience program.  

\appendix

\section{Counting statistics and optical analogies} \label{con082}

The question which naturally originates after consideration of the
shot noise is the following: Can we obtain some information about the
higher moments of the current? Since, as we have seen, the shot noise
at zero frequency contains more information about the transmission
channels than the average conductance, the studies of the higher
moments may reveal even more information. Also, we have seen that in
the classical theories of shot noise the distribution of the Langevin
sources (elementary currents) is commonly assumed to be Gaussian, in
order to provide the equivalence between the Langevin and
Fokker-Planck equations \cite{Kampen}. An independent analysis of the
higher moments of the current can reveal whether this equivalence
in fact exists, and thus how credible the classical theory is. 

A natural quantity to study is the $k$-th cumulant of the number of
particles $n(t)$ which passed through the barrier during the time $t$
(which is assumed to be large). In terms of the current $I(t)$, this
cumulant is expressed as
\begin{equation} \label{cum801}
\langle\langle n^k (t) \rangle\rangle = \frac{1}{e^k} \int_0^t dt_1
\dots dt_k \langle\langle I(t_1) \dots I(t_k) \rangle\rangle, 
\end{equation} 
where $\langle\langle \dots \rangle\rangle$ means the cumulant
(irreducible part). For the following, we only consider the
time-independent problems, {\em i.e.} the noise in the presence of a
dc voltage. Then the first cumulant is $\langle \langle n(t)
\rangle\rangle = \langle I \rangle t/e$, and the second one is
expressed through the zero-temperature shot noise power,
\begin{displaymath} 
\langle\langle n^2 (t) \rangle\rangle = \frac{St}{2e^2}.
\end{displaymath}
In particular, the ratio of $\langle \langle n^2 (t) \rangle\rangle$
and $\langle \langle n (t) \rangle\rangle$ gives the Fano factor. The
cumulants with $k > 2$ in Eq. (\ref{cum801}) contain additional
information about the statistics of current. Thus, if the distribution
of the transmitted charge is Poissonian, all the cumulants have the
same value; for the Gaussian distribution all the cumulants with $k >
2$ vanish. 

The general expression for the cumulants of the number of transmitted
particles was obtained by Lee, Levitov, and Yakovets \cite{LLY95}, who
followed the earlier paper\footnote{The paper \protect\cite{LLev93}
corrects Ref. \cite{LLev92}.} by Levitov and Lesovik \cite{LLev93}. 
We only give the results for zero temperature. Consider first one
channel with the transmission probability $T$. The probability that
$m$ particles pass through this channel during the time $t$ is given
by Bernoulli distribution, as found by Shimizu and Sakaki
\cite{Sakaki91} and Levitov and Lesovik \cite{LLev93},
\begin{equation} \label{Bern801}
P_m (t) = C^m_N T^m (1-T)^{N-m}, \ \ \ m \le N,
\end{equation} 
where $N(t)$ is the ``number of attempts'', on average given by
$\langle N \rangle = eVt/2\pi\hbar$. Actually, $N(t)$ fluctuates,
but in the long-time limit $N \gg 1$ these fluctuations are
insignificant \cite{LLev93}, and therefore $N$ in Eq. (\ref{Bern801})
must be understood as the average value. The expression
(\ref{Bern801}) gives the probability that out of
$N$ attempts $m$ particles go through (with the probability $T$) and
$N-m$ others are reflected back. 

The next step is to define the characteristic function, 
\begin{equation} \label{char801}
\chi(\lambda) = \sum_{m=0}^N P_m \exp (im\lambda) = \left[
Te^{i\lambda} + 1 - T \right]^N.
\end{equation}
If we have several independent\footnote{In the sense that we can
diagonalize the matrix $t^{\dagger} t$ and define the transmission
eigenvalues.} channels, the characteristic function is
multiplicative,
\begin{equation} \label{char802}
\chi(\lambda) = \prod_j \left[ T_je^{i\lambda} + 1 - T_j \right]^N, 
\end{equation}
where the product is taken over all the transmission channels. The
coefficients in the series expansion of $\ln \chi$ are the 
cumulants that we are looking for,
\begin{equation} \label{chi801}
\ln \ \chi_{\lambda} = \sum_{k=1}^{\infty} \frac{(i\lambda)^k}{k !}
\langle\langle n^k (t) \rangle\rangle,
\end{equation}
and Lee, Levitov, and Yakovets \cite{LLY95} obtain in this way the
explicit expression,
\begin{equation} \label{cum802}
\langle\langle n^k (t) \rangle\rangle = \left. N \sum_j \left[ T \left(
1 - T \right) \frac{d}{dT} \right]^{k-1} T \right\vert_{T = T_j}.  
\end{equation}
 
We see from Eq. (\ref{cum802}) that $\langle\langle n(t)
\rangle\rangle = N \sum T_j$, and $\langle\langle n^2(t)
\rangle\rangle = N\sum T_j (1-T_j)$, which are the results for the
average current and the zero frequency shot noise power. However,
higher cumulants do not vanish at all. Though they generally cannot
be calculated in a closed form, the distributions are studied for many
systems, and we give a brief overview below.  

For the tunnel barrier, when $T_j \ll 1$ for any $j$, all the
cumulants are equal, $\langle\langle n^k(t) \rangle\rangle = N\sum
T_j$, and thus the distribution of the transmitted charge is
Poissonian. 

De Jong \cite{Jong961} analyzes the counting statistics for
double-barrier structures using the concept of the distribution
function for the transmission probabilities. For the symmetric
case ($\Gamma_L = \Gamma_R = \Gamma$) he finds $\langle \ln \chi
\rangle = 2N_{\perp} \Gamma t (\exp(i\lambda/2) - 1)$, and explicitly
for the cumulants $\langle\langle n^k(t) \rangle\rangle = N_{\perp}
\Gamma t /2^{k-1}$. Here $N_{\perp}$ is the number of transverse
channels at the Fermi surface. The cumulants decrease exponentially
with $k$, and thus the statistics are closer to Gaussian than to 
Poissonian. De Jong \cite{Jong961} was also able to obtain the same
results classically, starting from the master equation. 

The case of metallic diffusive wires is considered by Lee, Levitov,
and Yakovets \cite{LLY95}, and subsequently by Nazarov
\cite{Nazarov99}. They find that the disorder-averaged logarithm
of the characteristic function is\footnote{The characteristic function
is not self-averaging; the expansion of $\ln \langle \chi (\lambda)
\rangle$ yields different expression for the cumulants, as found by
Muttalib and Chen \protect\cite{Muttalib96}. As Levitov, Lee, and
Yakovets \protect\cite{LLY95} argue, the correct quantity to average
is $\ln \chi(\lambda)$, rather than $\chi(\lambda)$, since it is
linearly related to the cumulants of the transmitted charge.}
\begin{equation} \label{diff801} 
\langle \ln \chi (\lambda) \rangle = \frac{\langle I \rangle t}{e}
\mbox{\rm arcsinh}^2 \sqrt{e^{i\lambda} - 1}.
\end{equation}
The expressions for the cumulants cannot be found in a closed
form. As expected, from Eq. (\ref{diff801}) we obtain $\langle\langle
n^2 \rangle\rangle = \langle I \rangle t/3e$, in agreement with the
fact that the Fano factor is $1/3$. For the following cumulants one
gets, for instance, $\langle\langle n^3 \rangle\rangle = \langle I
\rangle t/15e$ and $\langle\langle n^4 \rangle\rangle = -\langle I
\rangle t/105e$. For high $k$ the cumulants $\langle\langle n^k
\rangle\rangle$ behave as $(k-1)!/((2\pi)^{k}k^{1/2})$, {\em i.e.}
they diverge! Moreover, Lee, Levitov, and Yakovets \cite{LLY95}
evaluate the sample-to-sample fluctuations of the cumulants, and
find that for the high-order cumulants these fluctuations become
stronger than the cumulants themselves. Thus, the far tails of the
charge distributions are strongly affected by disorder. Nazarov
\cite{Nazarov99} generalizes the approach to treat weak localization
corrections. 

For the transmission through the symmetric chaotic cavity,
Ref. \cite{BSB99} finds    
\begin{equation} \label{diff802} 
\langle \ln \chi (\lambda) \rangle = \frac{4\langle I \rangle t}{e}
\ln \frac{e^{i\lambda/2} + 1}{2},
\end{equation}
with the explicit expression for the cumulants 
\begin{equation} \label{diff803} 
\langle\langle n^{2l} \rangle\rangle = \frac{\langle I \rangle t}{e}
\frac{2^{2l} -1}{2^{2l-1}l} B_{2l},
\end{equation}
and $\langle\langle n^{2l+1} \rangle\rangle = 0$ ($l \ge 1$). Here
$B_k$ are the Bernoulli numbers ($B_2 = 1/6$, $B_4 = -1/30$). Indeed,
for the second cumulant we obtain $\langle\langle n^2 \rangle\rangle =
\langle I \rangle t/4e$, in accordance with the $1/4$--shot noise
suppression in symmetric chaotic cavities. Eq. (\ref{diff803}) can be
also obtained classically \cite{BSB99}, using the generalization of
the minimal correlation approach.   

Muzykantskii and Khmelnitskii \cite{MKsup} investigate the counting
statistics for the NS interface and find 
\begin{equation} \label{diff804} 
\chi (\lambda) = \prod_j \left[ T_{Aj} e^{2i\lambda} + 1 -
T_{Aj} \right]^N,  \ \ \ T_{Aj} \equiv \frac{2T_j^2}{(2 - T_j)^2}.
\end{equation}
It is clearly seen from the comparison with Eq. (\ref{char802}) that
the particles responsible for transport have effective charge $2e$. 

Other important developments include the generalization to the
multi-terminal (Levitov and Lesovik \cite{LLev93}) and time-dependent
(Ivanov and Levitov \cite{Ivanov93}; Levitov, Lee, and Lesovik
\cite{LLL96}; Ivanov, Lee, and Levitov \cite{Ivanov97}) problems, and
numerical investigation of the counting statistics for the
non-degenerate ballistic conductors (Bulashenko {\em et al}
\cite{Bulashenko98}).  

Thus, the counting statistics certainly reveal more information about
the transport properties of conductors than is contained in either the
conductance or the second order shot noise. The drawback is that it is
not quite clear how these statistics can be measured. A proposal, due
to Levitov, Lee, and Lesovik \cite{LLL96}, is to use the spin-$1/2$
galvanometer, precessing in the magnetic field created by the
transmission current. The idea is to measure the charge transmitted
during a certain time interval through the evolution of the spin
precession angle. However, the time dependent transport is a
collective phenomenon (Section \ref{freq}), and thus the theory of
such an effect must include electron-electron interactions. In
addition, this type of experiments is not easy to realize.   

On the other hand, measurements of photon numbers are routinely
performed in quantum optics. In this field concern with counting
statistics has already a long history. However, typical mesoscopic
aspects --- disorder, weak localization, chaotic cavities --- and
effects particular to optics, like absorption and amplification, make
the counting statistics of {\em photons} a promising tool of
research. Some of these aspects (relating to disorder and chaos, where
the random matrix theory may be applied), have been recently
investigated by Beenakker \cite{Beenlas}; however, there are still
many unsolved problems.   

\section{Spin effects and entanglement} \label{con083}

A notion which mesoscopic physics recently borrowed from quantum
optics is entanglement. States are called entangled, if they cannot
be written simply as a product of wave functions. For our purpose, we
will adopt the following definition. Imagine that we have two leads,
$1$ and $3$, which serve as sources of electrons. The {\em entangled
states} are defined as the following two-particle states described in
terms of the creation operators, 
\begin{equation} \label{entang1}
\vert \pm \rangle = \frac{1}{\sqrt{2}} \left( \hat
a^{\dagger}_{3\downarrow} (E_2) \hat a^{\dagger}_{1\uparrow} (E_1) \pm
\hat a^{\dagger}_{3\uparrow} (E_2) \hat a^{\dagger}_{1\downarrow}
(E_1) \right) \vert 0 \rangle, 
\end{equation} 
where $\hat a^{\dagger}_{\alpha\sigma} (E)$ is the operator which
creates an electron with the energy $E$ and the spin projection
$\sigma$ in the source $\alpha$. The state corresponding to the lower
sign in Eq. (\ref{entang1}) is the spin singlet with the symmetric
orbital part of the wave function, while the upper sign describes a
triplet (antisymmetric) state.   

Such entangled states are very important in the field of quantum
computation. Condensed matter systems are full of entangled states:
there is hardly a system for which the ground state can be
expressed simply in terms of a product of wave functions. The key
problem is to find ways in which entangled states can be generated and
manipulated in a controlled way. Optical experiments on noise have
reached a sophisticated stage since there exist optical sources of
entangled states (the production of twin photon-pairs through down
conversion). It would be highly desirable to have an electronic
equivalent of the optical source and to analyze to what extent such
experiments can be carried out in electrical conductors
\cite{Vinc,buscience}. An example of such source is a p-n junction
which permits the generation of an electron-hole pair and the
subsequent separation of the particles. The disadvantage of such an
entangled state is that the electron and hole must be kept apart at
all times. Similarly, a Cooper pair entering a normal conductor,
represents an entangled state. But in the normal conductor it is
described as an electron-hole excitation and there is to our knowledge
no deterministic way to separate the electron and the hole. To date
most proposals in condensed matter related to quantum computation 
consider entangled states in closed systems.   

Theoretically, entanglement opens a number of interesting
opportunities. One of the questions is: Provided we were able to
prepare entangled states, how do we know the states are really
entangled? Since we deal with two-particle states, it is clear that
entanglement can only be measured in the experiments which are
genuinely two-particle. Burkard, Loss, and Sukhorukov
\cite{Burkard99,Burkard991} investigate the multi-terminal
noise. Indeed, add to the structure two more reservoirs (electron
detectors) $2$ and $4$, and imagine that there is no reflection back
to the sources (the geometry of the exchange HBT experiment,
Fig.~\ref{HBTfig}(b), with the additional ``entangler'' creating the
states (\ref{entang1})). The system acts as a three-terminal device,
with an input of entangled electrons and measuring the current-current
correlation at the two detectors.   

Since the shot noise is produced by the motion of the electron charge,
it is plausible that the noise measurements are in fact sensitive to
the symmetry of the {\em orbital part} of the wave function, and not to
the whole wave function. Thus the noise power seen at a single contact
is expected to be enhanced for the singlet state (symmetric orbital
part) and suppressed for the triplet state  (anti-symmetric orbital
part). It is easy to quantify these considerations by repeating the
calculation of Section \ref{scat} in the basis of entangled states
(\ref{entang1}). Assuming that the system is of finite size, so that
the set of energies $E$ is discrete, and the incoming stream of
entangled electrons is noiseless, Burkard, Loss, and Sukhorukov
\cite{Burkard99} obtained the following result, $\langle I_2 \rangle =
\langle I_4 \rangle$, $S_{22} = S_{44} = -S_{24}$, with 
\begin{equation} \label{entang2}
S_{22} = 2eT(1-T)\left( 1 \mp \delta_{E_1,E_2} \right).
\end{equation}
Thus, indeed, the shot noise is suppressed for the triplet state and
enhanced for the singlet state, provided the electrons are taken at
(exactly) the same energy. For the singlet state this suppression is
an indication of the entanglement, since there are no other singlet
states. One can also construct the triplet states which are not
entangled, 
\begin{equation} \label{entang3}
\vert \uparrow\uparrow \rangle = \hat a^{\dagger}_{3\uparrow}
(E_2) \hat a^{\dagger}_{1\uparrow} (E_1) \vert 0 \rangle, 
\end{equation} 
and an analogous state with spins down. These states, as shown by
Burkard, Loss, and Sukhorukov \cite{Burkard99}, produce the same noise
as the entangled triplet state. Thus, the noise suppression in this
geometry is not a signature of the entanglement. 

Another proposal, due to Loss and Sukhorukov \cite{Loss99}, is that
the entangled states prepared in the double quantum dot can be probed
by the Aharonov-Bohm transport experiments. The shot noise is
Poissonian in this set-up, and both current and shot noise are
sensitive to the symmetry of the orbital wave function. 

More generally, one can also ask what happens if one can operate with
spin-polarized currents separately. (Again, presently no means are
known to do this). Burkard, Loss, and Sukhorukov \cite{Burkard99}
considered a transport in a two-terminal conductor where the chemical
potentials are different for different spin projections. In
particular, if $V_{\uparrow} = -V_{\downarrow}$, the total average
current is zero (the spin-polarized currents compensate each
other). Shot noise, however, exists, and may be used as a means to
detect the motion of electrons in this situation. 

\section{Noise induced by thermal transport} \label{con084}

Sukhorukov and Loss \cite{SLlong} consider shot noise in metallic
diffusive conductors in the situation when there is no voltage applied
between the reservoirs, and the transport is induced by the difference
of temperatures. To this end, they generalize the Boltzmann-Langevin
approach to the case of non-uniform temperature. For the simplest
situation of a two-terminal conductor, when one of the reservoirs is
kept at zero temperature, and the other at the temperature $T$, their
result reads
\begin{equation} \label{thermal1}
S = \frac{4}{3} (1 + \ln 2) G k_BT 
\end{equation}
for the purely elastic scattering, with $G$ being the Drude
conductance. This shows, in particular, that the noise induced by
thermal transport is also universal --- the ratio of the shot noise 
power to the thermal current does not depend on the details of the
sample. This is an experiment that would be interesting to
realize. 

Another prospective problem concerning the noise induced by the
non-uniform temperature, is that the applied temperature gradient
would cause not only the transport of electrons, but also transport of
phonons. Thus, in this kind of experiments one can study shot noise
(and, possibly, also counting statistics) of phonons. This really
looks very promising, and, to our knowledge, by now has never been
discussed.

\end{document}